\documentclass[traditabstract]{aa} 
\usepackage{graphicx,xcolor}

\usepackage{natbib}
\bibpunct{(}{)}{;}{a}{}{,}

\usepackage{txfonts}
\newcommand{\alphaco} {$\alpha_{\rm CO}$}
\newcommand{\alphacosb} {$\alpha_{\rm CO,SB}$}
\newcommand{\alphacoms} {$\alpha_{\rm CO,MS}$}
\newcommand{\alphacogal} {$\alpha_{\rm CO,Gal}$}
\newcommand{\fsb} {$f_{\rm SB}$}

\newcommand{\faper} {\hbox{$f_{{\rm aper}}$}}

\newcommand{\herschel}{{\it Herschel} }
\newcommand{\ico}{$I_{\rm CO}$}
\newcommand{\icoone}{$I_{\rm CO(1-0)}$}
\newcommand{\icotwo}{$I_{\rm CO(2-1)}$}
\newcommand{\lsun}{$L_\odot$}
\newcommand{\msun}{$M_\odot$}
\newcommand{\mi}{$\mu$m}
\newcommand{\kms}{km~s$^{-1}$}
\newcommand{\ks}{K$_{\rm s}$}

\newcommand{\lco} {\hbox{$L^\prime_{\rm CO}$}}
\newcommand{\lir} {\hbox{$L_{\rm IR}$}}

\newcommand{\lk} {\hbox{$L_{\rm K}$}}
\newcommand{\lksun} {\hbox{$L_{\odot,\rm K}$}}

\newcommand{\mhtwo} {\hbox{$M_{{\rm H}_2}$}}
\newcommand{\mmol} {\hbox{$M_{{\rm H}_2}$}}
\newcommand{\mmolcenter} {\hbox{$M_{{\rm H_2, center}}$}}

\newcommand{\mdust} {\hbox{$M_{{\rm dust}}$}}
\newcommand{\mgas} {\hbox{$M_{{\rm gas}}$}}
\newcommand{\mstar} {\hbox{$M_{{\rm *}}$}}

\newcommand{\mhi}   {\hbox{$M_{\rm HI}$}}

\newcommand{\rkron}{$r_{\rm kron}$}
\newcommand{\diso}{$d_{\rm 25}$}
\newcommand{\riso}{$r_{\rm 25}$}
\newcommand{\rtwoone}{$R_{\rm 21}$}
\newcommand{\sfegas}{SFE$_{\rm gas}$}
\newcommand{\sfehi}{SFE$_{\rm HI}$}

\newcommand{\wise}{WISE}

\newcommand{\figuredir}{figures}
\newcommand{\figuredirspectra}{figures}
\newcommand{\figuredirpython}{figures}

\begin{document}

   \title{CO observations of major merger pairs at z=0: Molecular gas mass and star formation}


   \author{Ute Lisenfeld
          \inst{1}
          \and
          Cong Kevin Xu \inst{2,3}  
          \and
          Yu Gao\inst{4}
          \and
          Donovan L. Domingue\inst{5 }
          \and 
          Chen Cao\inst{6,7}
          \and 
          Min S. Yun\inst{8}
          \and
          Pei Zuo\inst{2,9}
}

   \institute{Departamento de F\'isica Te\'orica y del Cosmos, Universidad de Granada, Spain and Instituto Carlos I de F\'isica T\'eorica y Computacional, 
   Facultad de Ciencias, 18071 Granada, Spain,             
    \email{ute@ugr.es}
         \and
    National Astronomical Observatories, Chinese Academy of Sciences, Beijing 1000012, China
    \and
    South American Center for Astronomy, CAS, Camino El Observatorio 1515, Las Condes, Santiago, Chile
    \and
    Purple Mountain Observatory, Chinese Academy of Sciences, 2 West Beijing Road, Nanjing 210008, China
    \and
    Georgia College \& State University, CBX 82, Milledgeville, GA 31061, USA 
    \and
    School of Space Science and Physics, Shandong Univrsity, Weihai, Shandong 264209, China
    \and 
     Shandong Provincial Key Laboratory of Optical Astronomy \& Solar-Terrestrial Environment, Weihai, Chandong 264209, China
     \and 
    Department of Astronomy, University of Massachusetts, Amherst, MA 01002, USA
    \and 
     University of Chinese Academy of Sciences, Beijing 100049, China     
             }

   \date{Received September 15, 1996; accepted March 16, 1997}

 
  \abstract
   {We present CO observations of 78 spiral galaxies in local merger pairs. These galaxies represent
   a subsample of a  \ks-band selected sample consisting of 88 close major-merger pairs (HKPAIRs), 44 
   spiral-spiral (S+S) pairs and 44 spiral-elliptical (S+E) pairs, with separation $< 20\, h^{-1}$ kpc and mass ratio $<$ 2.5. 
  For all objects, the star formation rate (SFR) and dust mass were derived from 
    \herschel PACS and SPIRE data, and the atomic gas mass, \mhi, from the Green Bank Telescope HI observations.  The complete data set allows us to 
   study the relation between the gas (atomic and molecular) mass, dust mass and SFR in merger galaxies.
    We derive the molecular gas fraction (\mmol/\mstar), molecular-to-atomic gas mass ratio (\mmol/\mhi),
    gas-to-dust mass ratio and SFE (=SFR/\mhtwo) and study their dependences on pair type (S+S compared to S+E),
    stellar mass and the presence of morphological interaction signs.
    We find an overall moderate enhancements ($\sim 2\times$) in both molecular gas fraction (\mhtwo/\mstar), and molecular-to-atomic gas ratio 
    (\mhtwo/\mhi)  for star-forming galaxies in  major-merger pairs compared to non-interacting comparison samples, whereas
    no enhancement was found for the SFE nor for the total gas mass fraction ((\mhi+\mmol)/\mstar).
    When divided into S+S and S+E, low mass and high mass, and with and without interaction signs, there is 
    a small difference in 
    SFE, moderate difference in \mhtwo/\mstar, and strong differences in \mhtwo/\mhi\  between subsamples. 
    For the molecular-to-atomic gas ratio \mhtwo/\mhi, the difference between S+S and S+E subsamples is $0.69\pm0.16$ dex and between pairs with and without interaction signs is $0.53\pm 0.18$ dex. Together, our results suggest (1) star formation enhancement in close major-merger pairs occurs mainly in S+S pairs 
after the first close encounter (indicated by interaction signs)  because the HI gas is compressed into star-forming molecular gas by the tidal torque;
(2) this effect is much weakened in the S+E pairs. 
       }


   \maketitle
%

\section{Introduction}

Gravitational interaction is an important process for the evolution of galaxies 
in clusters \citep{dressler80, moore96}, groups \citep{hickson92, lisenfeld17}, 
triplets \citep{duplancic15, argudo15} and pairs \citep{ellison10, argudo15}.
It is now well established that galaxy interactions in pairs can cause an enhancement of the 
star formation rate (SFR). The amount of the enhancement depends on
parameters of the galaxies (mass ratio, gas fraction), on the orbital parameters
of the interacting galaxies \citep[e.g.,][]{kennicutt87, xu90}.
and on the phase of the interaction \citep[e.g.,][]{dimatteo07, cox08, scudder12}.
The largest enhancement of the SFR occurs in equal-mass mergers (major mergers) after the first pericenter
and, much stronger, during coalescene \citep[e.g.,][]{nikolic04,scudder12}.
In most cases the SFR enhancement  is only moderate ($<$ factor of 5 in 85\% of the cases, di Matteo et al. 2008),
but when all parameters are favourable (gas rich, equal mass galaxies during coalescence) short
episodes of very high SFR can occur, the most extreme examples being (Ultra) Luminous Infrared galaxies \citep{sanders96,ellison13}.

Simulations of galaxy mergers have helped to improve our understanding of the 
processes that affect the SFR during a merger. It has long been  recognized that tidal forces can produce an increase of the
SFR \citep[see][]{barnes92}. Gravitational torques produced by asymmetric tidal forces cause the gas at large radii
to lose angular momentum and fall into the central regions where the high gas surface density 
produces a central starburst. High resolution simulations have shown, that, in addition, changes in
the substructure of the Interstellar Medium (ISM) can favor the collapse of molecular clouds and
thus enhance the SFR \citep{teyssier10}. Parsec-resolution simulations of \citet{renaud14} find that
during a typical galaxy merger tidal compression can increase and modify turbulence, leading to an excess of
dense gas and an enhancement in the SF activity and the SF efficiency (SFE = SFR/\mmol).

In order to observationally better understand the question of when and how SF is enhanced during the merging process,
\citet{domingue09} selected, based on the \ks-band, a sample of close major mergers (KPAIR sample).
\citet{xu10} studied the specific SFR (sSFR=SFR/\mstar) enhancement in this sample and found
enhancement in galaxies in spiral-spiral (S+S) pairs, but none in spiral-elliptical (S+E) pairs. This result was confirmed
by \citet{cao16} for an extended sample (H-KPAIR sample) based on \herschel PACS and SPIRE data. These data allowed them
to derive the SFR and dust mass which can be used as  an indicator of the total
gas mass, assuming a constant dust-to-gas mass ratio. They found
an increase in the \sfegas (= SFR/\mgas) of spirals in S+S pairs, whereas the value in spirals in S+E pairs
is the same as for a control sample. 
The difference between the gas fraction (\mgas/\mstar) in star forming galaxies in S+S and S+E pairs is weak ($\sim$40\%) and insignificant,
 indicating that the amount  of gas is not the reason
for this difference. Using \wise\ and \herschel\ data,  \citet{domingue16}  found that
spirals in S+S pairs exhibit a significant enhancement in the interstellar radiation
field and dust temperature, while spirals in S+E pairs do not.
\citet{zuo18} observed the sample in HI with the Green Bank Telescope and  found a difference in \sfehi\ (=SFR/\mhi) between
S+S and S+E pairs.

Molecular gas is more closely related to the process of SF than the HI gas.
The SFR can be enhanced by a larger amount of molecular gas from which stars form,
or by making the process of star formation from gas more efficient (i.e. increasing SFE), 
e.g. by increasing the gas density. Both scenarios can
be distinguished observationally by measuring the molecular gas mass (compared to the stellar or atomic
gas mass) and the SFE.
The role of the molecular gas in galaxy interactions has been investigated in numerous studies. 
There is a general consensus that the molecular gas content is enhanced in interacting galaxies.
This has already been seen in the first studies with samples ranging from $\sim$ 10 to 1000 galaxies \citep{braine93,
combes94, casasola04}, and has been 
confirmed in  more recent studies.  \citet{violino18} found for  a small sample of 9 nearby
galaxies which they compared to a well-matched comparison sample that both the molecular gas fraction (\mmol/\mstar)
and  the  SFE are  enhanced in the
interacting sample. Both parameters are, however, consistent with non-mergers of similarly enhanced 
SFR. \citet{pan18} investigated a sample of 58 pairs and found an enhancement of the SFR, SFE, \mmol\
and \mmol/\mstar. Whereas the enhancement of the SFR, \mmol and \mmol/\mstar\ increases with decreasing
pair separation, the SFE is only enhanced in close (separation $<$ 20 kpc) pairs and equal mass systems.
An enhancement of the SFE was found in the earlier studies only in strongly interacting galaxies
\citep{solomon88, sofue93} whereas more weakly interacting pairs did not show any enhancement 
\citep{solomon88}. No enhancement in the SFE was found in studies with mixed interacting samples
\citep{combes94, casasola04}, whereas the more recent studies \citep{violino18, pan18} did find a 
small ($\lesssim$  factor of 2) increase in the SFE.

In the present paper we present new CO data for a subsample of H-KPAIR, which includes only close major mergers (r$< 20 h^{-1}$ kpc, mass ratio $<$ 2.5).
These data allow
us to calculate and analyse the molecular gas content in a homogenous sample with a large
ancillary data set. Together with the data from \citet{cao16}
and \citet{zuo18} we now have a complete data set to study the relation between the
cold interstellar medium (atomic, molecular gas and dust) and SF.
This sample allows us to  analyse this relation as a function of pair type  (S+S or S+E), stellar mass and
interaction stage classified by the morphology.  The goal of the present paper is to better 
understand how SF is enhanced in the merging process, and what role the gas, in particular
the molecular component, plays.

Throughout this paper, we adopt the $\Lambda$-cosmology with 
$\Omega_{\rm m} = 0.3$ and $\Omega_{\rm \Lambda} = 0.7$
and $H_{\rm 0} = 70$ km s$^{-1}$ Mpc$^{-1}$.


\section{The sample}

The local  galaxy pair sample used in this  work was constructed from
the  KPAIR sample which is a complete and unbiased \ks  -band (2.16~\mi) selected sample of 
170 close major-merger galaxy pairs \citep[see details in ][]{domingue09,xu12}.
Their projected distance, $r$, is in the range of 5 -- 20 $h^{-1}$  kpc and the mass ratio of the 
pair components $\le 2.5$.
\citet{cao16} selected a subsample of 88 galaxy pairs for observations with
\herschel\  from this sample (hereafter H-KPAIR) by excluding
(i) elliptical+elliptical (E+E) pairs; (2) pairs with only one measured redshift and (3) pairs 
with recession velocities < 2000 \kms. This sample includes 44 spiral+spiral (S+S) and
44 spiral+elliptical (S+E) pairs. All galaxies have $z<0.1$ with a median of z = 0.04.
The galaxies were  classified by visual inspection as not showing any merger signs
(labeled "JUS"), galaxies with interactions signs (labeled "INT") and pairs in the process of merging
(labeled "MER")

From the H-KPAIR sample we selected  a subsample of spirals in both S+S and S+E pairs that were
observed in \mbox{CO(1-0)}.  We did not observe elliptical galaxies because they generally have a low
molecular gas content and are not actively star-forming. 
In order to enhance the probability of detection we restricted the sample to
close-by (redshift $<$ 0.055) and relatively infrared bright objects, i.e. detected at  70~\mi .
Following these criteria, we observed 78 spiral galaxies out of the H-PAIR sample,
55 of them in S+S pairs, and 23 in S+E pairs.

{ Among 176 galaxies in the HKPAIR sample, 12 (8 of them with CO data) contain an AGN according to  optical spectroscopy 
\citep{cao16}.
Most of them do not show any significant difference in their mid- or far-infrared (FIR) emission compared to other galaxies in the sample. 
Only two galaxies, J13151726+4424255 and J12115648+4039184, show possible AGN contributions in their 
WISE colors \citep{domingue16}. This is consistent with \citet{nordon12} and \citet{lam13} who found that for most AGN 
hosting galaxies, the contribution from AGN to the infrared luminosity is insignificant. The inclusion of these AGNs shall not affect our main results. }


\section{Data}

\subsection{CO observations and data reduction}

The observations were carried out  between July 2015 and May 2018
with the Institut de Radioastronomie Milimetrique (IRAM) 30-meter telescope on Pico Veleta.
We observed the redshifted $^{12}$CO(1-0) and $^{12}$CO(2-1) lines in parallel in the central position of each galaxy.
We used  the dual polarization receiver EMIR in combination  with 
the autocorrelator FTS at  a   frequency resolution of 0.195 MHz  (providing a velocity resolution of $\sim$ 0.5 \kms\ at CO(1--0))
and with the autocorrelator  WILMA with a frequency resolution of  2MHz (providing a velocity resolution of  $\sim$ 5 \kms\  at CO(1--0)).
The observations were done in wobbler switching mode with a
wobbler throw between 40 and 120\arcsec\ in azimuthal direction. The wobbler throw was chosen individually in order to 
ensure that the off-position was well outside the partner galaxy.

The  broad bandwidth of the receiver (16 GHz) and backends (8 GHz for the FTS and 4 GHz for WILMA) allowed grouping the observations
of galaxies into similar redshifts. We organized the groups giving priority to the CO(1-0) line, and accepted that in some cases
CO(2-1) was not covered by the narrower velocity bandwidth of the E230 receiver.
The observed frequencies, taking into account
the redshift of the objects, range between108 and 113~GHz  for CO(1-0)  and between 218 and 227~GHz for CO(2-1). 
Each object was observed until it was detected with a S/N  ratio of at least 5 or
until a root-mean-square noise (rms) $ < 3$ mK (T$_{\rm A}^*$) was achieved for a velocity resolution of 20  km s$^{-1}$.
The integration times per object ranged between 15 and 100 minutes.
Pointing was monitored on nearby quasars  every 60 -- 90  minutes.
During the observation period, the weather conditions were 
generally good, with a pointing accuracy better than 3-4~\arcsec.
The mean system temperature for the observations
was 170~K for CO(1-0) and 360~K  for CO(2-1) 
on the $T_{\rm A}^*$ scale.
At 115 GHz (230 GHz), the  IRAM forward
efficiency, $F_{\rm eff}$, was 0.95 (0.91) and the 
beam efficiency, $B_{\rm eff}$, was 0.77 (0.58).
The half-power beam size for CO(1-0) ranges between  21.8$^{\prime\prime}$ (for 113 GHz)  and 22.8$^{\prime\prime}$ (for 108 GHz),
and the values for CO(2-1) are a factor 2 smaller.
All CO spectra and luminosities are
presented on the main beam temperature scale ($T_{\rm mb}$) which is
defined as $T_{\rm mb} = (F_{\rm eff}/B_{\rm eff})\times T_{\rm A}^*$.

The data were reduced in the standard way via the CLASS software
in the GILDAS package\footnote{http://www.iram.fr/IRAMFR/GILDAS}.
We first discarded poor scans and then subtracted a constant or linear baseline.
Some observations taken with  the FTS backend  were affected by platforming,
i.e. the baseline level changed abruptly at one or two positions along the band. This effect could be
reliably corrected because the baselines in between these (clearly visible) jumps were linear and could
be subtracted from the different parts individually, using the procedure {\it FtsPlatformingCorrection5.class}
provided by IRAM.
We then averaged the spectra and 
smoothed them  to resolutions of  21~\kms.

We present the detected spectra  in the appendix (Figs. A1 and A2).
For each spectrum, we determined visually the zero-level line widths, if detected. 
The velocity integrated spectra were calculated by summing the individual channels in between
these limits. 
 For non-detections we set an upper limit as

\begin{equation}
I_{\rm CO} < 3 \times {\rm rms} \times \sqrt{\delta \rm{V} \ \Delta V},
\end{equation}

\noindent { where $\delta \rm{V}$ is the channel width (in \kms), $\Delta$V the zero-level line width (in \kms), and rms the root mean square noise (in K).}
For the non-detections, we assumed a linewidth of $\Delta$V = 400  \kms\ which is close to the mean velocity width found
for CO(1-0) in the sample (mean $\Delta \rm{V}$ = 435 \kms\ with a standard deviation of 210 \kms).
We treated tentative detections, with a S/N ratio between 3-5, as upper limits in the statistical analysis.
The results of our CO(1-0)  observations are listed in Table~\ref{tab:ico}.
In addition to the statistical error of the velocity integrated line intensities, a calibration error of 
15 \% for CO(1-0) and 30\% for CO(2-1) has to be taken into account. These errors were determined 
by comparing the observations of  several strong sources (J0823+2120A, J0823+2120B, J1315+4424a and J1444+1207A) 
on different days.

\begin{table*}
\caption{\label{tab:ico} Velocity integrated CO intensities}
\begin{tabular}{llllllll}
\noalign{\smallskip} \hline \noalign{\medskip}
Galaxy name & Pair name &  rms\tablefootmark{a} &  $I_{\rm CO(1-0)}$\tablefootmark{b}  & $\rm \Delta \rm{V}_{CO(1-0)}$\tablefootmark{c} & rms\tablefootmark{a} & $I_{\rm CO(2-1)}$\tablefootmark{b,d}  & $\rm \Delta \rm{V}_{CO(2-1)}$\tablefootmark{c}  \\
   & & [mK] & [K \kms]  & [\kms] & [mK] & [K \kms]  & [\kms] \\
\noalign{\smallskip} \hline \noalign{\medskip}
 J00202580+0049350  &  J0020+0049   &    6.19   &    6.54 $\pm$    0.47   &  280   &   27.16   &   12.78 $\pm$    2.06   &  275  \\
 J01183417-0013416  &  J0118-0013   &    3.49   &    6.86 $\pm$    0.38   &  558   &       -   &                     -   &    -  \\
 J01183556-0013594  &  J0118-0013   &    3.93   &    2.63 $\pm$    0.22   &  150   &       -   &                     -   &    -  \\
 J02110638-0039191  &  J0211-0039   &    4.32   &    7.53 $\pm$    0.43   &  465   &   13.24   &   22.42 $\pm$    1.31   &  465  \\
 J03381222+0110088  &  J0338+0109   &    3.00   &    4.07 $\pm$    0.27   &  380   &    9.92   &    8.59 $\pm$    0.80   &  310  \\
.....  &    .....    &      .....    &  .....   &  .....    &   .....   &               .....   &  .....  \\
\noalign{\smallskip} \hline \noalign{\medskip}
\end{tabular}
\tablefoot{
\tablefoottext{*}{Tentative detections.}
\tablefoottext{a}{Root-mean-square noise at a velocity resolution of 21 \kms.}
\tablefoottext{b}{Velocity integrated intensity of the CO(1-0) and the CO(2-1) line.}
\tablefoottext{c}{Zero-level line width. The uncertainty is roughly given by the velocity resolution (21 \kms).}
\tablefoottext{d}{For some objects the bandwidth does not entirely cover the expected CO(2-1) frequency so that no value for $I_{\rm CO(2-1)}$
can be given.}
The full table is available online at the CDS.
}
\end{table*}

\subsection{Aperture correction and molecular gas mass}
\label{sec:molecular_gas_mass}

We observed the galaxies only in their central pointing  which in most cases
covers only a fraction of the entire galaxies. This fraction is furthermore different for each galaxy
depending on its size. Therefore,  we need to apply a correction for emission
outside the beam.
We carried out this aperture correction in the same way as described in \citet{lisenfeld11}, 
assuming an exponential distribution of the CO flux:
\begin{equation}
S_{\rm CO}(r) = S_{\rm CO,center}\propto \exp(-r/r_{\rm e}) .
\label{eq:Ico_r}
\end{equation}
where $S_{\rm CO,center}$ is the CO(1-0) flux in the central position and derived from
the mesured \ico\ applying the $T_{\rm mB}$ -to-flux  conversion factor of the IRAM 30m telescope (5 Jy/K).
\citet{lisenfeld11} derived a  scale length of $r_{\rm e}$ = 0.2$\times$\riso, where \riso\ is 
the major optical isophotal radius at 25 mag arcsec$^{-2}$,  from different  studies  of  local spiral galaxies
 \citep{2001PASJ...53..757N, 2001ApJ...561..218R, leroy08}
 and from their own CO data.  
 { Very similar values for $r_{\rm e}$/\riso\ have been found by \citet{boselli14a} ($r_{\rm e}$/\riso $\sim
  0.2$)
 and \citet{casasola17} ($r_{\rm e}$/\riso = 0.17$\pm$ 0.03) from an analysis of nearby mapped galaxies.}

For our sample \riso\
 is only available for 45 out of the 178 HKPAIR galaxies. In addition,  an isophotal
  radius as \riso\ can be affected by confusion in close pairs. 
 We therefore use the Kron radius, \rkron, derived from the \ks -band and obtained from the
 2MASS archive.
 We compared \riso\  and \rkron\ for the 45 galaxies
 where both radii are available and obtained a mean value \riso/\rkron = 1.33, with a standard deviation of 0.66,
 and a median value of 1.2. The large scatter is mostly due to few objects with a relatively 
 high \riso/\rkron\  ($\gtrsim 2$) where \riso\ most likely  is affected by confusion.
 We therefore use the median value of \riso/\rkron\ = 1.2 as the more robust estimate.
 Thus, we adopt $r_{\rm e} = 0.2 \times 1.2 \times$ \rkron = 0.24 $\times$ \rkron\ in 
 eq.~\ref{eq:Ico_r} and use this
 distribution to calculate the expected CO flux  from the entire disk, $S_{\rm CO,tot}$, taking the galaxy inclination into account,
by { 2-dimensional integration over the exponential galaxy disk} \citep[see][for more details]{lisenfeld11}.
We derive the inclination of the galaxy from the \ks -band major-to-minor axis ratio, obtained from the 2MASS archive.
{ \citet{boselli14a} generalized this method to 3 dimensions by taken the finite thickness of galaxy disks  
into account.
Except for edge-on galaxies ($i>80^\circ$) the 3-dimensional method gives basically the same result as the 2-diminsional approximation,
and also for edge-on galaxies the difference is $<5\%$ for $z_{\rm CO}/\Theta < 0.1$ ($z_{\rm CO}$ being the scale height of the CO perpendicular to
the disk and $\Theta$ the beam size). We therefore  consider the 2-dimensional aperture correction sufficient.
}

The resulting aperture correction factors, \faper, defined as the ratio between $S_{\rm CO,center}$ and the total,
aperture corrected flux $S_{\rm CO,tot}$,
lie between 1.0 and 5.0 with a mean (median) value of 1.5 (1.4). 
The values for the  extrapolated molecular gas mass and \faper\ are
listed in Table~\ref{tab:mh2}.

We calculate the molecular gas mass 
from the CO(1-0) luminosity, \lco\ \citep[following][]{solomon97}, derived as:
\begin{equation}
L^\prime_{\rm CO} [{\rm K \, km\, s^{-1} pc^{-2}}]= 3.25 \times 10^7\, S_{\rm CO,tot} \nu_{\rm rest}^{-2} D_{\rm L }^{2} (1+z)^{-1},
\label{eq:lco}
\end{equation}
where  $S_{\rm CO, tot}$ is the aperture corrected CO line  flux (in Jy \kms), 
$D_{\rm L }$ is the luminosity distance in Mpc, $z$ the redshift
and $\nu_{\rm rest}$ is the rest frequency of the line in GHz.
We then calculate the molecular gas mass as 
\begin{equation}
M_{\rm H_2} [M_\odot]= \alpha_{\rm CO} L^\prime_{\rm CO} 
\end{equation}
\label{eq:mh2}
We adopt the Galactic value,  \alphaco =  \alphacogal = 3.2 \msun/(K \, km s$^{-1}$ pc$^{-2}$) \citep{bolatto13},  
and do not  include helium or heavy metals in the mass.
This conversion factor corresponds to 
 X=$N_{\rm H_{2}}/I_{\rm CO}$ = $2\times 10^{20}\rm cm^{-2}$ (K km s$^{-1})^{-1}$.

\begin{figure}
   \centering
 \includegraphics[width=8cm,trim=0.cm 0.cm 0cm 0.cm,clip]{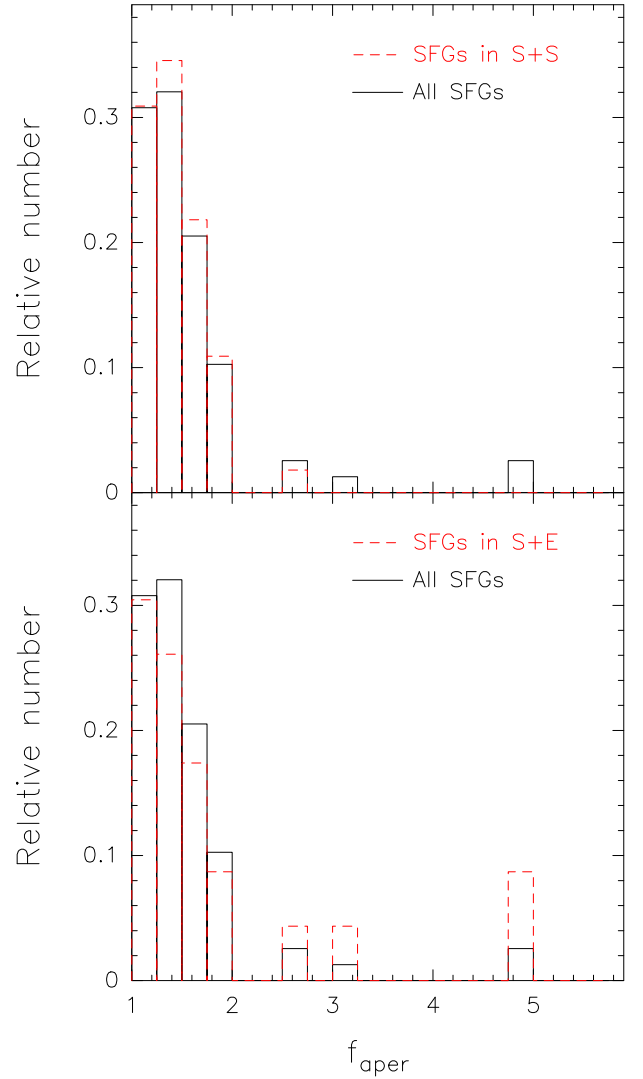}
      \caption{Histogram of the distribution of the aperture correction, \faper , for the
      different subsamples (red dashed line) compared to the full sample (full black line).}
         \label{fig:histo_faper}
   \end{figure}

This empirical aperture correction potentially could introduce a bias in our analysis.
We therefore investigate in 
Figure~\ref{fig:histo_faper}  the distribution of \faper, separate for  
 S+S pairs and S+E pairs. Two features are visible that are relevant for this question.
Firstly, for most of the objects, the aperture
 correction is below 2 (91\%, and 60\% below 1.5) which means that possible errors introduced by this correction are small.
Secondly,  the distribution of \faper\ for spirals in S+S and in S+E pairs
 is very similar and no features indicating the presence of a bias are obvious.

\begin{table}
\caption{\label{tab:mh2} Molecular gas mass}
\begin{tabular}{llll}
\hline
Galaxy name &Distance   &log(\mmol)\tablefootmark{a} & \faper\tablefootmark{b} \\
   & [Mpc]  &  [\msun] &   \\
\hline
   J00202580+0049350  &    64.4  &    9.44  &    2.63\\
   J01183417-0013416  &   200.9  &   10.10  &    1.22\\
   J01183556-0013594  &   201.7  &    9.71  &    1.29\\
   J02110638-0039191  &    77.0  &    9.52  &    1.92\\
   J03381222+0110088  &   172.9  &    9.80  &    1.38\\
   .... & .... & .... & .... \\
\hline 
\end{tabular}
\tablefoot{
\tablefoottext{a}{Cold molecular gas mass, extrapolated from the central pointing to the entire disk, calculated as
described in Sect.~\ref{sec:molecular_gas_mass}.}
\tablefoottext{b}{Aperture correction \faper = \mmol/\mmolcenter.}
The full table is available online at the CDS.
}
\end{table}

\subsection{Dust emission, SFR and stellar mass}

All objects in our sample were imaged with the Herschel instruments PACS \& SPIRE \citep{cao16}.
These authors fitted the dust SED with the model of \citet{draine07}  and derived the total IR luminosity, \lir,  as well as the dust mass 
from it. They calculated  the SFR from  \lir\ using the expression given in Kennicutt (1998), adapted to
a Kroupa IMF, as SFR (\msun yr$^{-1}) = 2.2 \times 10^{-10}$ (\lir/\lsun).
{ This derivation of the SFR misses the contribution from unobscured UV radiation, which is  on the order of 20\% for 
KPAIR galaxies \citep{yuan12}. It can also be affected by the contribution of old stars to the dust heating which is
about 30\% for spiral galaxies \citep{buat96} and even lower \citep[$0.17\pm0.11$,][]{buat11} for massive, actively star-forming galaxies.
Thus, both affects are not expected to be severe for our actively star-forming sample.
In addition, since we use the same formalism for both the H-KPAIR and the AMIGA control sample, these possible
biases are not expected to  affect the results involving the SFR.}

The stellar mass was calculated from the 2MASS \ks-band luminosity as 
 \mstar (\msun)  =  0.54 \lk/\lksun\ \citep{xu12}.
{  The near-infrared \ks\ band luminosity, mainly from old stars that dominate the stellar mass, is insensitive to dust 
 extinction and star formation \citep{bell01}. The calibration for the conversion from \lk\ to \mstar\ was derived from a comparison between \lk
and \mstar\ for normal galaxies in \citet{kauffmann13}.}
We use the data provided by \citet{cao16}, in particular the 70~\mi\  PACS flux, SFR, dust and stellar masses.

In addition, 
we derived the central 70~\mi\  and \ks\ band fluxes within the IRAM
CO(1-0) beam directly from the corresponding images
in order to locally compare the SFR and stellar mass with the measured molecular gas
mass in the center of the galaxies, \mmolcenter. 
{ To achieve
this, we multiplied the  70~\mi\ image ($Im_{\rm 70\mu m}(x,y)$) and the \ks\ image ($Im_{\rm Ks}(x,y)$) with the IRAM beam
pattern (approximated as a normalized Gaussian beam) placed at
the position where the CO beam was pointed  during the observations ($x_0,y_0)$:
\begin{equation}
Im_{\rm 70\mu m, beam}(x,y) = Im_{\rm 70\mu m}(x,y) 
  \cdot \exp{\left(\frac{(x-x_0)^2+(y-y_0)^2)}{2\sigma^2}\right)}
\end{equation}
A corresponding expression  is used for the \ks\ band image. 
The Gaussian standard deviation $\sigma$ is related to the FWHM as $\sigma$ = FWHM/$(2\sqrt{2\ln(2)})$ =  FWHM/2.35.
For the \ks\ band image we use the FWHM of the IRAM 30m CO(1-0) beam, FWHM = FWHM(IRAM) = 21.34\arcsec(1+z).
For the 70~\mi\ PACS image we use a slightly smaller value,
FWHM =$\sqrt{\rm FWHM(IRAM)^2-FWHM(PACS)^2}$,  in order to take into account that the 70~\mi\ image is already
convolved to the PACS resolution, FWHM(PACS)$\sim$ 6\arcsec.
From  the resulting maps $Im_{\rm 70\mu m, beam}(x,y)$ and $Im_{\rm Ks, beam}(x,y)$ we then measured the
total 70 \mi\ flux,  $S_{\rm70\mu m, center}$,  and   \ks\ flux, $S_{\rm Ks, center}$.
}
We derived the central stellar mass within the IRAM beam from $S_{\rm Ks, center}$ with the same relation as for the
total stellar mass (see above). For the central SFR we assumed that it is well traced by the  70~\mi\
emission and derived it as ${\rm SFR_{center}}={\rm SFR}\cdot S_{\rm70\mu m, center}/S_{\rm70\mu m}$ where
SFR and $S_{\rm70\mu  m}$ are the total SFR and 70~\mi\ flux from \citet{cao16}, respectively.

\subsection{Atomic gas}

We use the  data for the atomic hydrogen (HI) data for 70 galaxy pairs (34 S+S pairs and 36 S+E pairs) from the H-KPAIR sample, 
from \cite{zuo18}. { These authors observed
 58 pairs  with the Green Bank  Telescope (GBT) and retrieved the data for an additional 12 pairs 
from the literature, giving a total of 70 pairs with HI data}. We have CO data for 38 of these pairs.

\section{Results}

In the following section, we mainly study the relations between molecular gas mass, atomic gas
mass, dust mass, stellar mass and SFR. The main goal is to search for differences between
spiral galaxies in S+S and in S+E pairs, for trends with the stellar mass,
and to investigate the influence of the stage of the merger process 
 by distinguishing galaxies classified  as not showing any merger signs
(labeled "JUS") from galaxies with interactions signs (labeled "INT") and pairs in the process of merging
(labeled "MER").
Following \citet{cao16}, we exclude from our sample galaxies with a low sSFR $< -11.3$ yr$^{-1}$ (7 objects, only 1 detection
 in CO(1-0)) which belong to the red sequence and are thus not actively star-forming galaxies. 

Since the beam of the GBT (9~\arcmin\ for HI) is too large to separate the emission from the individual galaxies, the analysis 
taking into account HI data is done for each pair as a whole. 
For S+S pairs we sum the values for the SFR, stellar mass, dust mass and molecular gas
for both components. 
We then divide these values, as well as the HI mass, by two in order to obtain values 
typical for one galaxy.  
If only one member of a pair is observed in CO (2 cases), we flag the total molecular gas mass
and the total gas mass as a lower limit.
For S+E pairs we assume, following \citet{zuo18}, that the atomic gas is mostly associated to the spiral component.
\citet{zuo18}, tested this assumption using the gas mass derived from the dust mass 
\citep{cao16} and found that the mean gas mass in elliptical is only about 10\% of the total mass in a S+E pair.
We thus make a 10\% correction and assume that \mhi\ of the spiral component in S+E pairs is 90\% of the 
\mhi\ observed in the pair.
The other variables, \mmol,  \mstar, SFR and \mdust\ are taken for the spiral component only.

We list the mean and median values
for different ratios, separated for the different groups,  in Table~\ref{tab:means}.
In this table, we present the values for the entire galaxies (or pairs).
As a check, we also examined all relations involving
\mhtwo, \mstar\ and SFR for the central values within the IRAM beam. We obtained very similar results, with the
mean values agreeing within the errors.  This supports the robustness of our conclusions,

\subsection{Comparison sample}

In order to search for differences between our  merger sample and non-interacting galaxies,
we need to compare our results to a suitable comparison sample. The comparison sample
used in \citet{cao16} which was selected  from the SDSS and matched to the HKPAIR sample
does not include data for the molecular gas and can therefore only be used for a comparison 
of the SFR or \mdust, which is a good indicator for the total gas mass \citep[e.g.,][]{eales12, corbelli12}.

{ Different catalogues of non-interacting galaxies containing CO, HI, SFR and \mstar\ exist in the literature, 
e.g. the AMIGA sample of isolated galaxies \citep{verdes-montenegro05, lisenfeld11}, the COLDGASS sample of mass-selected nearby galaxies 
\citep{saintonge11a,saintonge11b}
and a catalogue of the ISM of normal galaxies \citep{bettoni03}.
Here, we use the AMIGA and the COLD GASS sample for comparison because  the CO observations
for both samples have been taken with the IRAM 30m telescope and have been processed in a similar way as for our interacting
sample.}
We present the relevant mean and median values
in Table~\ref{tab:mean_comp}. We adapt all values to the Kroupa IMF,
the Galactic \alphaco\  and do not include
 helium and heavy metals in the gas masses, and we limit the objects to those with
 sSFR $> -11.3$ yr$^{-1}$, as in our pair sample.

The AMIGA  sample consisting of 1050 local  isolated galaxies \citep{verdes-montenegro05}.
Out of this sample, a volume-limited (recession velocities between 1500 and 5000 \kms) subsample 
of 173 object possess CO data and their molecular gas properties were analyzed in
\citet{lisenfeld11}.  The CO data were observed, as in the present paper,  for the central position of the galaxy
and the total molecular mass was calculated with the same extrapolation procedure as in
the present paper. In \citet{lisenfeld11}  the K-band luminosities, \lk,  are presented
and we use  them to calculate the stellar mass in the same way as here (see Sect. 3.3).
Also the SFR is calculated in a very similar way as here from the total FIR luminosity based on IRAS data, \lir,  following the expression from \citet{kennicutt98}.
The AMIGA sample comprises a stellar mass range of (log(\mstar) $\sim$ 9-11 
which extends to lower masses than our sample.  Therefore, we recalculated all mean and
median values restricting the sample to high-mass spirals (\mstar $> 10^{10}$ \msun), and using the Kaplan-Meier
estimator to take  upper limits into account.

The  COLD GASS galaxy sample \citep{saintonge11a,saintonge11b}
is a mass-selected (\mstar $> 10^{10}$ \msun)  local sample of $\sim$ 350 galaxies. The HI fluxes 
were obtained from the GASS survey \citep{catinella18}. The COLD GASS sample matches 
the HKPAIR sample in stellar mass (log(\mstar) $\sim$ 10-11.4).
{ The CO (1-0) was observed with the IRAM 30m telescope for the central position.
In order to obtain the total CO flux,
 an aperture correction  was derived by the authors  based on CO maps of nearby spiral galaxies \citep{kuno07} 
for which the impact of the observation with the IRAM beam was simulated for different redshifts.
In this way, the flux that would be measured by a 22\arcsec\ Gaussian beam ($S_{\rm CO-beam}$) was calculated and
compared to the total flux of the maps, $S_{\rm tot}$. The dependence of the aperture correction factor 
$S_{\rm tot}/S_{\rm CO-beam}$ on the optical radius, \diso, was fitted and provided an aperture correction
as a function of \diso\ \citep[see eq.~2 in][]{saintonge11a}. 
For galaxies with \diso $\geq$ 40\arcsec\ they 
improved the scatter in the relation between the aperture correction factor and \diso\ by performing an
offcenter pointing along the major axis and used an empirical relation  based 
on the ratio of the offcenter and central CO fluxes  \citep[see eq.~3 in][]{saintonge11a}.}
The stellar mass and the SFR were calculated from optical/UV spectral energy
distribution (SED) based on  a Chabrier IMF, which is very similar to the Kroupa IMF used in the present paper.
In order to match the COLD GASS sample to ours, we furthermore made two restrictions:
(i) We selected late-type galaxies based on the concentration index, C (the ratio of the r-band Petrosian radii encompassing 90\% and 50\% of the flux, 
$C = r_{90,r}/r_{50,r}$). We selected those galaxies with $C> 2.85$ corresponding to galaxies with morphological
types of Sa or later \citep{yamauchi05}.
(ii) We excluded those galaxies that were found to be in pairs  by \citet{pan18}.

The results for the two comparison samples are reasonably consistent.  The mean values have
differences of less than $2\sigma$ except for  the molecular gas fraction where the 
AMIGA sample has a mean value which is $0.22 \pm 0.06$ dex lower than that of the COLDGASS sample.
The reason for this difference is not completely clear. The AMIGA sample was selected with strict isolation criteria
and has  a low molecular gas content \citep{lisenfeld11} and a low
SFR \citep{lisenfeld07} compared to other samples. However, if this were the reason for the difference with COLDGASS,
we would  expect not only a difference in \mmol/\mstar\ but a similar difference
in \mmol/\mhi\ which is not the case.
In the following, we use the AMIGA sample for quantitative comparisons, but we note that all the
conclusions are also valid compared to the COLDGASS sample.

\begin{table}
\caption{\label{tab:mean_comp} Mean and median values of the comparison samples}
\begin{tabular}{l | lll}
\noalign{\smallskip} \hline \noalign{\medskip}
Parameter & AMIGA & ColdGASS & Cao et al. (2016)\\
    & mean & mean & mean \\
    & median & median & median\\
        & $n/n_{\rm up}$\tablefootmark{a} & $n/n_{\rm up}$\tablefootmark{a} & $n$\tablefootmark{a} \\
\noalign{\smallskip} \hline \hline \noalign{\medskip}

log(\mmol/\mhi) & -0.63$\pm$0.06  &  -0.49$\pm$0.05 \\
   &   -0.58 & -0.46 & \\
   &  76/12 & 81/10& \\
   \hline
log(\mmol/\mstar) & -1.48$\pm$0.05& -1.26$\pm$0.04 & \\
   &  -1.35 & -1.20&\\
   &  78/14 & 168/15&\\
   \hline
log(\mgas/\mstar) & -0.76$\pm$0.05& -0.82$\pm$0.04 & \\
   &  -0.71& -0.80&\\
   &  77/13 & 81/10&\\
   \hline
log(SFE) &-9.07$\pm$0.05 &  -8.97$\pm$0.04 &  \\
   & -9.13  & -8.98&\\
   &  62/7 & 168/15&\\
   \hline
   log(\sfegas) & -9.73$\pm$0.05 &   -9.63$\pm$0.07 & -9.81$\pm$ 0.05\tablefootmark{a}\\
   &  -9.79&-9.66 & -- \\
   &  61/6 &81/10 & 132/0\\
\noalign{\smallskip} \hline \noalign{\medskip}
\end{tabular}
\tablefoot{
\tablefoottext{a} {Total number of galaxies (n) and number of upper limits ($n_{\rm up}$).
\tablefoottext{b} The value from \citet{cao16} is adapted to our
mean gas-to-dust mass ratio of 138.
}
}
\end{table}

\subsection{CO line ratio}

\begin{figure}
   \centering
\includegraphics[width=8.cm,trim=0.cm 0.cm 0cm 0cm,clip]{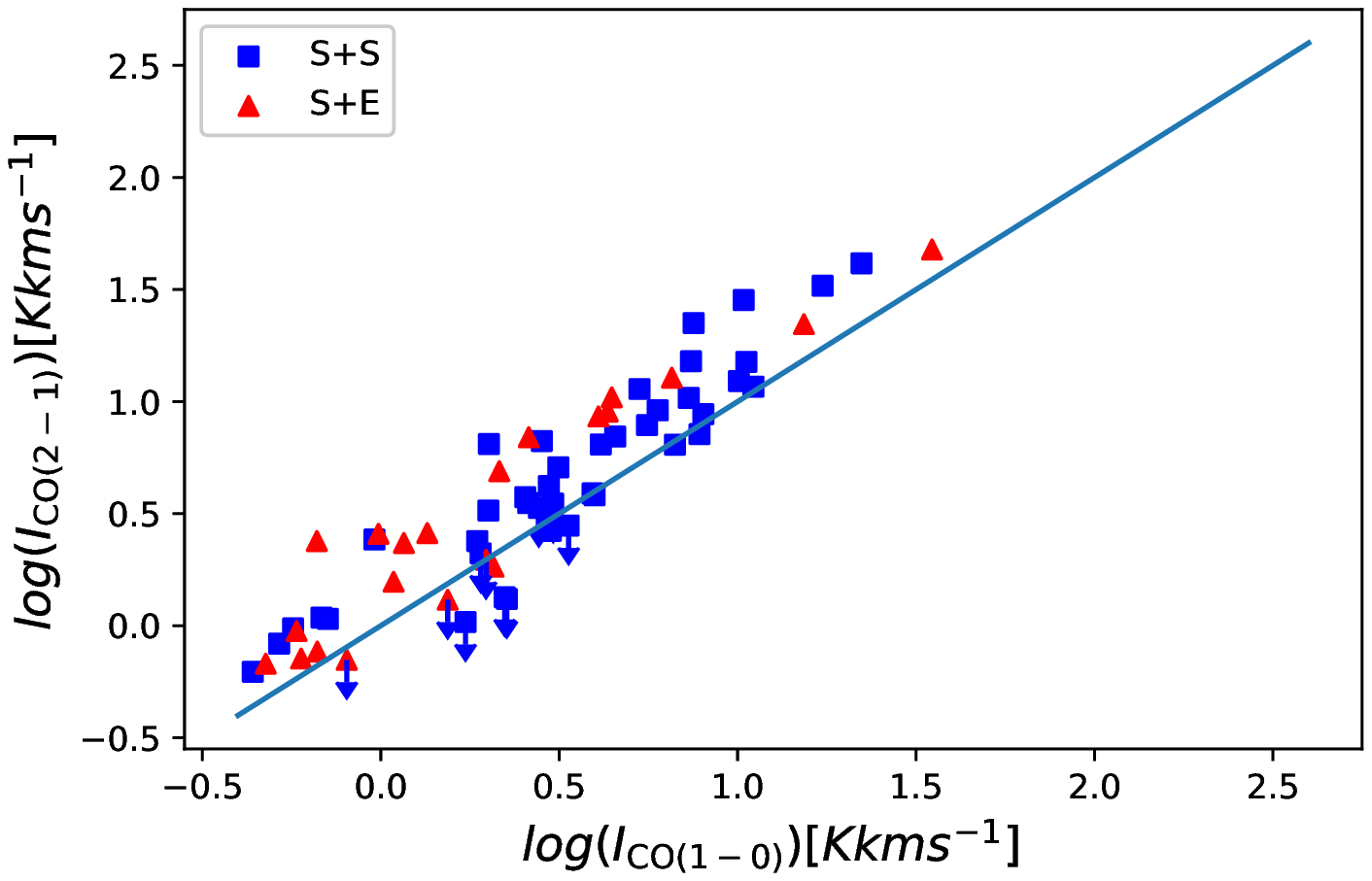}
\includegraphics[width=8.cm,trim=0.cm 0.cm 0cm 0cm,clip]{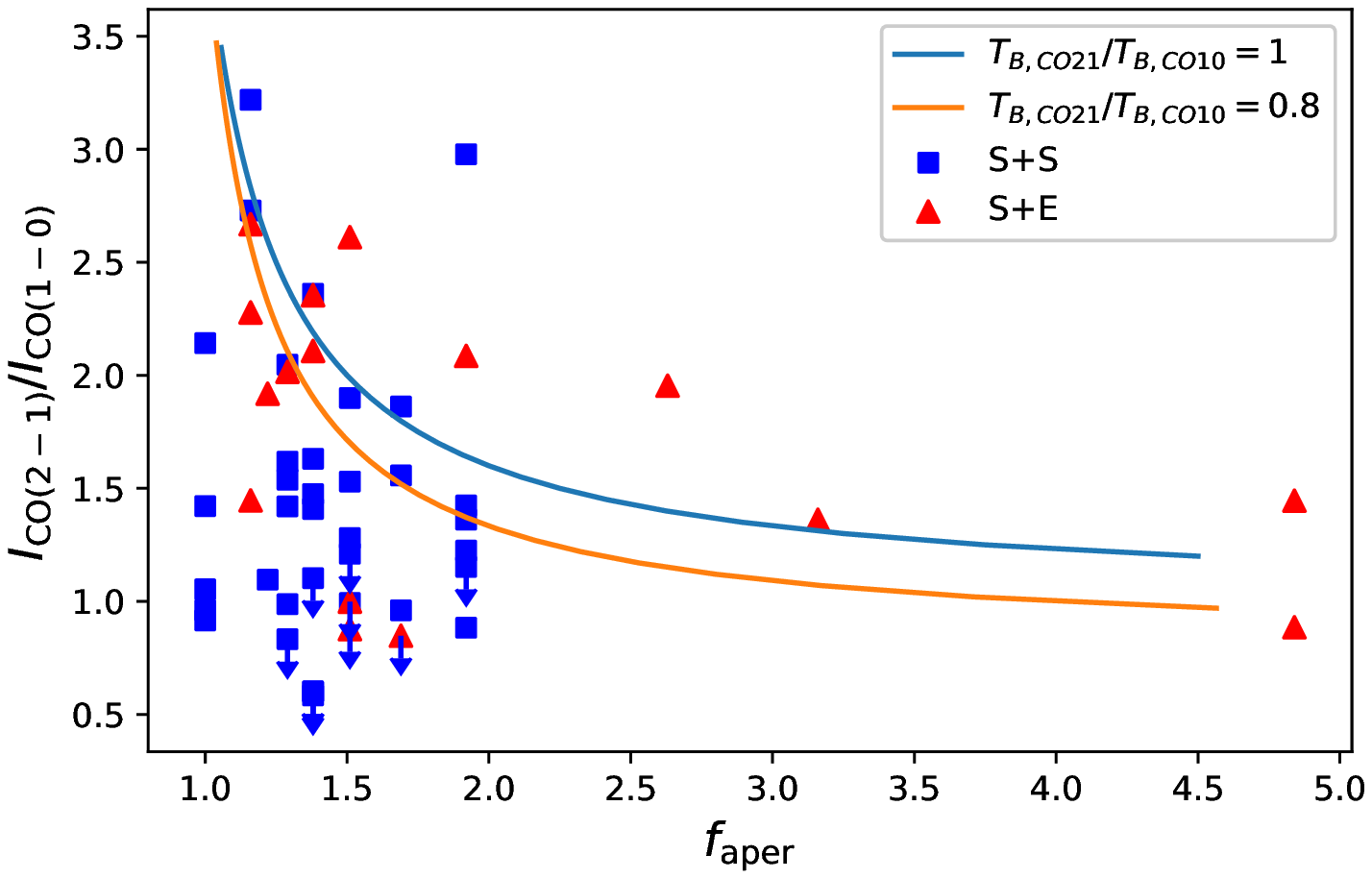}
      \caption{\icotwo\ as a function of   \icoone (top panel). The line indicates unity. In the lower panel we 
      show the line ratio as a function of the aperture correction, \faper, together with the prediction of a simple model assuming a 
      Gaussian source and beam size. The model results are shown for two values of the ratio of the intrinsic source 
      brightness temperatures.
      }
               \label{fig:line-ratio}
   \end{figure}

In our sample we have 53 objects for which measurements for both \icoone\ and \icotwo\
are available and which allow us to derived their line ratios. We show the relation between both
lines in Fig.~\ref{fig:line-ratio}.
We derive (taking upper limits in \icotwo\ into account) a mean value of 
\rtwoone=\icotwo/\icoone = 1.5 $\pm$ 0.1.
{ If we only take objects with detections in both \icoone\ and \icotwo\ into account (42 galaxies) the mean value is
\rtwoone = 1.7 $\pm$ 0.1.
Both values are not aperture corrected.}

{ To interpret \rtwoone\ one has to consider, { apart from the  excitation  temperature of the gas},
two main parameters: the source size relative to the beam and the opacity of the molecular gas. For
optically thick, thermalized emission with a point-like distribution
we expect a ratio \rtwoone\ = $\Theta^2_{\rm B(CO10)}/\Theta^2_{\rm B(CO21)} = 4$
(with 
$\Theta_{\rm B}$ being the FWHM of the beams). On the
other hand, for  a source that is more extended than the beams  we expect
\rtwoone$ \sim 0.6 -1 $ for optically thick gas in thermal equilibrium, where \rtwoone\  depends on the 
temperature of the gas, and \rtwoone$> 1$  for optically thin gas.

We derive for our sample a mean \rtwoone $>1$ which could be interpreted as a source extension smaller than at least 
the CO(1-0) beam or an extended source and optically thin emission. 
Based on our data alone, we cannot distinguish between these two cases. 
We can, however, consider results 
for  spatially resolved observations from the literature which allows
to derive the emission of CO(1-0) and CO(2-1) from the same area
so that the source-to-beam size is irrelevant.

Matched aperture observations of  CO(1-0) and CO(2-1) give  line ratios
of \rtwoone$=0.89\pm0.6$ \citep[][for a small sample of nearby spiral  galaxies]{braine93},
\rtwoone $\sim 0.8$ \citep[][for the SINGS sample]{leroy09}, 
and \rtwoone $\sim 0.6-0.9$ \citep[][for 4 low-luminosity AGNs from the NUGA survey]{casasola15}.
\rtwoone $\sim 0.8$ is consistent with optically thick gas with an excitation temperature of $\sim$10 K \citep{leroy09}.
It is therefore very likely that a similar situation holds in our sample so that \rtwoone $>1$ can be
interpreted as optically thick,  thermalized gas which a spatial extension smaller than (at least) the 
CO(1-0) beam.
}

{ Based on these assumptions, we  can quantify the relation between \rtwoone\ and the source size  in a simple model.}
The relation between the velocity integrated intensity and the intrinsic source brightness temperature, $T_{\rm B}$, is
\citep{solomon97}:
\begin{equation}
I_{\rm CO} \Omega_{\rm S*B}(1+z) = T_{\rm B}\triangle V \Omega_{\rm S},
\end{equation}
where $\Omega_{S*B}$ is the solid angle of the source convolved with the beam. Adopting for simplicity a source
with a Gaussian distribution and neglecting the redshift dependence we obtain:
\begin{equation}
I_{CO} = T_B\triangle V \frac{\Omega_S}{\Omega_{S*B}} = T_B\triangle V \frac{\Theta^2_S}{\Theta^2_B+\Theta^2_S},
\end{equation}
where $\Theta_S$ 
is the FWHM of the source. 
With these simplifications, and furthermore assuming that the intrinsic brightness temperature $T_{\rm B}$ is the same for 
both lines we can describe the line ratio as
\begin{equation}
R_{\rm 21}= \frac{I_{\rm CO(2-1)}}{I_{\rm CO(1-0)} }=  \frac{\Theta^2_{\rm B(CO10)}+\Theta^2_{\rm S}}{\Theta^2_{\rm B(CO21)}+\Theta^2_{\rm S}}.
\label{eq:ratio}
\end{equation}
For point-like source ($\Theta^2_{\rm S} \ll \Theta^2_{\rm B}$) the line ratio is $R=4$. 

Again adopting Gaussian distributions for both the source and the beam we can derive the aperture correction as a function
of beam and source sizes:
\begin{equation}
f_{\rm aper} = \frac{S_{\rm CO, tot}}{S_{\rm CO, center}} =   \frac{\Theta^2_{S}}{\Theta^2_{\rm B(CO10)}} +1,
\label{eq:faper}
\end{equation}
where $S_{\rm CO, tot}$ is the total flux of the source and ${S_{\rm CO, center}}$ is the flux observed within the Gaussian CO(1-0) beam.
Together, eq.~\ref{eq:ratio} and \ref{eq:faper} yield a relation between the aperture correction and the line ratio, both derived
within the simplifying assumptions made here.
We show the resulting relation in Fig. ~\ref{fig:line-ratio}. It described the observed line ratios reasonably well, supporting the
correctness of the aperture relation that we use here.

\subsection{Molecular gas, atomic gas and stellar mass}

\begin{figure}
   \centering
\includegraphics[width=8.cm,trim=0cm 0.cm 0cm 0cm,clip]{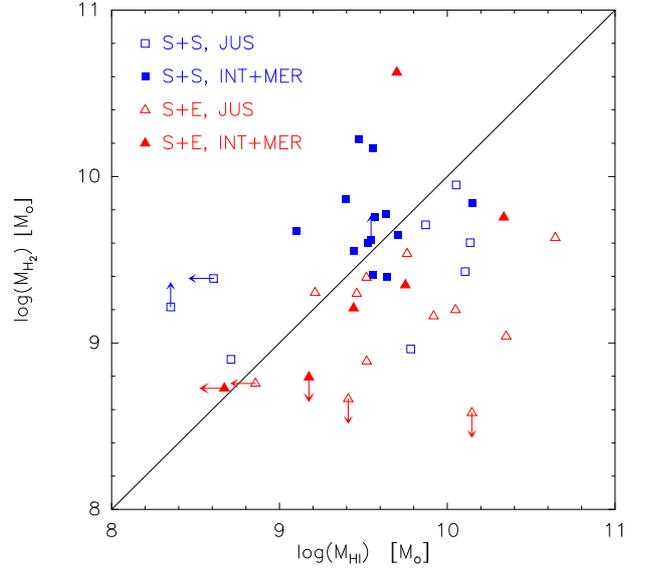}
      \caption{Molecular gas mass vs. atomic gas mass of galaxy pairs. 
            The values for \mhi\ and \mmol\ are divided by 2 for the S+S pair.
            { To guide the eye, we include the line of unity (\mhi = \mmol, black line)}.
}
         \label{fig:mhi_mmol}
   \end{figure}

\begin{figure}
   \centering
\includegraphics[width=8.cm,trim=0.cm 0.cm 0cm 0cm,clip]{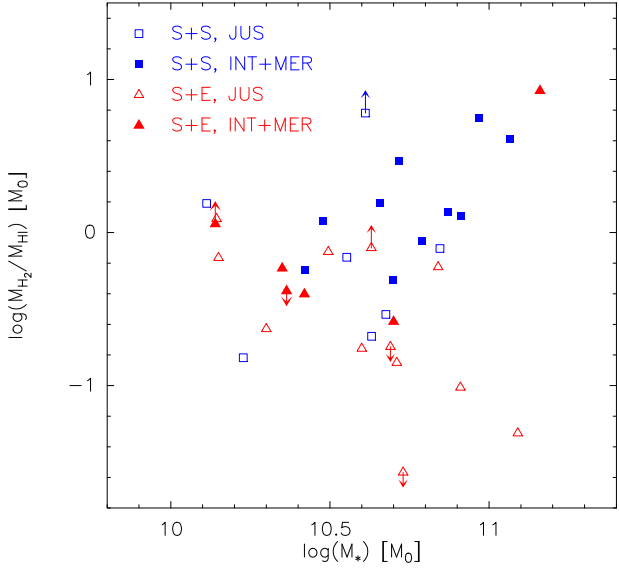}
      \caption{The ratio between molecular gas mass and  atomic gas mass of galaxy pairs as a function
      of stellar mass. 
             The values for \mhi\ and \mmol\ are divided by 2 for the S+S pair.
}
         \label{fig:mmol_over_mhi}
   \end{figure}

 Figure~\ref{fig:mhi_mmol} shows the  molecular and atomic gas mass of the pairs.
 { There is only a  weak correlation (the Spearman's rank correlation coeficient is 0.27 and the
significance is  0.096)}.
 The ratio between molecular and atomic gas mass is shown in Figure~\ref{fig:mmol_over_mhi} 
 as a function of stellar mass.
The mean molecular-to-atomic gas mass ratio  (see  Table~\ref{tab:means}) is significantly higher
than the value for the AMIGA comparison sample (by $0.38\pm 0.12$ dex).
When inspecting the mean values of different subsamples  (see  Table~\ref{tab:means}), we find
the largest and most significant differences between galaxies with and 
without morphological signs of interaction. Whereas the JUS sample has a mean log(\mmol/\mhi)
compatible within the errors with the comparison samples, the INT+MER galaxies have a value
which is higher by $0.69\pm0.16$. This difference is more significant for
S+S galaxies than for S+E galaxies although the number of galaxies  in the corresponding subsamples
are very low.
There is a difference of $0.55\pm0.18$ dex in log(\mmol/\mhi) between S+S and S+E pairs, with the value for
S+E pair being compatible with the comparison sample and that for S+S pairs being higher.
As for subsamples with different stellar mass,  there is no difference for the entire sample. Only for the subsample of S+S pairs
there is a difference  of $0.51\pm0.16$ dex between low stellar mass (log(\mstar) $< 10.7$ \msun) and 
high stellar mass (log(\mstar) $>10.7$ \msun) but the subsamples contain a low number of galaxies.
 
      \begin{figure}
   \centering
\includegraphics[width=8cm,trim=0.cm 0.cm 0cm 0cm,clip]{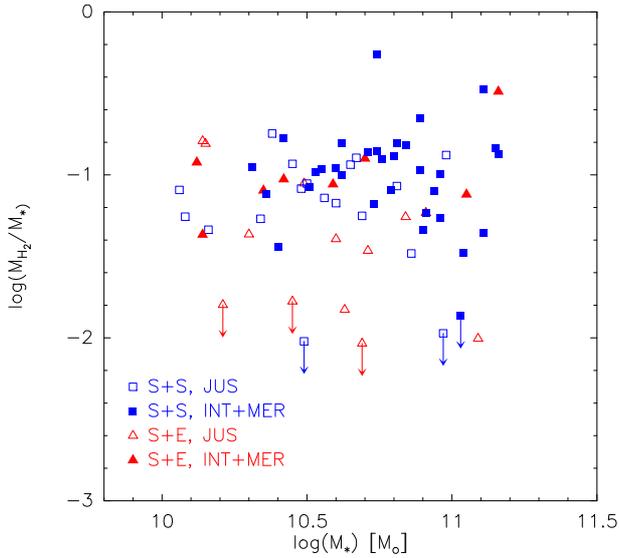}
      \caption{Molecular gas mass fraction (= \mmol/\mstar) vs. stellar mass.}
         \label{fig:fract_mmol_mstar}
   \end{figure}

   Figure~\ref{fig:fract_mmol_mstar} 
   shows the molecular gas   fraction (defined as \mmol/\mstar) as a function of
   stellar mass. 
   The mean logarithmic molecular gas fraction is above  the values found for 
  the comparison sample by $0.34\pm 0.07$ dex for the AMIGA and  by $0.12\pm 0.06$ for the COLDGASS sample.
   Also here, there is a large and significant difference  ($0.32\pm 0.9$ dex) between
  the mean values of the JUS and INT+MER sample.
    This difference is also present when considering the S+S and S+E galaxies separately.
   The molecular gas fracion is also smaller for star-forming galaxies in S+E galaxies compared to S+S galaxies,
   with a difference of difference is $0.21\pm 0.11$ dex (1.9$\sigma$).
   There is no  trend with the stellar mass.
   
\begin{figure}
   \centering
\includegraphics[width=8cm, trim=0.cm 0.cm 0cm 0cm,clip]{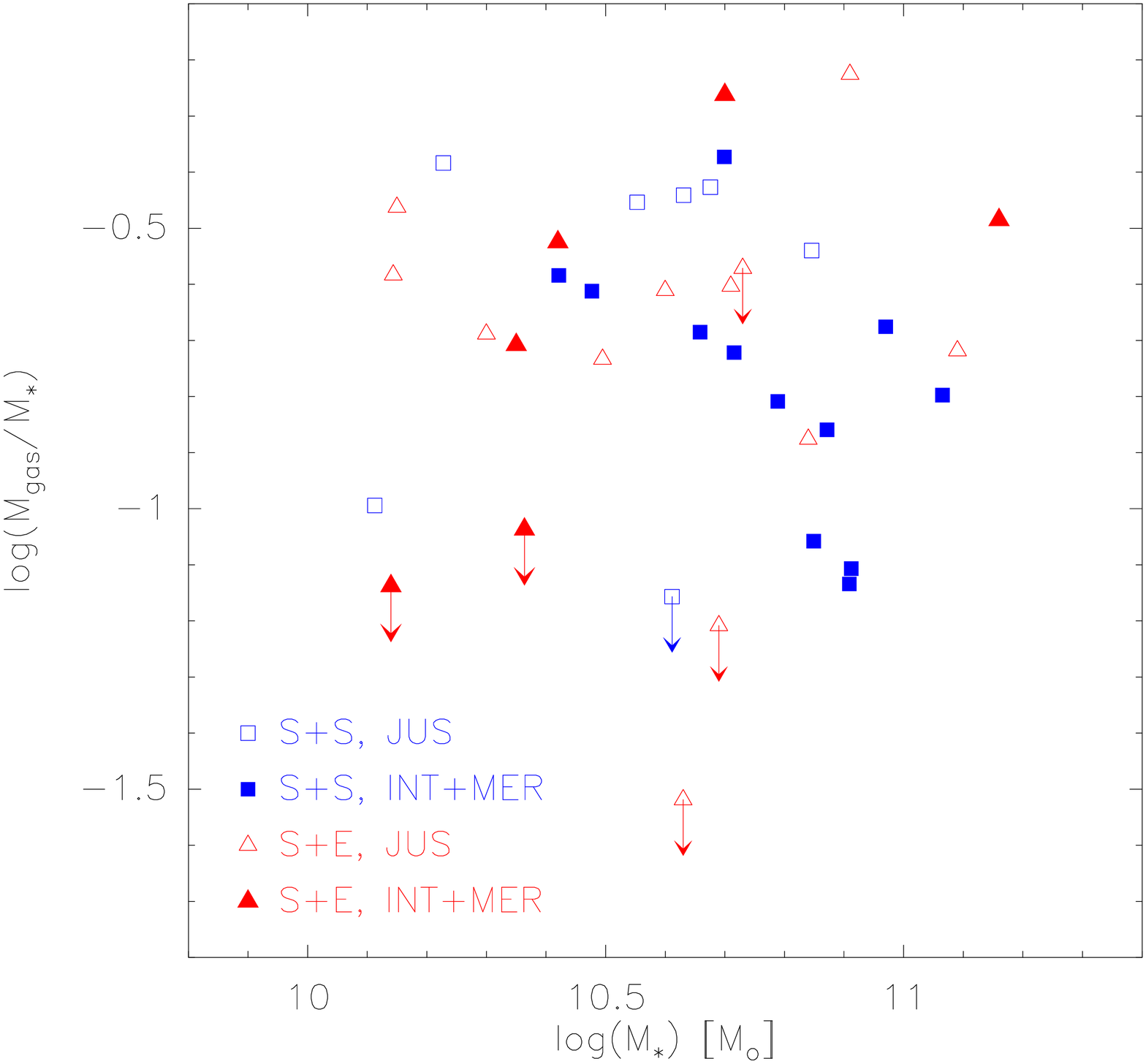}
      \caption{Total gas mass fraction ((\mmol+\mhi)/\mstar) vs. stellar mass of galaxy pairs.  
                  The values for \mdust\ and (\mhi+\mmol) are devided by 2 for the S+S pair.
}
         \label{fig:fract_mgas_mstar}
   \end{figure}

   The total gas mass fraction (\mgas/\mstar = (\mmol+\mhi)/\mstar), shown in 
   Figure~\ref{fig:fract_mgas_mstar}, does not show any trends with stellar mass, pair type (S+S or S+E)
   nor morphological feature of interaction.

\subsection{Molecular, atomic and  dust mass}

\begin{figure}
   \centering
\includegraphics[width=8.cm,trim=0.cm 0.cm 0cm 0cm,clip]{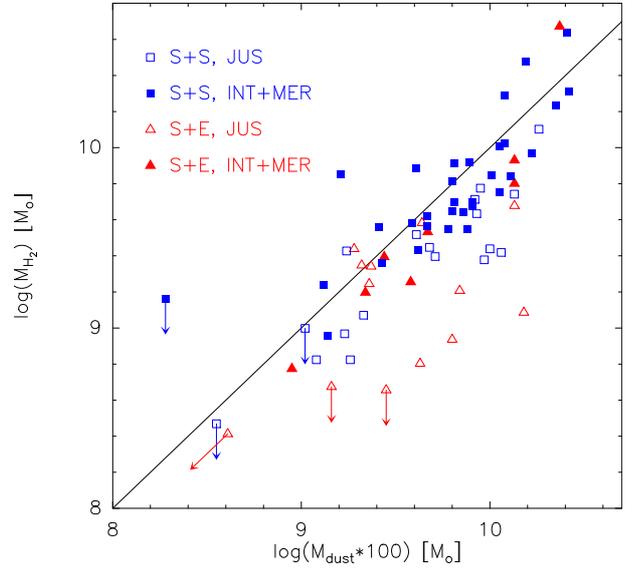}
      \caption{Molecular gas mass vs. $100\times$ dust mass. 
      The line shows unity.
}
         \label{fig:mmol_mdust}
   \end{figure}

\begin{figure}
   \centering
\includegraphics[width=8.cm,trim=0.cm 0.cm 0cm 0cm,clip]{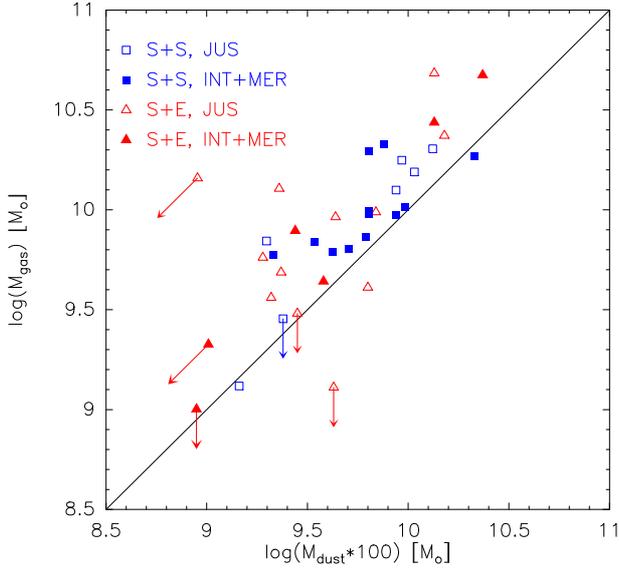}
      \caption{Total gas  (molecular + atomic) mass vs. total $100\times$  dust mass of galaxy 
      pairs.  
      The values for \mdust\ and (\mhi+\mmol) are devided by 2 for the S+S pair.
      The line shows unity.
}
         \label{fig:mgas_mdust}
   \end{figure}

The molecular gas mass shows a good correlation with the
dust mass from \citet{cao16} (Figure~\ref{fig:mmol_mdust}).
The Spearman's rank correlation coeficient is 0.73 and the
significance is  $8.4 \times 10^{-12}$.
The correlation is even tighter  
with the total gas mass (Figure~\ref{fig:mgas_mdust})
(Spearman's rank correlation coeficient of 0.84 and 
significance $3.7 \times 10^{-9}$).

\begin{figure}
   \centering
\includegraphics[width=8.cm,trim=0.cm 0.cm 0cm 0cm,clip]{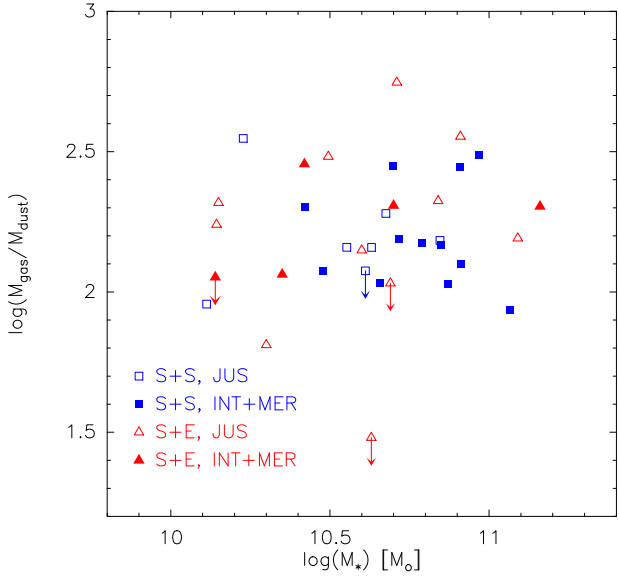}
      \caption{The gas-to-dust mass ratio vs stellar mass   in galaxy pairs.  
      The values for \mdust, \mstar and (\mhi+\mmol) are devided by 2 for the S+S pair.
}
         \label{fig:gd}
   \end{figure}

Figure~\ref{fig:gd} shows the ratio between the total gas mass  and the dust mass
as a function of stellar mass. The mean value is $138$ with an error of 12~\%. No trend with the stellar mass is visible,
nor a  significant difference between S+S or S+E pairs or as a function of interaction morphology.
The mean value is very close to the mean value for nearby galaxies \citep[\mgas/\mdust =  137, ][their Tab. 2]{draine07}, 
with values ranging  between $\sim 100$ and 400.
The relatively constant gas-to-dust mass ratio confirms the correctness of the analysis of \citet{cao16} who used the dust
mass as a gas mass tracer.

\subsection{Gas mass, stellar mass and star formation rate}

\begin{figure}
   \centering
\includegraphics[width=8.cm,trim=0.cm 0.cm 0cm 0cm,clip]{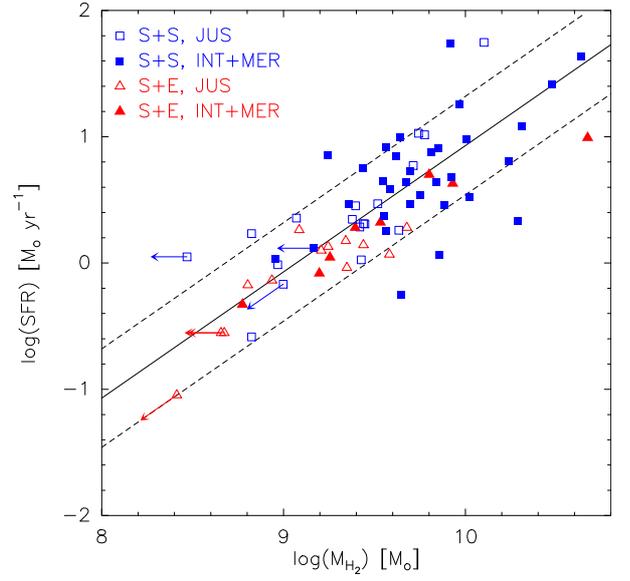}
      \caption{Star formation rate as a function of molecular gas mass.   The lines  are not a fit to the
      data, they show the
      mean value (full line) and dispersion (dashed lines) of the AMIGA comparison sample.
}
         \label{fig:mmol_sfr}
   \end{figure}

Figure~\ref{fig:mmol_sfr} shows the SFR as a function of \mmol.
We also show the relation between SFR and \mmol\ found for the AMIGA sample which
is consistent with linearity \citep{lisenfeld11}, together with the standard deviation.
The galaxies in S+S pairs follow this relation reasonably well, whereas spirals in S+E pairs 
lie below.

\begin{figure}
   \centering
\includegraphics[width=8.cm,trim=0.cm 0.cm 0cm 0cm,clip]{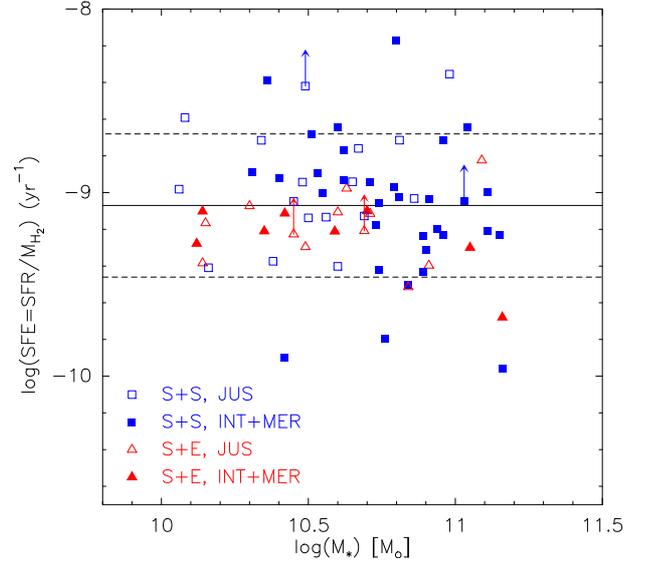}
      \caption{Star formation efficiency (= (SFR)/\mmol) vs stellar mass.        The lines are not a fit to the
      data, they  show the
      mean value (full line) and dispersion (dashed lines)  of the AMIGA comparison sample.
      }
         \label{fig:sfe_mstar}
   \end{figure}

Figure~\ref{fig:sfe_mstar} shows the relation between the SFE, defined as the 
ratio between SFR and molecular gas mass (SFE = SFR/\mmol) and the stellar mass.
The value of log(SFE) for spirals in S+E pairs is
 below that of spirals in S+S pairs by $0.18\pm0.06$ dex (see Table~\ref{tab:means}).
 
\begin{figure}
   \centering
\includegraphics[width=8.cm,trim=0.cm 0.cm 0cm 0cm,clip]{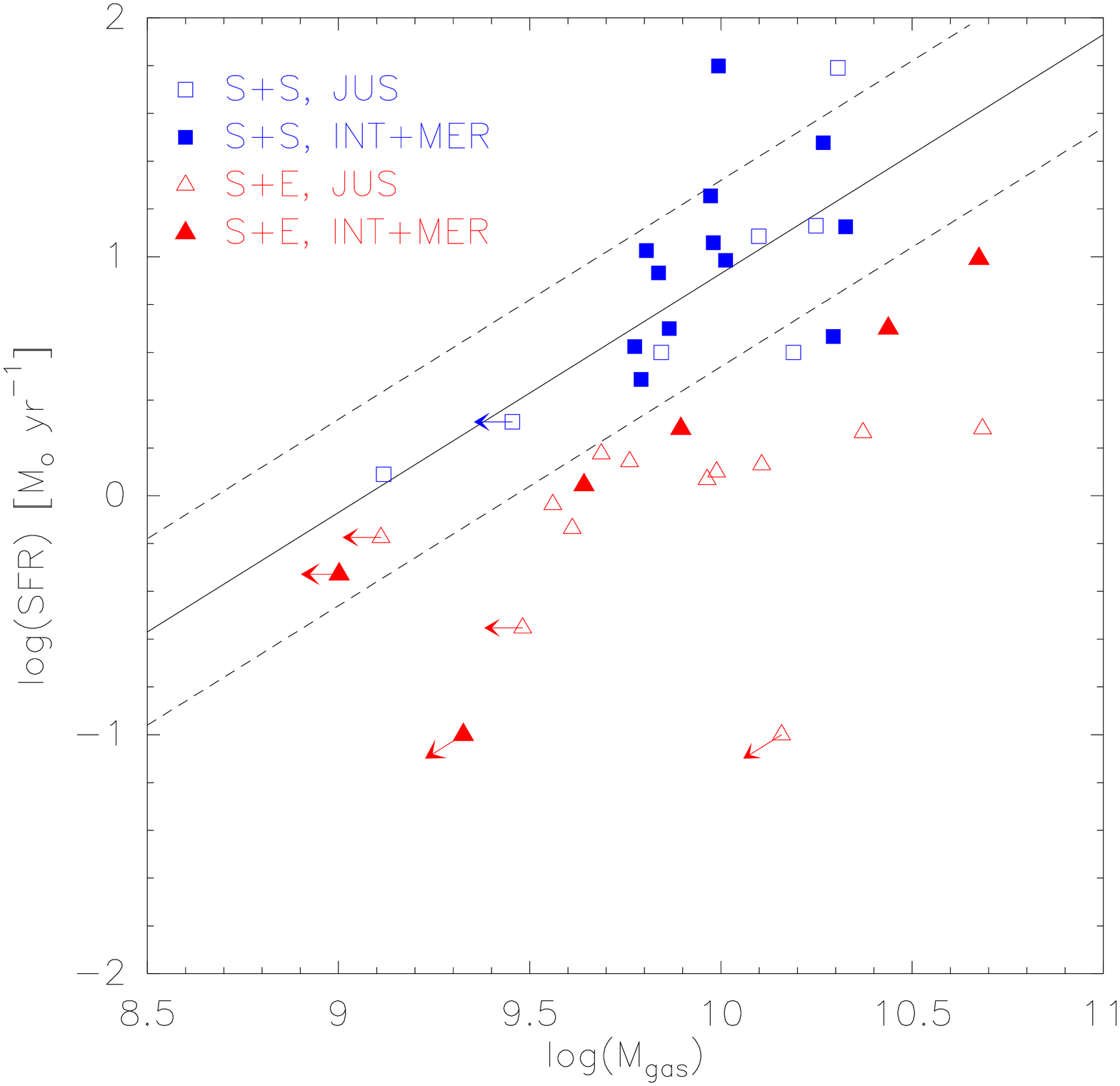}
      \caption{Star formation rate as a function of total (molecular+atomic) gas mass.    To guide the eye,
      the lines are the same as in Fig.~\ref{fig:mmol_sfr} and show the
      mean value (full line) and dispersion (dashed lines) of the AMIGA comparison sample.
}
         \label{fig:mgas_sfr}
   \end{figure}

\begin{figure}
   \centering
\includegraphics[width=8.cm,trim=0.cm 0.cm 0cm 0cm,clip]{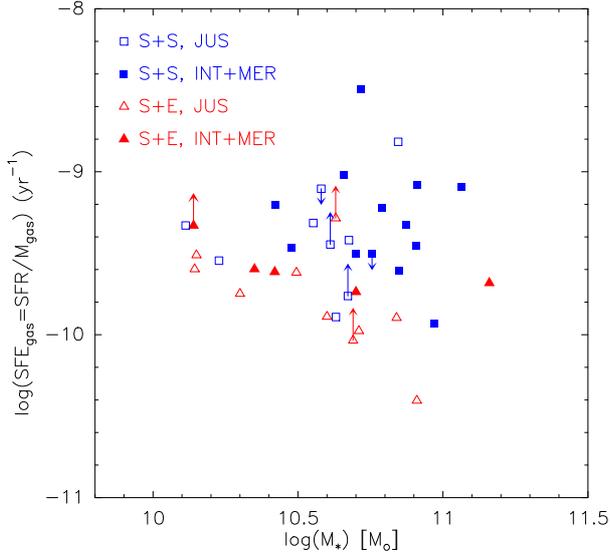}
      \caption{Star formation efficiency per total gas mass (\sfegas = (SFR)/\mgas) vs. stellar mass.  }
         \label{fig:sfe_tot_mstar}
   \end{figure}

Figure~\ref{fig:mgas_sfr} shows the SFR as a function of the total (molecular + atomic) gas mass,
and Figure~\ref{fig:sfe_tot_mstar} the ratio between the SFR and the gas mass (\sfegas = SFR/\mgas).
The difference between S+S and S+E spirals is even more pronounced  than for the SFE;
the value of log(\sfegas) is higher by $0.42\pm0.11$  dex for spirals in S+S compared to S+E pairs. 
Again, no trend with \mstar\ or the interaction phase is found.

This difference in \sfegas\ between S+S and S+E pairs is  in agreement with the results of Cao et al. (2016) who studied the
\sfegas\  using  the dust mass as a tracer of the total gas mass for the H-KPAIR sample. They found that  log(\sfegas) is
 $0.3\pm 0.1$   higher for spirals in S+S pairs than for S+E pairs.

\begin{table*}
\caption{Mean  and median   values of parameters for the total sample and subsamples. The X-factor was chosen to vary with
the distance from the galaxy MS as described in Sect. 5.1. {\bf Not to be shown in the paper, just for our comparison}.}
\begin{tabular}{l|cc|cc|cc|cc|cc|cc|cc|c}
\hline
 & S+S  &  S+E & log(\mstar)$< 10.7$  & log(\mstar)$> 10.7$ & JUS & INT+MER  & Total \\
 & mean & mean  & mean & mean & mean  & mean & mean\\
 & median & median & median & median & median & median & median \\
 & n/n$_{\rm up}$\tablefootmark{a}  &n/n$_{\rm up}$\tablefootmark{a} & n/n$_{\rm up}$\tablefootmark{a}  & n/n$_{\rm up}$\tablefootmark{a}  & n/n$_{\rm up}$\tablefootmark{a}  & n/n$_{\rm up}$\tablefootmark{a}  & n/n$_{\rm up}$\tablefootmark{a} \\
\hline
\hline
  &   {\bf -8.93$\pm$0.07} &   {\bf -9.18$\pm$0.04}    &   -8.96$\pm$0.06  &  -9.04$\pm$0.10  & -8.95$\pm$0.08&  -9.02$\pm$0.08  &  -8.99$\pm$0.05  \\
SFE   &  -8.91 &  -9.18 &  -8.98 &  -9.11 &  -9.07 &  -9.05 &  -9.05 \\
 & 49/2 &  20/2  & 38/3 & 31/1  &  29/3 &40/1  &69/4 \\
\hline
  &   &     &   -8.88$\pm$0.08  &  -8.98$\pm$0.11  & { -8.84$\pm$0.10} &  {-8.97$\pm$0.09 } &  \\
SFE   & -- & --   & -8.87 &  -9.01 &  -8.97 &  -8.91 &   -- \\
S+S & &  & 24/1  & 25/1 & 17/1  &  32/1 & \\
\hline
  &    &      &   -9.14$\pm$0.03  &  -9.26$\pm$0.10  & { -9.16$\pm$0.06} &  { -9.20$\pm$0.05}  &   \\
SFE   &  -- & --  &  -9.11 &  -9.22 &  -9.11 &  -9.18 &  -- \\
S+E &  &   & 14/2 & 6/0  &  12/2 & 8/0  &  \\
\hline
\hline
 &  {\bf -9.27$\pm$0.09} &   {\bf -9.74$\pm$0.07}    &   -9.46$\pm$0.06  &  -9.49$\pm$0.14  & -9.60$\pm$0.11 &  -9.33$\pm$0.10  &  -9.46$\pm$0.08  \\
\sfegas  &  -9.29 &  -9.65 &  -9.47 &  -9.60 &  -9.64 &  -9.44 &  -9.47 \\
 & 19/1 &  16/3  & 20/4 & 15/0  &  18/3 &17/1  &35/4 \\
\hline
 &  -- &   --  &   -9.37$\pm$0.08  &  -9.18$\pm$0.16  & -9.34$\pm$0.13 &  -9.25$\pm$0.12  &  --  \\
\sfegas  &  -- & -- &  -9.40 &  -9.19 &  -9.34 &  -9.29 &  -- \\
 S+S&  -- &  --  & 10/1 & 9/0  &  7/1 &12/0  & -- \\
\hline
 &  -- &   --  &  {\bf  -9.60$\pm$0.06}  &  {\bf -9.96$\pm$0.11}  & {\bf  -9.81$\pm$0.09} &  {\bf  -9.59$\pm$0.05 } &  --  \\
\sfegas  &  -- & -- &  -9.63 &  -9.93 &  -9.76 &  -9.63 &  -- \\
 S+E&  -- &  --  & 10/3 & 6/0  &  11/2 &5/1  & -- \\
\hline 
\hline 
  &  {\bf -0.07$\pm$0.10} &   {\bf -0.51$\pm$0.15}  &   -0.34$\pm$0.08  &  -0.18$\pm$0.17  & {\bf  -0.56$\pm$0.12} &  {\bf 0.03$\pm$0.10}  &  -0.27$\pm$0.10  \\
\mhtwo/\mhi  & 0.06  & -0.55  & -0.36 & -0.13 & -0.64  & 0.06 & -0.16 \\
 & 18/0 &  16/3 & 18/2 & 16/1  &  17/2 &17/1  &34/4 \\
\hline 
  &  --  &   --  &   {\bf -0.31$\pm$0.12}  & {\bf  0.17$\pm$0.12}  & {\bf  -0.43$\pm$0.14} &  {\bf 0.11$\pm$0.10}  & -- \\
\mhtwo/\mhi  & --  &--  & -0.32 & 0.08 & -0.38  & 0.06 & --\\
S+S &   --& --& 9/0  &  9/0 &6/0 & 12/0 &-- \\
\hline 
  &  --  &   --  &   -0.35$\pm$0.10  &  -0.62$\pm$0.28  &  -0.64$\pm$0.16 &   -0.17$\pm$0.24  & -- \\
\mhtwo/\mhi  & --  &--  & -0.36& -0.77 & -0.68  & -0.36 & --\\
S+E  &  --& --& 9/2  &  7/1 &11/2 & 5/1 &-- \\

\hline
\hline
 &  -1.16$\pm$0.04 &   -1.31$\pm$0.09  &  -1.23$\pm$0.05 &  -1.18$\pm$0.07  & {\bf  -1.36$\pm$0.07} &  {\bf -1.09$\pm$0.04}  &  -1.21$\pm$0.04 \\
\mhtwo/\mstar &  -1.15& -1.24 & -1.15 & -1.2 & -1.28 & -1.06 & -1.17 \\
 & 50/3 &  21/3 & 39/4 & 32/2  &  31/5 &40/1  &71/6 \\
\hline
 &   &    & -1.17$\pm$0.05  &    -1.15$\pm$0.07 & {\bf  -1.28$\pm$0.07} &  {\bf -1.09$\pm$0.05}  &    \\
\mhtwo/\mstar &  & & -1.15 & -1.18 & -1.25 & -1.06 &  \\
S+S & &  & 24/1 &  26/2 &  18/2 & 32/1 & \\
\hline
 &   &    & -1.32$\pm$0.11  &   -1.30$\pm$0.17  & {\bf  -1.48$\pm$0.12} &  {\bf -1.05$\pm$0.07}  &    \\
\mhtwo/\mstar &  & &-1.1  & -1.24 & -1.39 & -1.06 &  \\
S+E & &  & 15/3 &  6/0 &  13/3 & 8/0 & \\
\hline
\hline
 &  -0.76$\pm$0.06&   -0.79$\pm$0.10  &  -0.79$\pm$0.09 &  -0.76$\pm$0.07  & -0.74$\pm$0.09 &  -0.79$\pm$0.06  &  -0.77$\pm$0.06 \\
\mgas/\mstar  & -0.76& -0.67 & -0.62 & -0.84 & -0.64 & -0.84 & -0.68 \\
 & 19/1 &  18/5 & 21/5 & 16/1  &  19/4 &18/2  &37/6 \\
\hline
\hline
 &  2.16$\pm$0.05 &   2.18$\pm$0.09  &  2.13$\pm$0.06 &  2.25$\pm$0.06 & 2.18$\pm$0.08 &  2.18$\pm$0.05  &  2.18$\pm$0.05  \\
\mgas/\mdust & 2.13 & 2.28 & 2.13 & 2.20 & 2.16 & 2.14 & 2.16 \\
 & 19/1 &  16/3 & 20/4 & 15/0  &  18/3 &17/1  &35/4 \\
\hline
\end{tabular}
\tablefoot{
Mean values for which the difference between the complementary subsamples are $\gtrsim 2\sigma$ are printed in boldface.
\tablefoottext{a} {total number of galaxies (n) and number of upper limits (n$_{\rm up}$)}.
}
\label{tab:means_varx}
\end{table*}

\subsection{``Holmberg effect"}

\begin{figure}
   \centering
\includegraphics[width=8.cm,trim=0.cm 0.cm 0cm 0cm,clip]{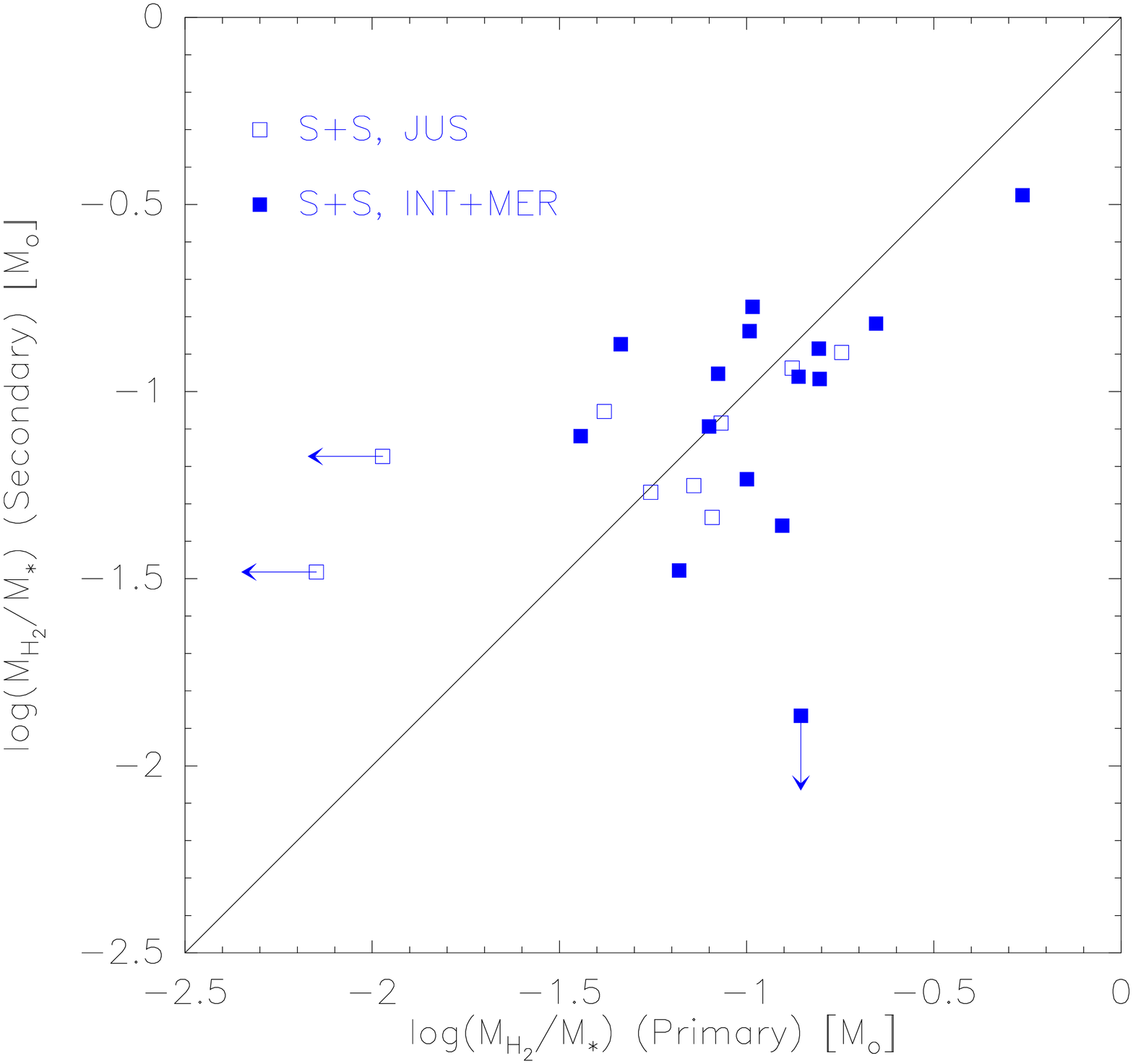}
      \caption{Molecular gas fraction of the secondary vs. primary galaxy in S+S pairs.  }
         \label{fig:holmberg_mh2_over_mstar}
   \end{figure}

"Holmberg effect" denotes usually any concordant behavior between the two components in galaxy pairs, 
after the discovery of \citet{holmberg37} that binary galaxies tend to have similar morphologies and optical colors. 
Figure~\ref{fig:holmberg_mh2_over_mstar} shows the correlation between the molecular gas fraction
(\mmol/\mstar) of the primary and secondary galaxies in S+S pairs. 
The Spearman's rank correlation coeficient is 0.55 and the
significance is  $9.4 \times 10^{-3}$.
A similar   "Holmberg effect" has been found for the sSFR both by
\citet{xu10}, \citet{cao16} and, for a larger sample of 1899 galaxies, by \citet{scudder12}.

\section{Discussion}

\subsection{Effect of a variable X-factor}

We base our analysis on a constant \lco-to-\mmol factor \alphaco. It is well accepted that  \alphaco\
can vary in different types of galaxies, mainly due to its dependence on  metallicity
and star formation activity  \citep[see][and references therein]{bolatto13}.
It can be considerably higher for low-metallicity galaxies { \citep[below 12+log(O/H) $\sim$ 8.4, e.g.,][]{leroy11,bolatto13, hunt15} } and
lower by a factor of 3-10 in extreme starbursts as ULIRGS \citep{downes98, downes03}.

\begin{figure}
   \centering
\includegraphics[width=8.cm,trim=0.cm 0.cm 0cm 0cm,clip]{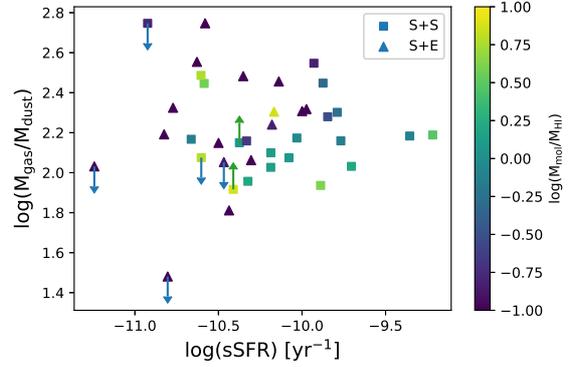}
      \caption{The gas-to-dust mass ratio as a function of the sSFR.
      }
               \label{fig:gd_ratio_ssfr}
   \end{figure}

Our sample does not contain low-metallicity galaxies because all galaxies are relatively
high mass ($>10^{10}$ \msun), and neither does it contain extreme starburst galaxies, so
that we do not expect \alphaco\ to vary considerably among them. 
We can test whether we find any indications of a relation with the SF activity by 
comparing the gas-to-dust mass ratio to the specific SFR (Fig.~\ref{fig:gd_ratio_ssfr}). 
If the molecular gas mass is correctly calculated  we expect no 
trend  with the sSFR. 
If, on the other hand,   \alphaco\  were decreasing with the SF activity we would expect to see
an overestimate of \mhtwo\  and thus an increase of the gas-to-dust mass ratio for increasing sSFR.
This is not the case, even for those objects where the molecular gas is dominating
the gas mass. At most, there might be a weak trend in the opposite direction  (decreasing gas mass
with sSFR).

In spite of this lack of evidence for a variation of \alphaco\ in our sample, we test
 the effect of  a possible variation  following  loosely the prescription proposed in \citet{violino18}
 and \citet{sargent14} who parametrized the variation of the conversion factor as:
 
 \begin{equation}
 \alpha_{\rm CO} = (1-f_{\rm SB}) \alpha_{\rm CO, MS} + f_{\rm SB}  \alpha_{\rm SB}
 \label{eq:alphaco}
 \end{equation}
 where \alphacoms\ is the conversion factor  for galaxies lying on the galaxy main-sequence, \alphacosb\ is the 
value for extreme starburst galaxies and \fsb\ is the probability of a galaxy being in a starburst
phase given its offset from the mean locus of the star-forming main
sequence in the SFR-\mstar\ plane.  
\citet{violino18}  take into account a metallicity dependence of \alphacoms,
which we do not consider necessary for our high-mass sample, so that we take \alphacoms\ as 
the Galactic X-factor,  \alphacoms\ = 3.2 \msun/(K \kms pc$^{-2}$) \citep{bolatto13}. We
assume that \fsb\ varies  linearly with the logarithmic distance  of a galaxy from the galaxy main sequence, 
$ \Delta_{\rm MS}$ = log(SFR) - log(SFR$_{\rm MS}$).
 We use the MS location following the prescription
 by \citet[][their eq. 5]{saintonge16} because their sample encompasses roughly the same
 mass range as ours. We normalize \fsb\  by taking Arp 220 as a prototypical extreme starburst galaxy.
Deriving the stellar mass from the K-band luminosity and the SFR from the far-infrared luminosity, we obtain
\mstar = $1.4 \times10^{10}$ \msun\ and SFR = 220 \msun\ yr$^{-1}$ and find that 
 Arp 220 lies  a factor  of 100 above the MS. We adopt  \alphacosb = 0.8 \msun/(K \kms pc$^{-2}$) (Downes \& Solomon 1998). 
 Together, this yields, for a linear relation between \fsb\ and 
 $\Delta_{\rm MS}$,  \fsb = $\Delta_{\rm MS}/\log(100) = \Delta_{\rm MS}/2$.
 Finally, we assume that galaxies that fall below the MS  have the Galactic \alphaco\  (i.e. \fsb = 0).
 
 We test all our results with this different prescription.
The results obtained with this variable \alphaco\ are very similar to the ones with a constant,
Galactic value. The differences are within the errors, and in particular all the differences and correlations
that we  found are also valid for a variable \alphaco. 
Therefore we conclude that our results are robust with respect to reasonable uncertainties
in the X-factor.

\subsection{Variations of the SFE}

Compared to AMIGA sample (Table 3), we find no enhancement in the SFE for the whole pair sample, nor for any of the subsamples
 (S+S and S+E. low mass and high mass, with and without interaction signs; see Table 4). 
 It is controversial in the literature on 
 whether paired galaxies have enhanced SFE compared to normal spirals. 
 { While \citet{violino18} and \citet{pan18} found
weakly enhanced SFE ($\lesssim 2\times$), \citet{solomon88}, \citet{combes94}, and \citet{casasola04}  found no
such enhancement.}
 
 We found a $\sim 3\sigma$ difference ($0.18\pm 0.6$) dex between the SFE of galaxies in S+S and in S+E pairs.
 Compared to the AMIGA sample the SFE in S+E is slightly  decreased (by $0.13\pm 0.06$) dex, indicating a lower
 capacity of the molecular gas to form stars in S+E pairs.  
 The difference with respect to the comparision sample is small and might be affected by systematic uncertainties, e.g.
by the differences in the calculation of the SFR. The difference between the S+S and S+E subsample
 is more robust. It holds at a 3$\sigma$ level when using a variable \alphaco\ ($0.25\pm 0.08$) dex and when 
 considering the central values for the molecular gas and SFR ($0.23\pm 0.7$) dex. It is unclear what the physical reason
 for this difference is. A possible reason could be hydrodynamical effects which are expect to be stronger in a
 S+S merger where the gas of both galaxies interacts compared to S+E mergers where the elliptical component has 
 very little cold gas. However, we would then expect to see a noticeable difference mainly in the later stages of the interaction, whereas
 in our study we find no differences in the SFE between galaxies with and without interaction signs.
 \citet{hwang11} suggested that the hot gas halos around elliptical galaxies could be responsible for the
  differences in the SF activity found between S+S and S+E. However, it is unclear how the hot gas can directly  affect the molecular gas
  without altering the total (\mhi + \mhtwo) gas content which we found equal for all subsample.

When considering the total gas, our results show significant enhancement in \sfegas\ for S+S pairs ($0.42\pm 0.10$ dex), pairs with 
 signs of interaction ($0.37\pm 0.10$ dex), and low mass pairs ($0.27\pm 0.08$ dex), but not for subsamples of S+E pairs, pairs without signs of 
 interaction, and high mass pairs. Simulations \citep[e.g.][]{renaud14} have predicted enhanced \sfegas\ 
 in interacting galaxies, which is particularly strong during the final stage of a merger, but also during the early phase just after the first pericenter passage. 
 Our results suggest that this scenario applies to S+S pairs, but not to S+E pairs. The mean \sfegas\
of S+E subsample is consistent with that of control galaxies, and is lower than that of S+S by $0.42\pm 0.11$ dex. This result agrees well with that of
\citep{cao16} who found a difference between the mean \sfegas\ of S+S and S+E pairs of $0.43\pm 0.13$ dex.

\subsection{Variation of the molecular gas fraction}

The molecular gas fraction (\mhtwo/\mstar) and molecular-to-atomic gas ratio (\mhtwo/\mhi) are enhanced for 
INT+MER galaxies whereas for pairs at the beginning of the interaction (JUS) no
enhancement is found. 
Significant enhancement is also present in S+S pairs but not in S+E pairs.
The total gas content (\mgas/\mstar), on the other hand,  does not show any
variation between the subsamples.

These results together give a consistent picture. From the total gas content in the
galaxy, a considerable fraction is converted from atomic to molecular gas during the
interaction. This conversion becomes stronger
in a later stage, when morphological signatures are  visible.
We would then expect a similar trend for the
sSFR, caused by the increase in the molecular gas fraction. Indeed, \citet{cao16} found a higher sSFR for
INT+MER galaxies compared to JUS galaxies.

{ Our results are in general consistent with numerous previous
observations that showed significant increase in the molecular gas
content in interacting galaxies (\citep{braine93, combes94, casasola04, violino18, pan18}. 
Our results, though, indicate that not all galaxies in close
major-merger pairs have an enhanced molecular gas content, but only those
in S+S pairs and pairs with interaction signs do.

With a statistically significant large and homogeneous sample of
paired galaxies, confined to close major-merger pairs that have small
separations and mass ratios, we pinpoint that the major enhancement
occurs in the molecular-to-atomic gas ratio. This conclusion disagrees 
with \citet{casasola04}. They
investigated the molecular-to-atomic gas ratios of interacting
galaxies and found that, for late types, these ratios are quite normal
compared to the control sample. Given that both the interacting galaxy
sample and the control sample of \citet{casasola04} were
constructed heterogeneously 
and the multi-band data collected from the
literature, their result most likely has larger uncertainties than ours, and,
most importantly,
their data does not allow a distinction between the pair type
and interaction phase which we found to be crucial to find a difference in \mmol/\mhi.}

\subsection{What drives the enhancement of the SFR in galaxy pairs?}

Our results show that the increase of the SFR in close major-merger pairs is mainly driven by an enhancement in the molecular-to-HI gas ratio. 
The total gas content, on the other hand, is constant so that inflow or loss of gas can be excluded. The SFE is not significantly enhanced compared to control galaxies, 
while the increase in the \sfegas\ can be explained by the enhancement of the 
molecular-to-HI gas ratio. Given that the enhancement is found only in pairs with signs of interaction and absent in pairs without, it is most likely triggered 
by the strong tidal torque after the first close encounter, which can compress the lower density HI gas into the higher density molecular gas. This scenario is 
different from that for (U)LIRGs which are in the final stage of coalescence. For (U)LIRGs, both the molecular gas content and SFE are strongly 
enhanced \citep{solomon88, mirabel89}. It seems that in paired galaxies after the first close encounter, while 
the tidal torque compresses more gas into star forming giant molecular clouds (GMCs), the SFE of individual GMCs is still comparable to the
standard value found in normal galaxies \citep{leroy08}. On the other hand, in (U)LIRGs such as Arp220, the GMCs in the 
central starburst are further compressed into a gas disk of very high density ($> 10^4$ cm$^{-3}$), resulting in a much higher SFE 
\citep{scoville97, scoville17}. 
{ \citet{gao99} found that, for 
infrared selected mergers, the \lir/\mhtwo\ ratio increases nearly 10$\times$
when projected separation decreases from $\sim 30$ to $\sim 2$~kpc,
consistent with a continuous transition of the SFE between close pairs and 
final stage mergers.}

This change of SFE of molecular gas along the merger sequence may provide a new and important constraint to simulations of galaxy mergers. 
Currently no simulation has separated molecular gas from the atomic gas. An indirect inference may be drawn from the results of \citet{renaud14}, 
who carried out parsec-resolution simulations with comprehensive treatment of turbulence and shocks. In their Fig.~2, they presented  predictions of gas 
density Probability Distribution Functions (PDFs) of different merger epochs, with the $t=35$ Myr epoch corresponding to the starburst after the first 
close encounter and the $t=170$ Myr epoch to the starburst in the final coalescence. For both epochs, strong enhancements are found both in the ratio 
between densities of $\sim 100$ cm$^{-3}$ and $\sim 1$ cm$^{-3}$, which corresponds roughly to the molecular-to-atomic gas ratio,  and in the ratio 
between densities of $\sim 10^4$ cm$^{-3}$ and $\sim 100$ cm$^{-3}$. The mass of dense gas of $\sim 10^4$ cm$^{-3}$  is linearly correlated to SFR 
\citep{Gao04}, therefore the enhancement in the ratio between densities of $\sim 10^4$ cm$^{-3}$ and $\sim 100$ cm$^{-3}$ indicates an 
SFE enhancement both in the  epoch after the first close encounter and in the final coalescence stage. This seems not fully consistent with our 
result which does not show strong SFE enhancement in paired galaxies.

We found differences in the SFE, \sfegas\ and \mmol/\mhi\ between S+S and S+E subsamples. In particular the mean  \mmol/\mhi\
ratio of the S+E is $0.55\pm 0.18$ dex lower than that of S+S (Table 4), and is consistent with those of the control samples (Table 3). The difference is 
caused by two factors: (1) only 31\% (5/16) of star-forming galaxies in S+E show signs of interaction while 67\% (12/18) of those in S+S do;
(2) the mean  \mmol/\mhi\ ratio of star-forming galaxies in S+E pairs without interaction signs is $0.35 \pm 0.22$ dex lower than that of their 
counterparts in S+S.
The first factor might be explained by the stabilizing effect that bulges can have during 
interaction, if star-forming galaxies in S+E are more likely to be earlier Hubble types with larger 
bulges compared to their counterparts in S+S, which is indeed expected according to the Holmberg effect 
\citep{holmberg58, hernandez01}.
Simulations have shown that large bulges can suppress 
the tidal effects during and after close encounters 
\citep{mihos96, dimatteo08, cox08}, making it more difficult for conspicuous 
tidal features to form. 
Interestingly, star-forming galaxies in S+E pairs with interaction signs show an enhancement of $0.55\pm0.30$ dex in $M_{H_2}/M_{HI}$ ratio compared 
to those in S+E pairs without interaction signs, suggesting that HI gas can also be compressed to molecular gas in S+E pairs when the tidal 
effects are sufficiently strong (indicated by interaction signs), similar to what is happening in S+S pairs with interaction signs. 
The factor (2), though with a low significance (1.6$\sigma$), might indicate 
that the progenitors of star-forming galaxies in S+E have lower   \mmol/\mhi\ than those in S+S.
Alternatively, the formation of the molecular gas due to tidal forces might have started early in the 
interaction and preferentially in the bulge-less S+S systems. 
x

\section{Conclusions and summary}

We presented new CO data for a sample of \ks-band selected local ($z<0.06$), 
close (projected separation $\lesssim 20\,h^{-1}$~kpc), major (mass ratio $\le 2.5$) merger pairs
which allows us to calculate the molecular gas mass.
These data, together with a large set of ancillary data, allows us to study the
molecular gas fraction, \mmol/\mstar, the molecular-to-atomic gas mass ratio, \mmol/\mhi,
SFE = SFR/\mmol, and \sfegas = SFR/\mgas\ as a function of galaxy mass, interaction type (S+S or S+E pair)
and interaction phase (undisturbed appearance "JUS", or with clear signs of tidal disturbance
or merging "INT+MER").
We compared the values of the merger sample to two comparison samples,
AMIGA \citep{lisenfeld11} and COLDGASS \citep{saintonge11a,  saintonge11b}. 
The main conclusions are:

\begin{enumerate}

\item We found no significant enhancement in SFE (=SFR/\mhtwo) for the whole pair sample, nor for any of the subsamples 
(S+S and S+E, low mass and high mass, with and without interaction signs).  The SFE in star-forming galaxies
in S+E pairs is $0.18\pm0.6$ dex lower than in S+S pairs.

\item When considering the total gas, \mgas = \mhi+\mhtwo, our
 results show significant enhancement in \sfegas\ for S+S pairs ($0.42\pm 0.09$ dex), pairs with signs of interaction
  ($0.37\pm 0.10$ dex), and low mass pairs ($0.27\pm 0.08$ dex), but not for subsamples of S+E pairs, pairs 
  without signs of interaction, and high mass pairs.

\item We found an enhancement of \mmol/\mhi\  from
JUS to INT+MER galaxies. The values of the JUS subsample are compatible with
those of the control samples. A similar, albeit less significant, trend was found for 
\mmol/\mstar. This indicates that the amount of molecular gas enhances
as the interaction proceeds.

\item We found differences in \mmol/\mhi\  and in \mmol/\mstar\ between S+S and 
S+E subsamples. The mean \mhtwo/\mhi\  ratio and mean \mhtwo/\mstar\ ratio of the S+E are $0.55\pm 0.18$ dex (2.5$\sigma$)
and $0.21 \pm 0.11$ dex (1.9$\sigma$)
lower than those of the S+S, and are consistent with those of control samples.

\end{enumerate}

Our results show that the star formation enhancement in close major-merger pairs is mainly driven by an accelerated conversion 
of atomic gas to molecular gas in pairs with interaction signs, likely triggered by the strong tidal effects after the first close encounter. 
Both the star formation and molecular gas content enhancements are significantly suppressed in star-forming galaxies in 
S+E pairs, probably due to the stabilizing effects of large bulges.

\begin{acknowledgements}
CKX acknowledges support by the National Key R\&D Program of China No. 2017YFA0402704 and National Natural Science Foundation of China No. Y811251N01. 
UL acknowledge support by the research projects
AYA2014-53506-P and AYA2017-84897-P from the Spanish Ministerio de Econom\'\i a y Competitividad,
from the European Regional Development Funds (FEDER)
and the Junta de Andaluc\'ia (Spain) grants FQM108.
YG acknowledges the NSFC grants\#11420101002 and 11861131007, the National Key R\&D 
Program of China grant \#2017YFA0402700, and the CAS Key Frontier Sciences Program.
This work is sponsored in part by the Chinese Academy of Sciences (CAS), 
through a grant to the CAS South America Center for Astronomy (CASSACA) in Santiago, Chile.
This work is based on observations carried out under project numbers  071-12 and 177-15 with the IRAM 30m telescope. 
IRAM is supported by INSU/CNRS (France), MPG (Germany) and IGN (Spain).
\end{acknowledgements}

\bibliographystyle{aa}
\bibliography{biblio_kpair}

\appendix

\section{Figures of spectra}


\begin{figure*}
\centerline{
\includegraphics[width=3.6cm,clip,trim = 0.cm 0.cm 0.cm 0.0cm,angle=-0]{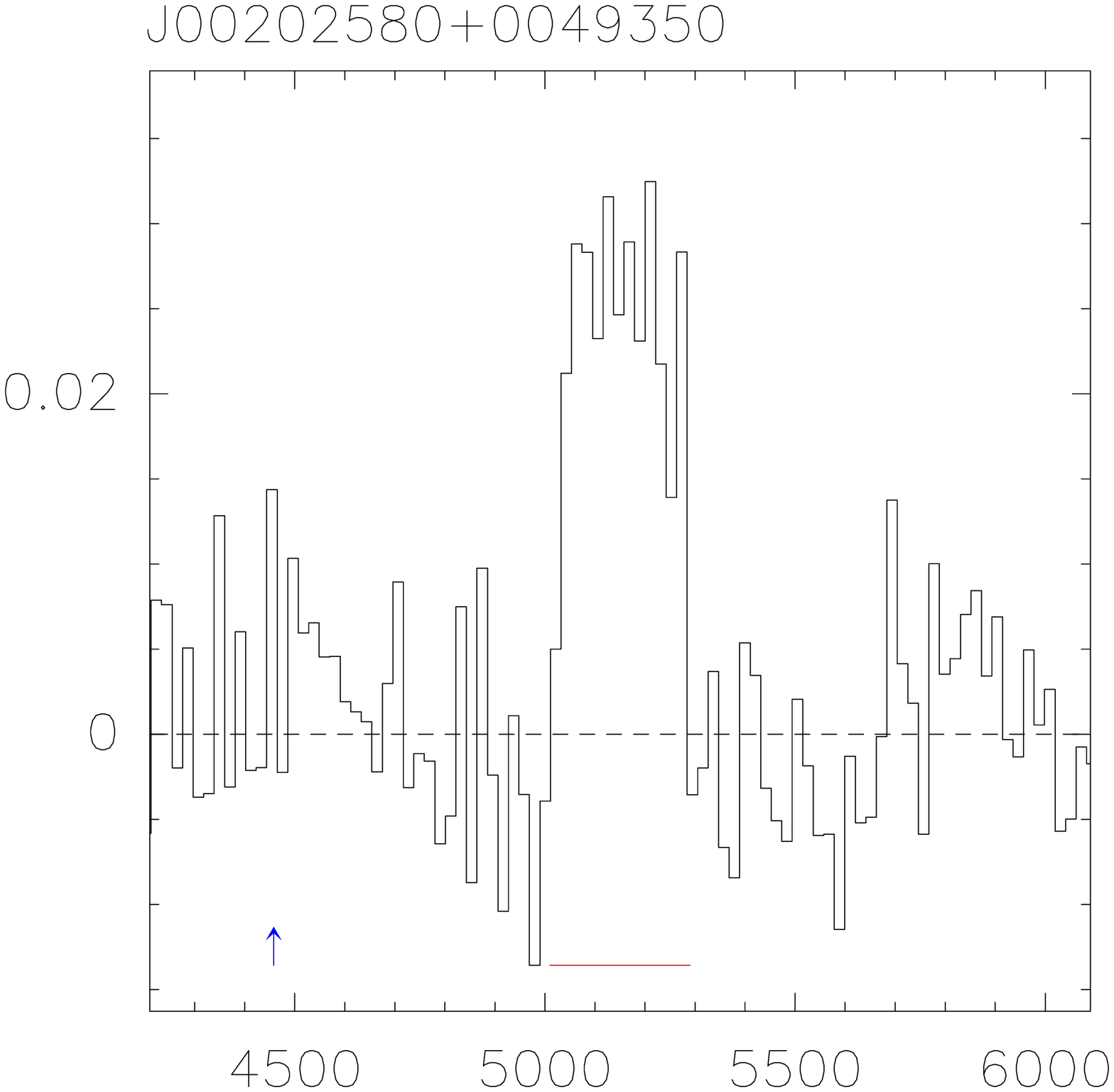}
\hspace{0.1cm}
\includegraphics[width=3.6cm,clip,trim = 0.cm 0.cm 0.cm 0.0cm, angle=-0]{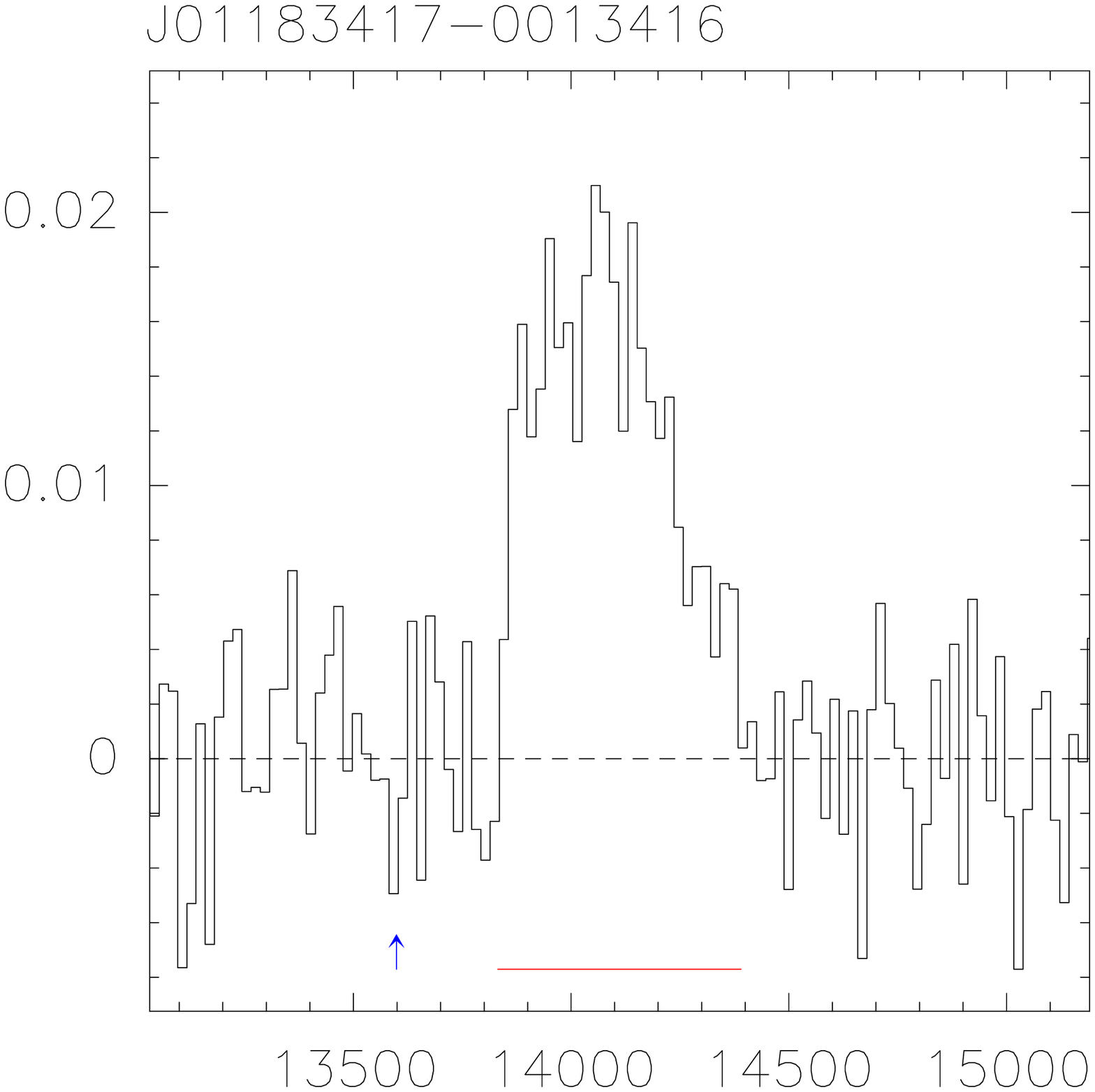}
\hspace{0.1cm}
\includegraphics[width=3.6cm,clip,trim = 0.cm 0.cm 0.cm 0.0cm, angle=-0]{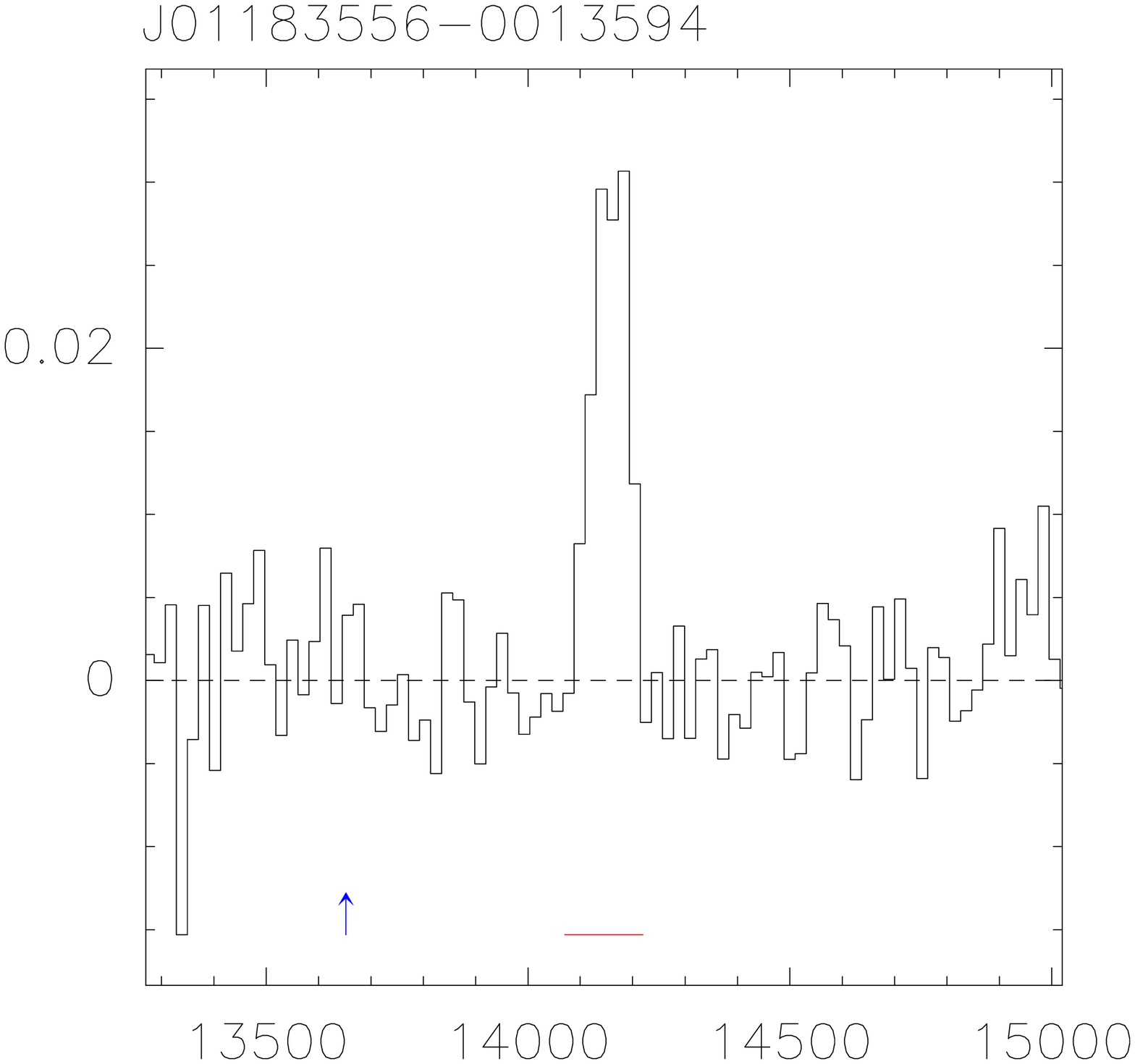}
\hspace{0.1cm}
\includegraphics[width=3.6cm,clip,trim = 0.cm 0.cm 0.cm 0.0cm, angle=-0]{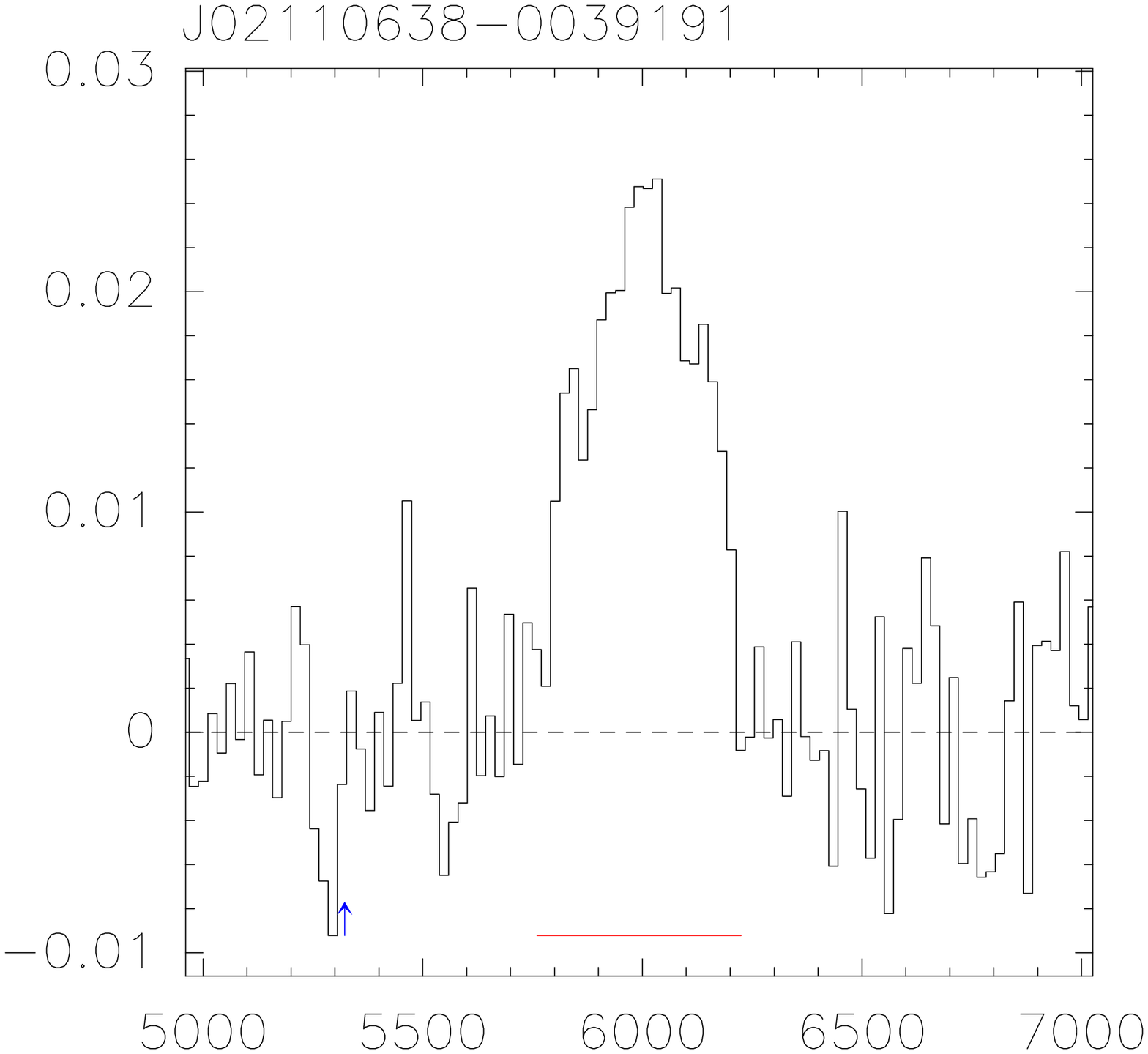}
}
\quad

\centerline{
\includegraphics[width=3.6cm,clip,trim = 0.cm 0.cm 0.cm 0.0cm,angle=-0]{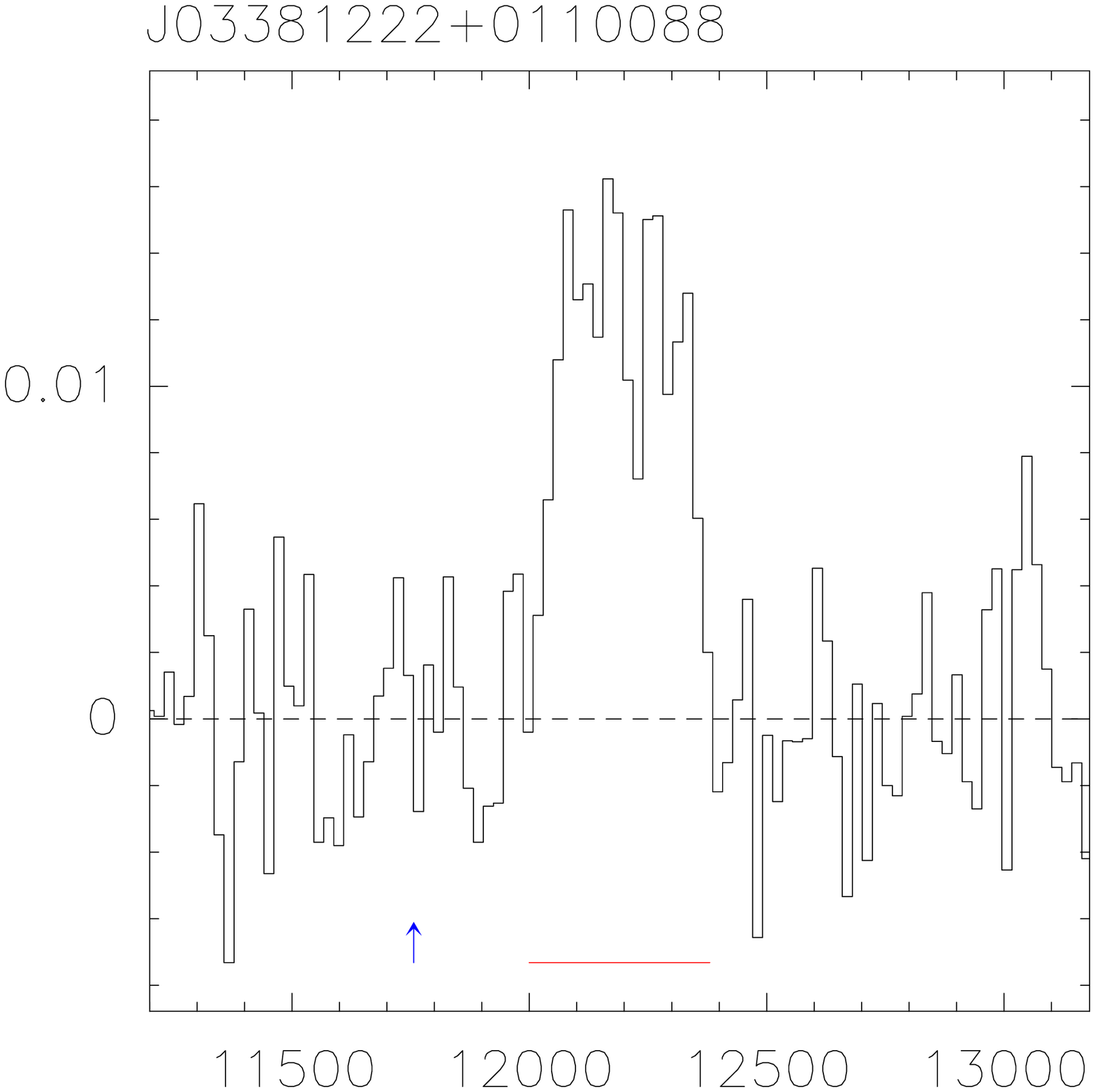}
\hspace{0.1cm}
\includegraphics[width=3.6cm,clip,trim = 0.cm 0.cm 0.cm 0.0cm, angle=-0]{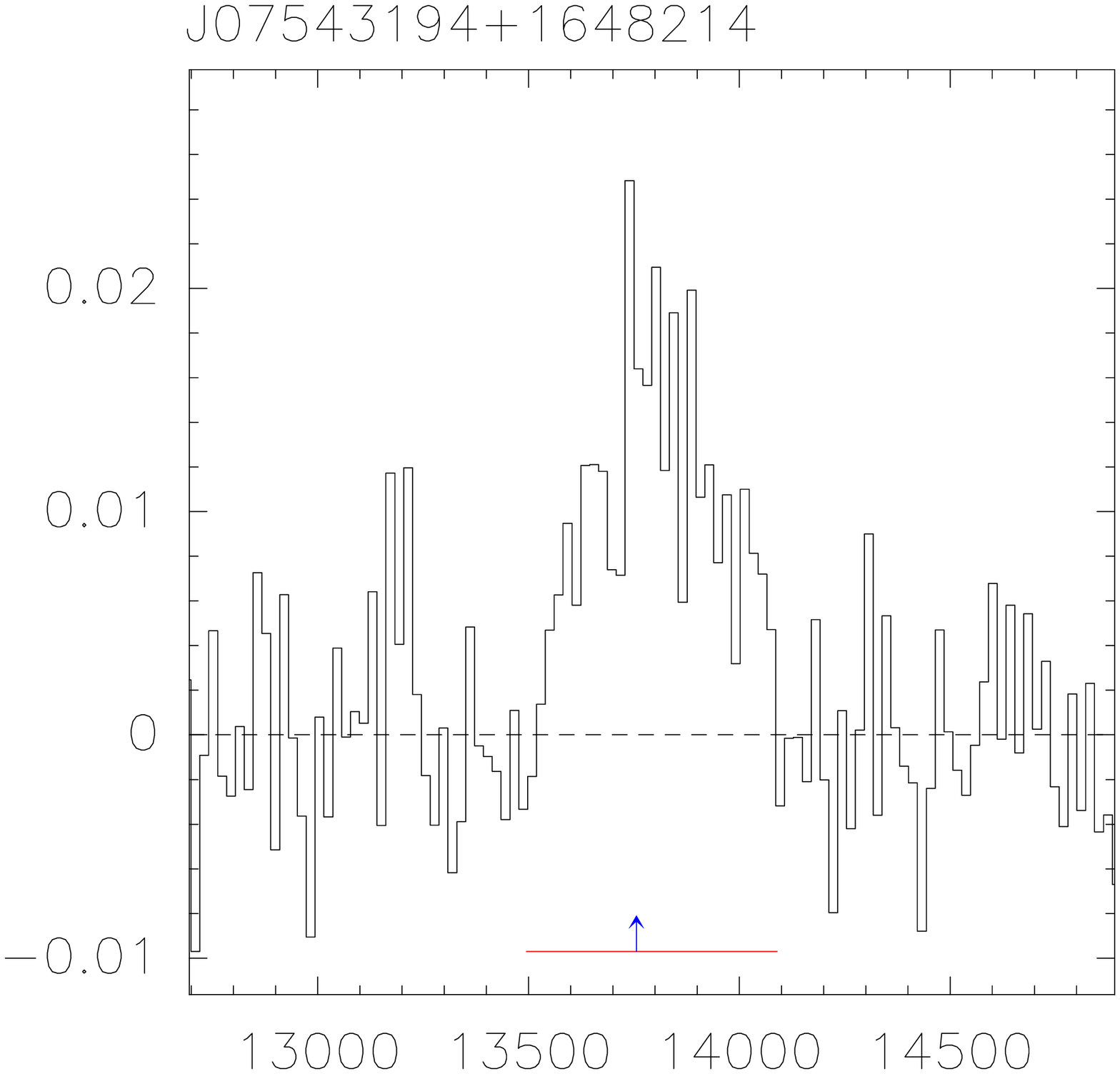}
\hspace{0.1cm}
\includegraphics[width=3.6cm,clip,trim = 0.cm 0.cm 0.cm 0.0cm, angle=-0]{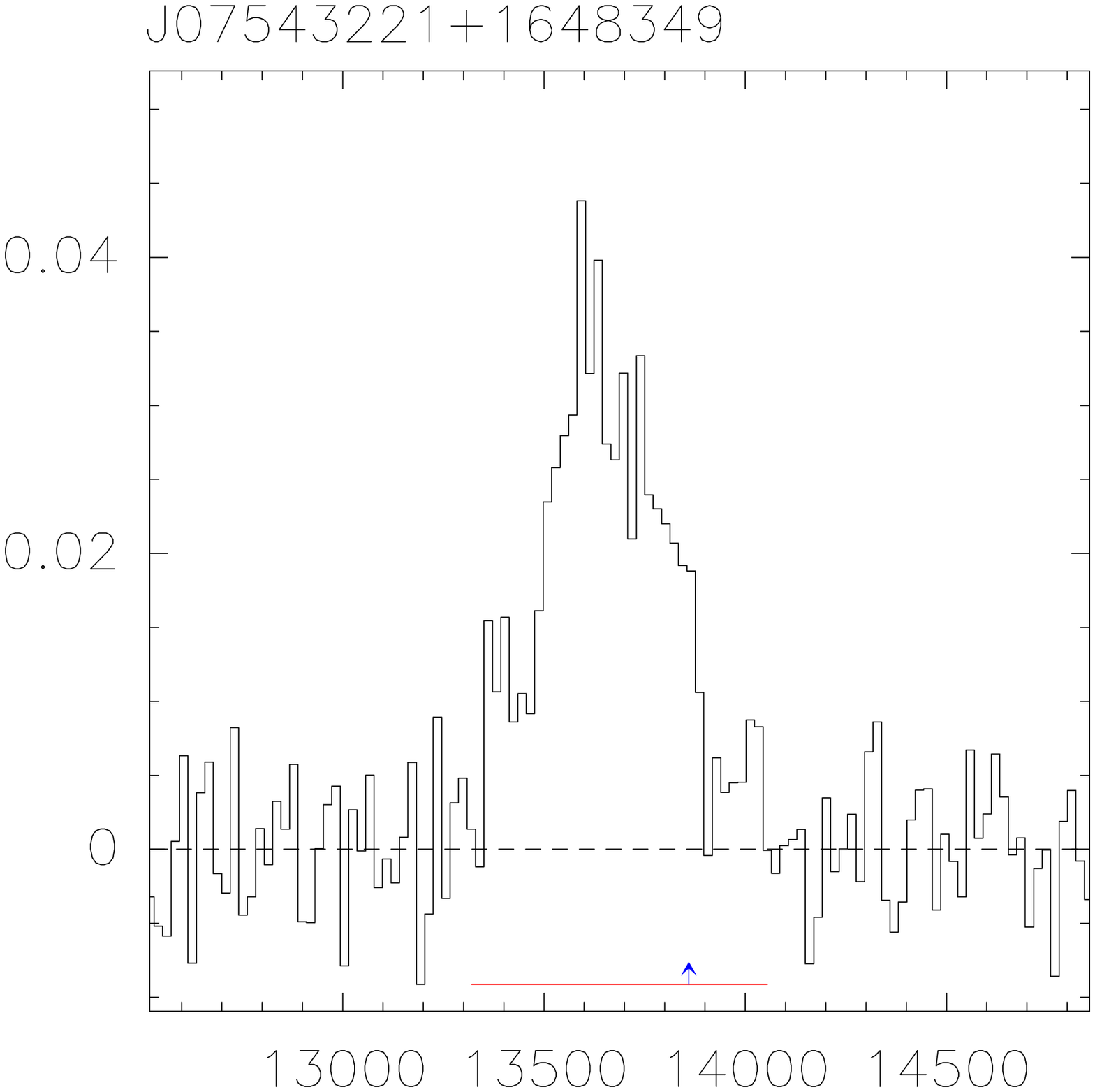}
\hspace{0.1cm}
\includegraphics[width=3.6cm,clip,trim = 0.cm 0.cm 0.cm 0.0cm, angle=-0]{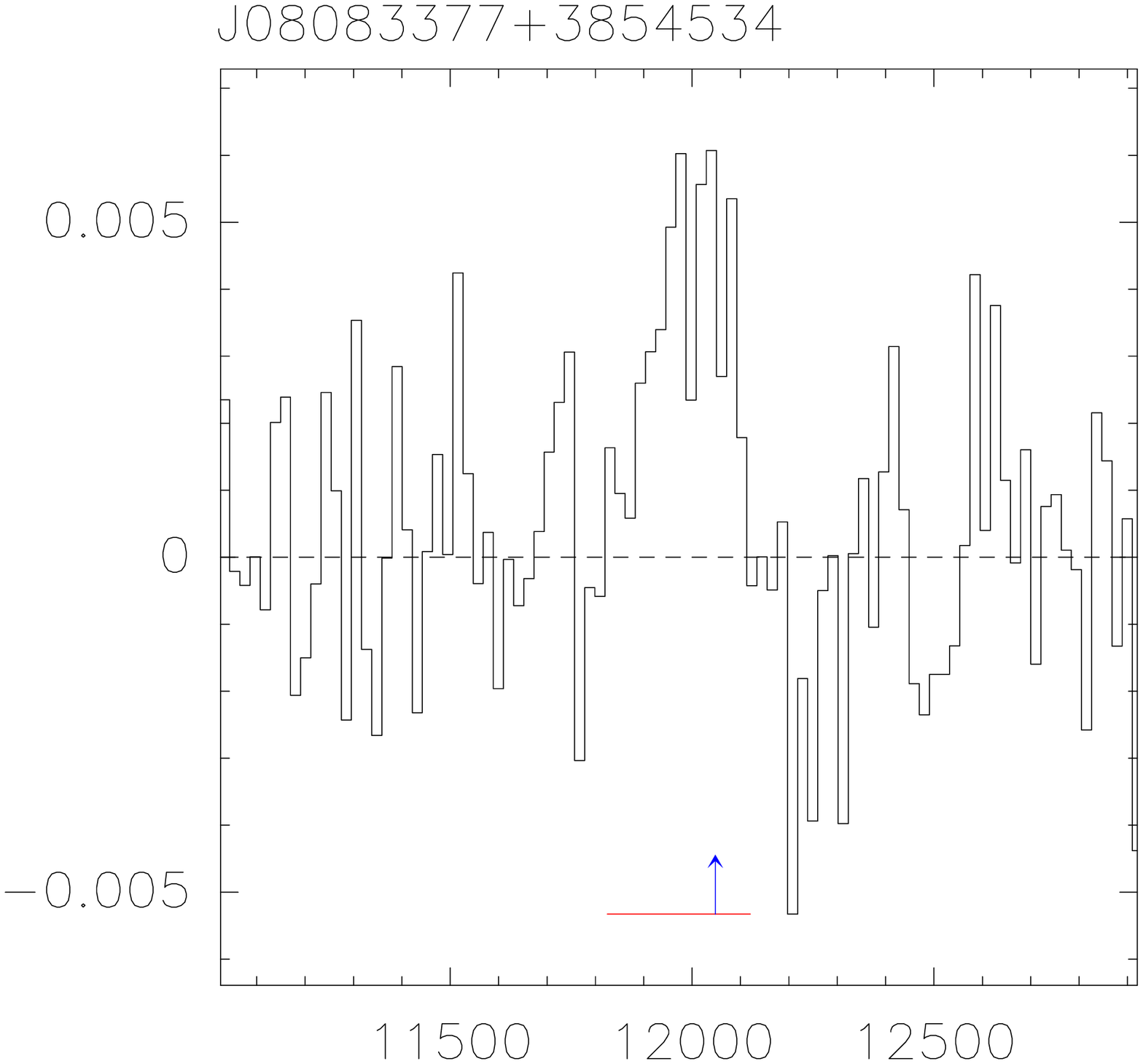}
}
\quad

\centerline{
\includegraphics[width=3.6cm,clip,trim = 0.cm 0.cm 0.cm 0.0cm,angle=-0]{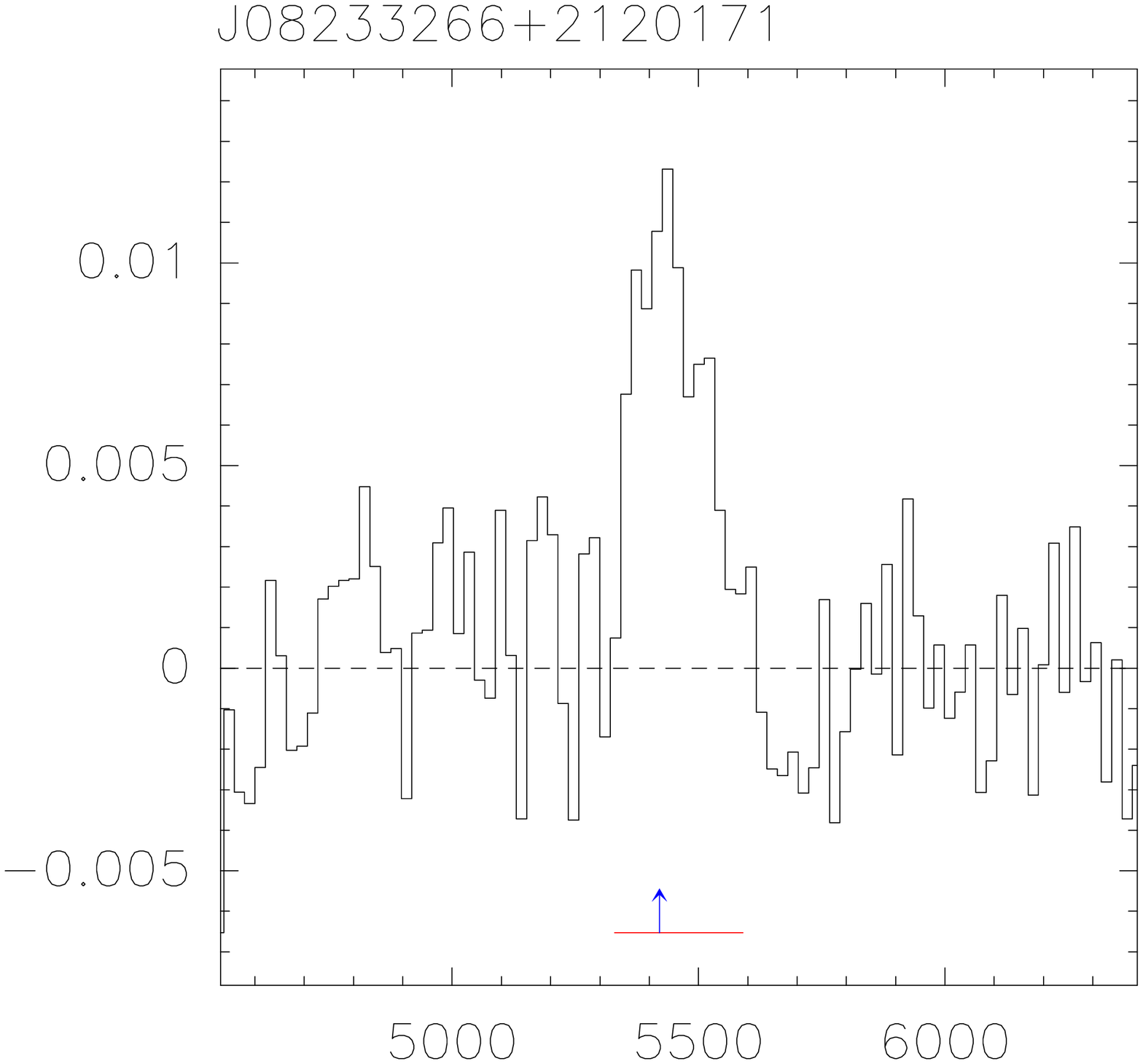}
\hspace{0.1cm}
\includegraphics[width=3.6cm,clip,trim = 0.cm 0.cm 0.cm 0.0cm, angle=-0]{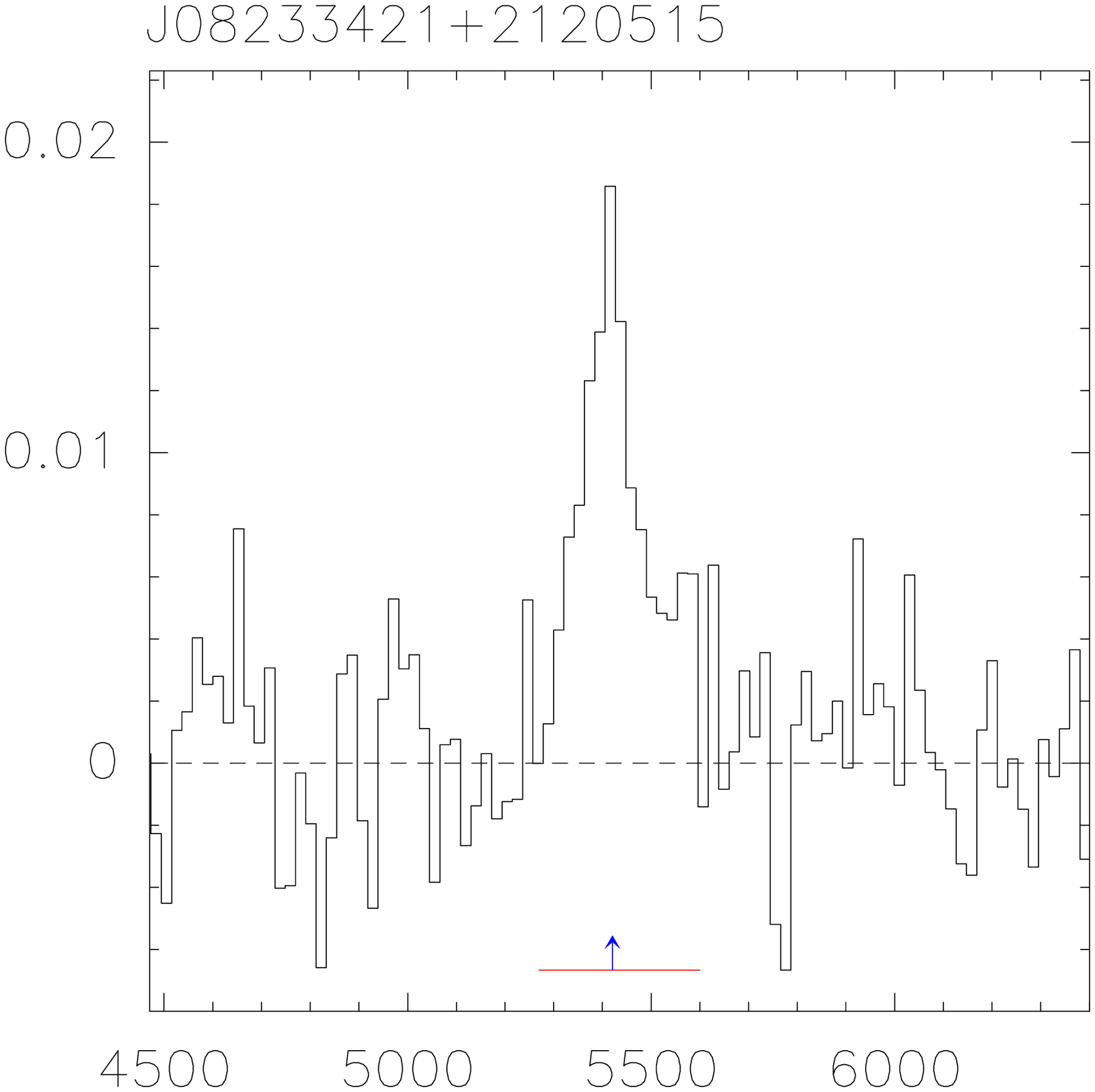}
\hspace{0.1cm}
\includegraphics[width=3.6cm,clip,trim = 0.cm 0.cm 0.cm 0.0cm, angle=-0]{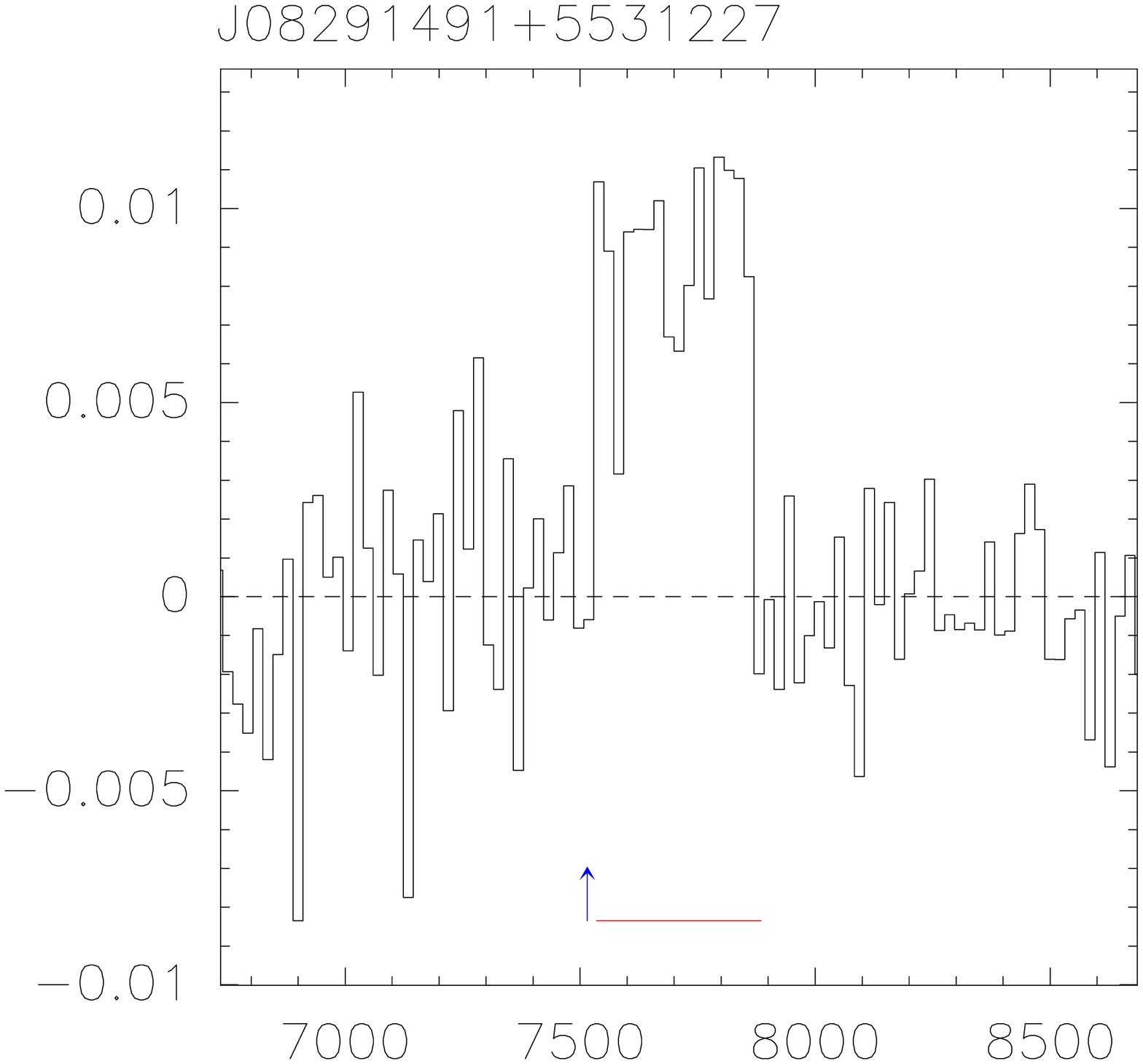}
\hspace{0.1cm}
\includegraphics[width=3.6cm,clip,trim = 0.cm 0.cm 0.cm 0.0cm, angle=-0]{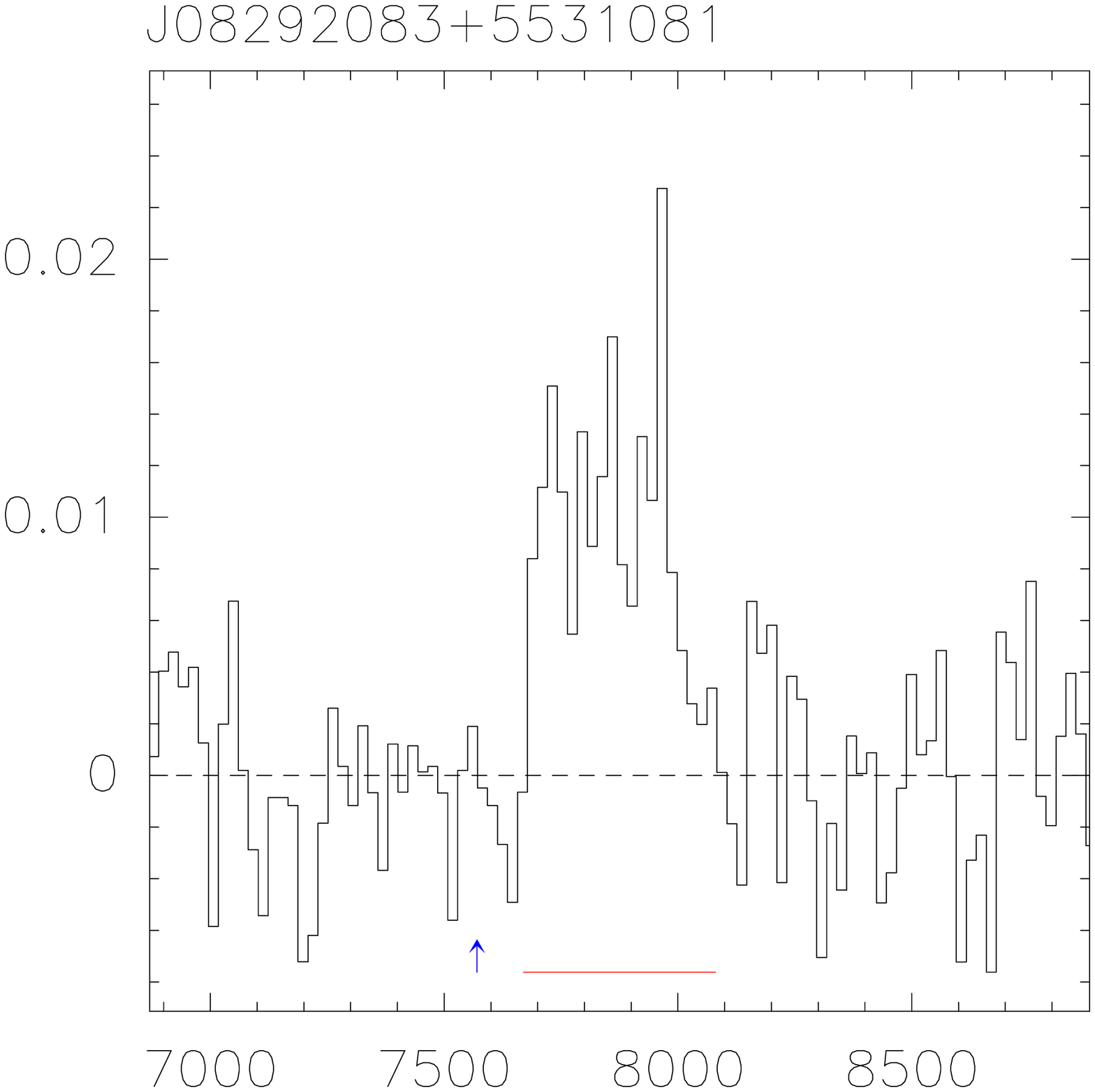}
}
\quad

\centerline{
\includegraphics[width=3.6cm,clip,trim = 0.cm 0.cm 0.cm 0.0cm,angle=-0]{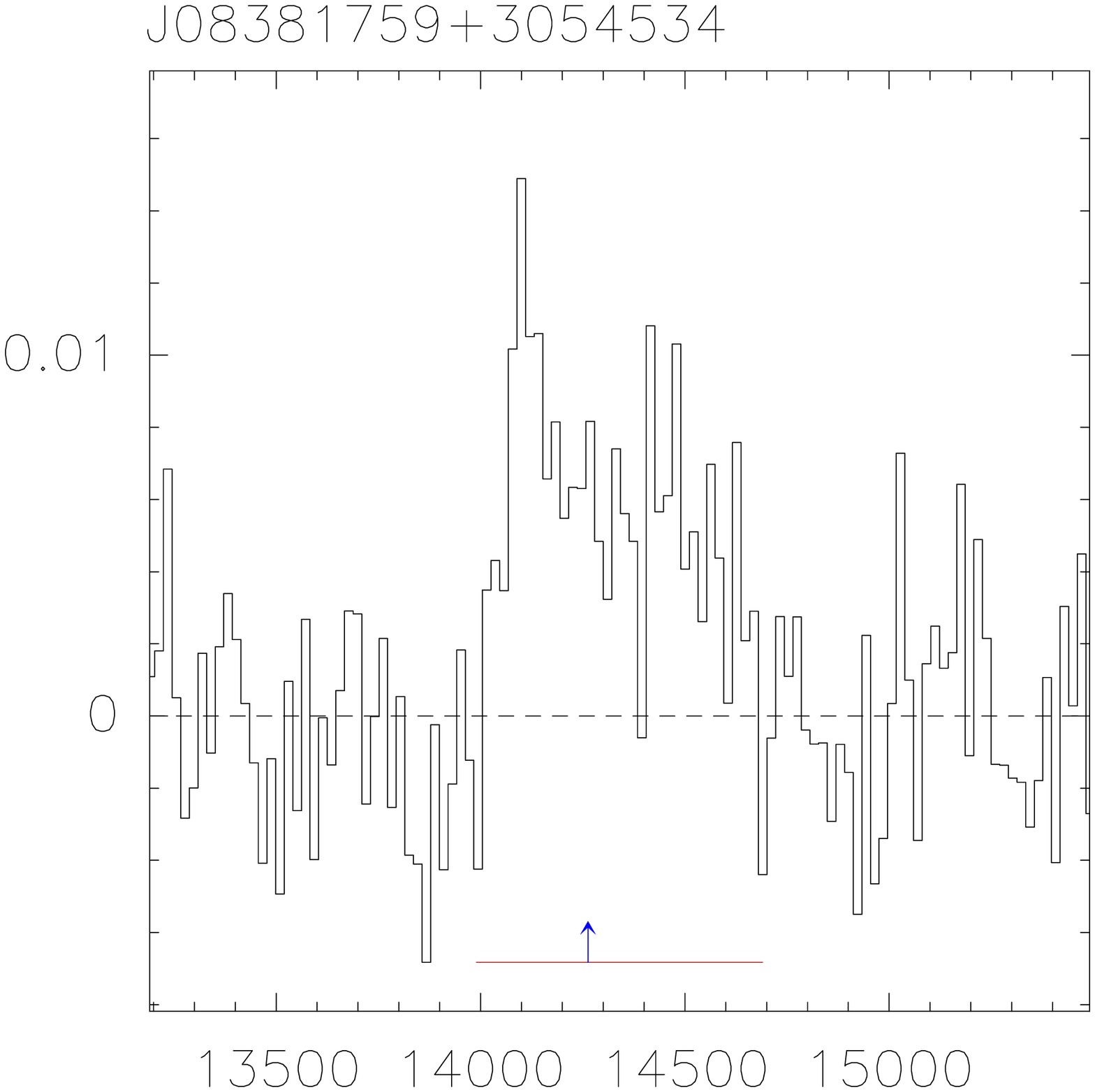}
\hspace{0.1cm}
\includegraphics[width=3.6cm,clip,trim = 0.cm 0.cm 0.cm 0.0cm, angle=-0]{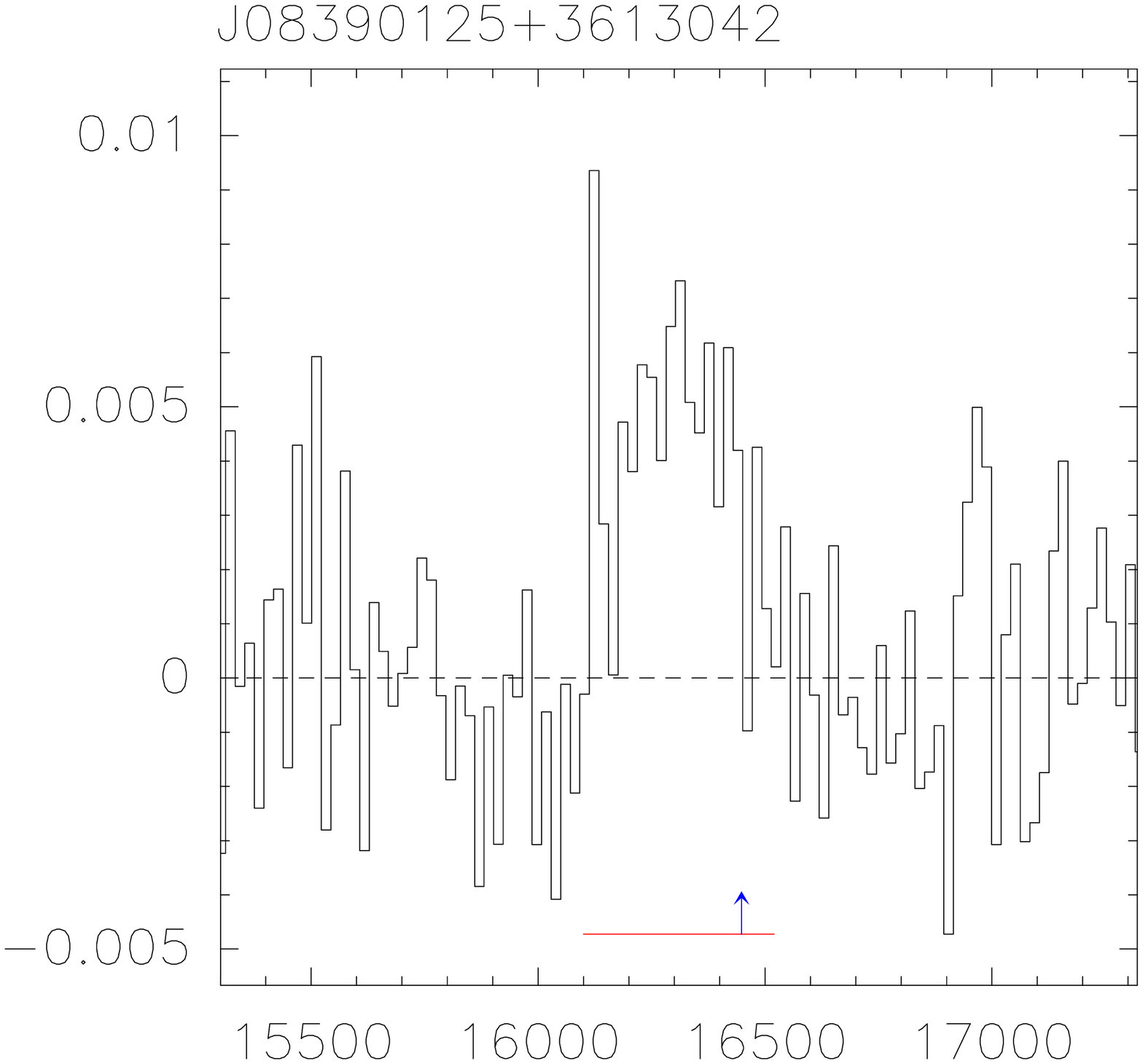}
\hspace{0.1cm}
\includegraphics[width=3.6cm,clip,trim = 0.cm 0.cm 0.cm 0.0cm, angle=-0]{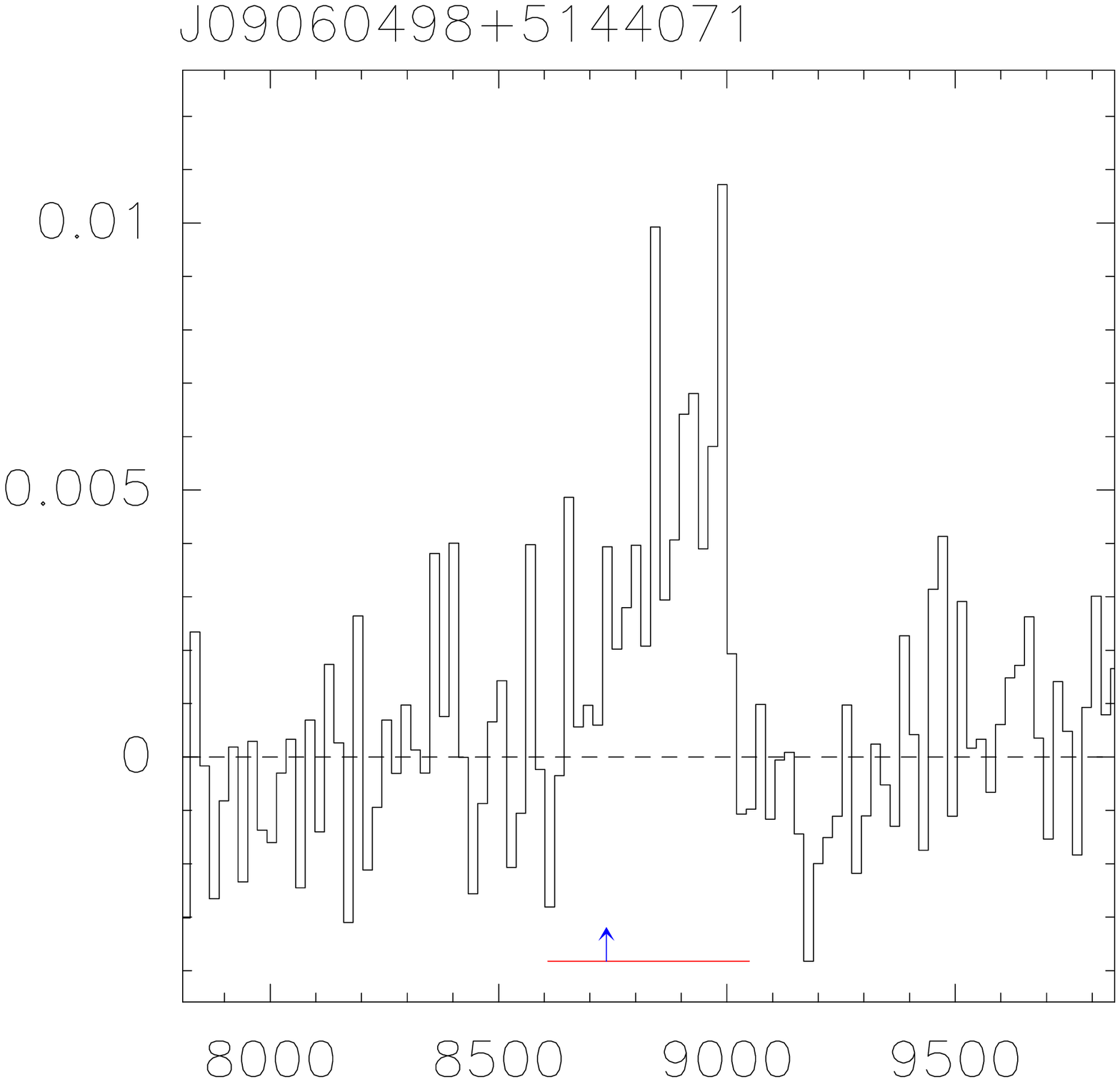}
\hspace{0.1cm}
\includegraphics[width=3.6cm,clip,trim = 0.cm 0.cm 0.cm 0.0cm, angle=-0]{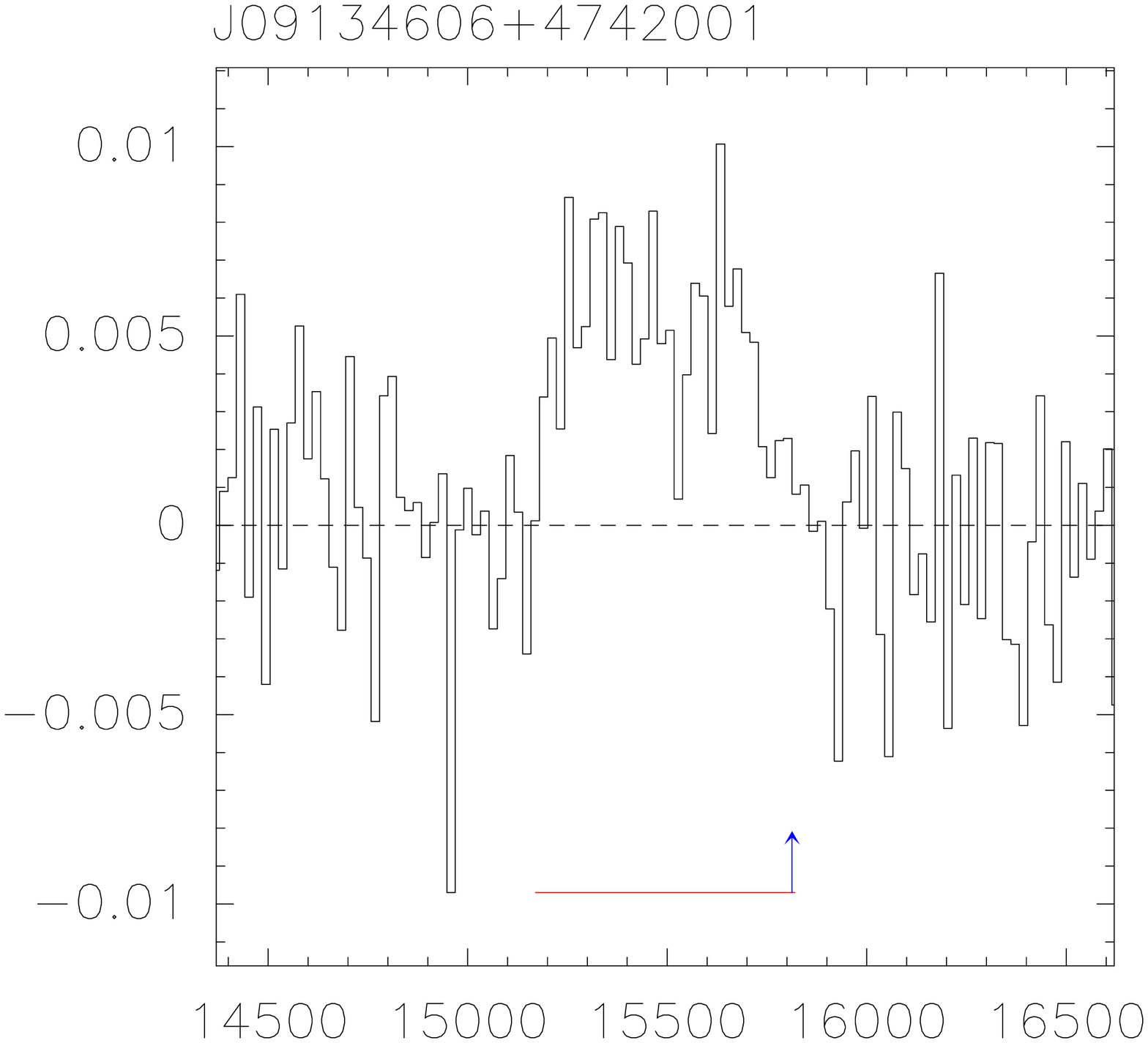}
}
\quad

\centerline{
\includegraphics[width=3.6cm,clip,trim = 0.cm 0.cm 0.cm 0.0cm,angle=-0]{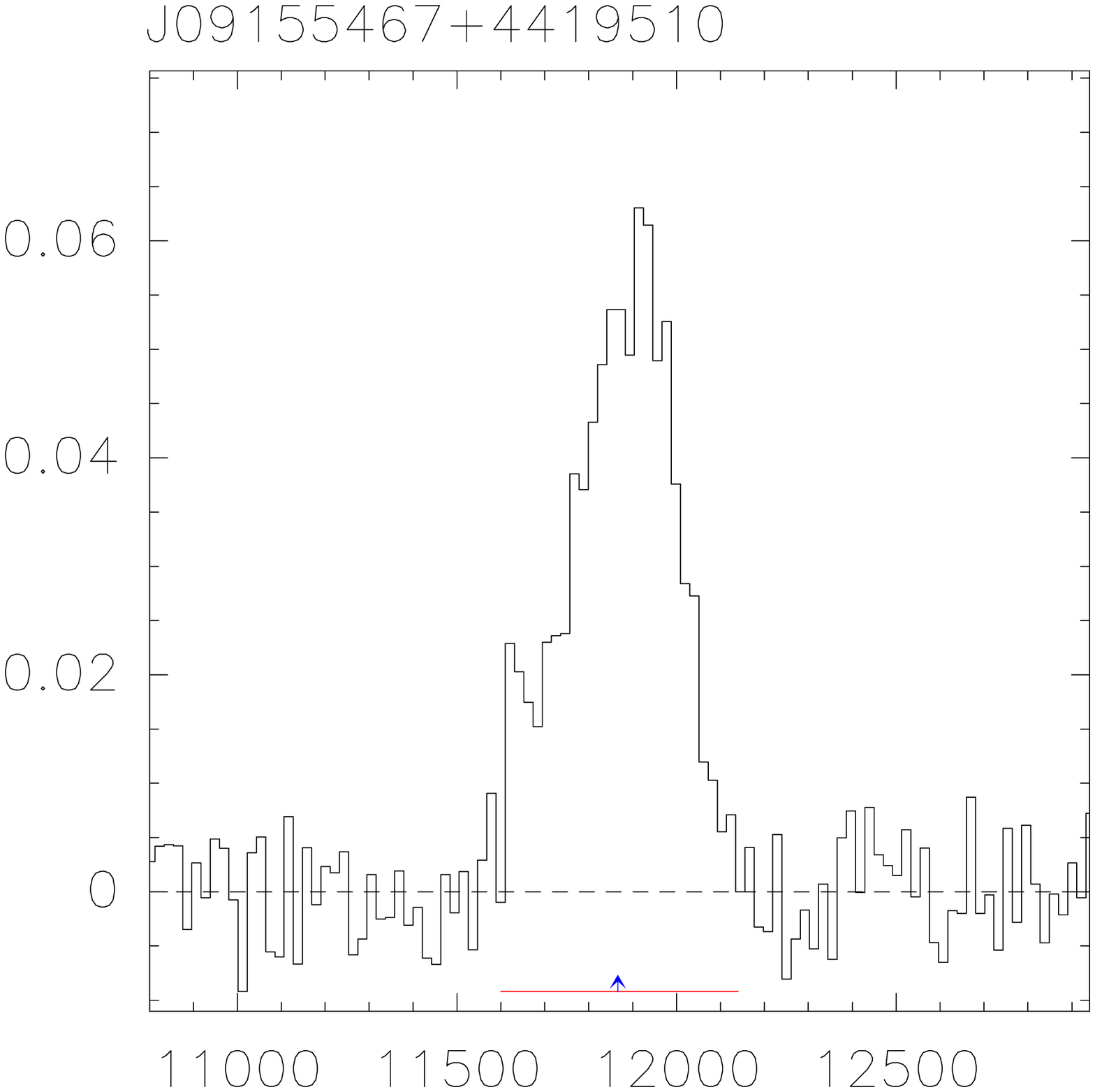}
\hspace{0.1cm}
\includegraphics[width=3.6cm,clip,trim = 0.cm 0.cm 0.cm 0.0cm, angle=-0]{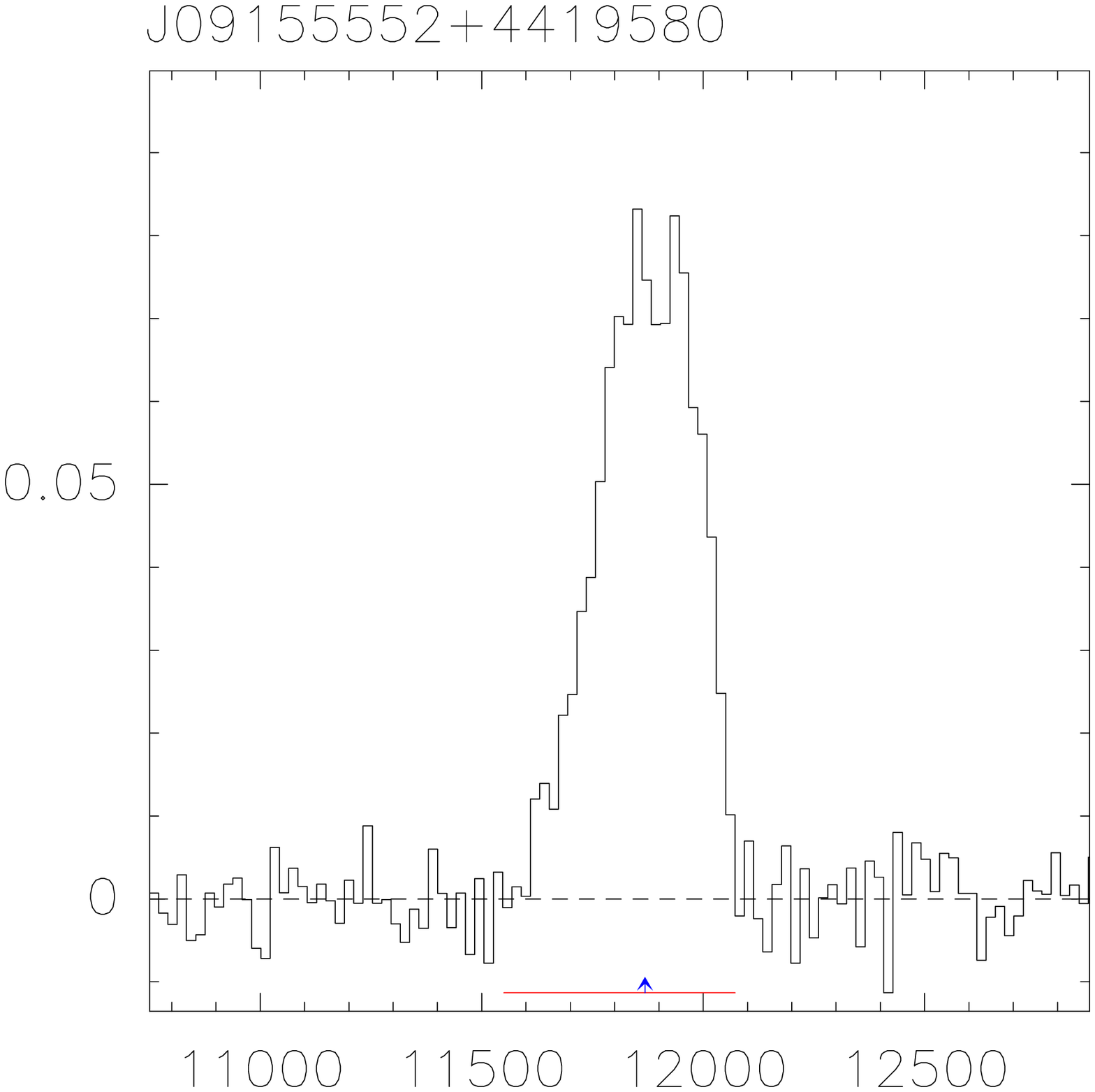}
\hspace{0.1cm}
\includegraphics[width=3.6cm,clip,trim = 0.cm 0.cm 0.cm 0.0cm, angle=-0]{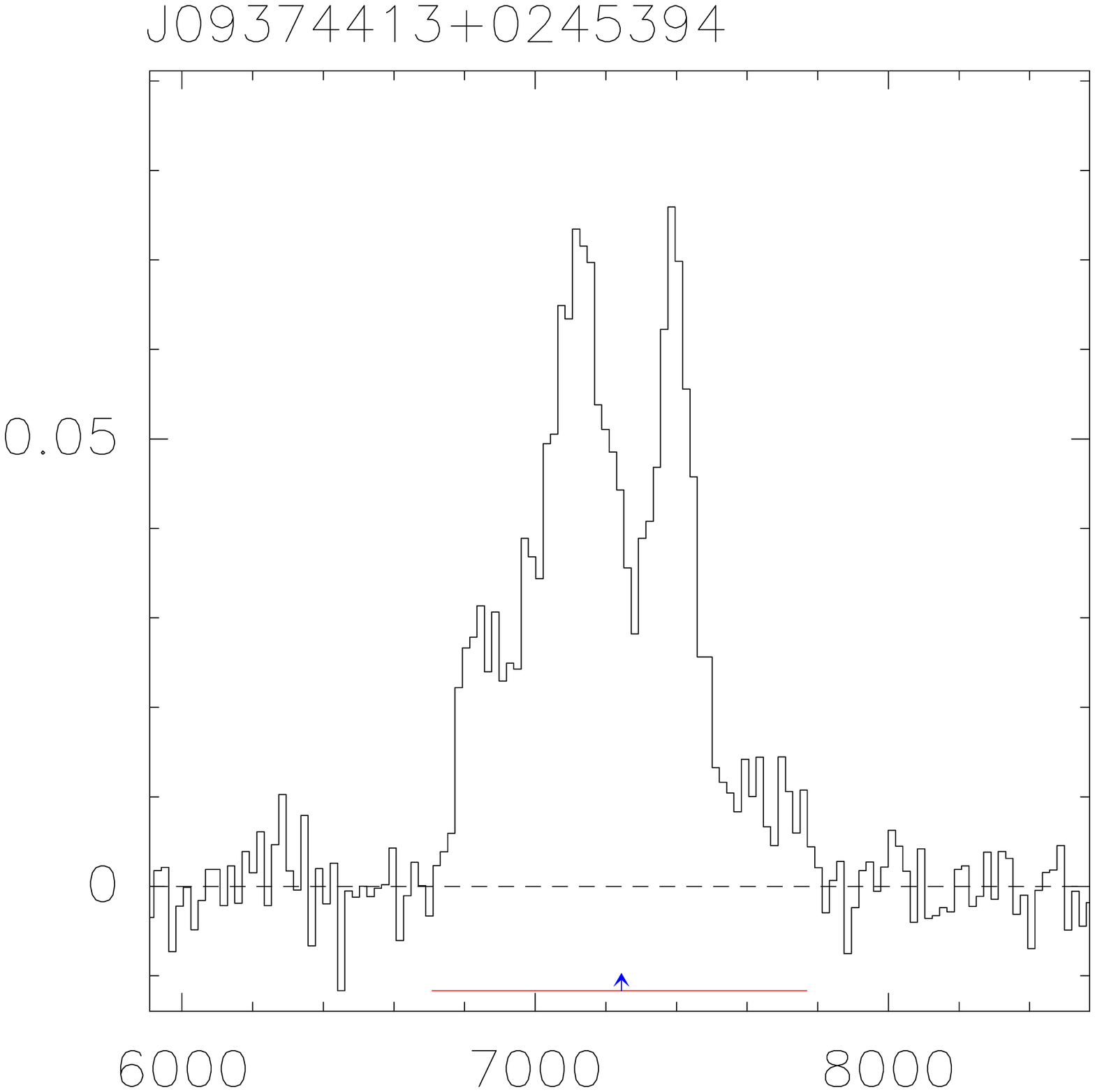}
\hspace{0.1cm}
\includegraphics[width=3.6cm,clip,trim = 0.cm 0.cm 0.cm 0.0cm,angle=-0]{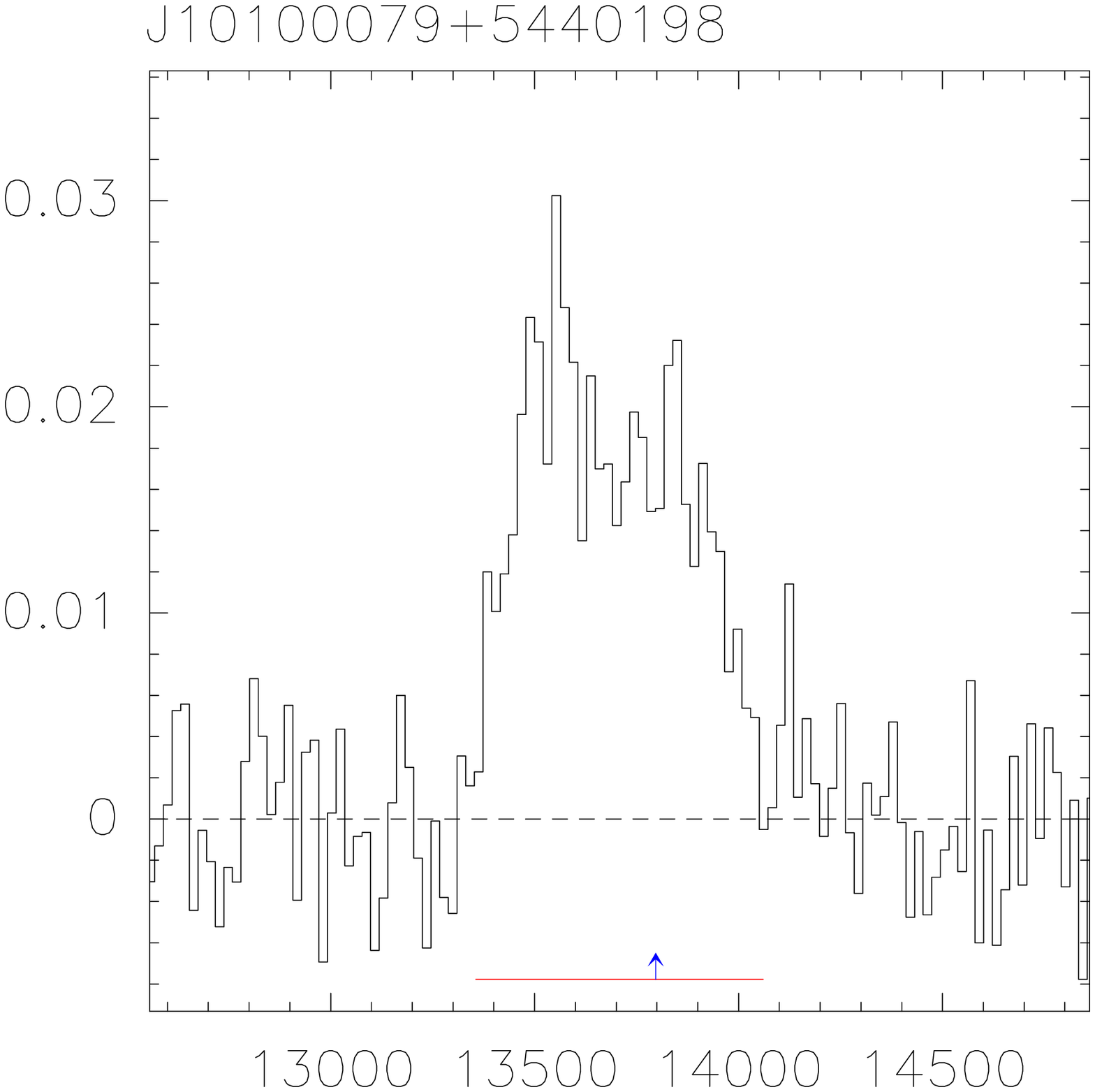}
}
\quad

\centerline{
\includegraphics[width=3.8cm,clip,trim = 0.cm 0.cm 0.cm 0.0cm, angle=-0]{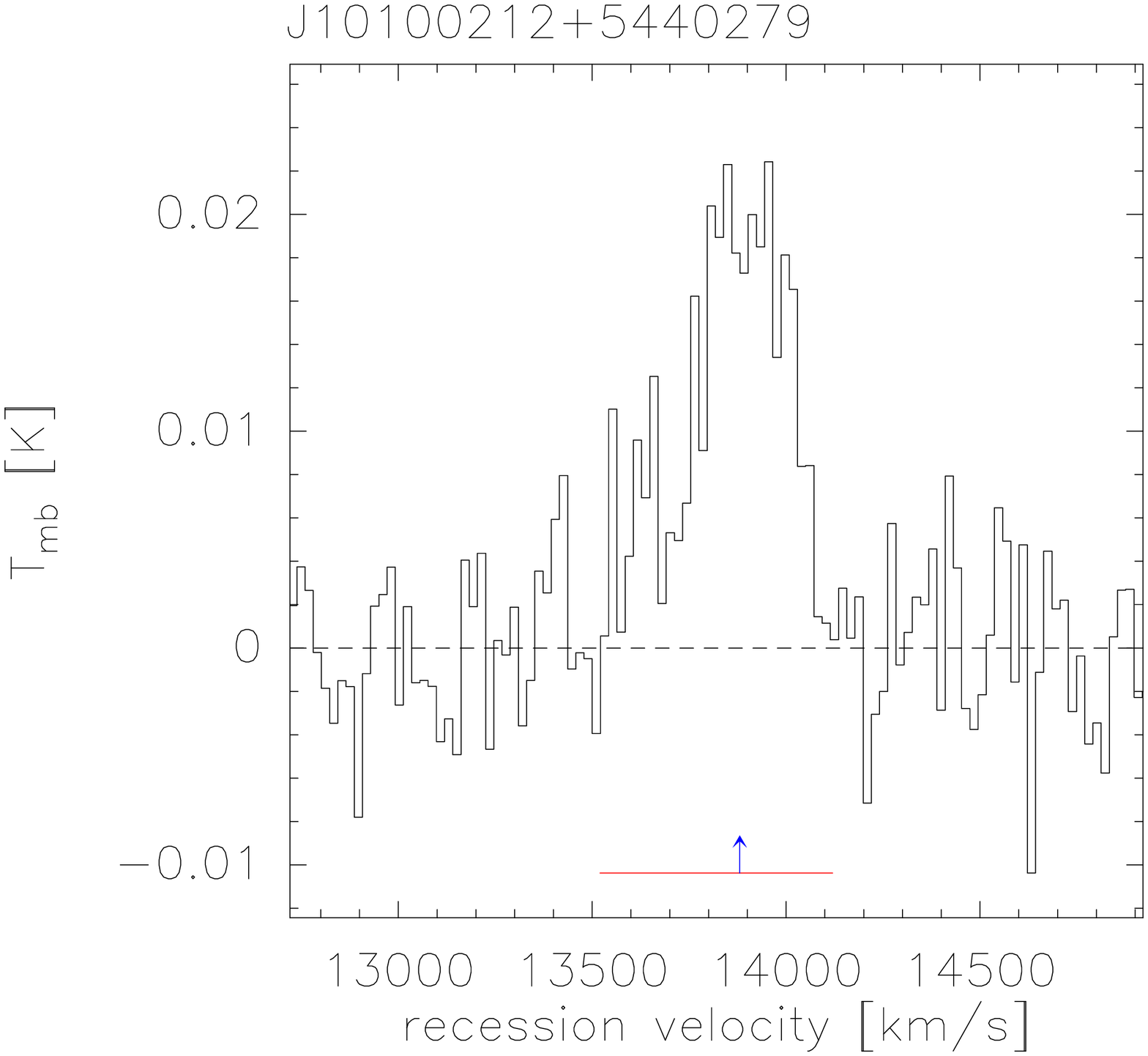}
\hspace{0.1cm}
\includegraphics[width=3.6cm,clip,trim = 0.cm 0.cm 0.cm 0.0cm,angle=-0]{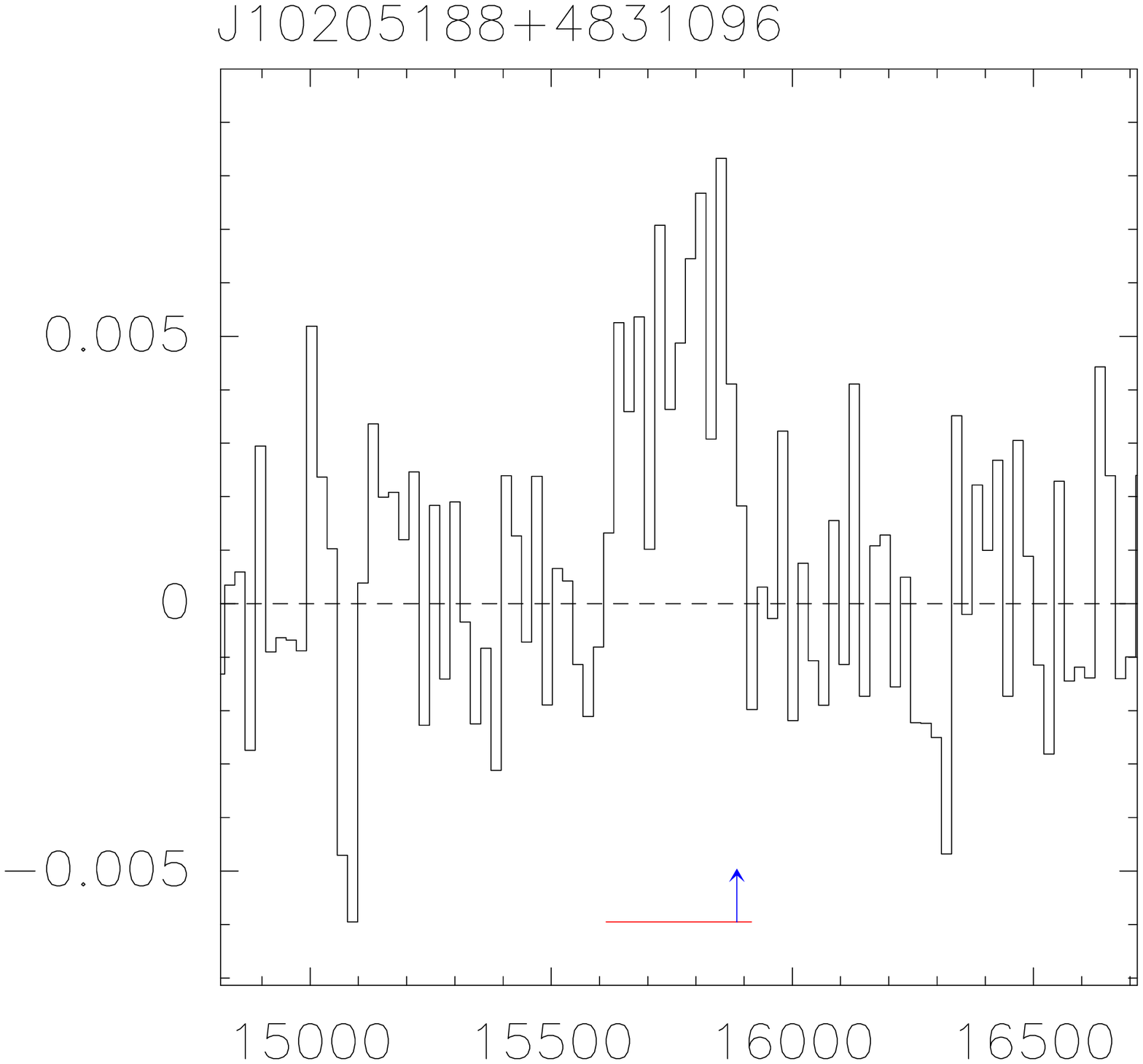}
\hspace{0.1cm}
\includegraphics[width=3.6cm,clip,trim = 0.cm 0.cm 0.cm 0.0cm,angle=-0]{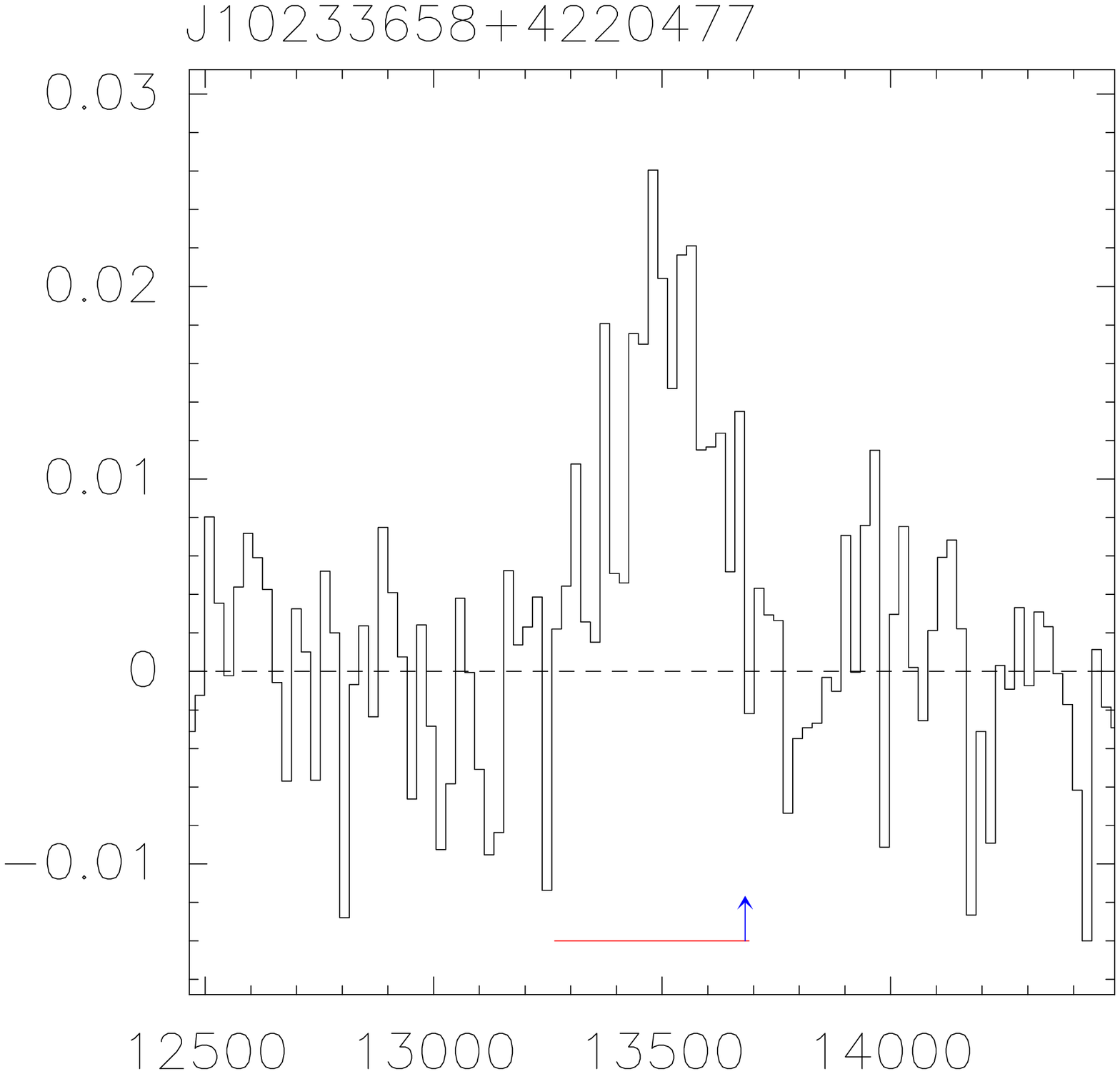}
\hspace{0.1cm}
\includegraphics[width=3.6cm,clip,trim = 0.cm 0.cm 0.cm 0.0cm, angle=-0]{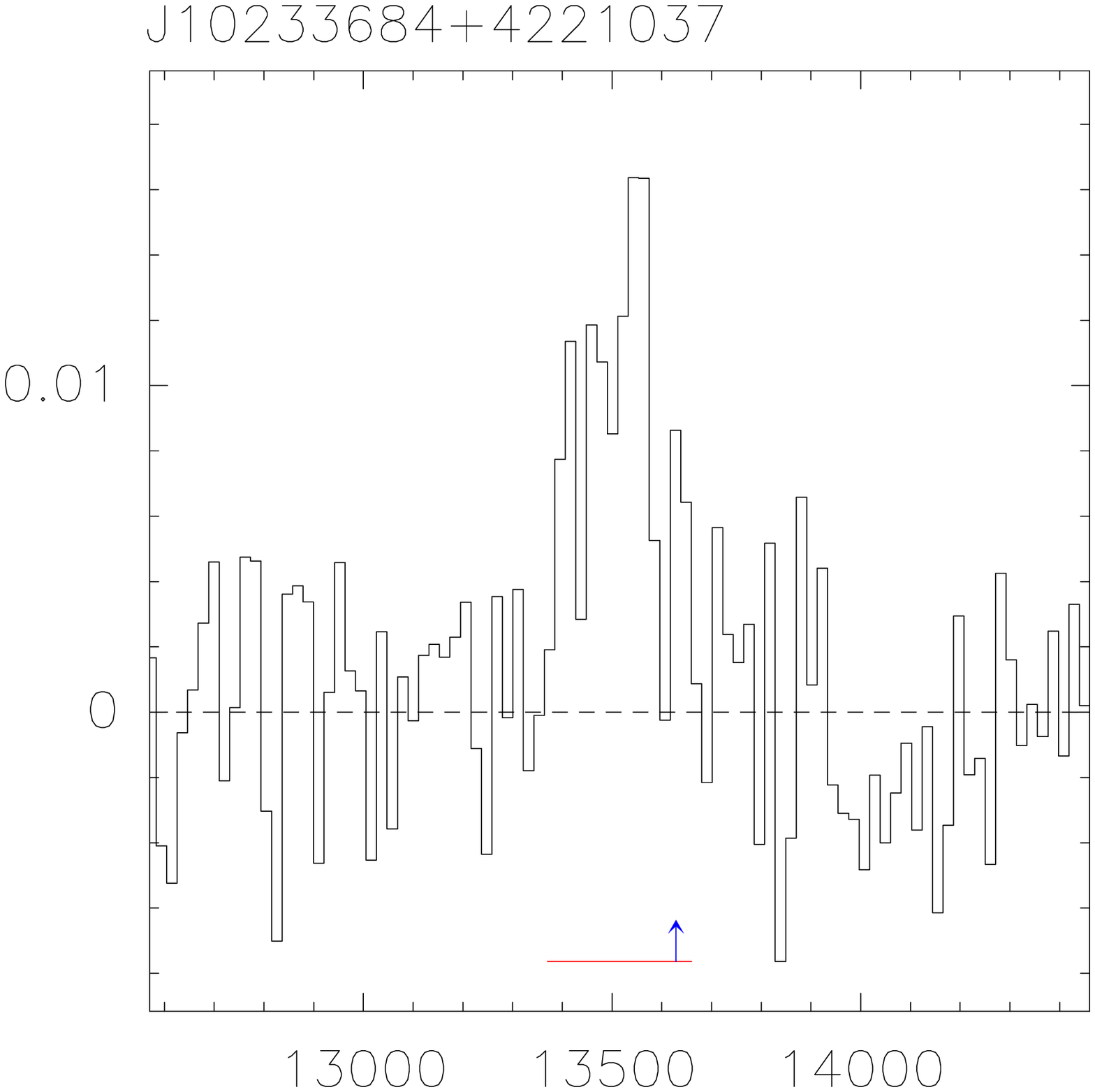}
}

\caption{CO(1-0) spectra of the detected  galaxies (including tentative detections). The velocity resolution
is $\sim$ 20 \kms\ for most spectra and $\sim$  40 \kms\ for some cases where a lower resolution was required
to clearly see the line. The red line segment shows the zero-level linewidth of the 
CO line adopted for the determination of the velocity integrated intensity. 
The blue upright arrow indicated the optical  heliocentric recession velocity. 
An asterisk next to the name indicates a tentative detection.
}
\label{fig:spectra_co10}
\end{figure*}

\setcounter{figure}{0} 
\begin{figure*}

\centerline{
\includegraphics[width=3.6cm,clip,trim = 0.cm 0.cm 0.cm 0.0cm, angle=-0]{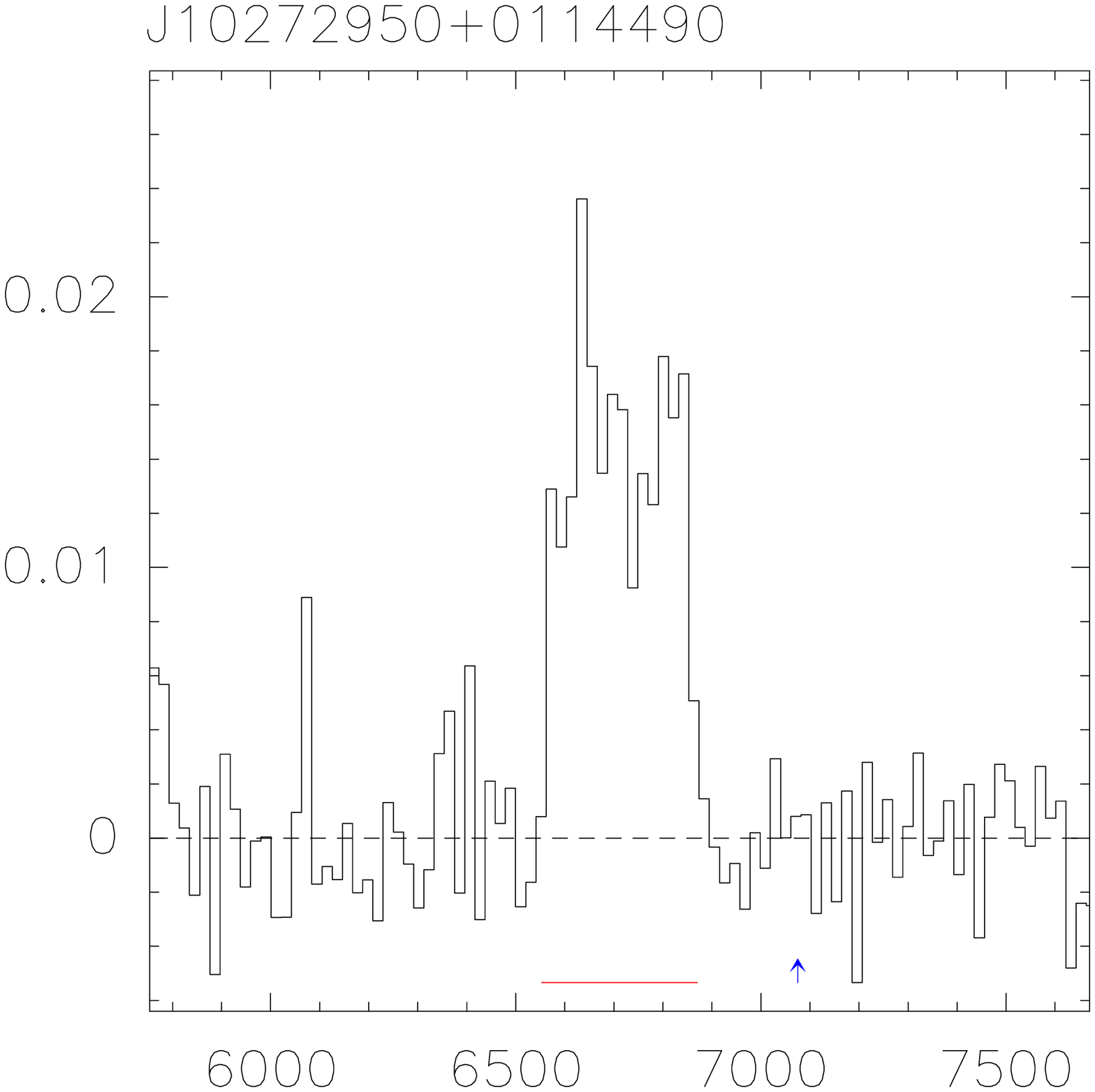}
\hspace{0.1cm}
\includegraphics[width=3.6cm,clip,trim = 0.cm 0.cm 0.cm 0.0cm, angle=-0]{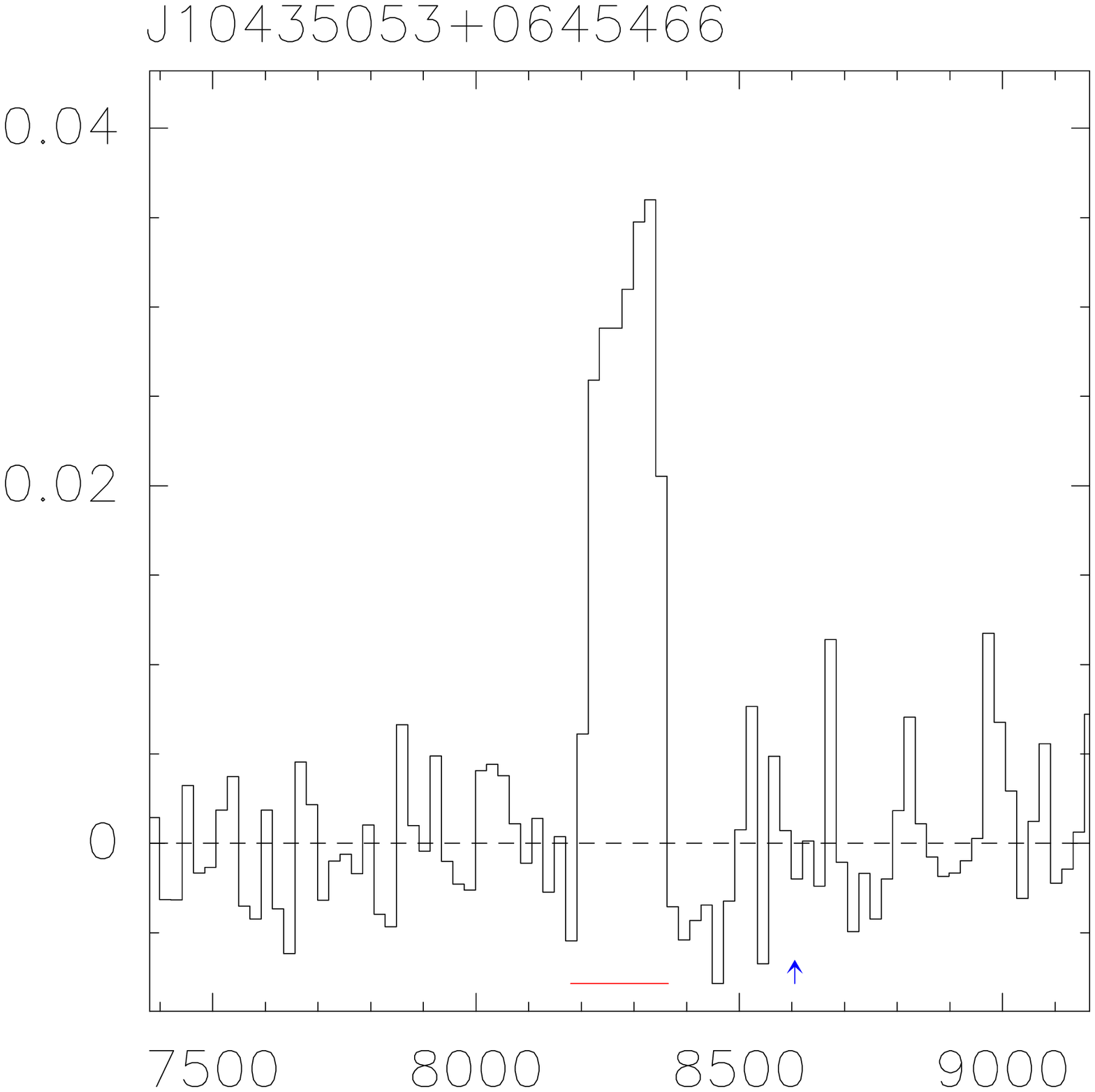}
\hspace{0.1cm}
\includegraphics[width=3.6cm,clip,trim = 0.cm 0.cm 0.cm 0.0cm,angle=-0]{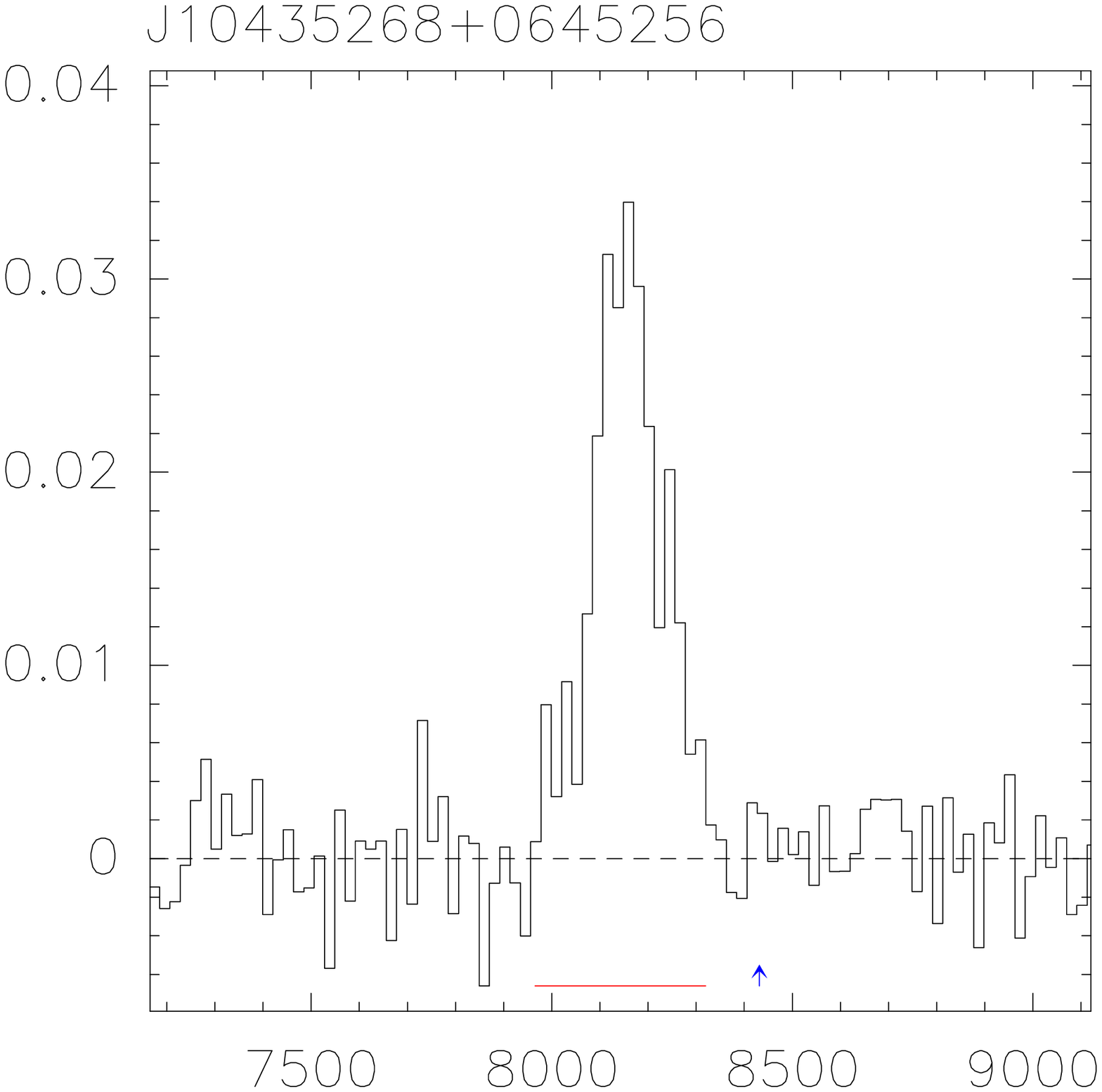}
\hspace{0.1cm}
\includegraphics[width=3.6cm,clip,trim = 0.cm 0.cm 0.cm 0.0cm,angle=-0]{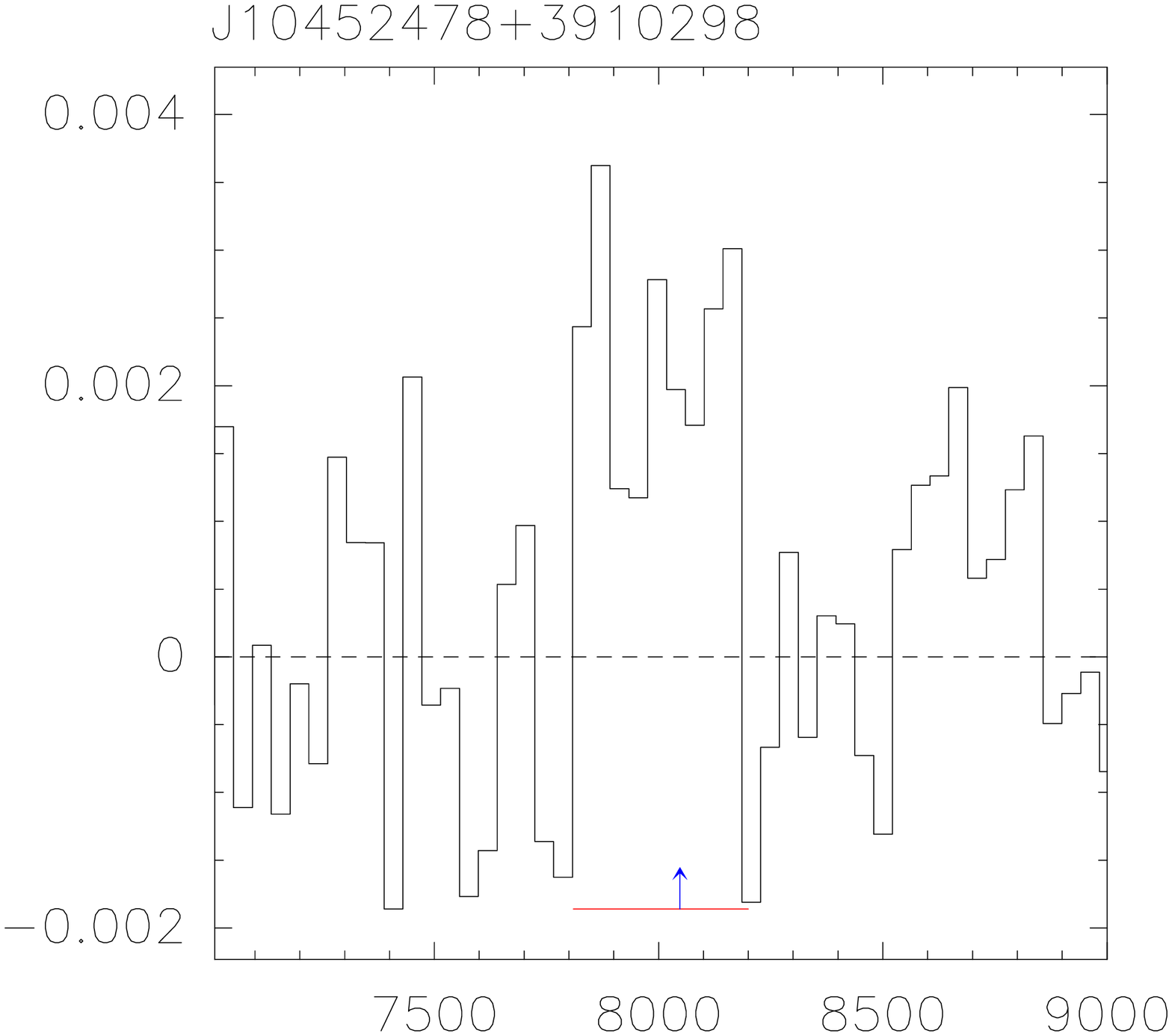}
}
\quad
\centerline{
\includegraphics[width=3.6cm,clip,trim = 0.cm 0.cm 0.cm 0.0cm, angle=-0]{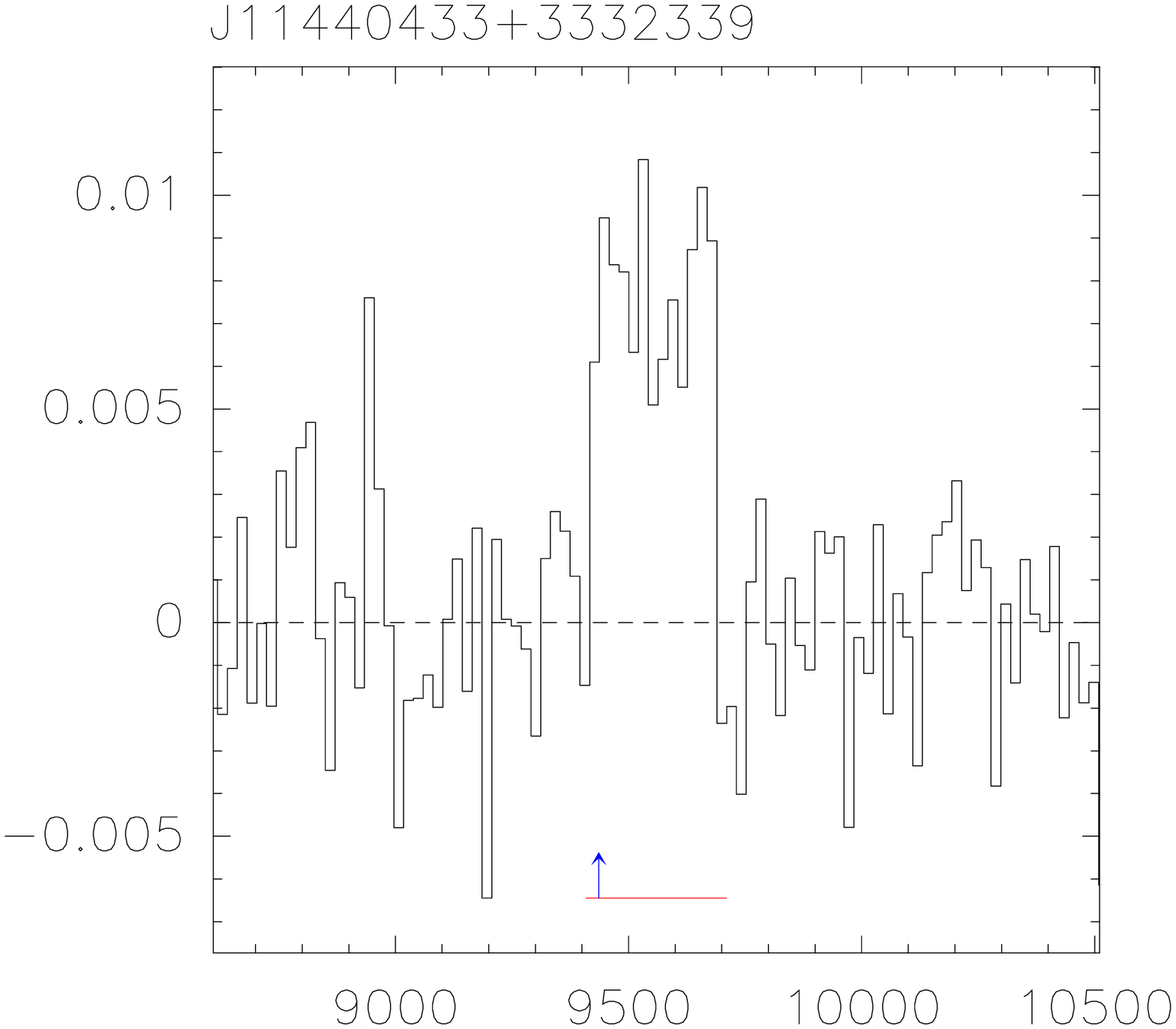}
\hspace{0.1cm}
\includegraphics[width=3.6cm,clip,trim = 0.cm 0.cm 0.cm 0.0cm, angle=-0]{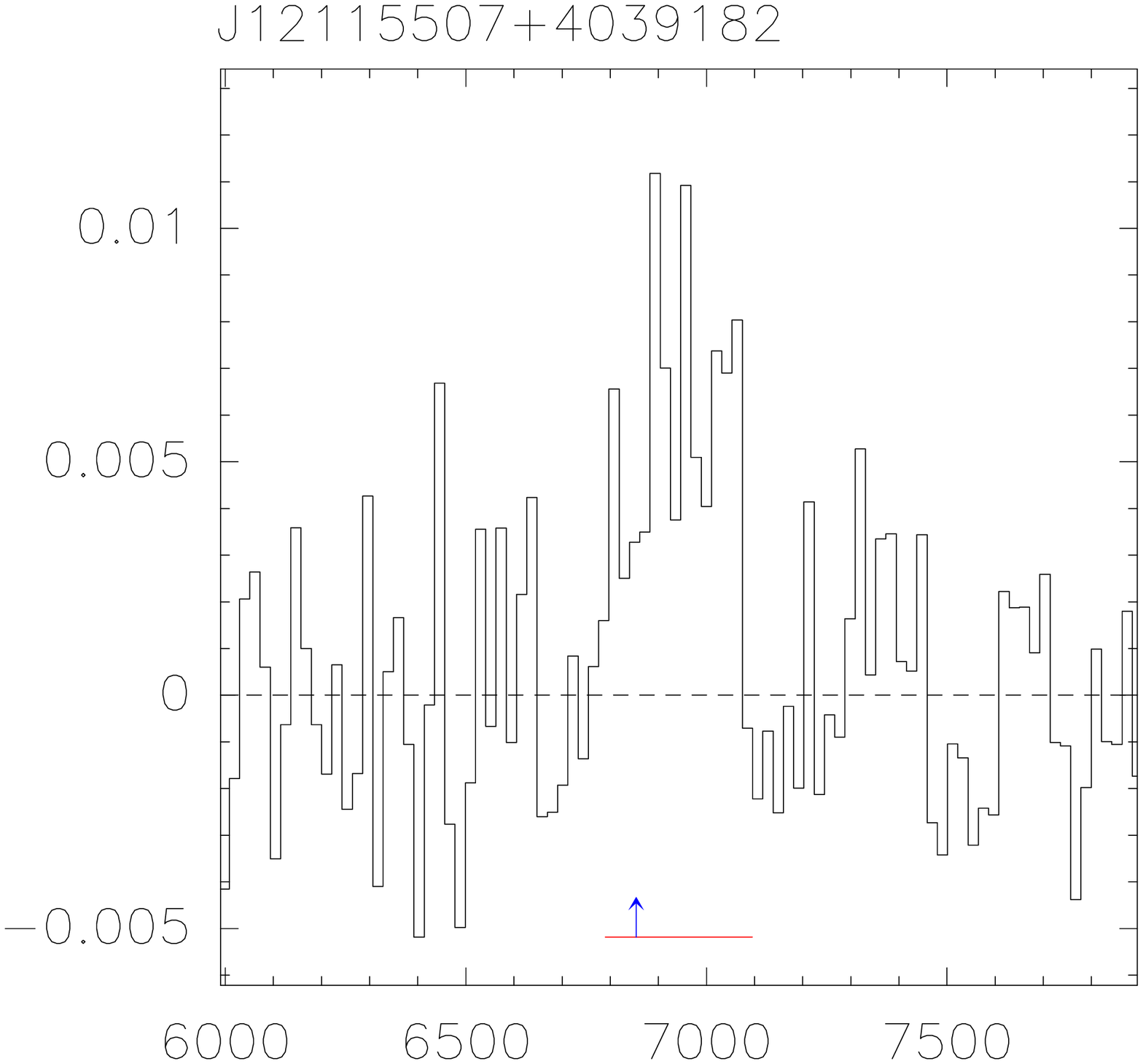}
\hspace{0.1cm}
\includegraphics[width=3.6cm,clip,trim = 0.cm 0.cm 0.cm 0.0cm, angle=-0]{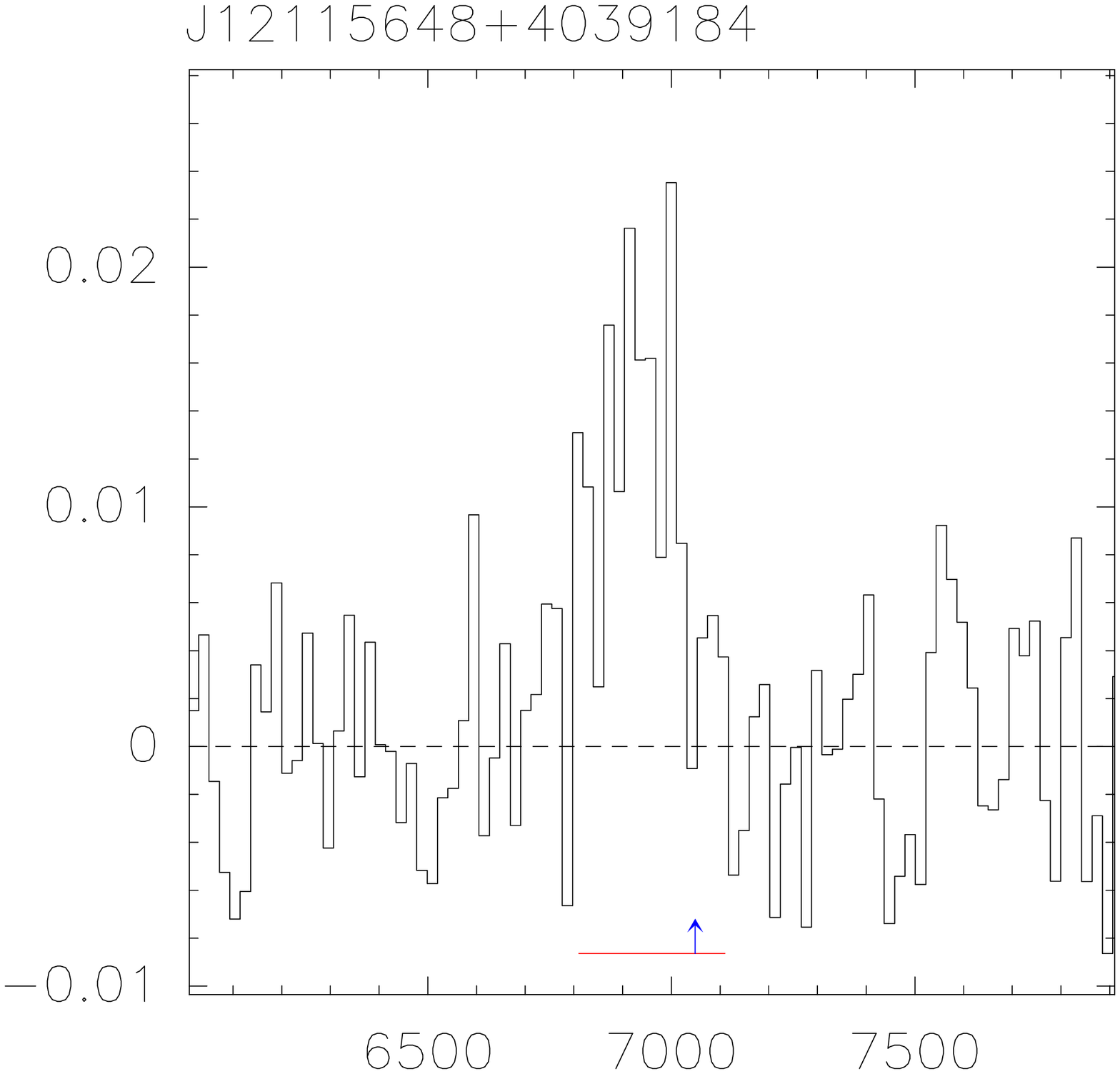}
\hspace{0.1cm}
\includegraphics[width=3.6cm,clip,trim = 0.cm 0.cm 0.cm 0.0cm,angle=-0]{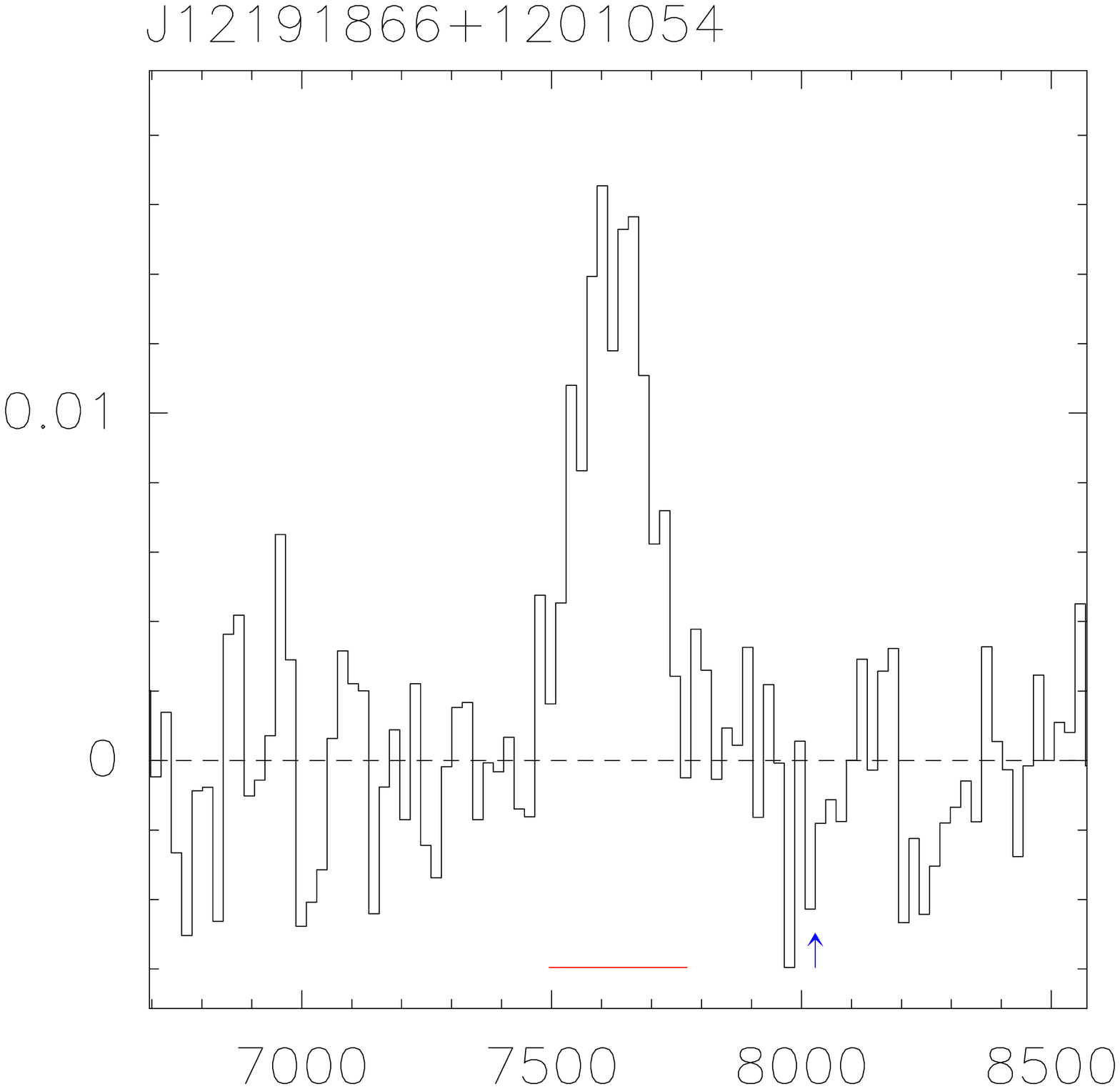}
}
\quad

\centerline{
\includegraphics[width=3.6cm,clip,trim = 0.cm 0.cm 0.cm 0.0cm, angle=-0]{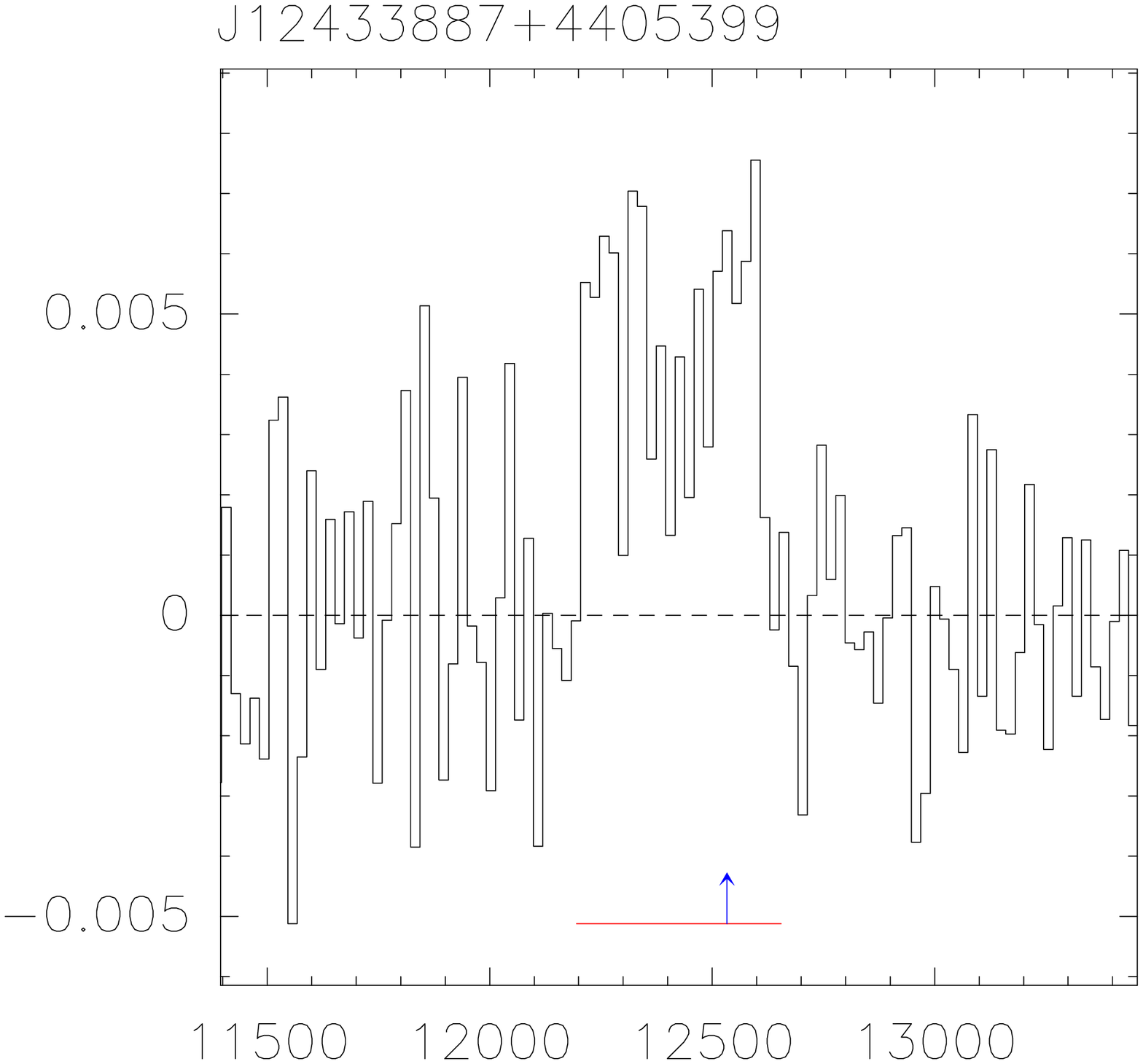}
\hspace{0.1cm}
\includegraphics[width=3.6cm,clip,trim = 0.cm 0.cm 0.cm 0.0cm, angle=-0]{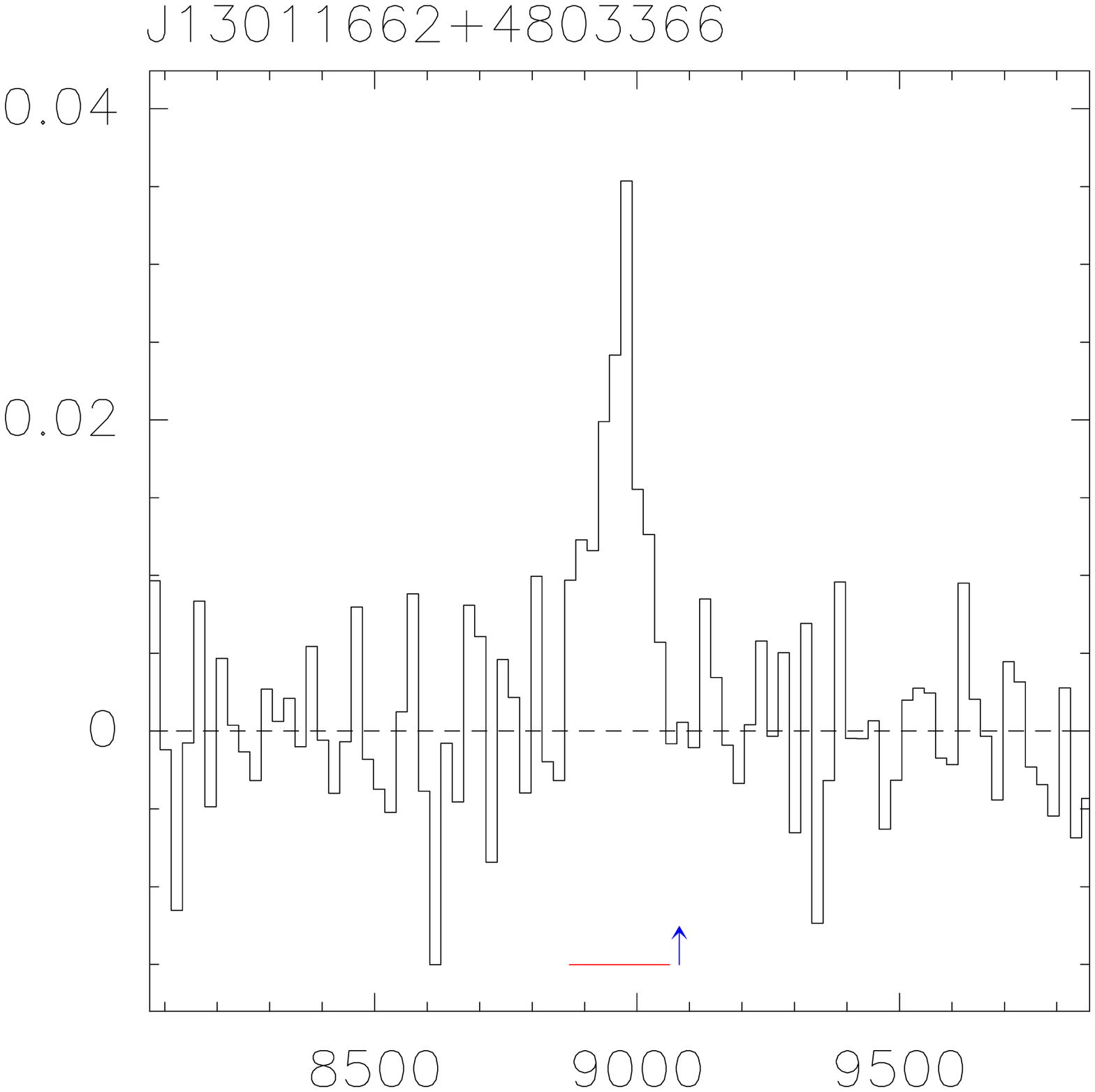}
\hspace{0.1cm}
\includegraphics[width=3.6cm,clip,trim = 0.cm 0.cm 0.cm 0.0cm, angle=-0]{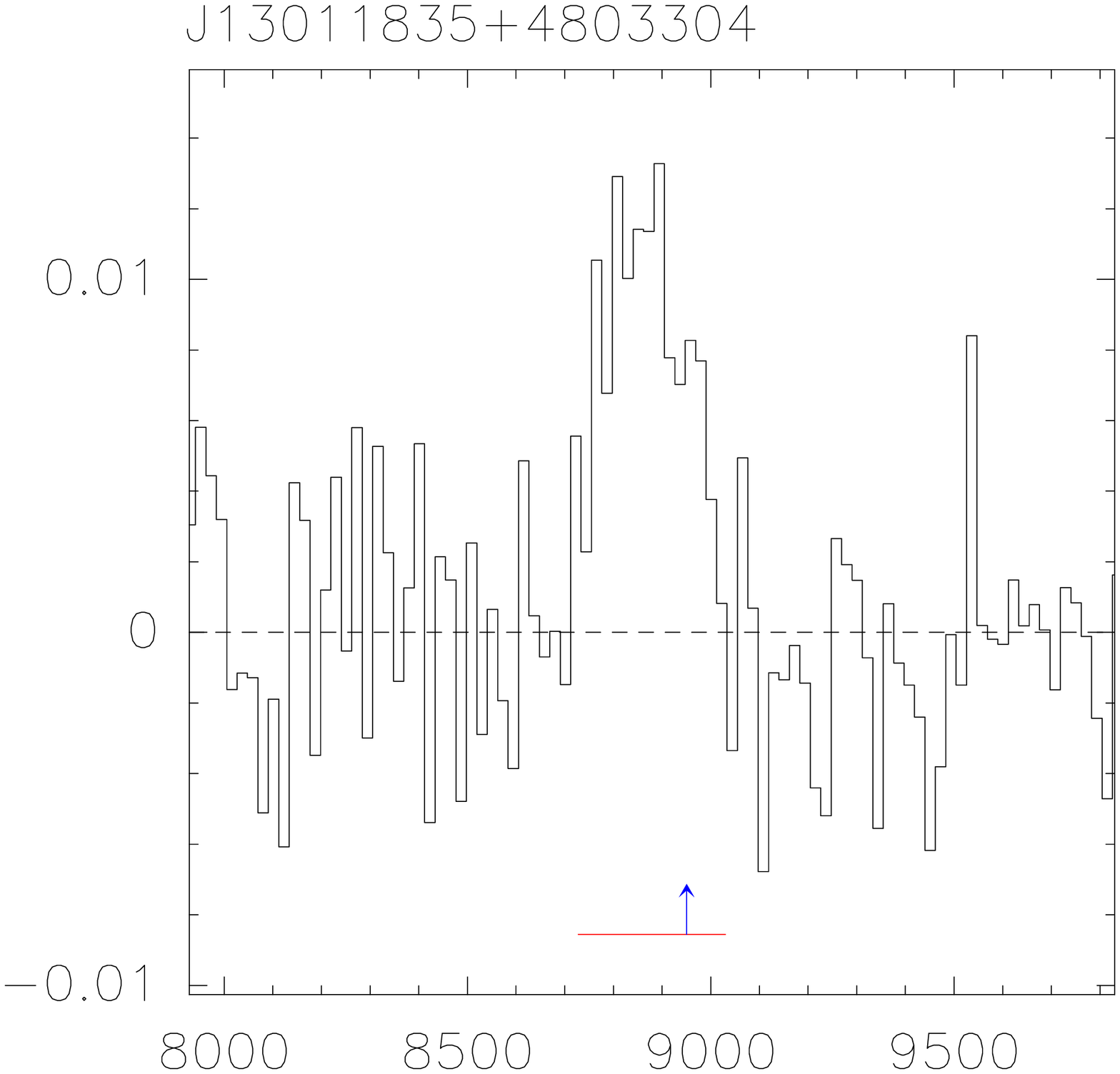}
\hspace{0.1cm}
\includegraphics[width=3.6cm,clip,trim = 0.cm 0.cm 0.cm 0.0cm,angle=-0]{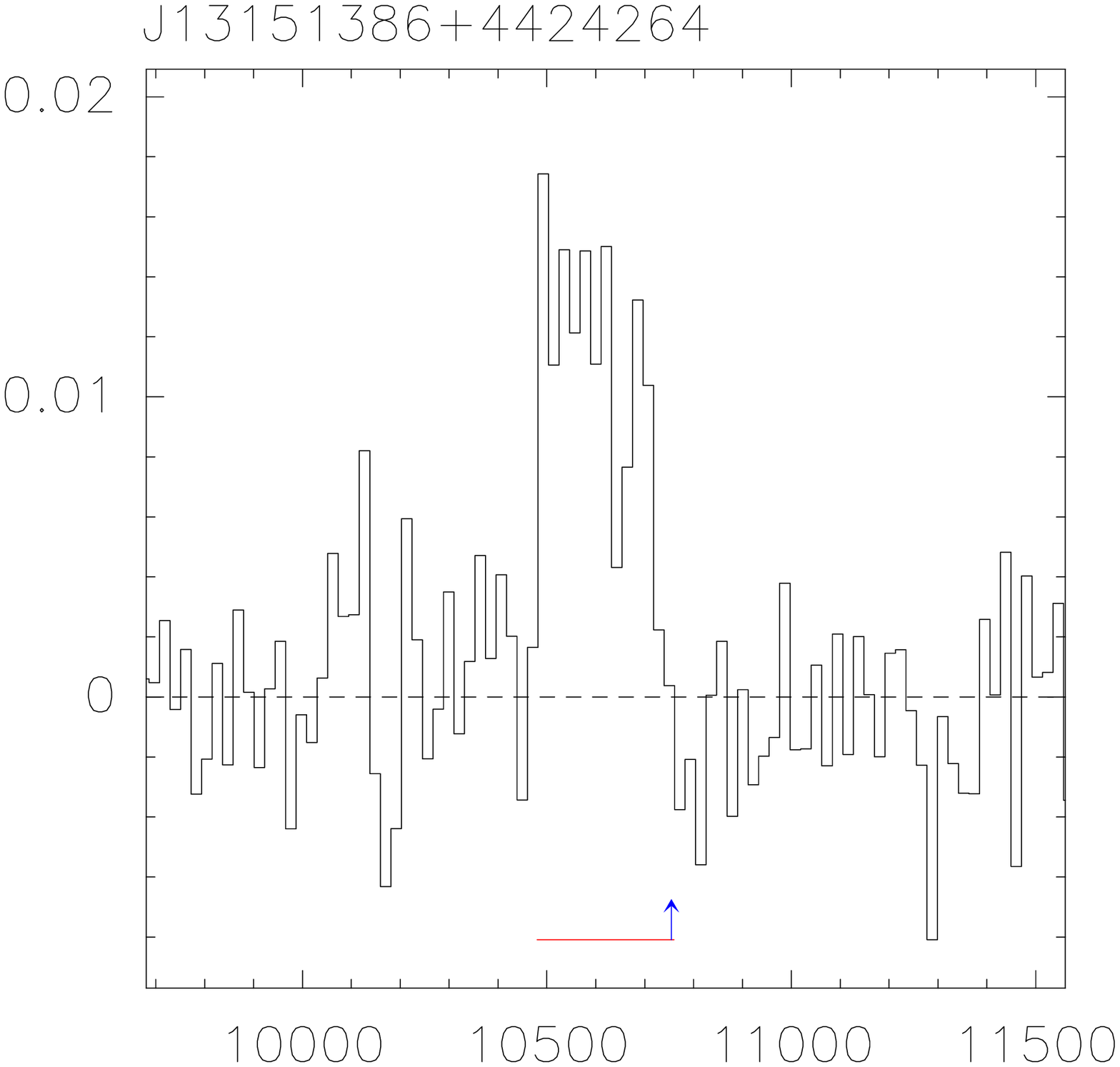}
}
\quad

\centerline{
\includegraphics[width=3.6cm,clip,trim = 0.cm 0.cm 0.cm 0.0cm, angle=-0]{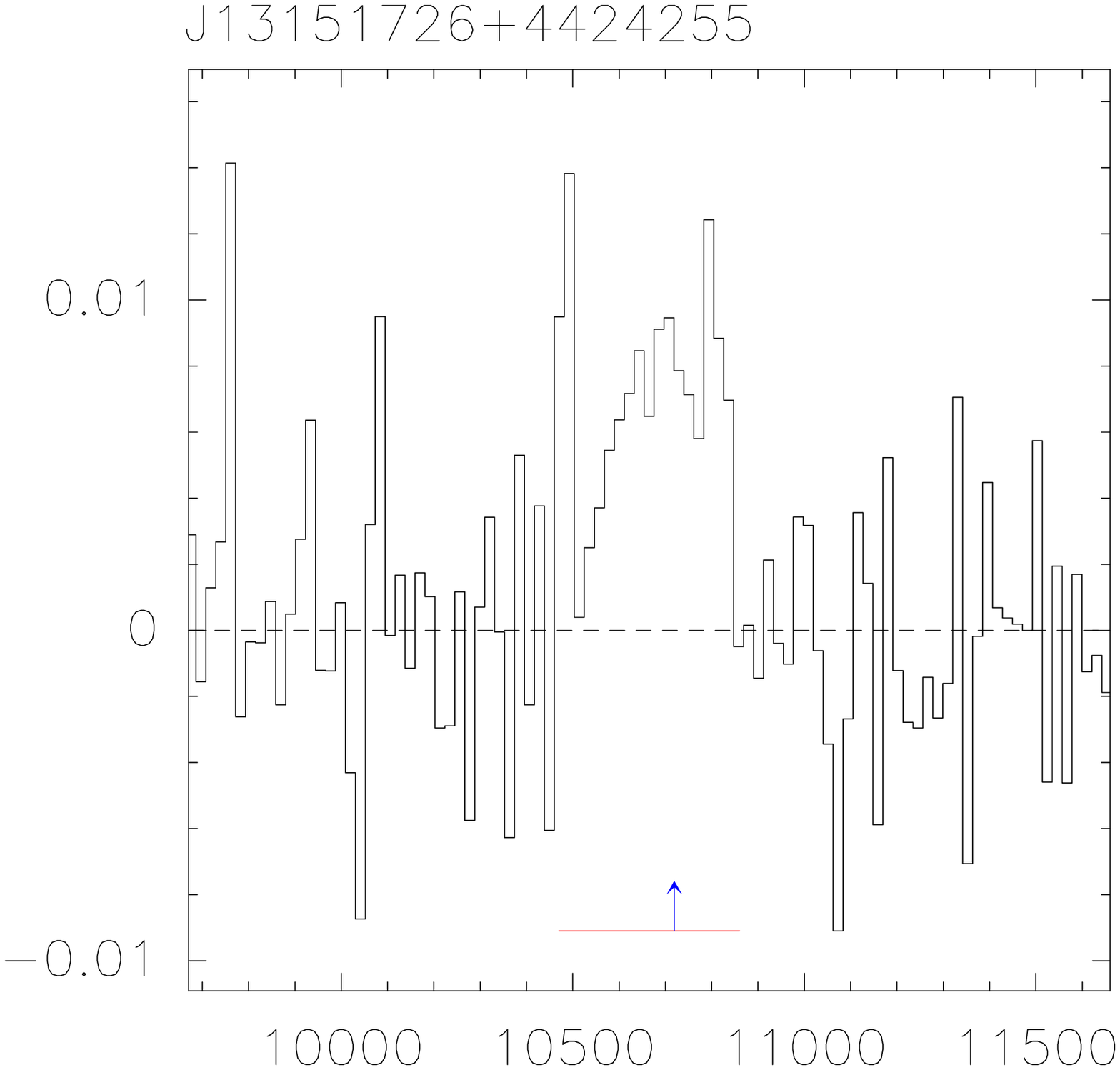}
\hspace{0.1cm}
\includegraphics[width=3.6cm,clip,trim = 0.cm 0.cm 0.cm 0.0cm, angle=-0]{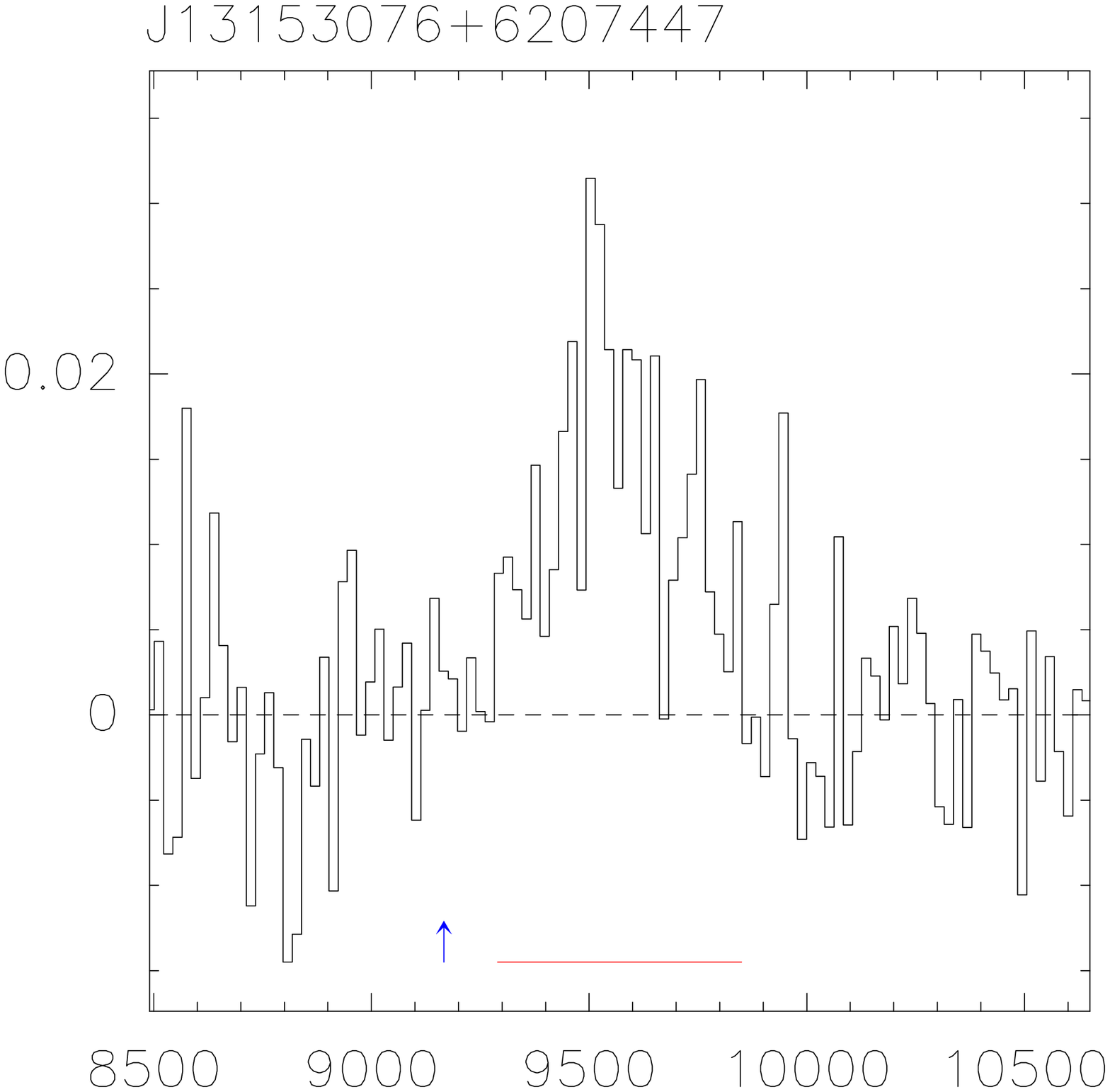}
\hspace{0.1cm}
\includegraphics[width=3.6cm,clip,trim = 0.cm 0.cm 0.cm 0.0cm, angle=-0]{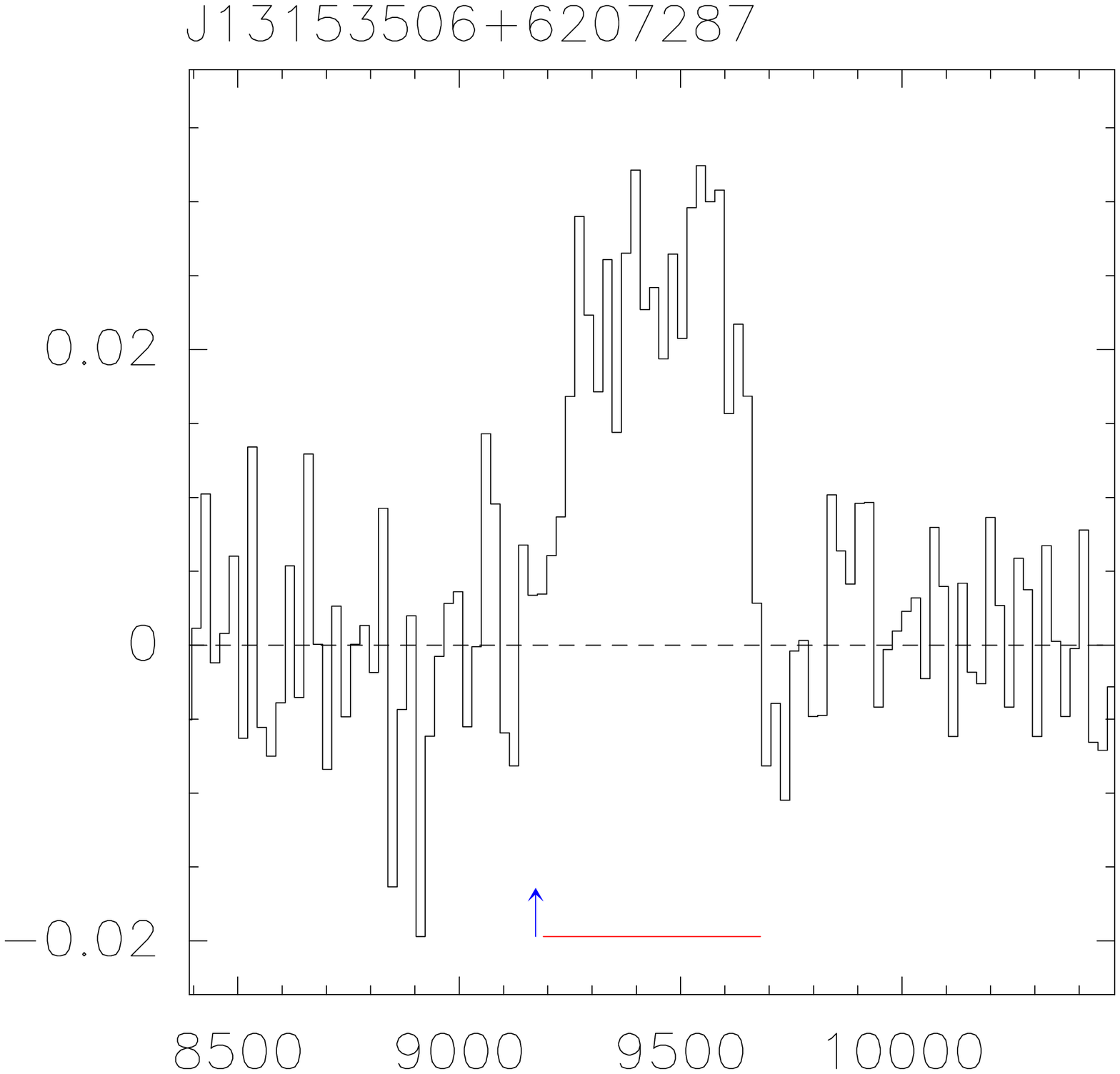}
\hspace{0.1cm}
\includegraphics[width=3.6cm,clip,trim = 0.cm 0.cm 0.cm 0.0cm,angle=-0]{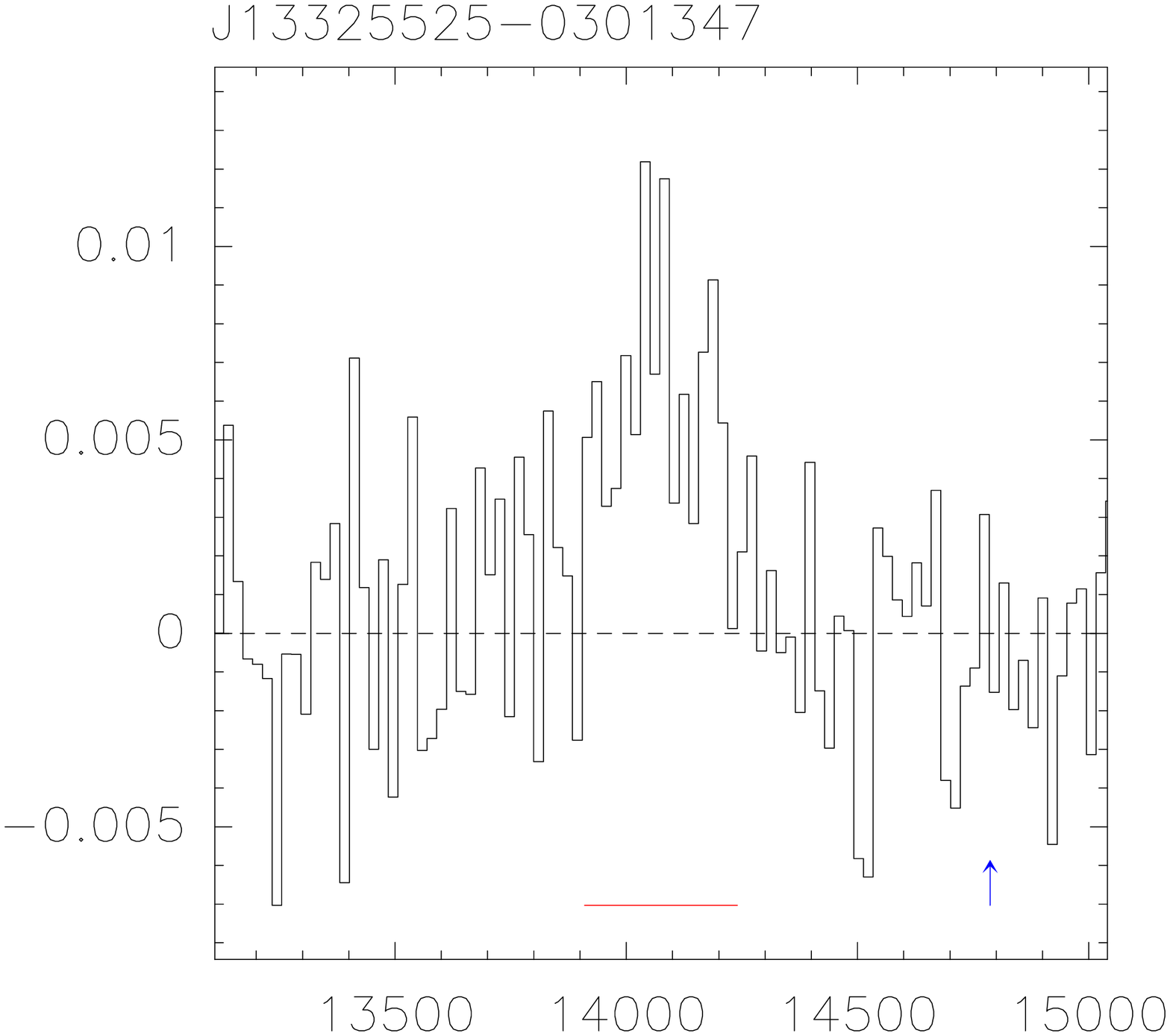}
}
\quad

\centerline{
\includegraphics[width=3.6cm,clip,trim = 0.cm 0.cm 0.cm 0.0cm, angle=-0]{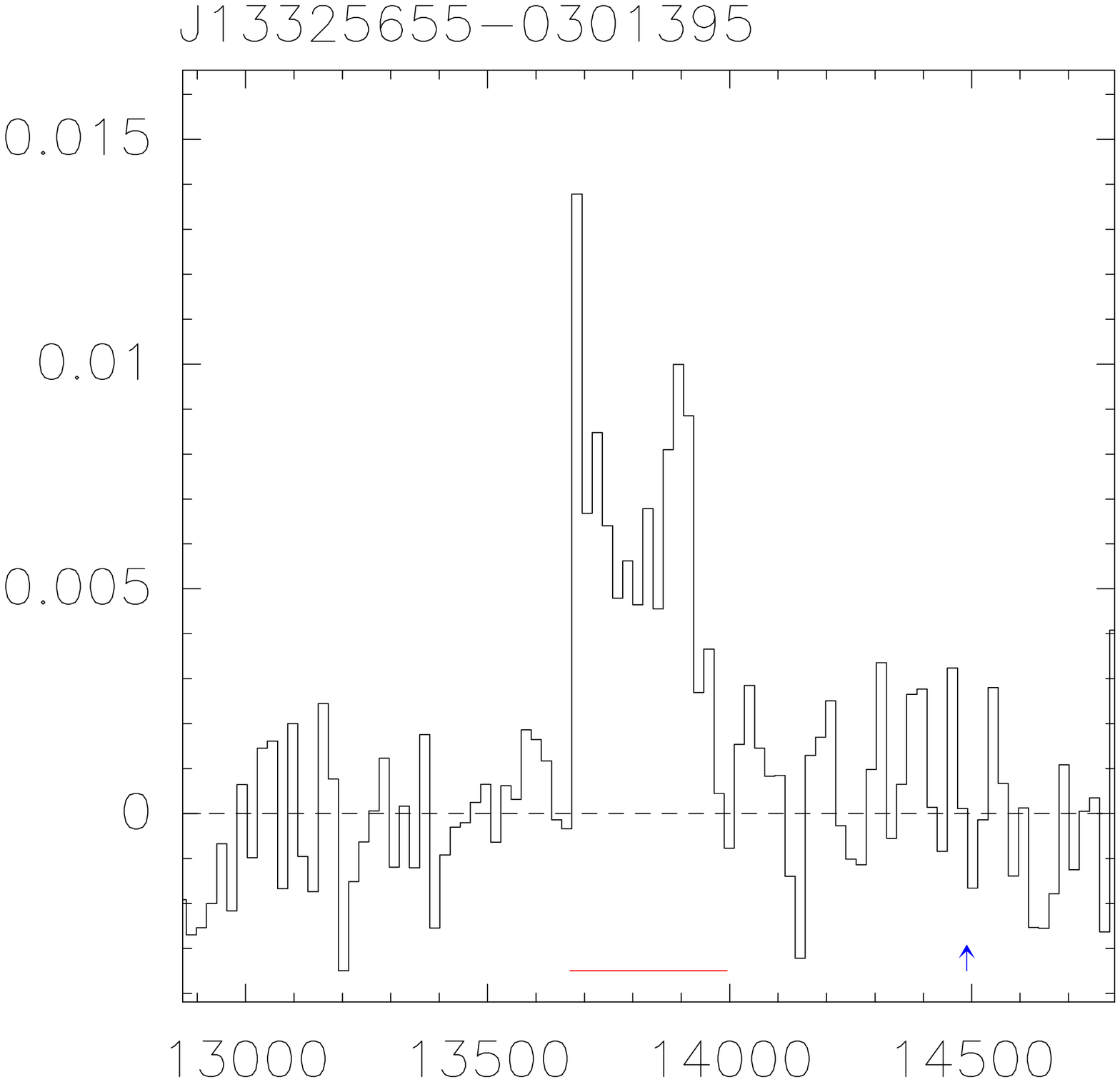}
\hspace{0.1cm}
\includegraphics[width=3.6cm,clip,trim = 0.cm 0.cm 0.cm 0.0cm, angle=-0]{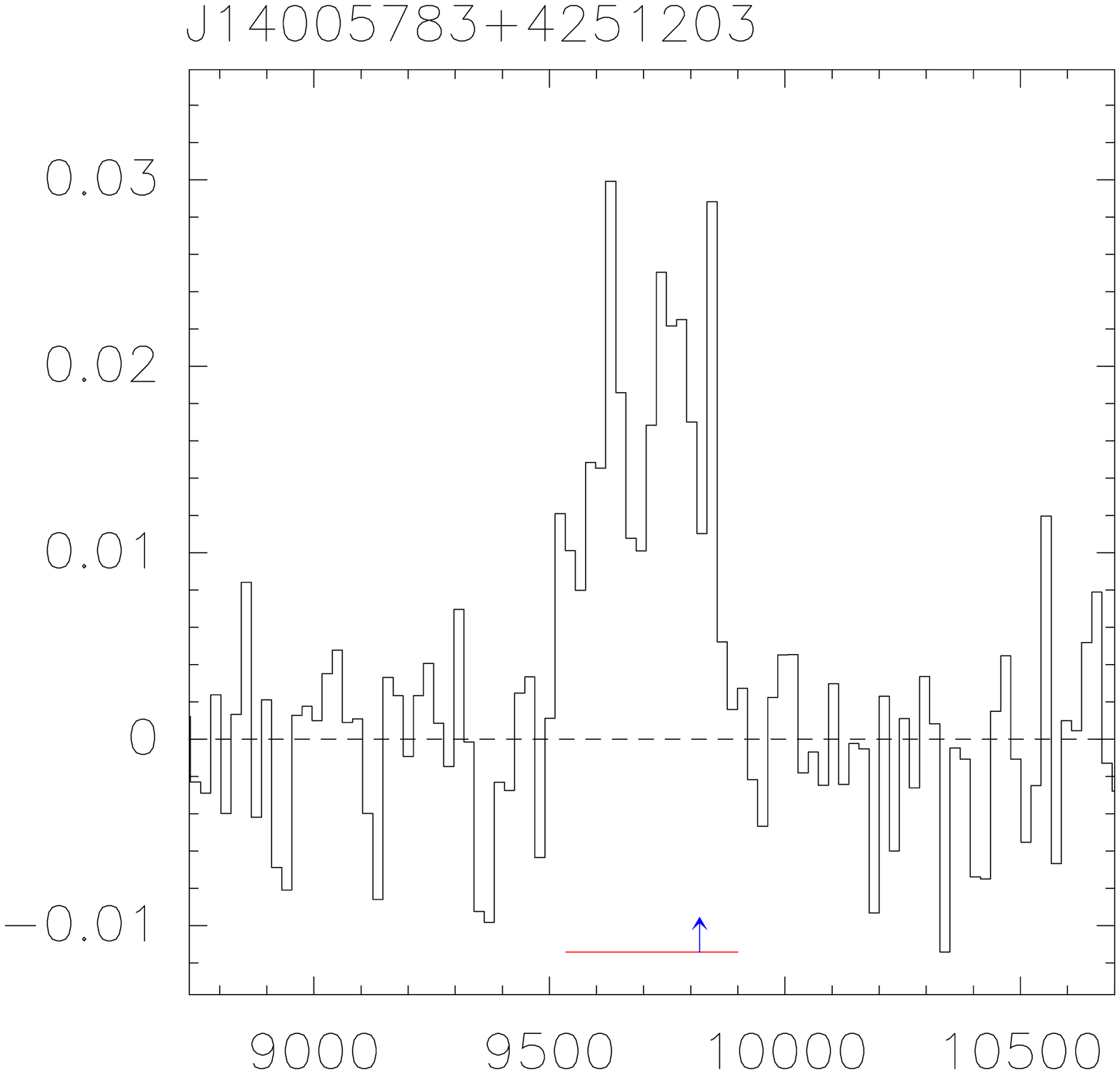}
\hspace{0.1cm}
\includegraphics[width=3.6cm,clip,trim = 0.cm 0.cm 0.cm 0.0cm, angle=-0]{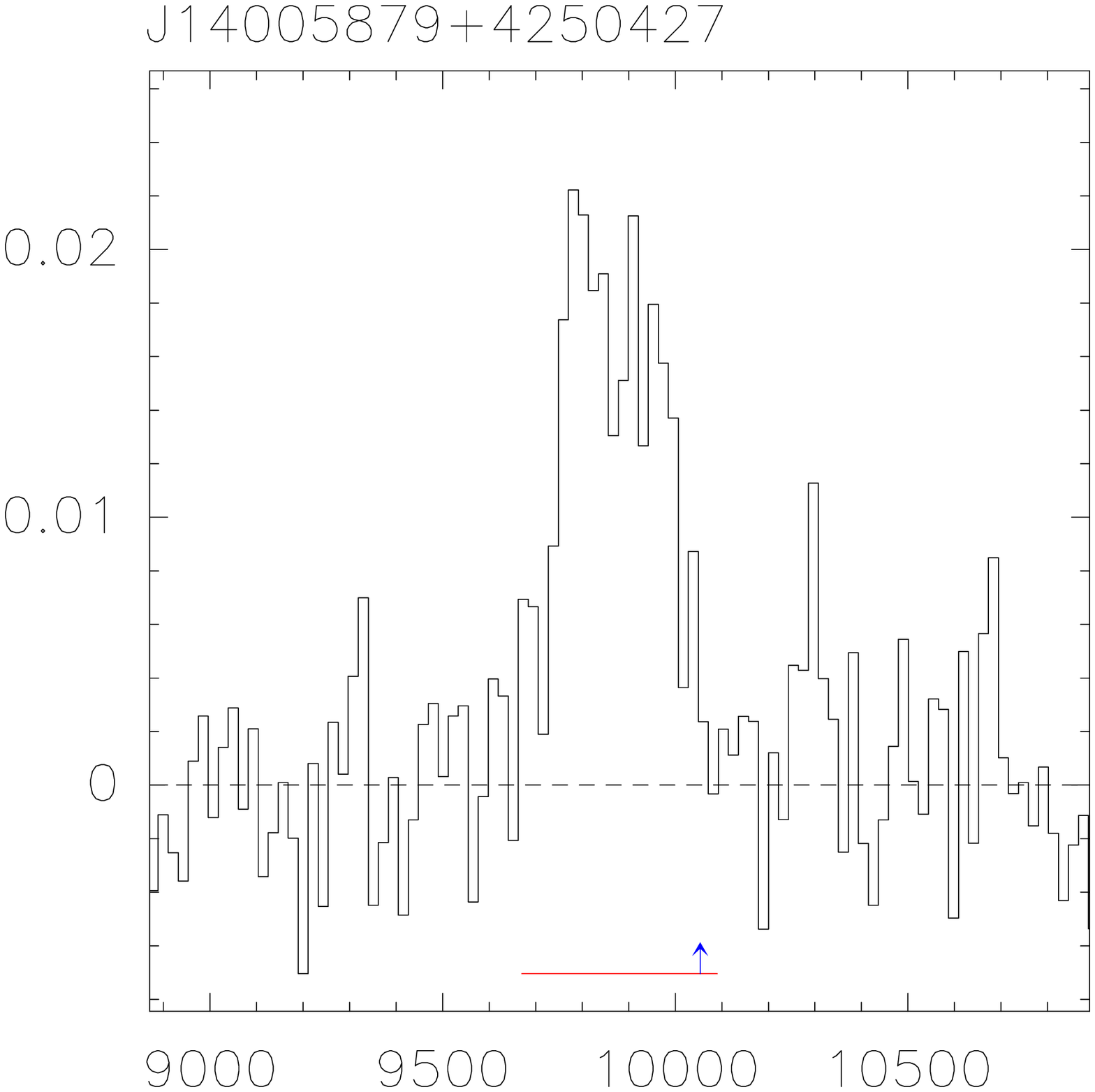}
\hspace{0.1cm}
\includegraphics[width=3.6cm,clip,trim = 0.cm 0.cm 0.cm 0.0cm,angle=-0]{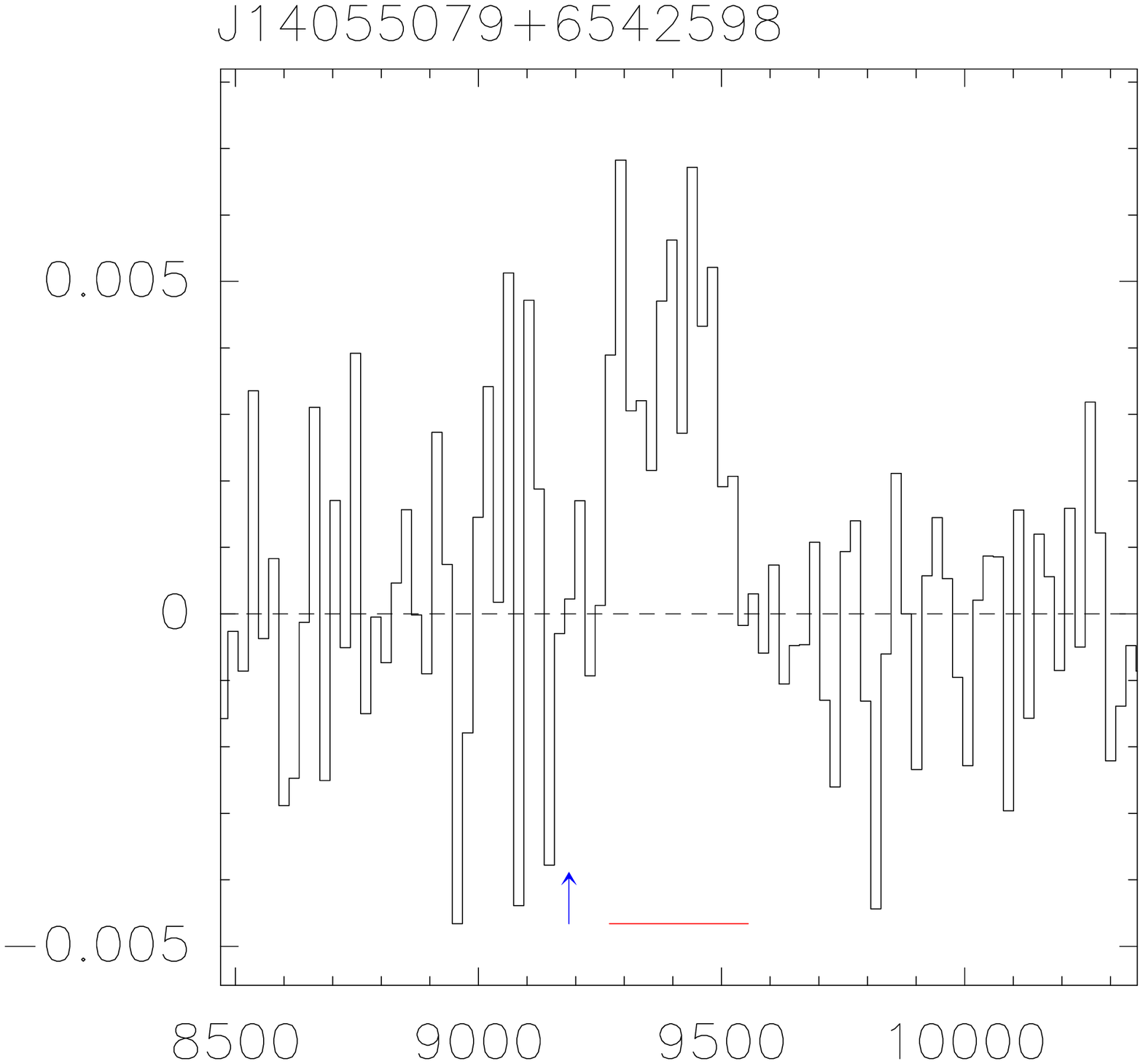}
}
\quad

\centerline{
\includegraphics[width=3.8cm,clip,trim = 0.cm 0.cm 0.cm 0.0cm, angle=-0]{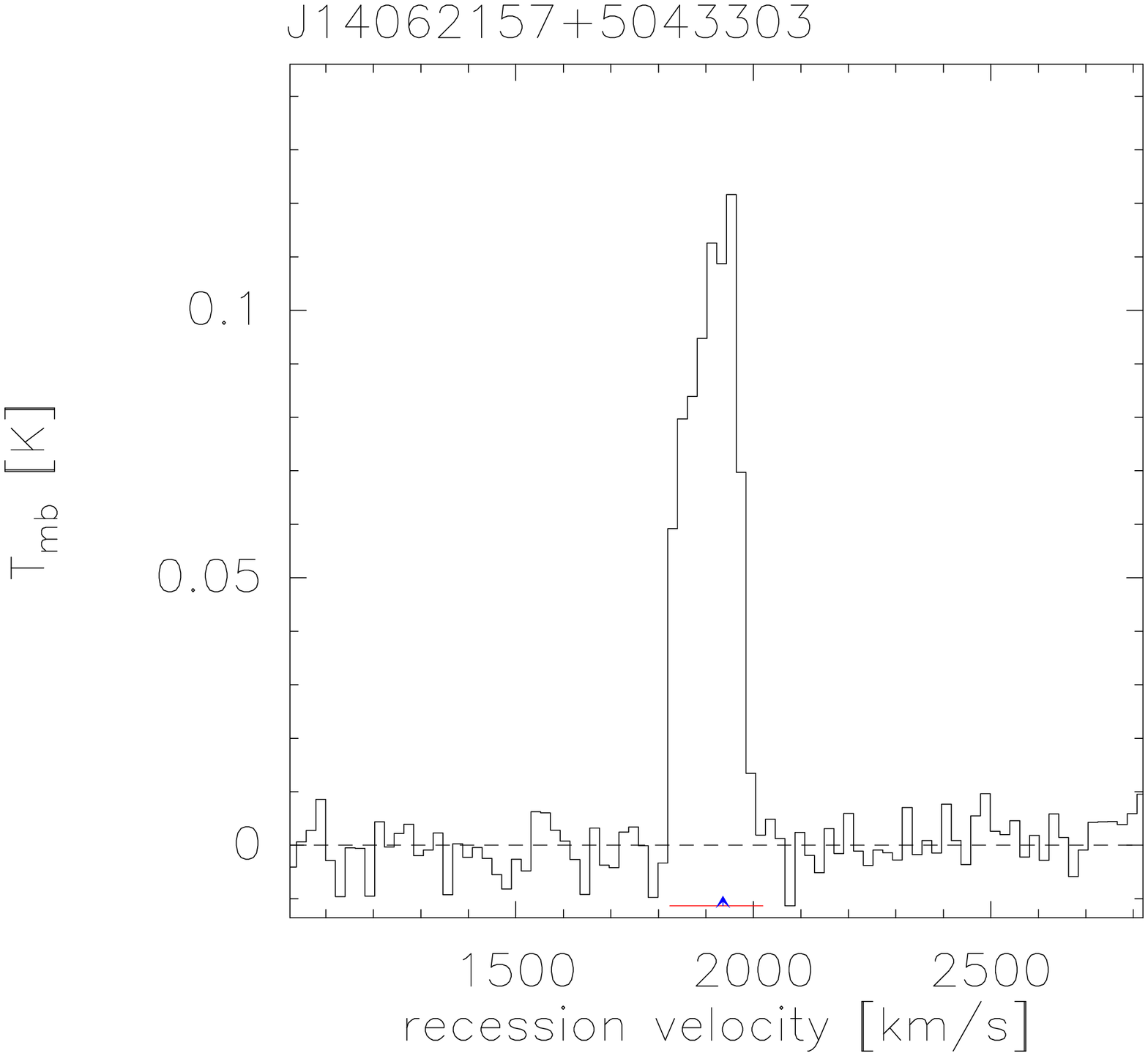}
\hspace{0.1cm}
\includegraphics[width=3.6cm,clip,trim = 0.cm 0.cm 0.cm 0.0cm, angle=-0]{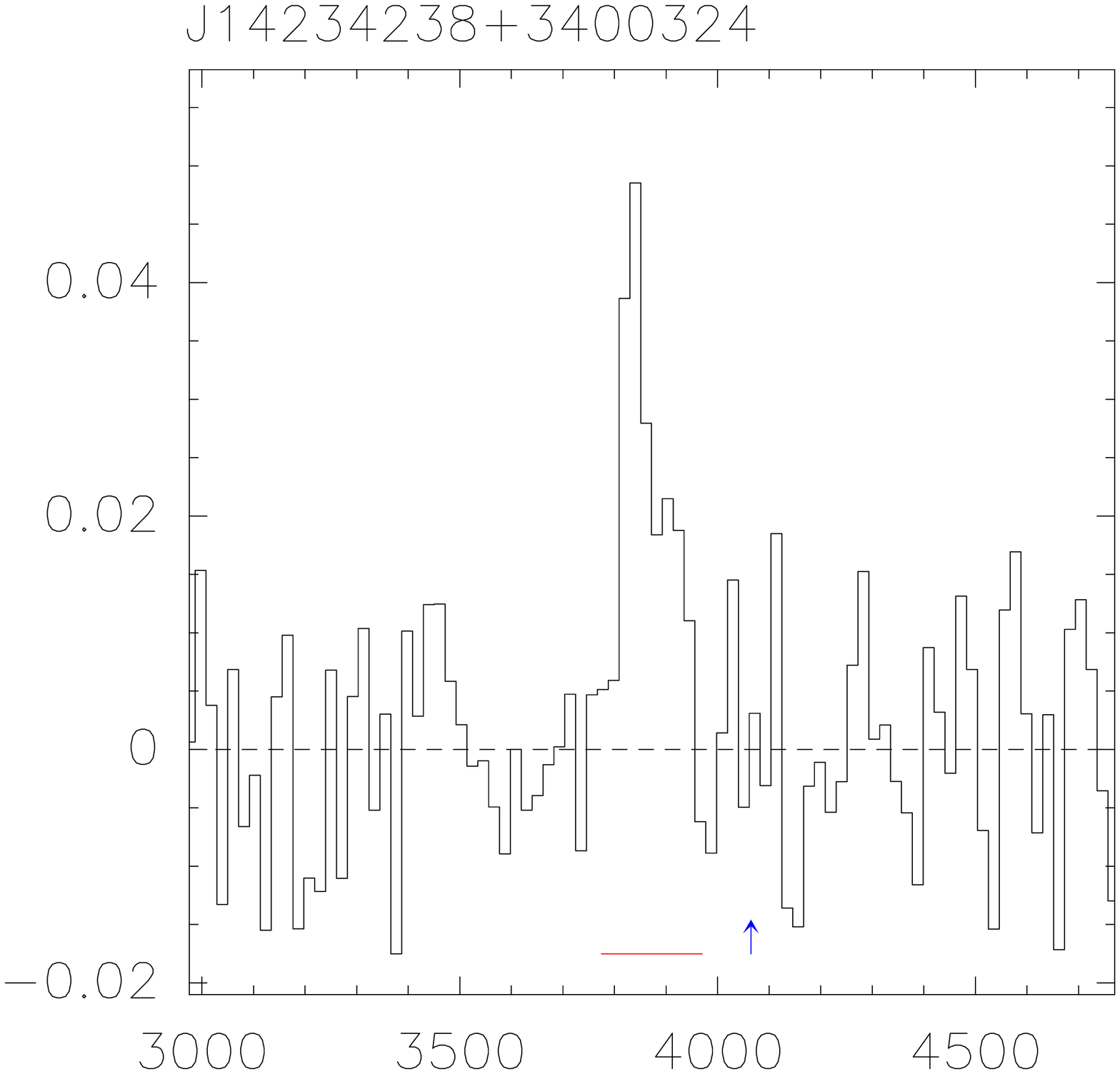}
\hspace{0.1cm}
\includegraphics[width=3.6cm,clip,trim = 0.cm 0.cm 0.cm 0.0cm, angle=-0]{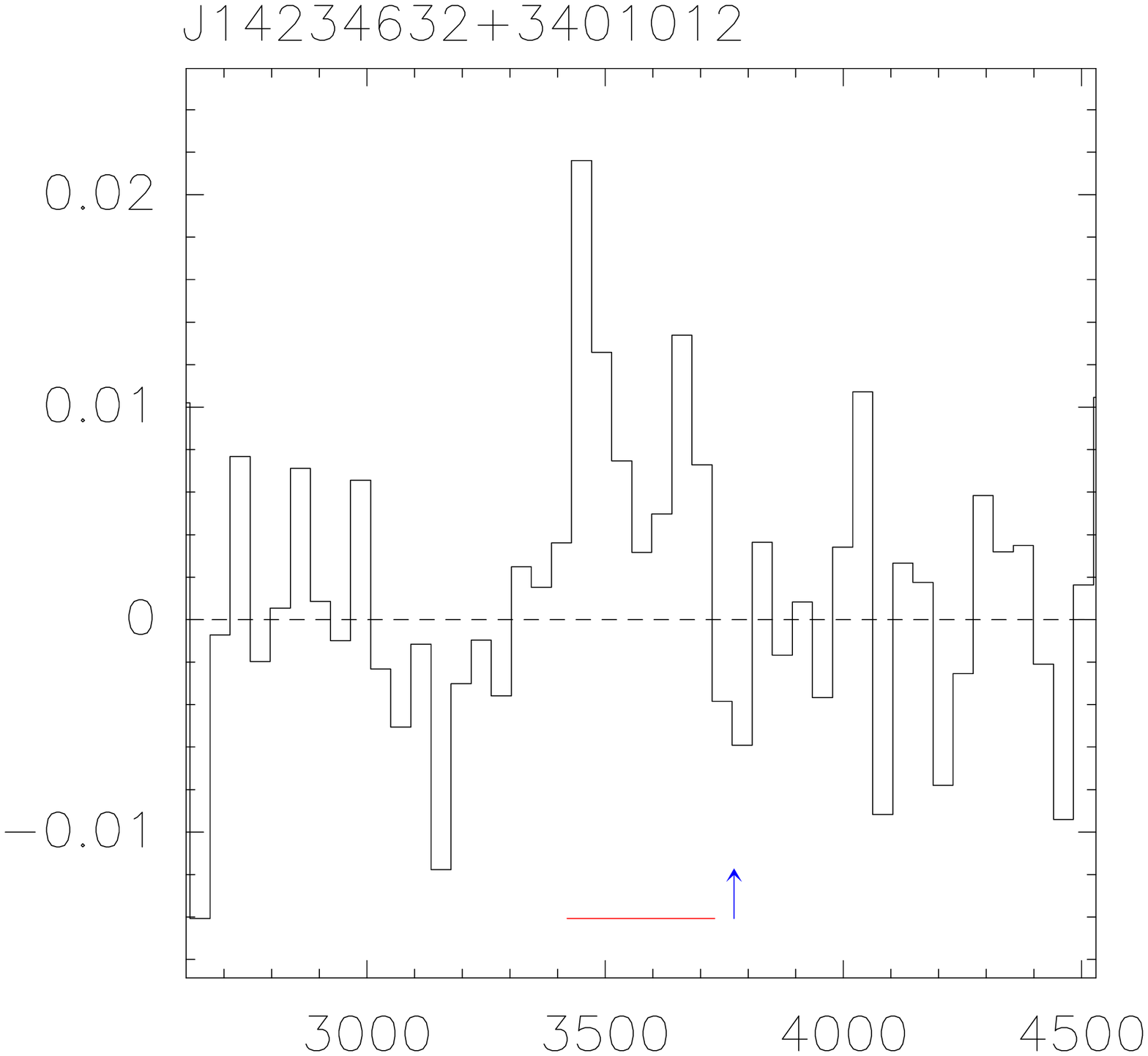}
\hspace{0.1cm}
\includegraphics[width=3.6cm,clip,trim = 0.cm 0.cm 0.cm 0.0cm,angle=-0]{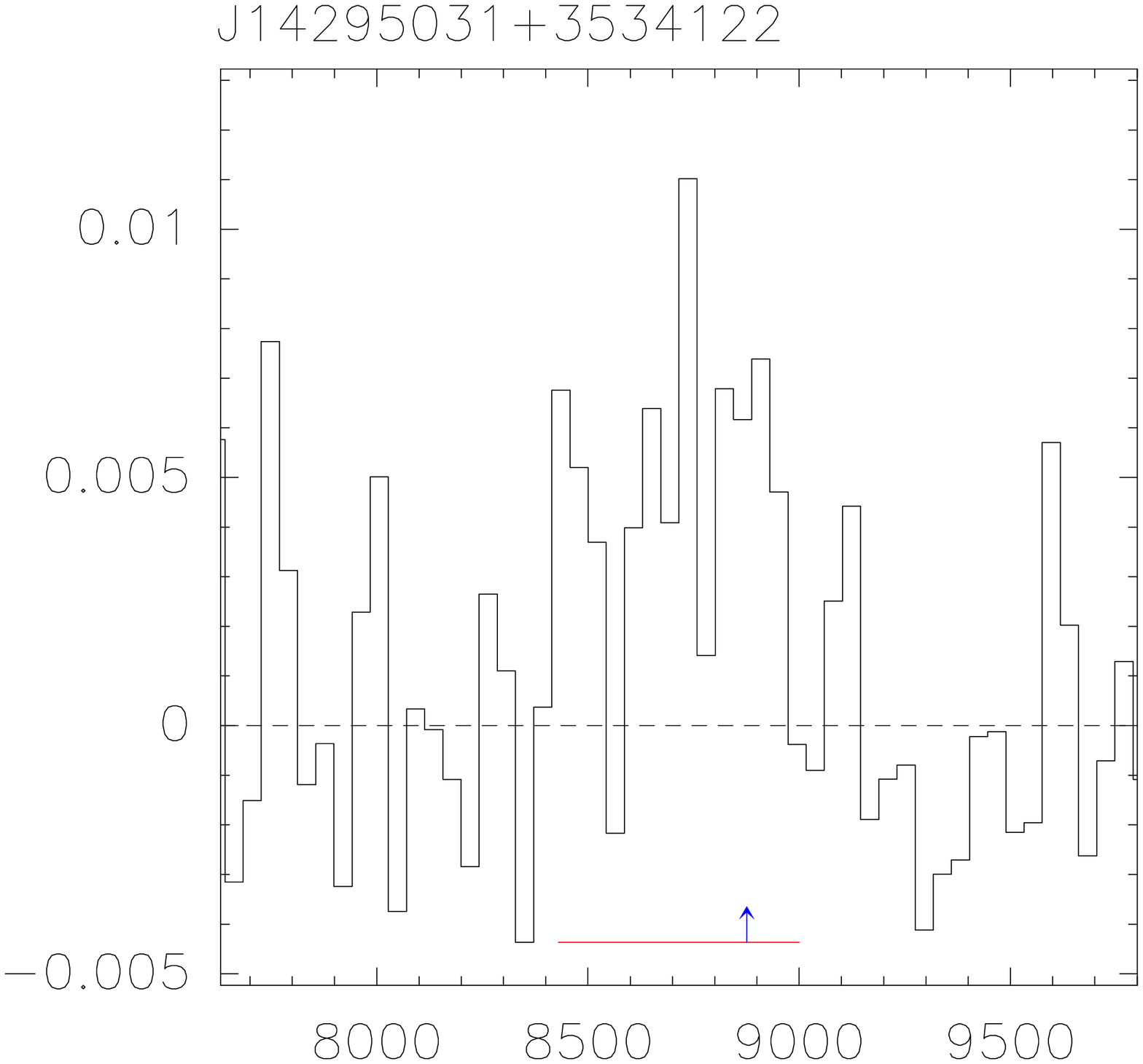}
}

\caption{Continued.}
\end{figure*}

\setcounter{figure}{0} 
\begin{figure*}

\centerline{
\includegraphics[width=3.6cm,clip,trim = 0.cm 0.cm 0.cm 0.0cm, angle=-0]{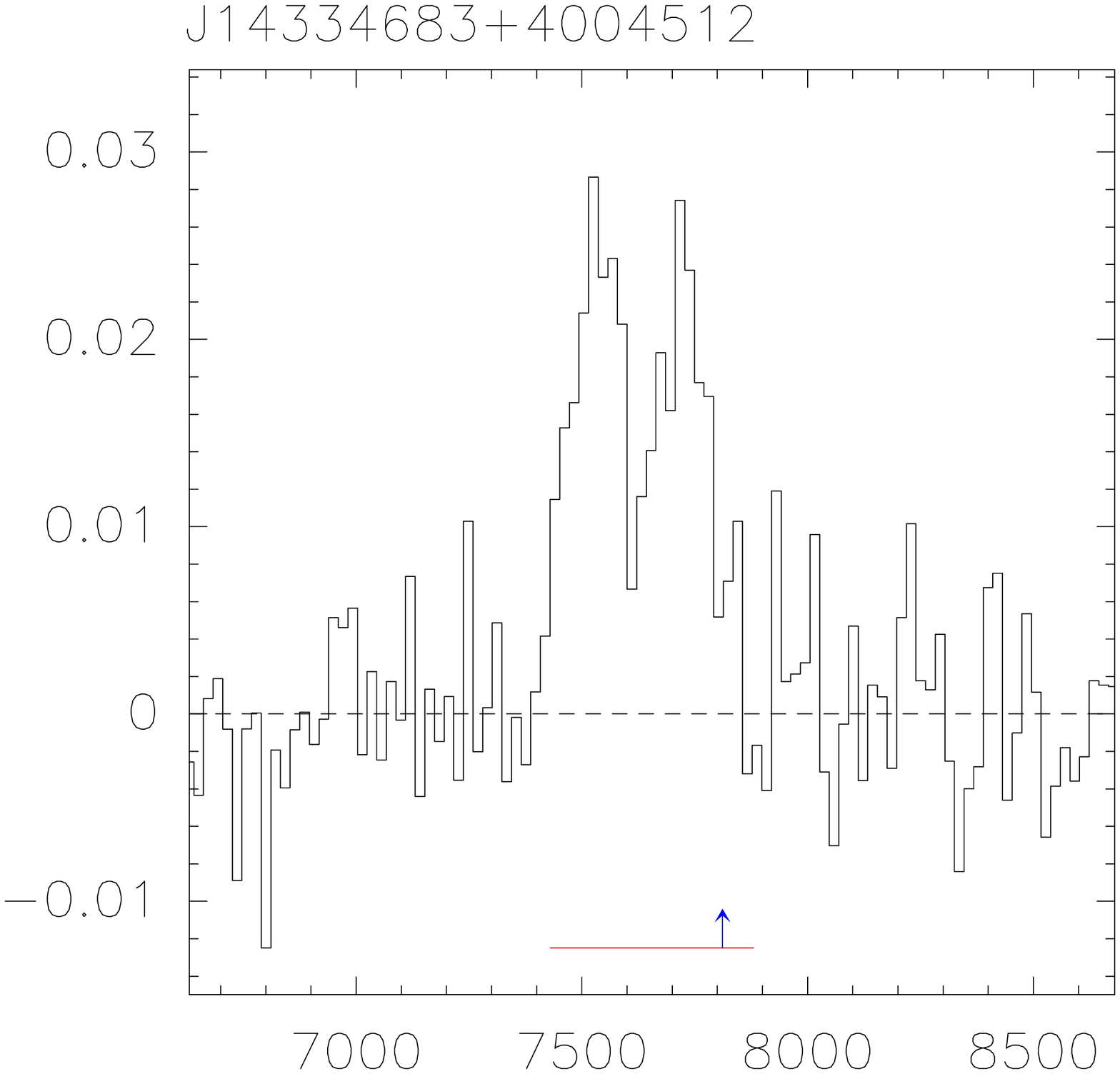}
\hspace{0.1cm}
\includegraphics[width=3.6cm,clip,trim = 0.cm 0.cm 0.cm 0.0cm, angle=-0]{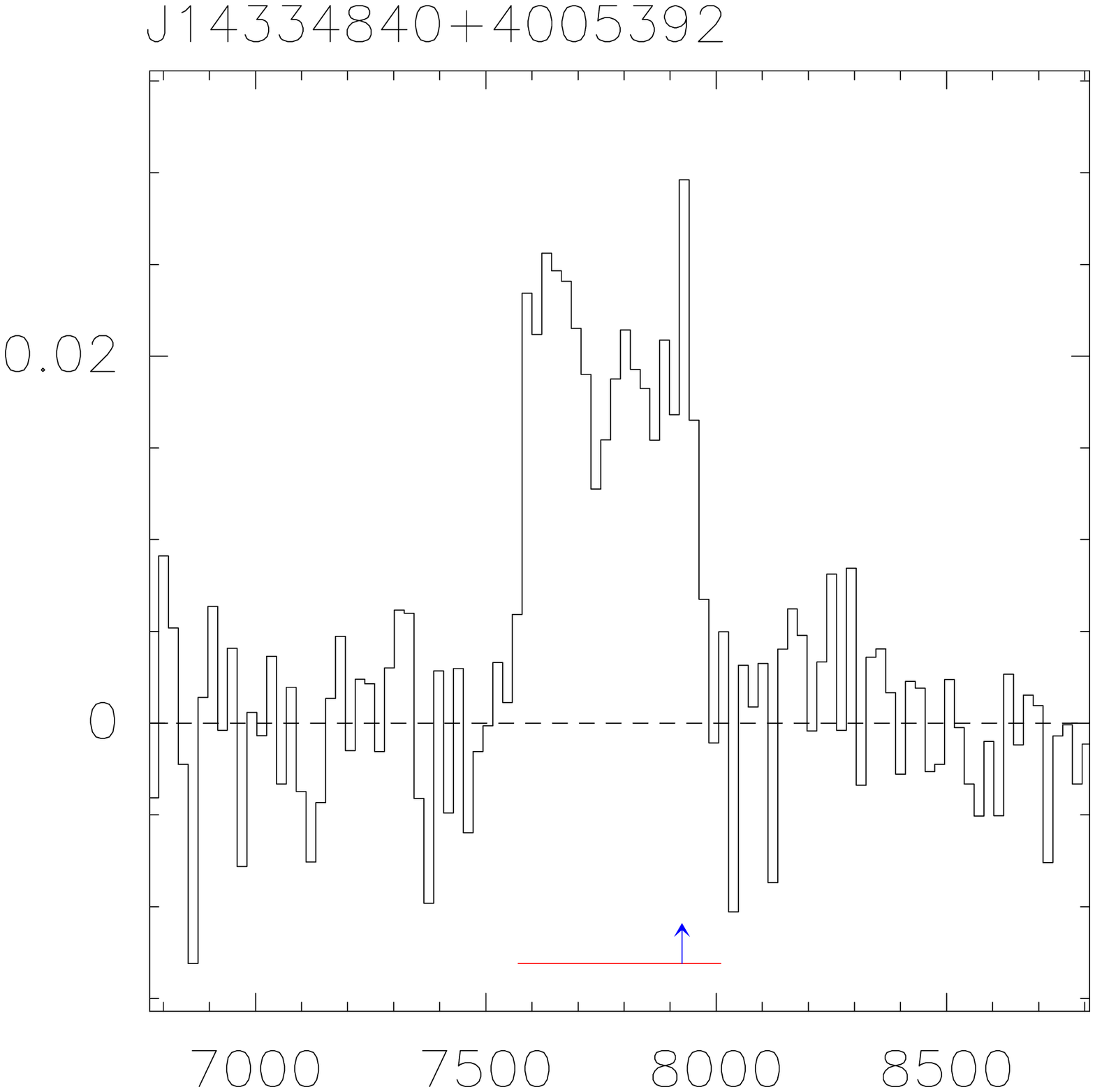}
\hspace{0.1cm}
\includegraphics[width=3.6cm,clip,trim = 0.cm 0.cm 0.cm 0.0cm, angle=-0]{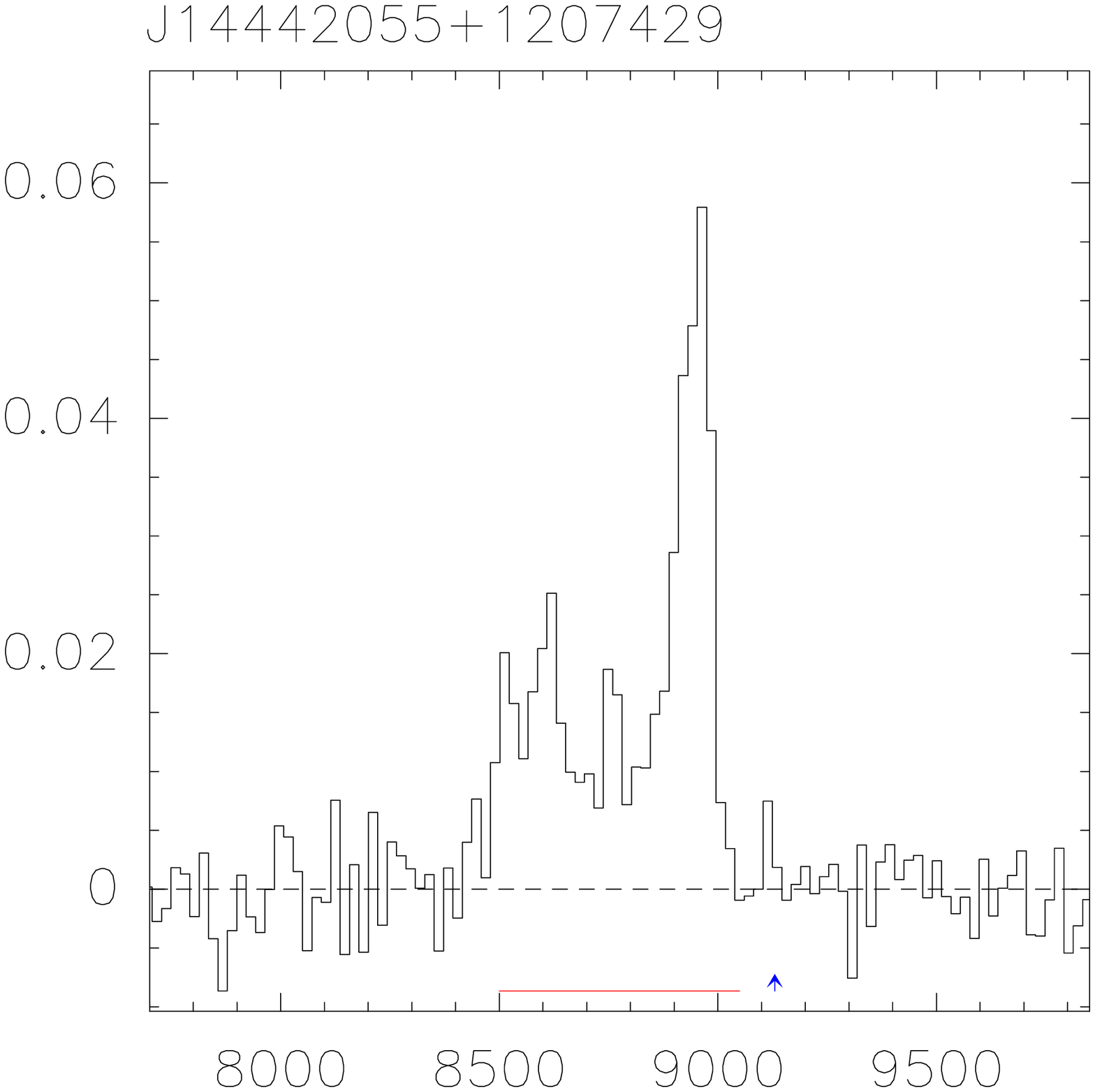}
\hspace{0.1cm}
\includegraphics[width=3.6cm,clip,trim = 0.cm 0.cm 0.cm 0.0cm,angle=-0]{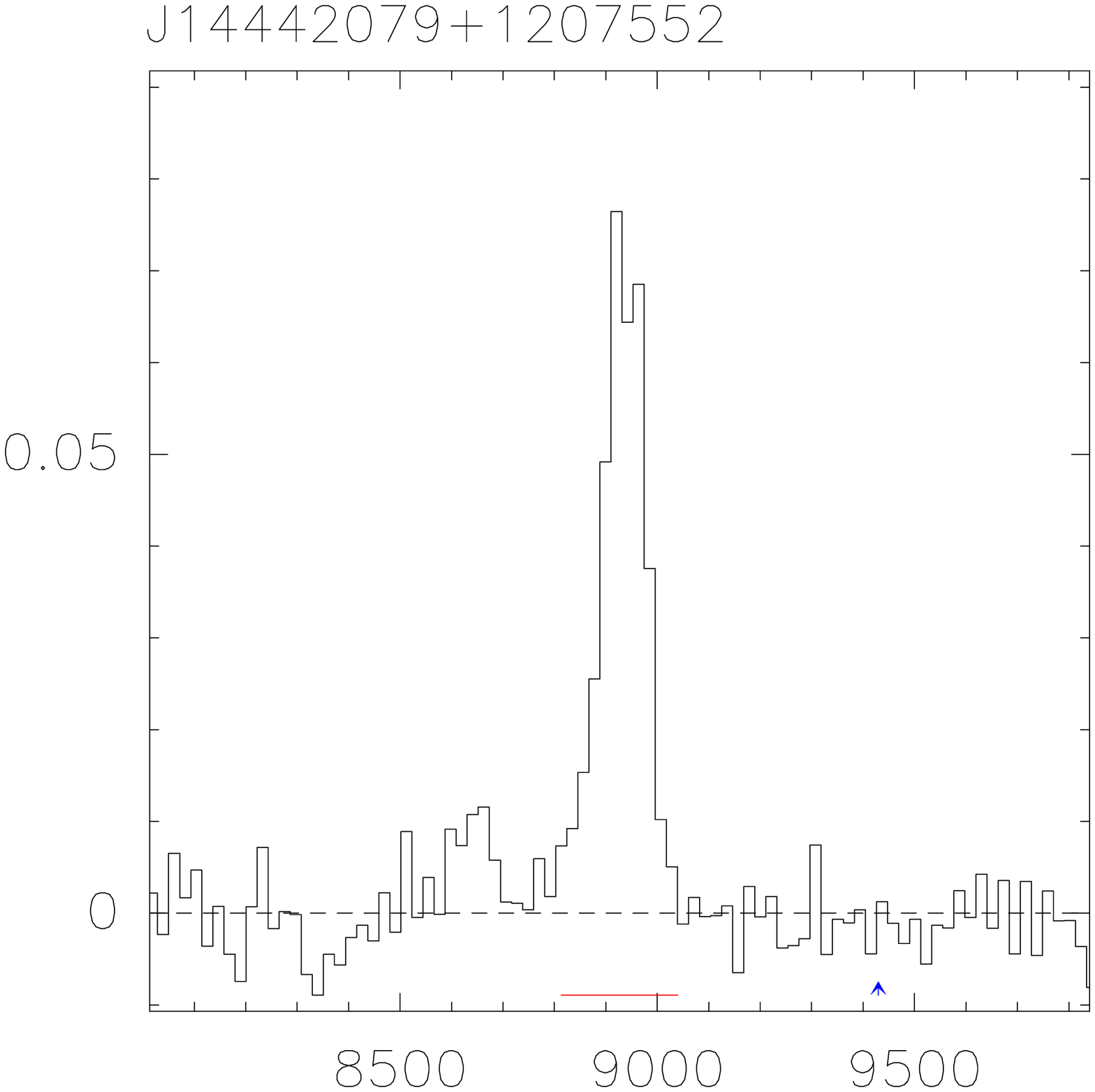}
}

\centerline{
\hspace{0.1cm}
\includegraphics[width=3.6cm,clip,trim = 0.cm 0.cm 0.cm 0.0cm, angle=-0]{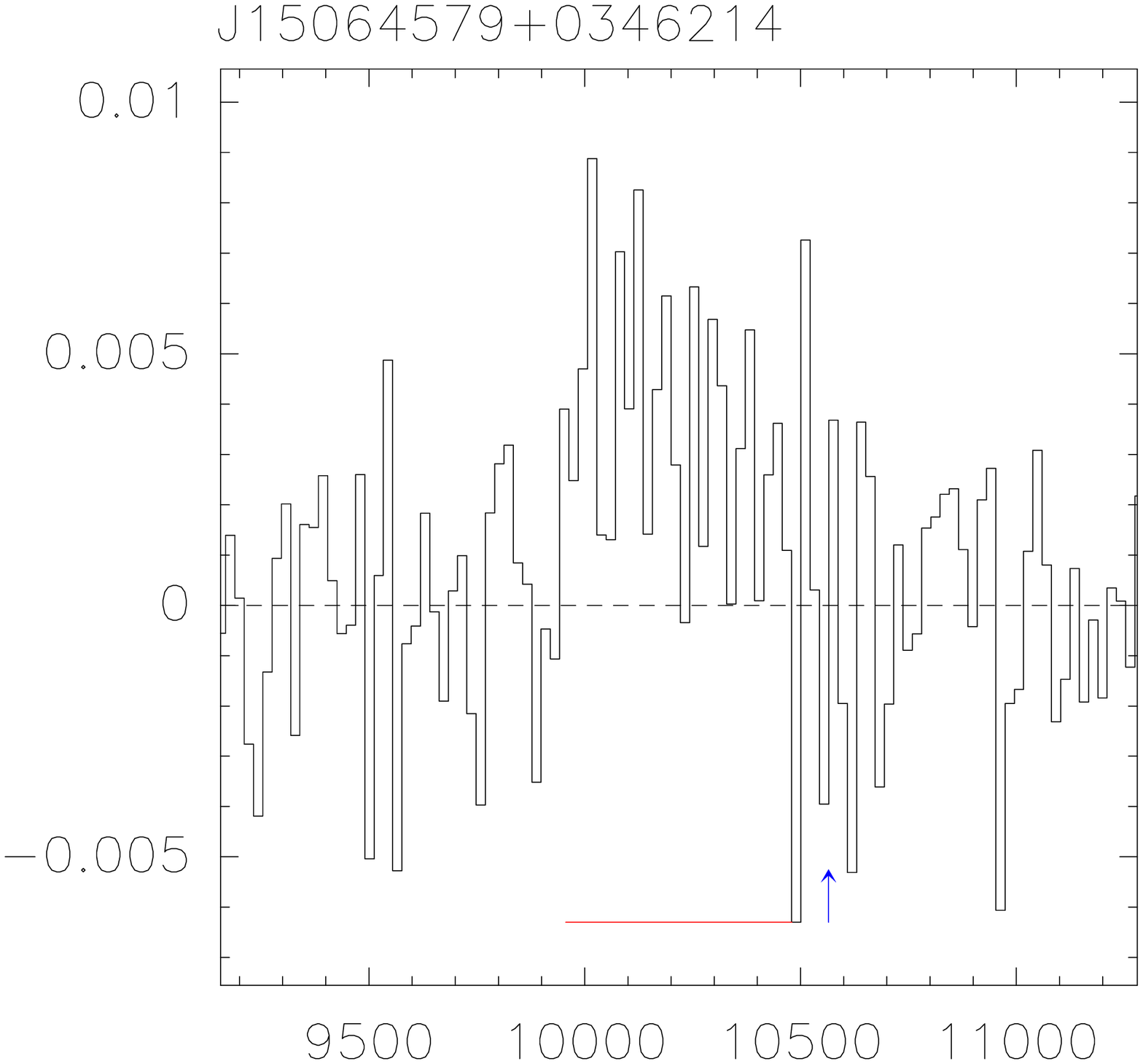}
\hspace{0.1cm}
\includegraphics[width=3.6cm,clip,trim = 0.cm 0.cm 0.cm 0.0cm, angle=-0]{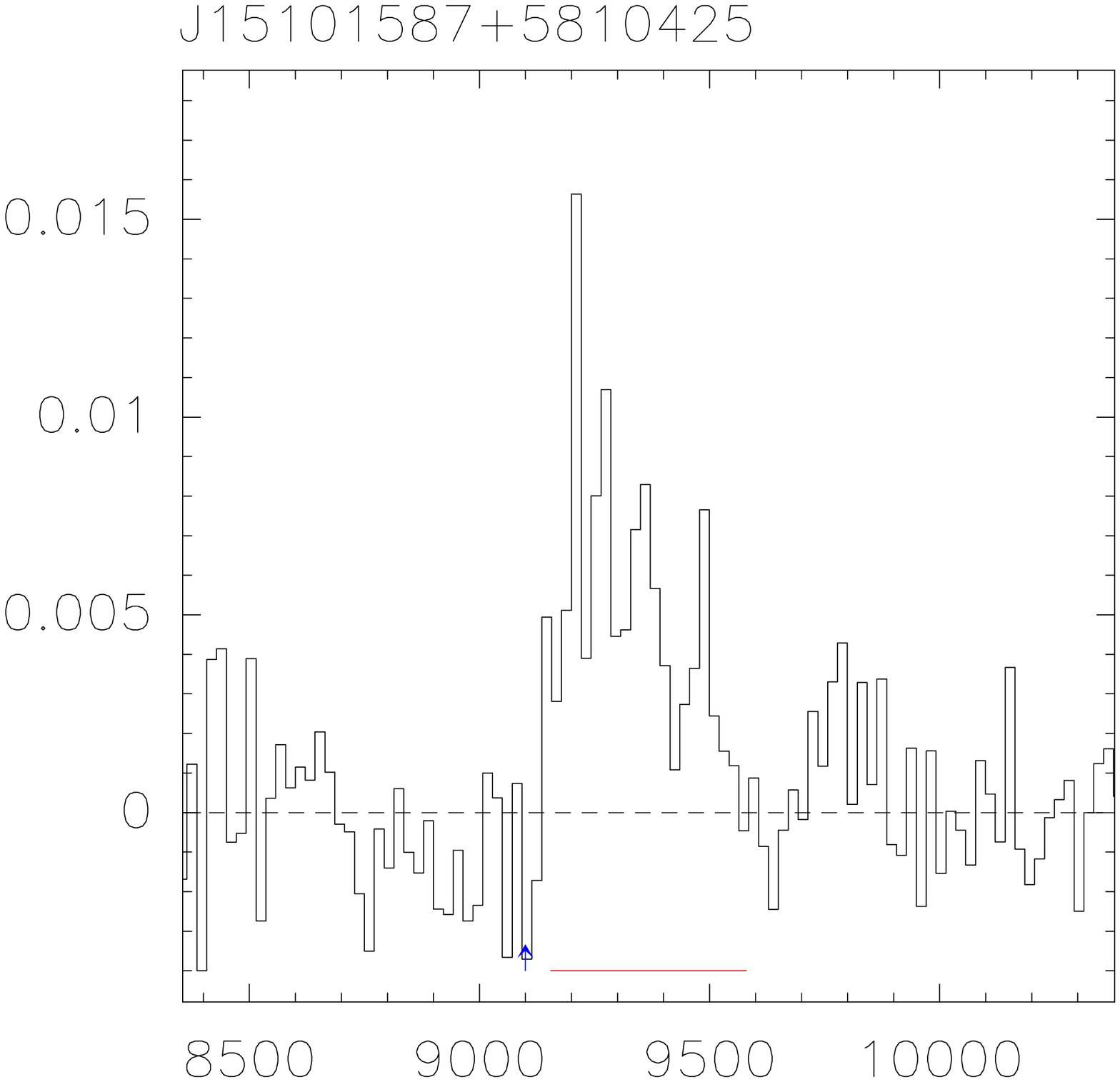}
\hspace{0.1cm}
\includegraphics[width=3.6cm,clip,trim = 0.cm 0.cm 0.cm 0.0cm,angle=-0]{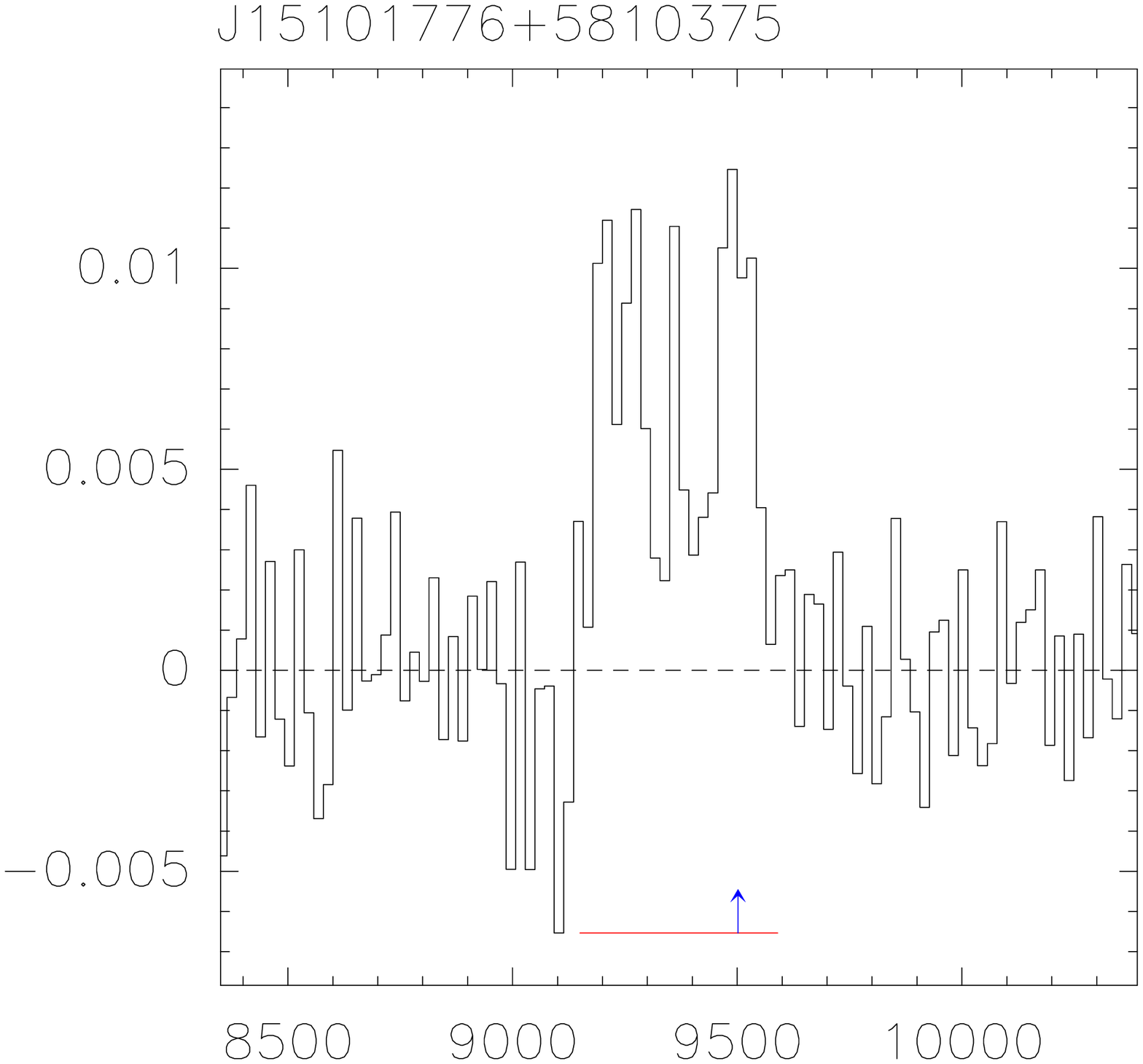}
\hspace{0.1cm}
\includegraphics[width=3.6cm,clip,trim = 0.cm 0.cm 0.cm 0.0cm, angle=-0]{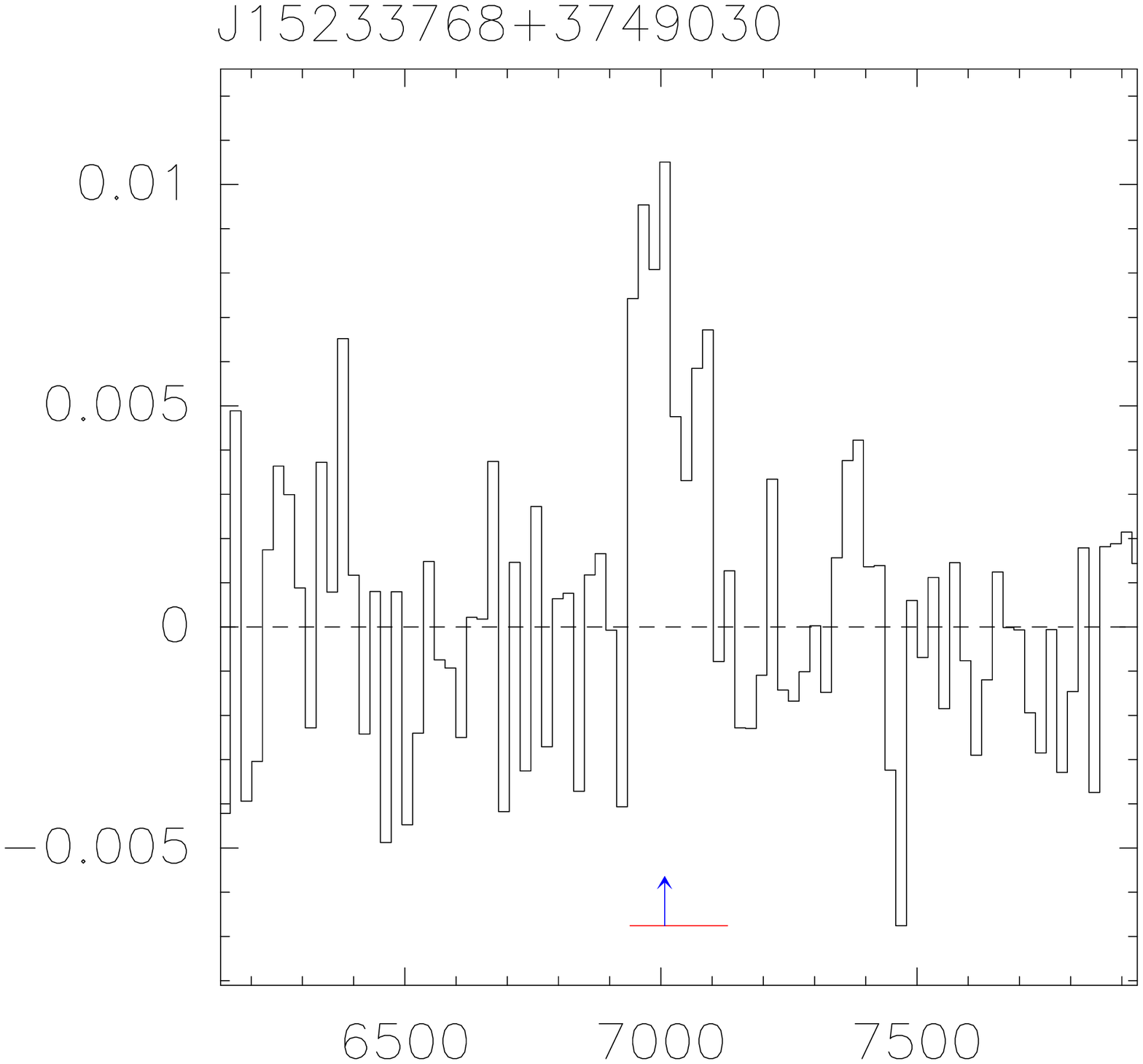}
}

\centerline{
\includegraphics[width=3.6cm,clip,trim = 0.cm 0.cm 0.cm 0.0cm, angle=-0]{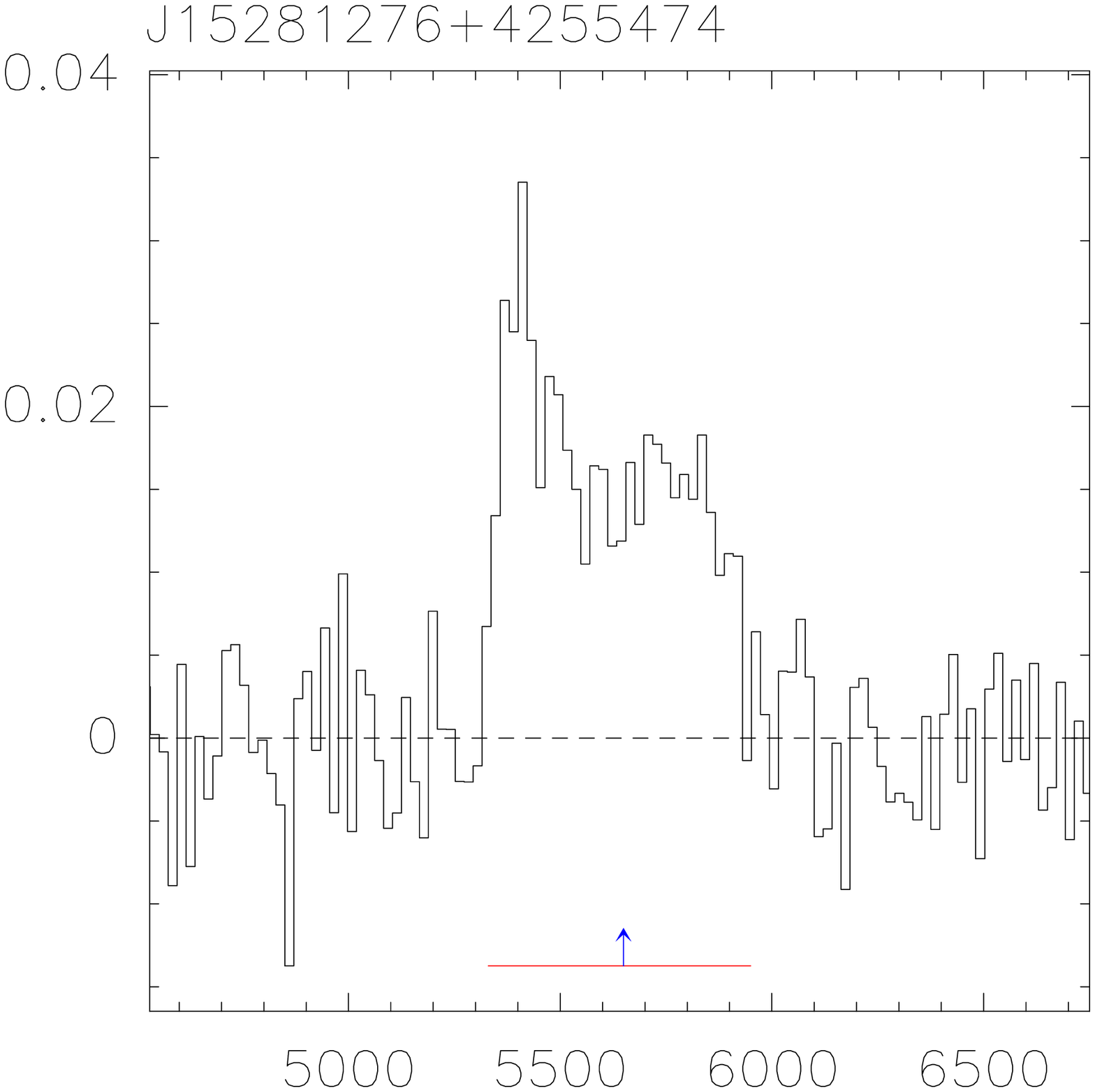}
\hspace{0.1cm}
\includegraphics[width=3.6cm,clip,trim = 0.cm 0.cm 0.cm 0.0cm, angle=-0]{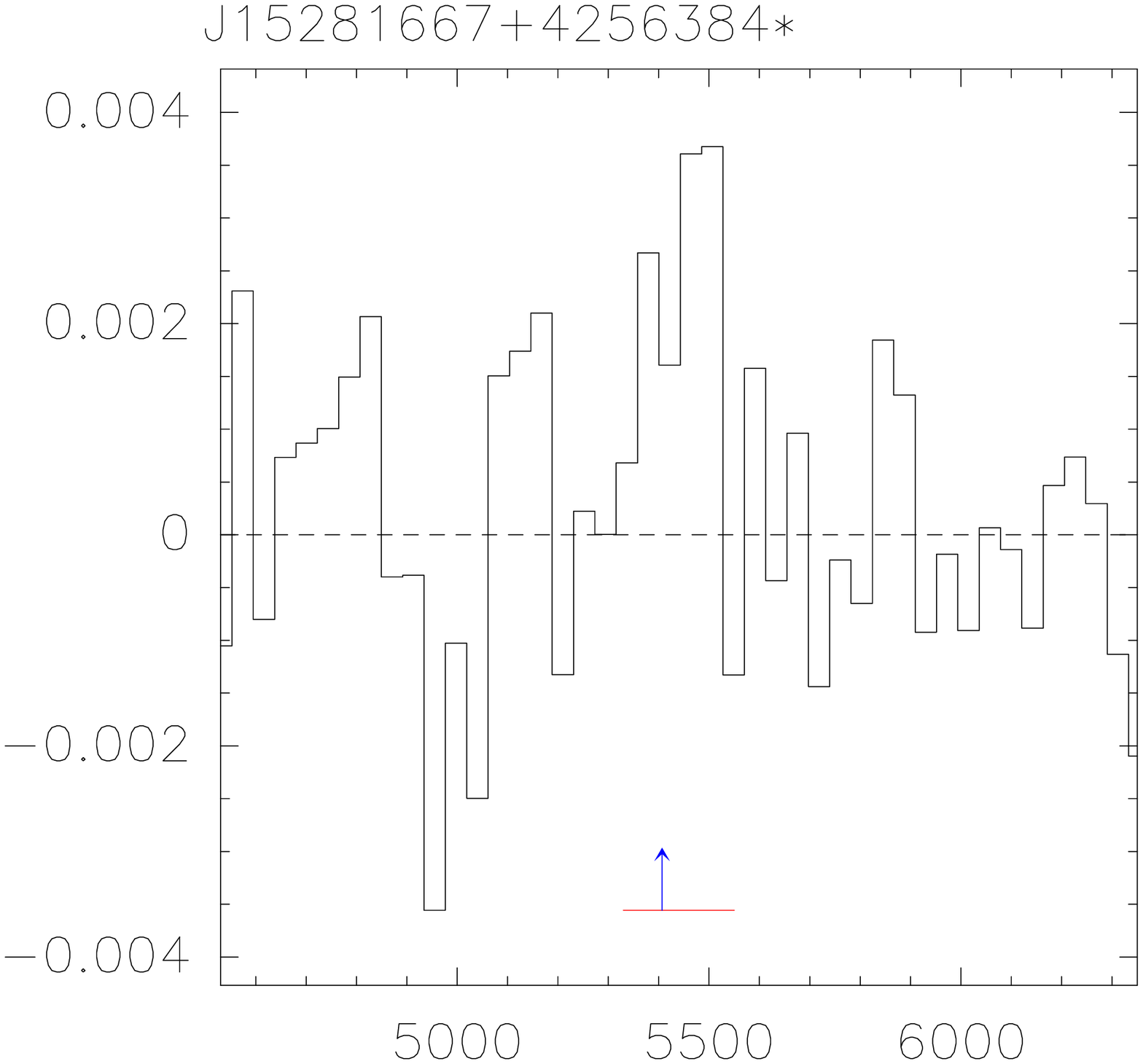}
\hspace{0.1cm}
\includegraphics[width=3.6cm,clip,trim = 0.cm 0.cm 0.cm 0.0cm,angle=-0]{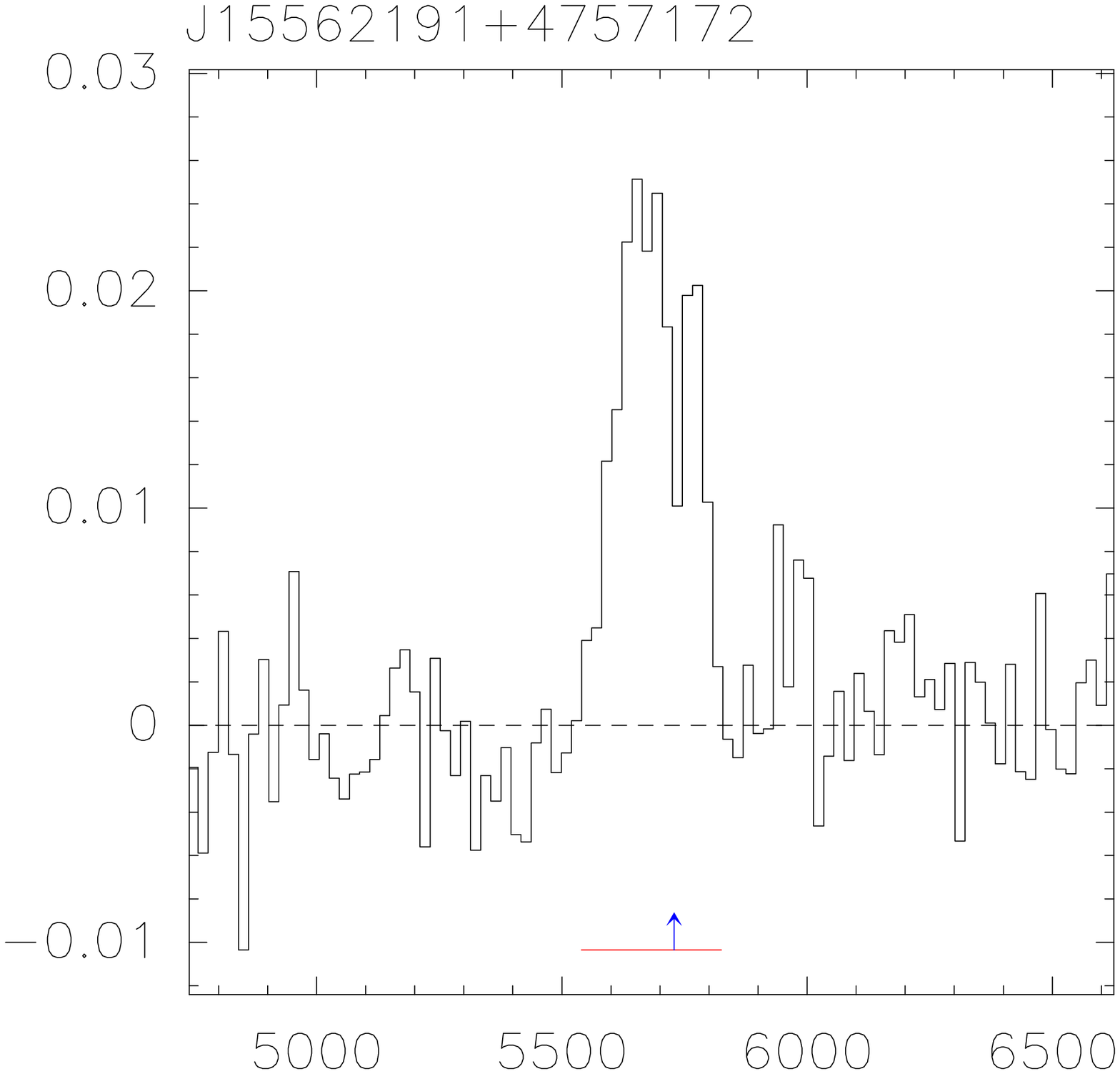}
\hspace{0.1cm}
\includegraphics[width=3.6cm,clip,trim = 0.cm 0.cm 0.cm 0.0cm, angle=-0]{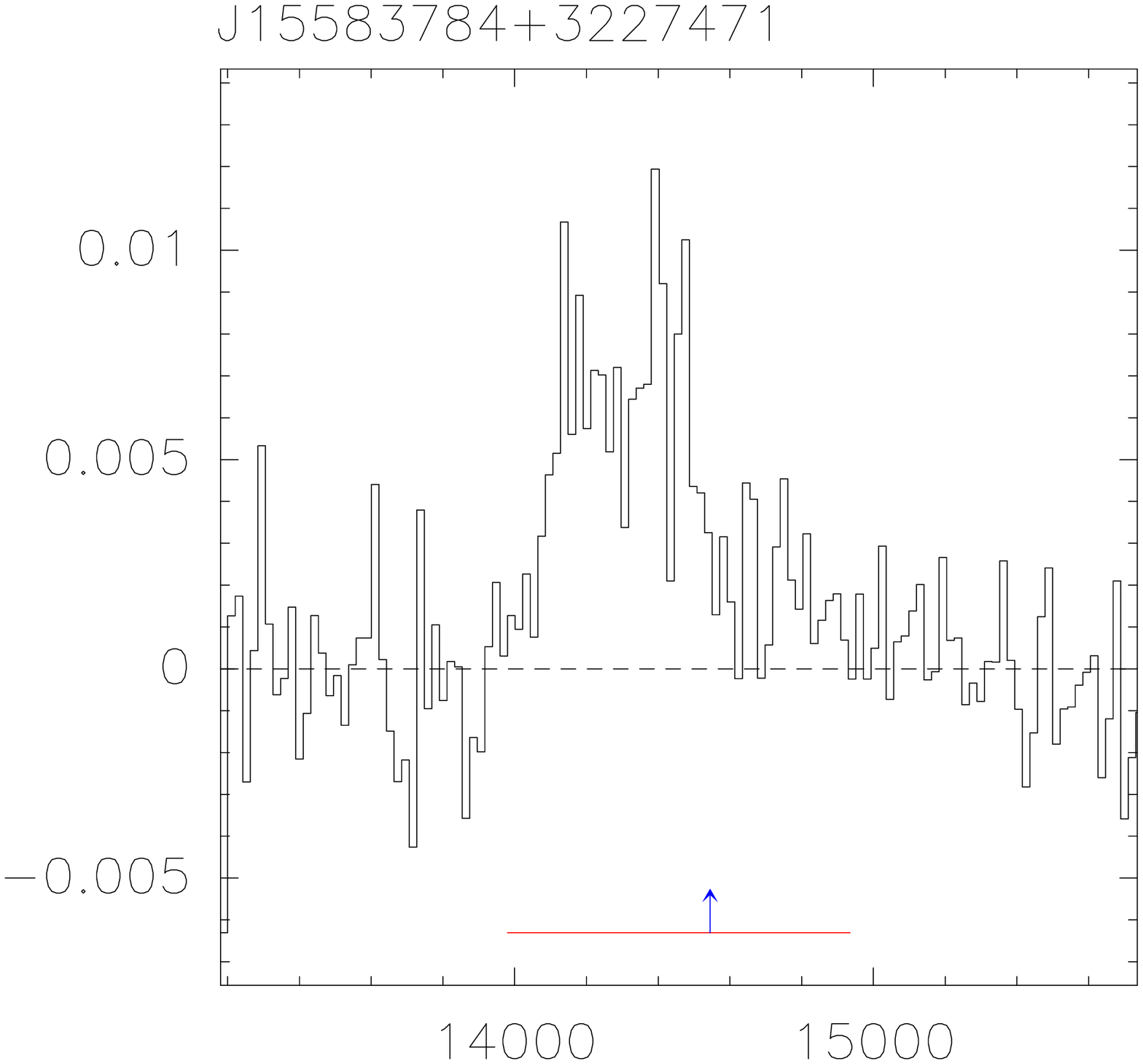}
}
\quad

\centerline{
\includegraphics[width=3.6cm,clip,trim = 0.cm 0.cm 0.cm 0.0cm, angle=-0]{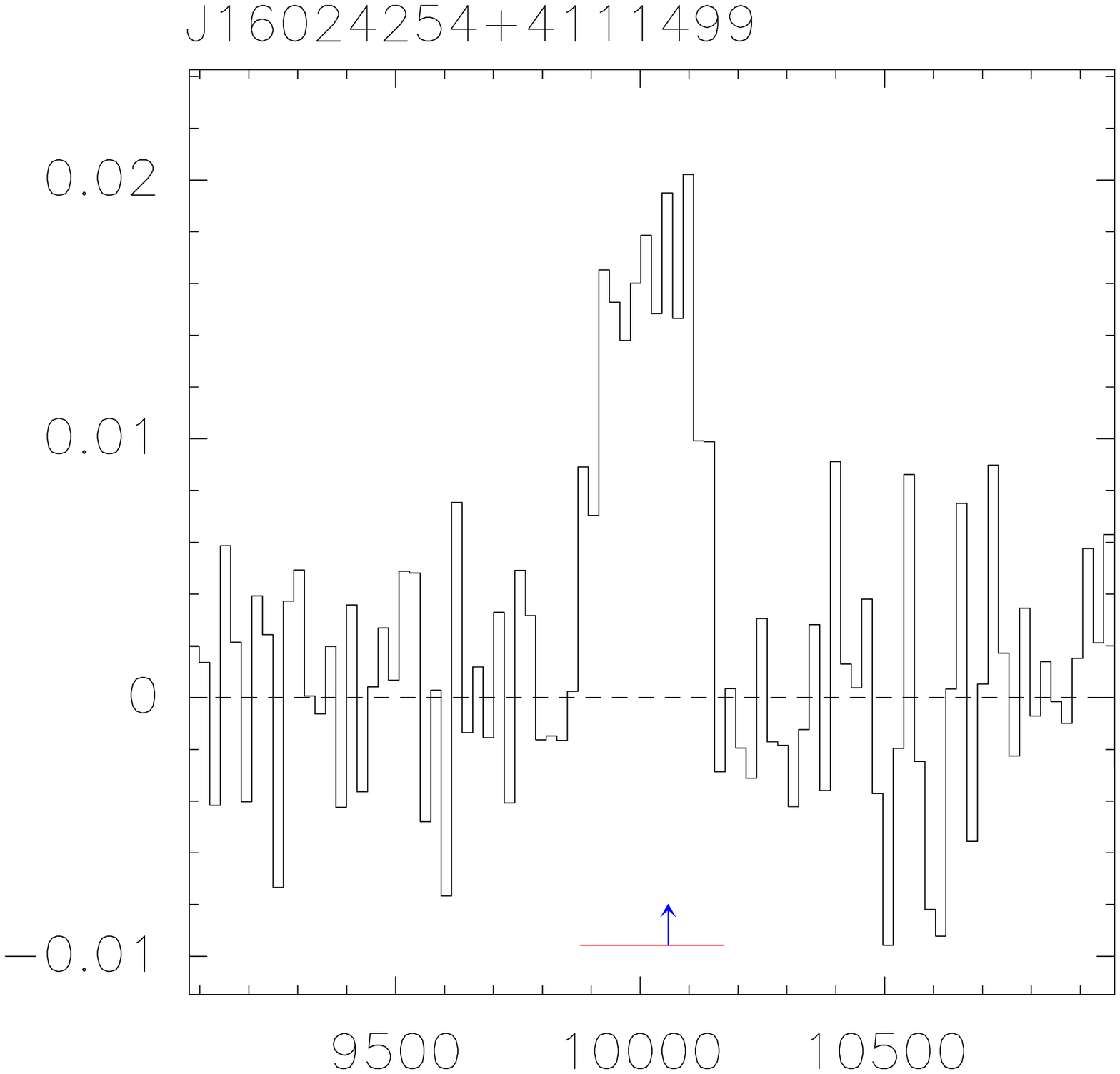}
\hspace{0.1cm}
\includegraphics[width=3.6cm,clip,trim = 0.cm 0.cm 0.cm 0.0cm, angle=-0]{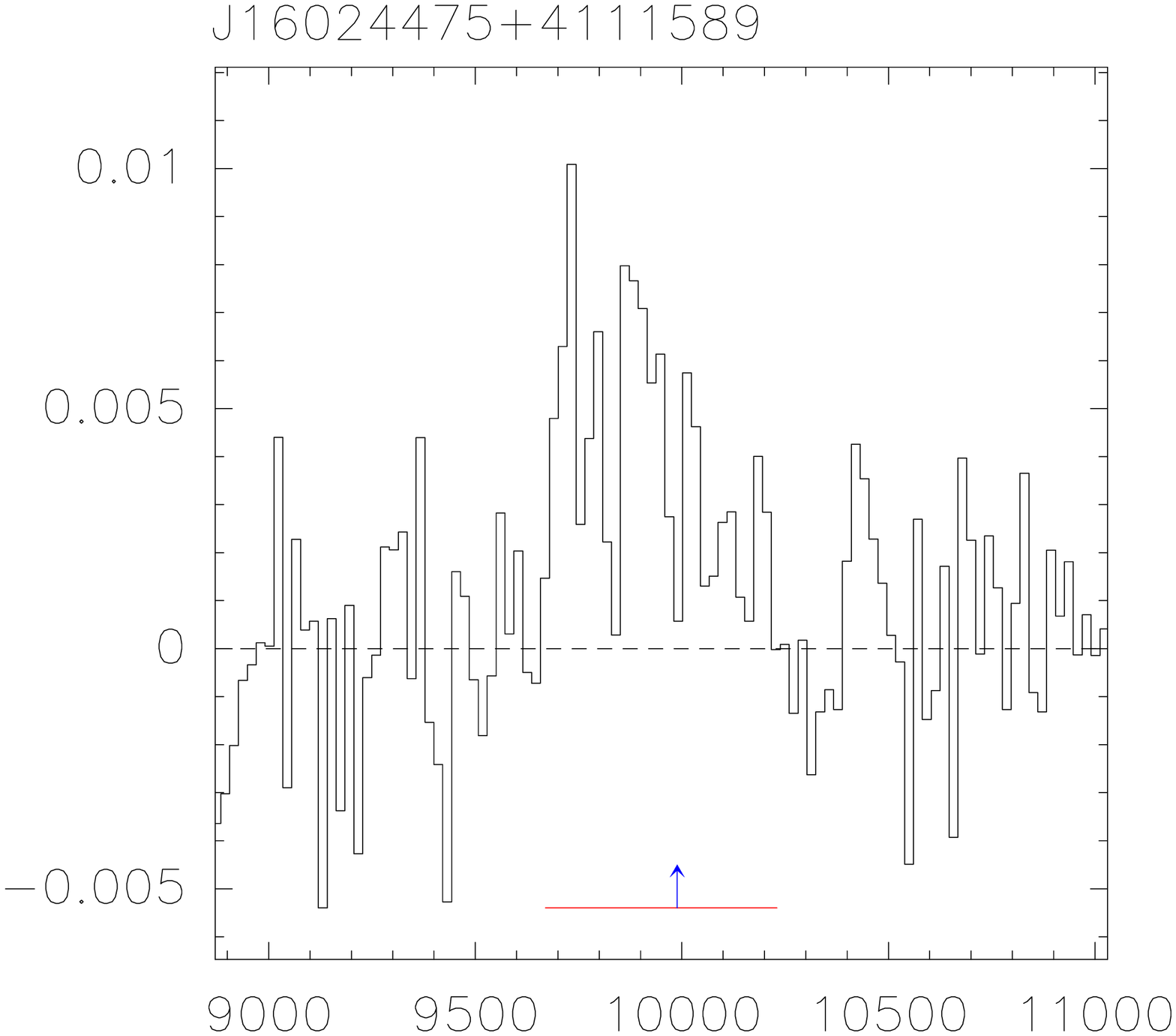}
\hspace{0.1cm}
\includegraphics[width=3.6cm,clip,trim = 0.cm 0.cm 0.cm 0.0cm,angle=-0]{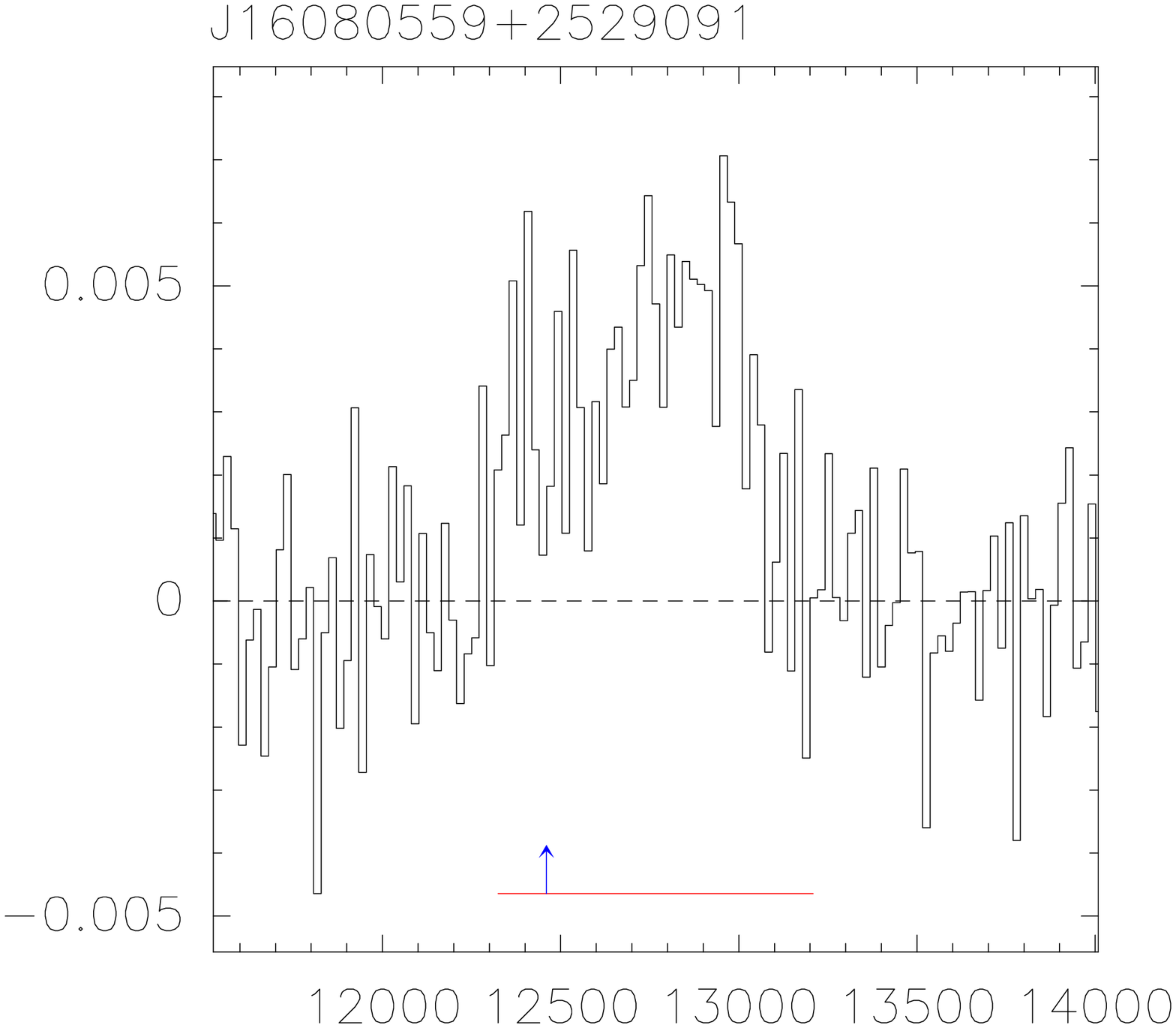}
\hspace{0.1cm}
\includegraphics[width=3.6cm,clip,trim = 0.cm 0.cm 0.cm 0.0cm, angle=-0]{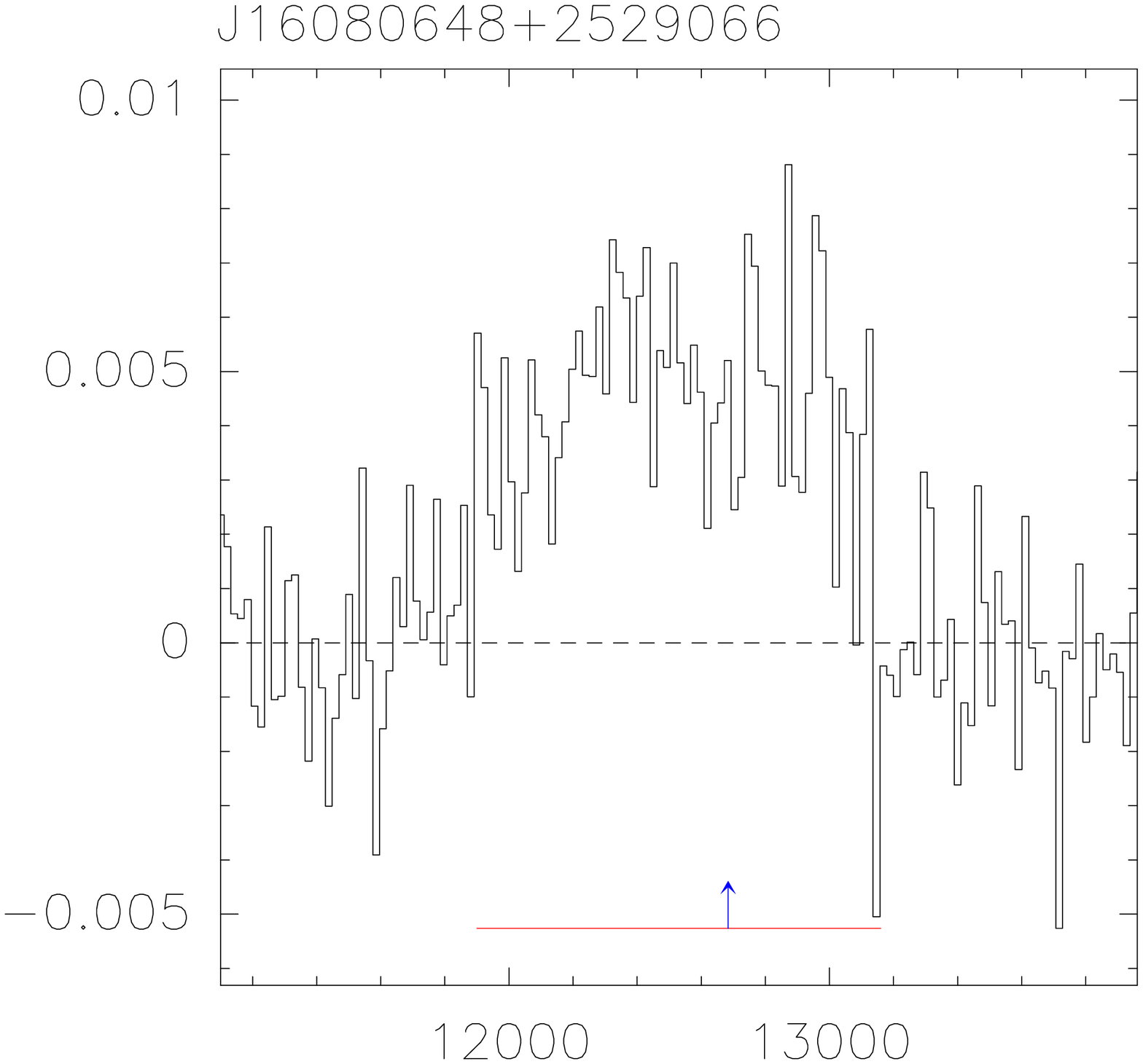}
}
\quad

\centerline{
\includegraphics[width=3.6cm,clip,trim = 0.cm 0.cm 0.cm 0.0cm, angle=-0]{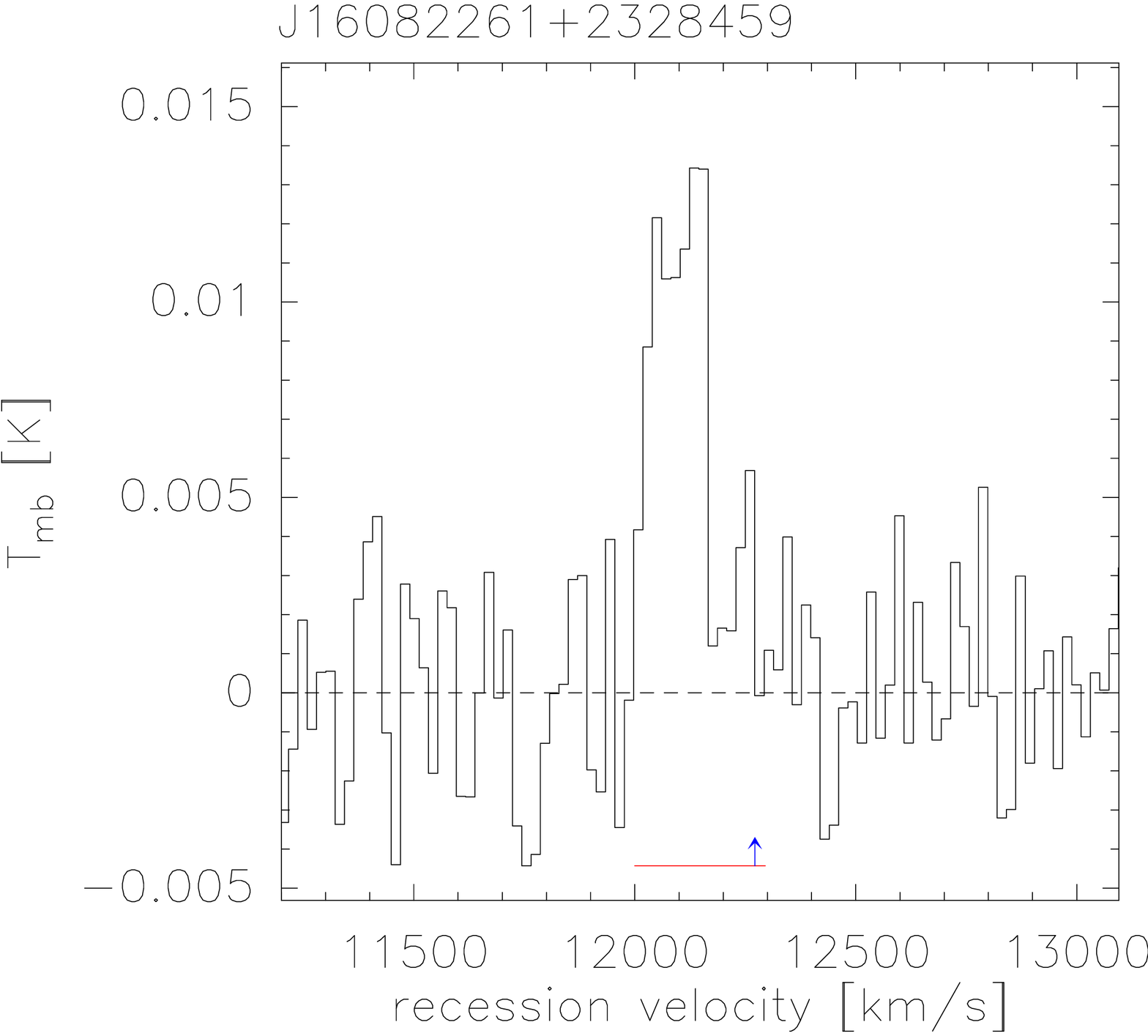}
\hspace{0.1cm}
\includegraphics[width=3.6cm,clip,trim = 0.cm 0.cm 0.cm 0.0cm, angle=-0]{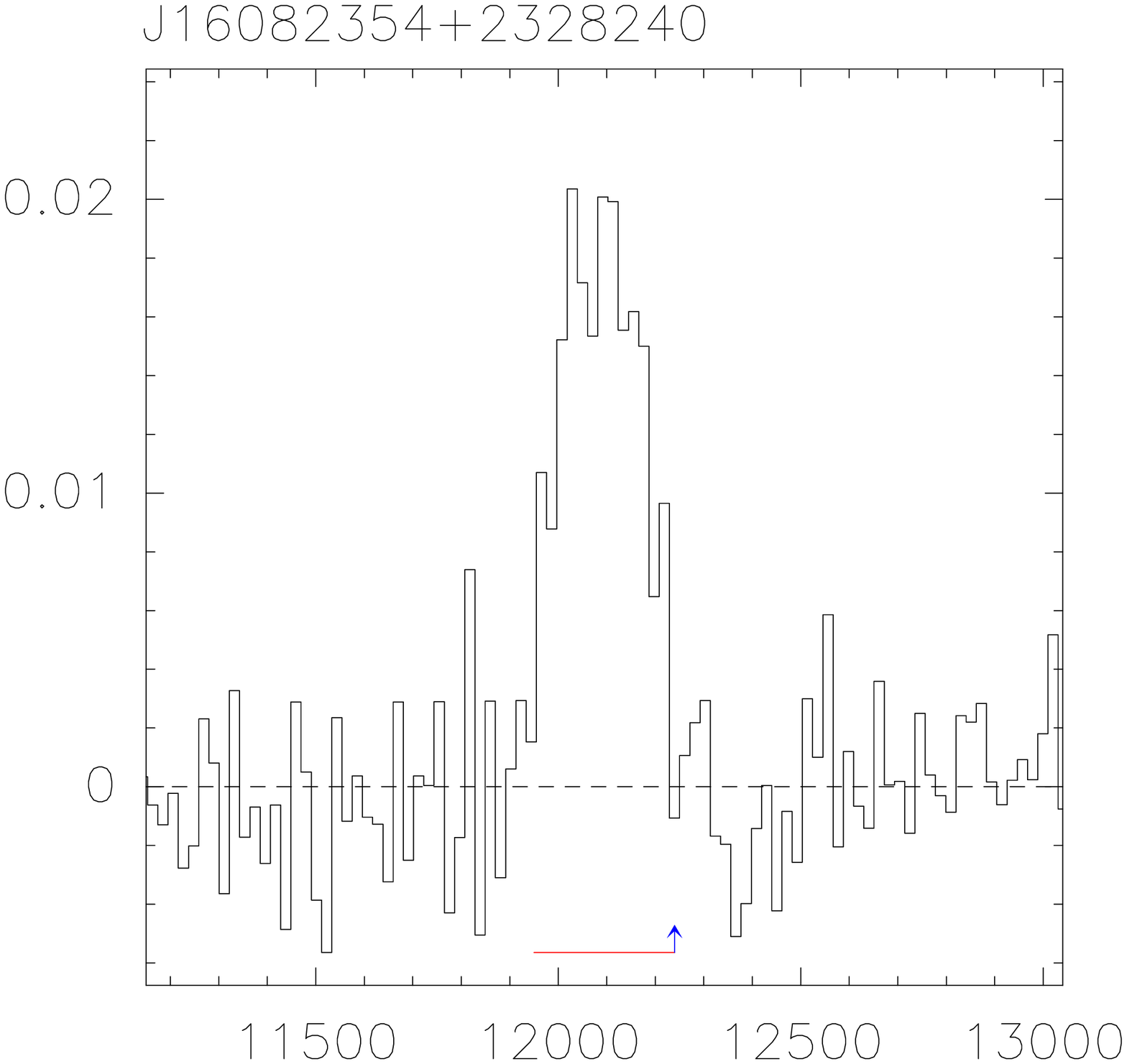}
\hspace{0.1cm}
\includegraphics[width=3.6cm,clip,trim = 0.cm 0.cm 0.cm 0.0cm,angle=-0]{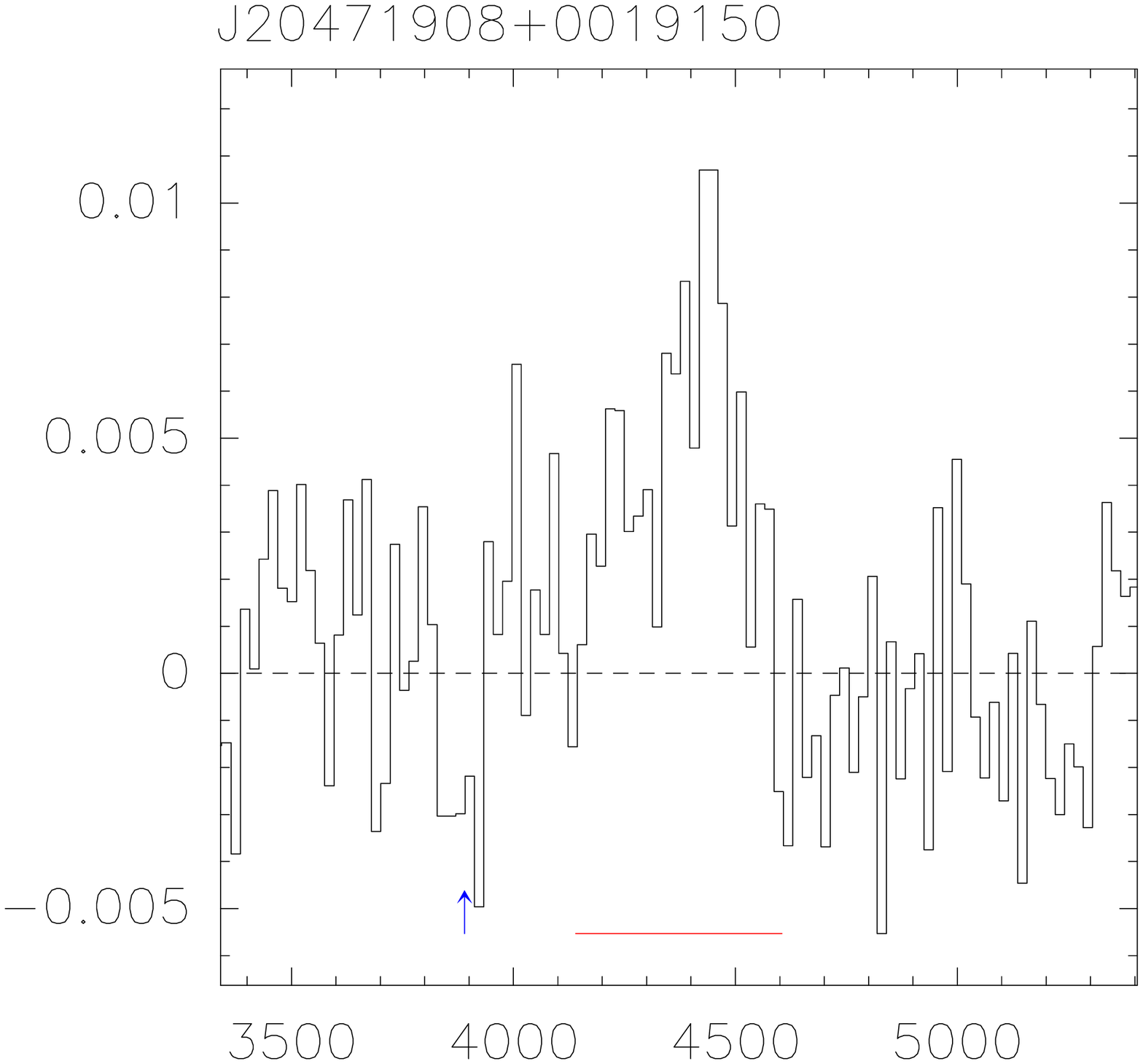}
}

\caption{Continued}
\end{figure*}



\begin{figure*}
\centerline{
\includegraphics[width=3.6cm,clip,trim = 0.cm 0.cm 0.cm 0.0cm,angle=-0]{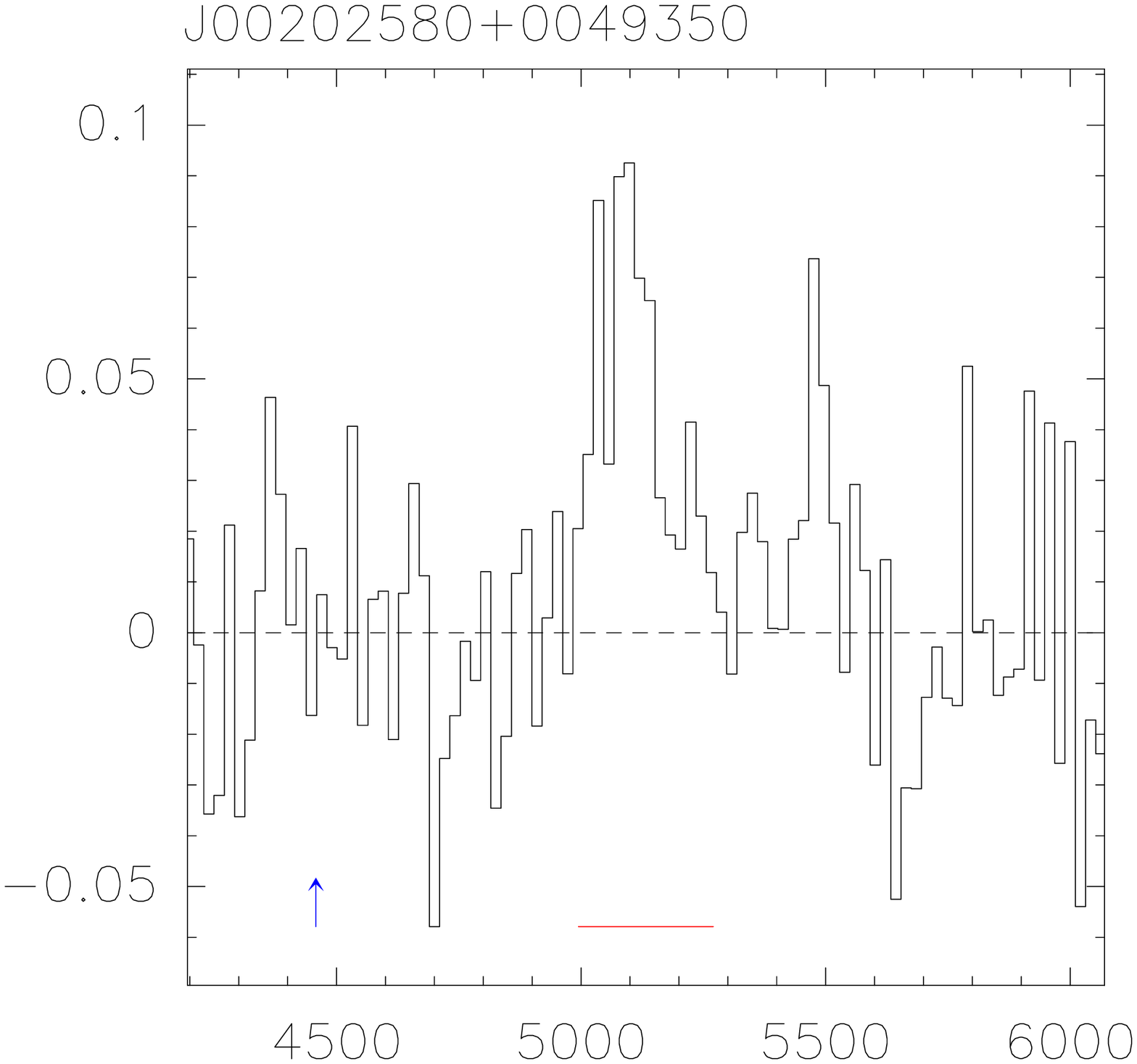}
\hspace{0.1cm}
\includegraphics[width=3.6cm,clip,trim = 0.cm 0.cm 0.cm 0.0cm, angle=-0]{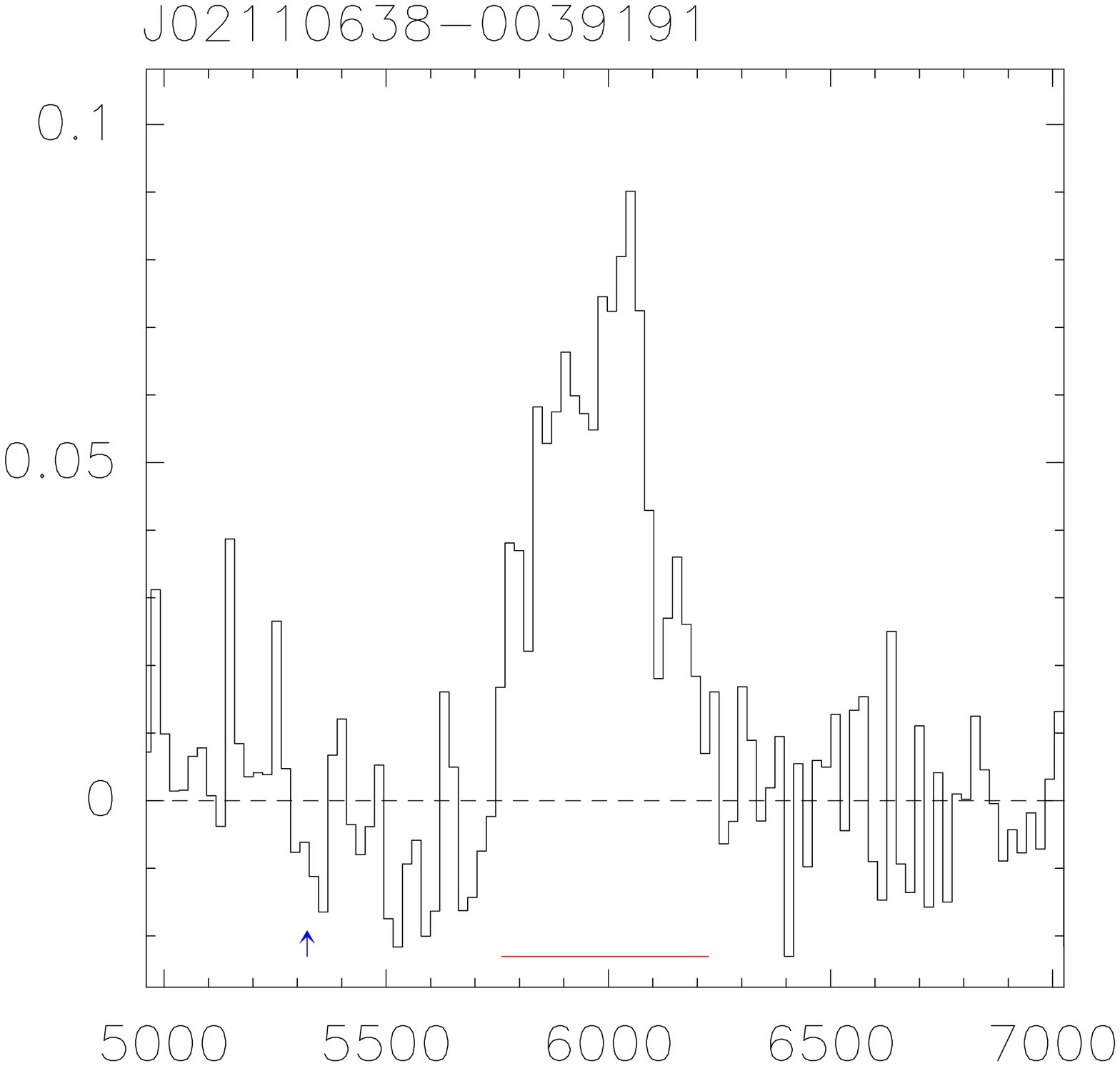}
\hspace{0.1cm}
\includegraphics[width=3.6cm,clip,trim = 0.cm 0.cm 0.cm 0.0cm, angle=-0]{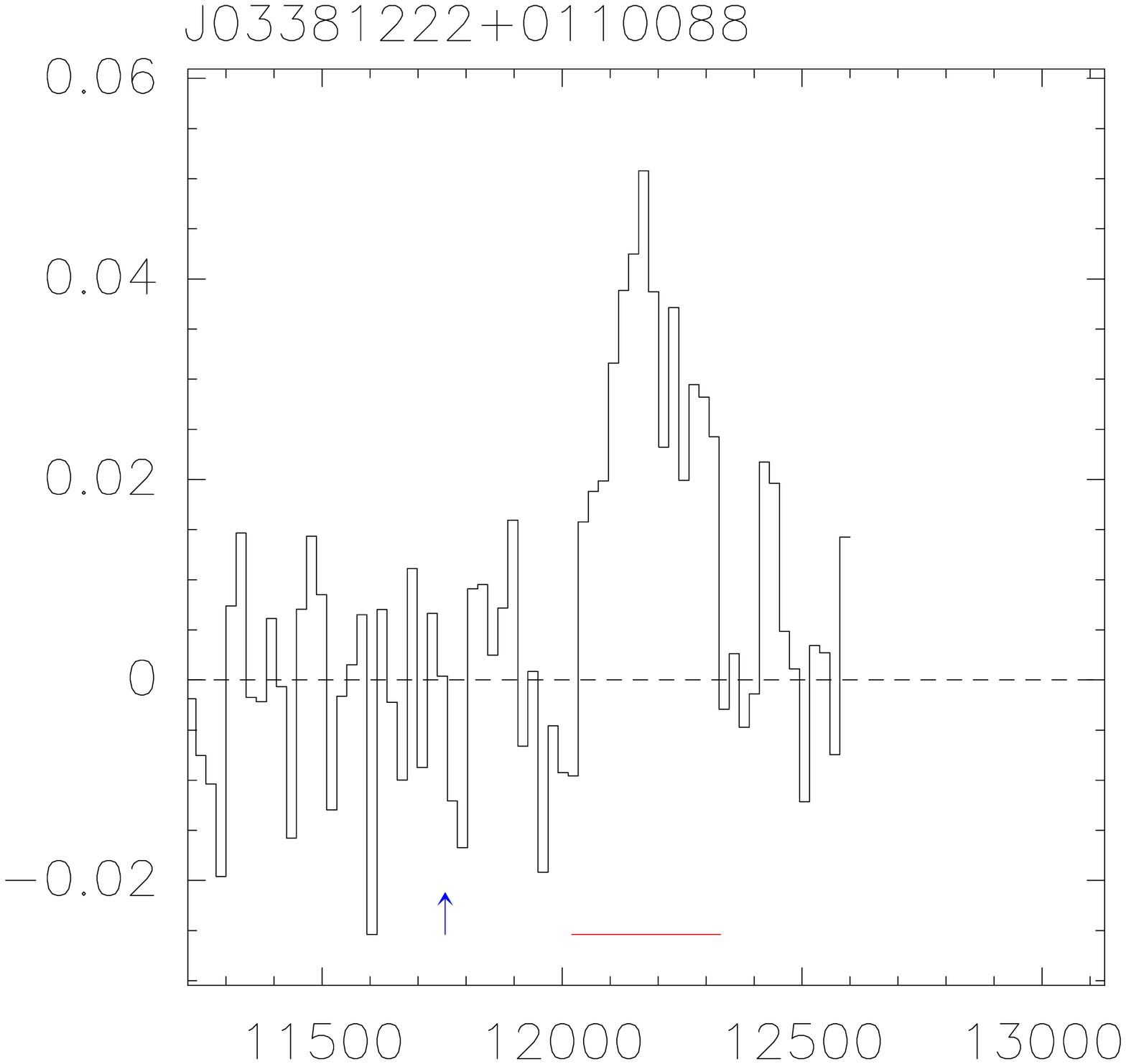}
\hspace{0.1cm}
\includegraphics[width=3.6cm,clip,trim = 0.cm 0.cm 0.cm 0.0cm, angle=-0]{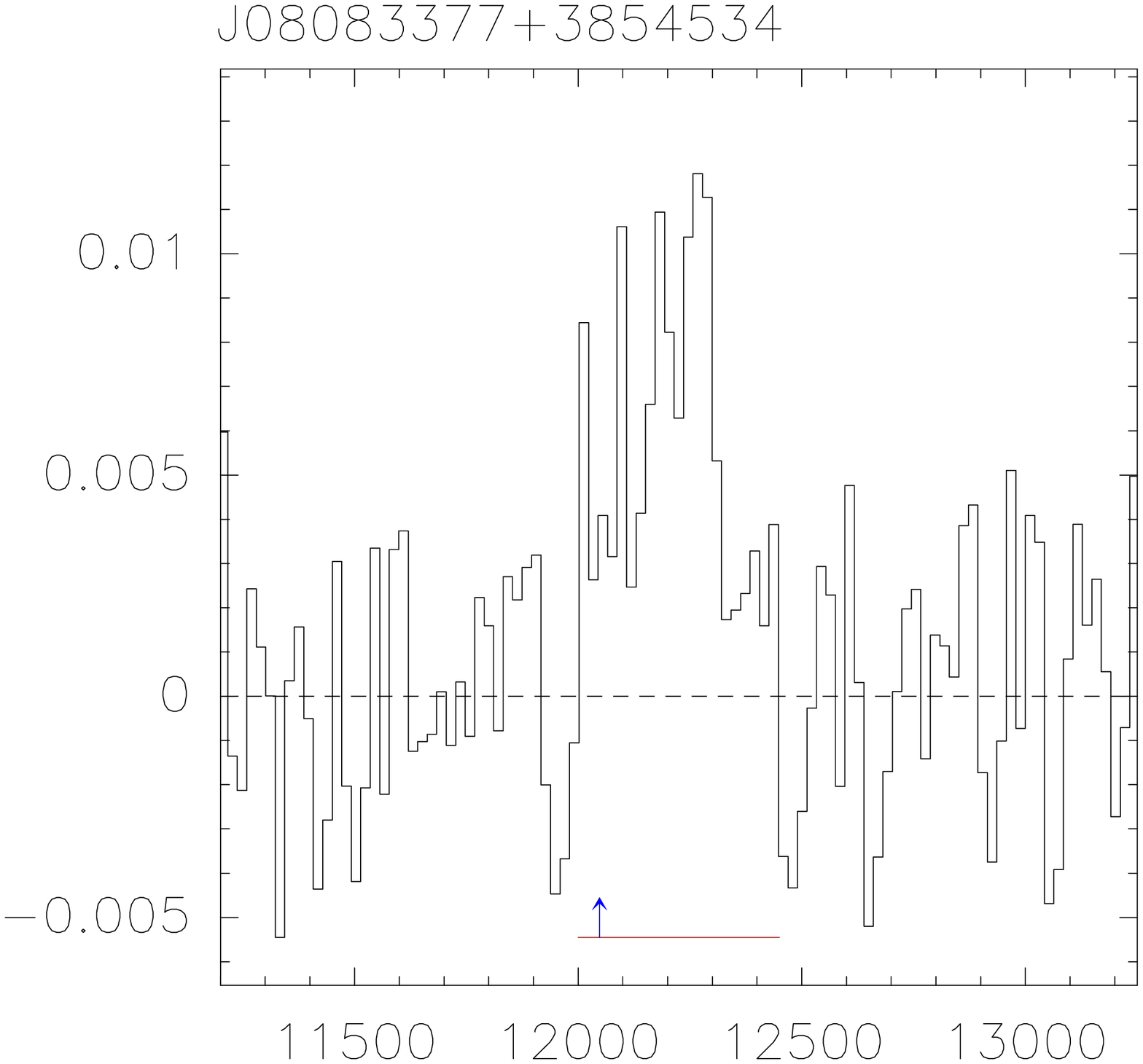}
}
\quad

\centerline{
\includegraphics[width=3.6cm,clip,trim = 0.cm 0.cm 0.cm 0.0cm,angle=-0]{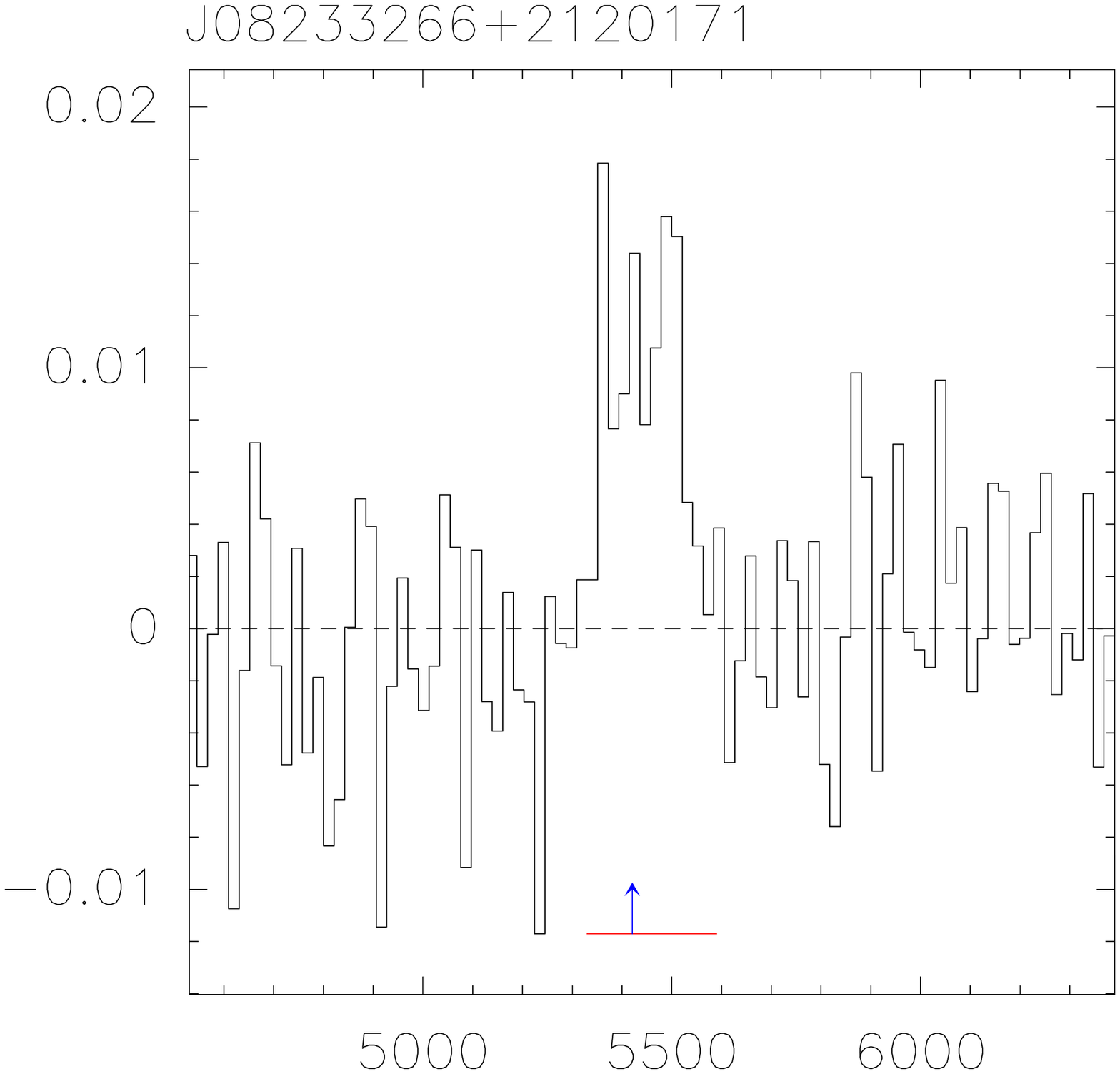}
\hspace{0.1cm}
\includegraphics[width=3.6cm,clip,trim = 0.cm 0.cm 0.cm 0.0cm, angle=-0]{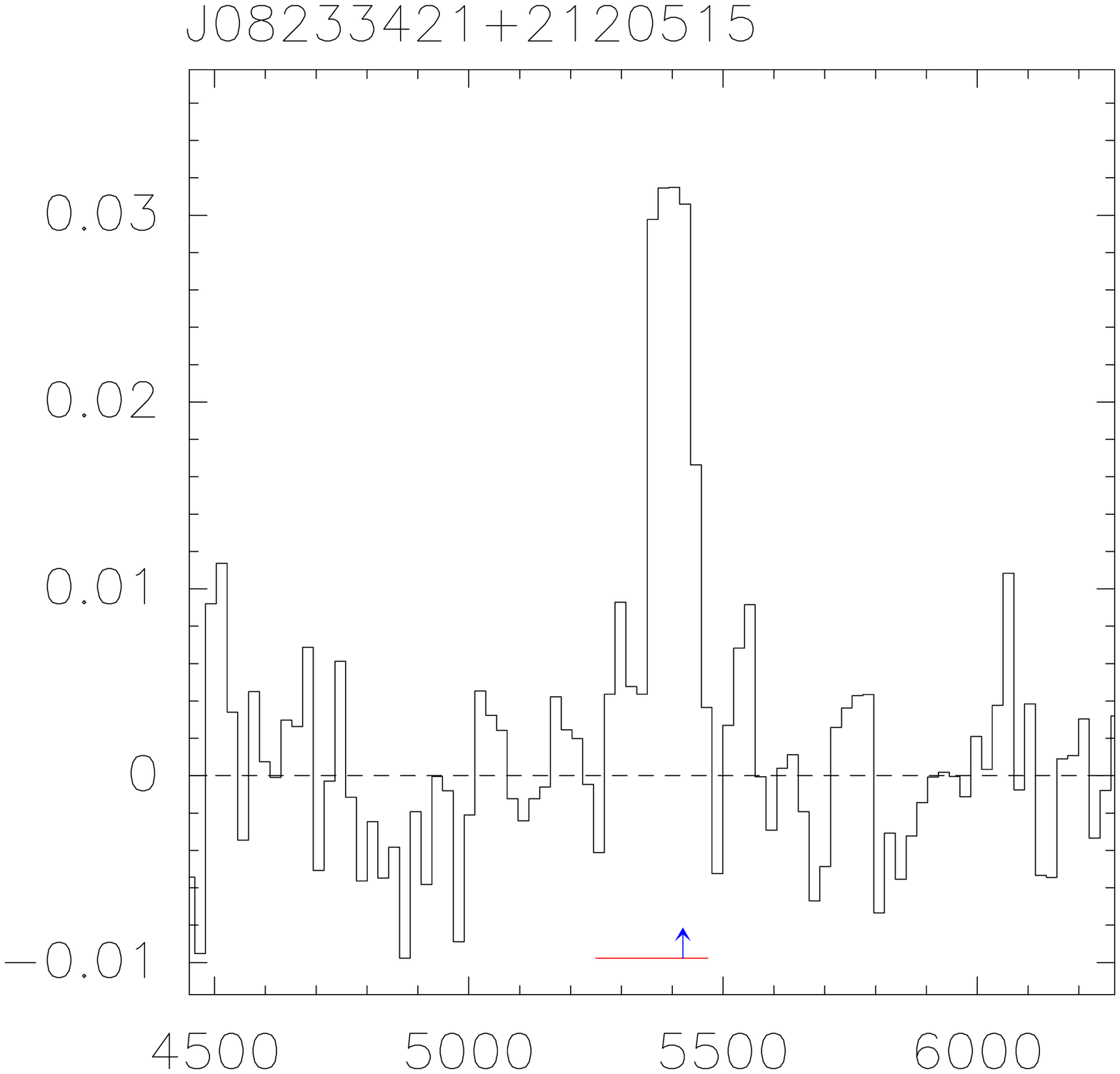}
\hspace{0.1cm}
\includegraphics[width=3.6cm,clip,trim = 0.cm 0.cm 0.cm 0.0cm, angle=-0]{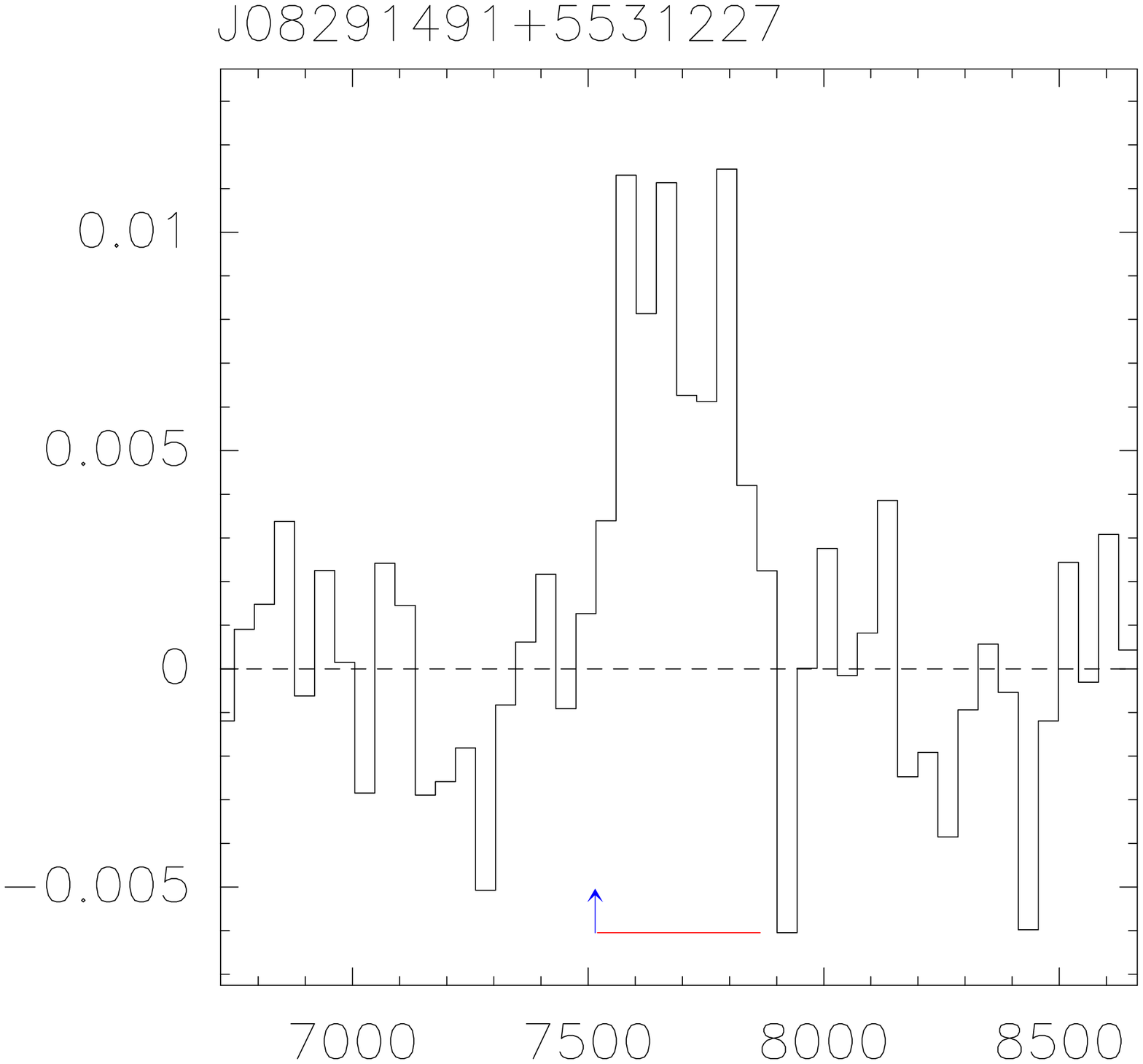}
\hspace{0.1cm}
\includegraphics[width=3.6cm,clip,trim = 0.cm 0.cm 0.cm 0.0cm, angle=-0]{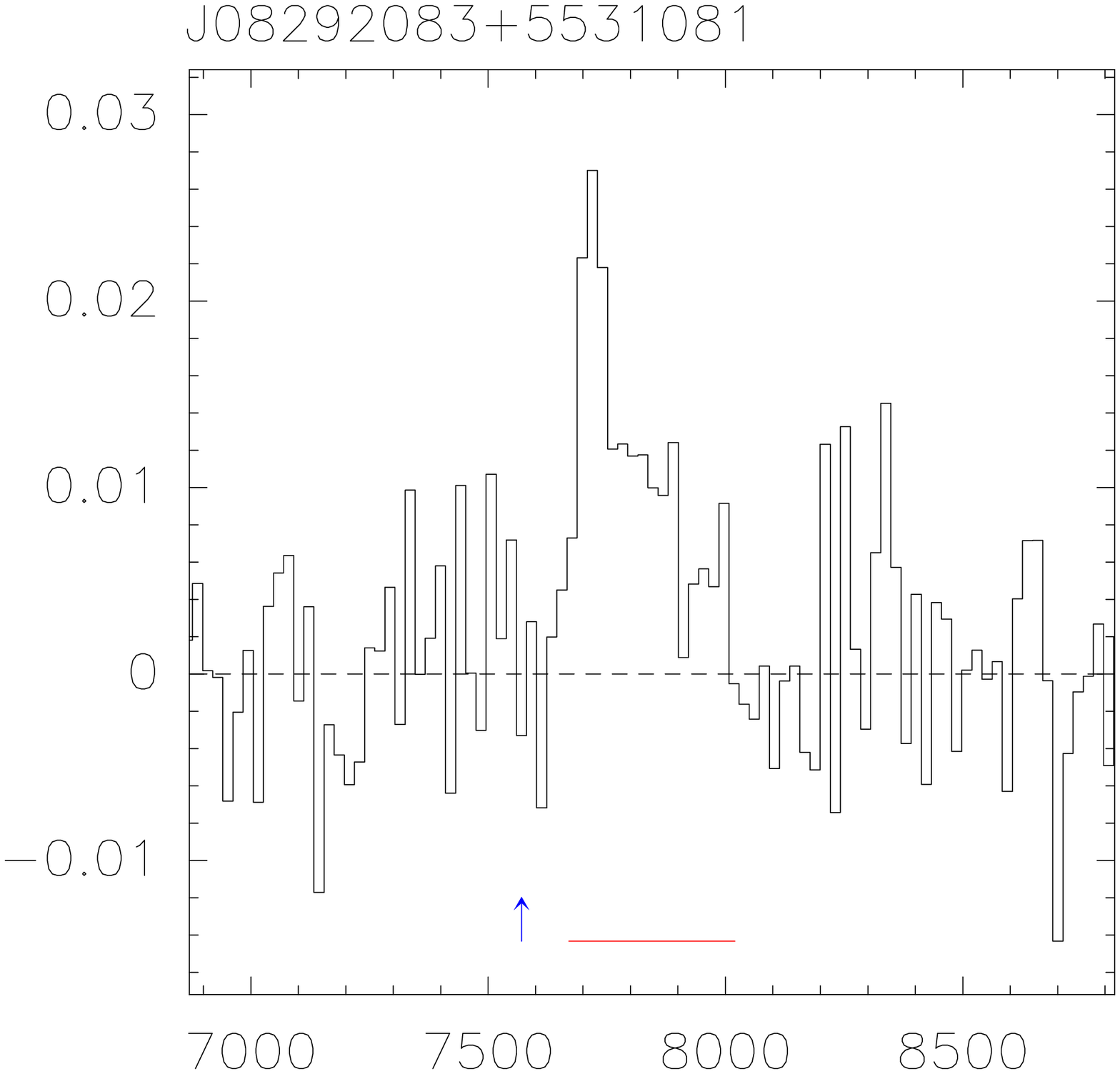}
}
\quad

\centerline{
\includegraphics[width=3.6cm,clip,trim = 0.cm 0.cm 0.cm 0.0cm,angle=-0]{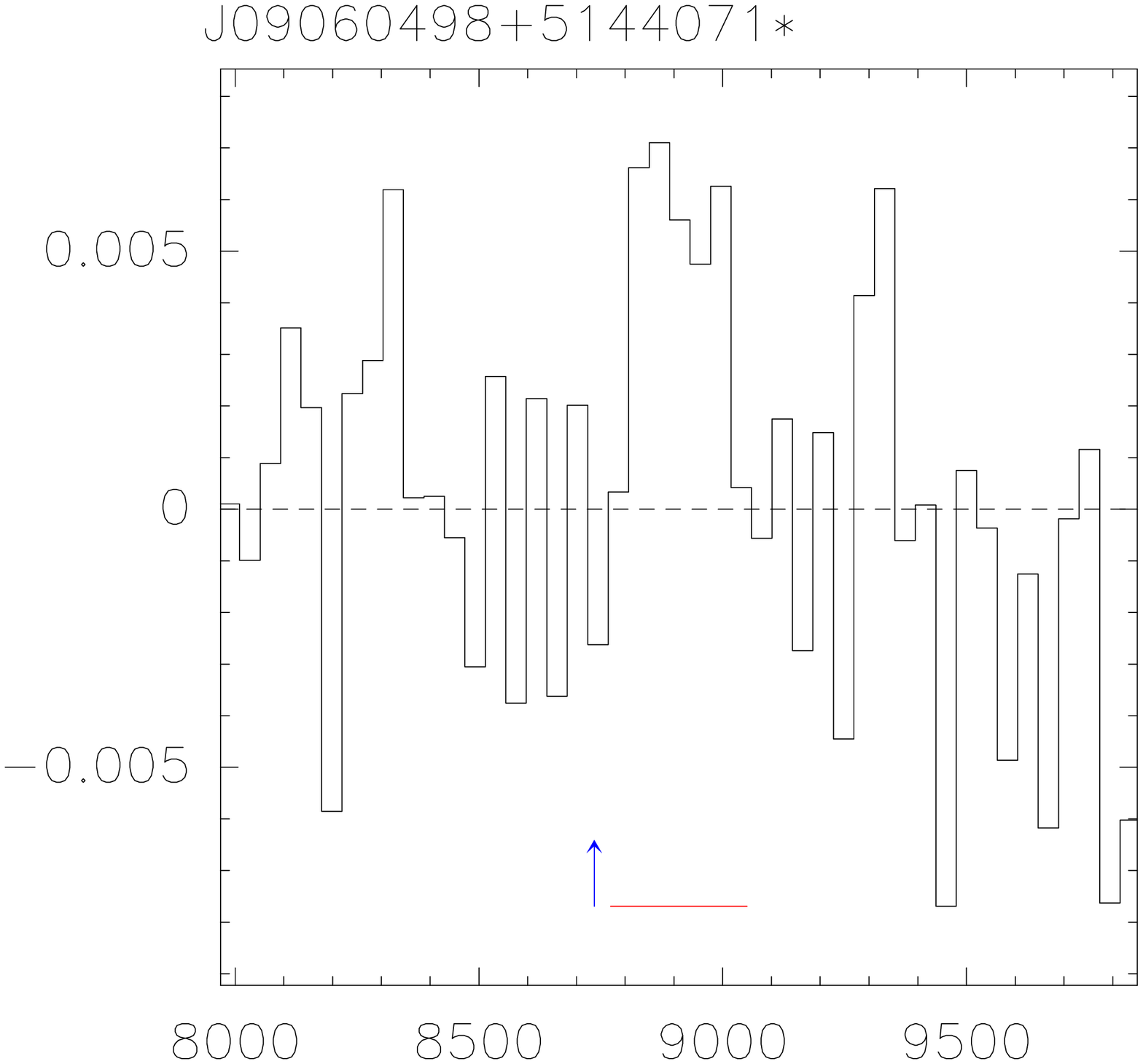}
\hspace{0.1cm}
\includegraphics[width=3.6cm,clip,trim = 0.cm 0.cm 0.cm 0.0cm, angle=-0]{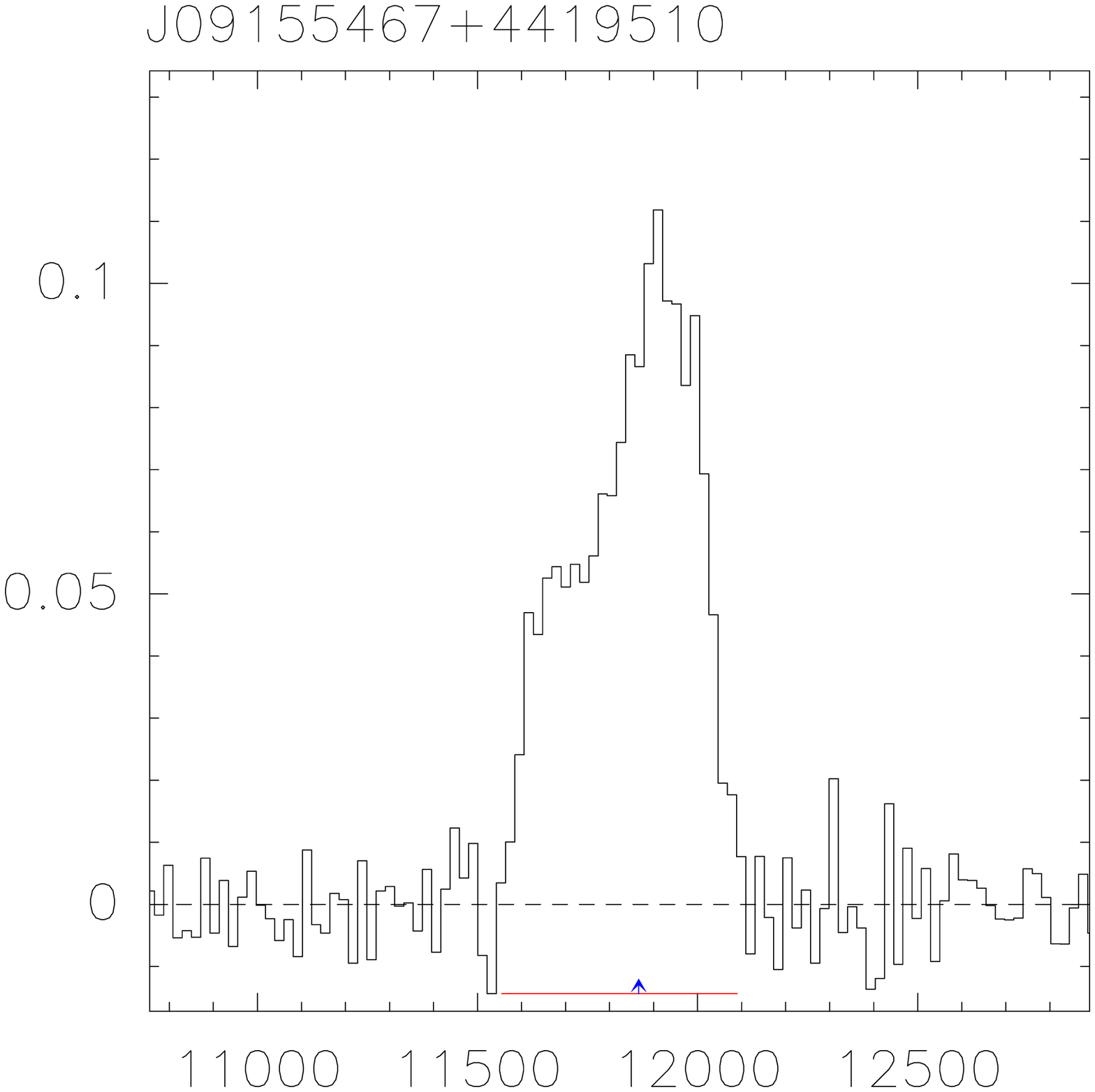}
\hspace{0.1cm}
\includegraphics[width=3.6cm,clip,trim = 0.cm 0.cm 0.cm 0.0cm, angle=-0]{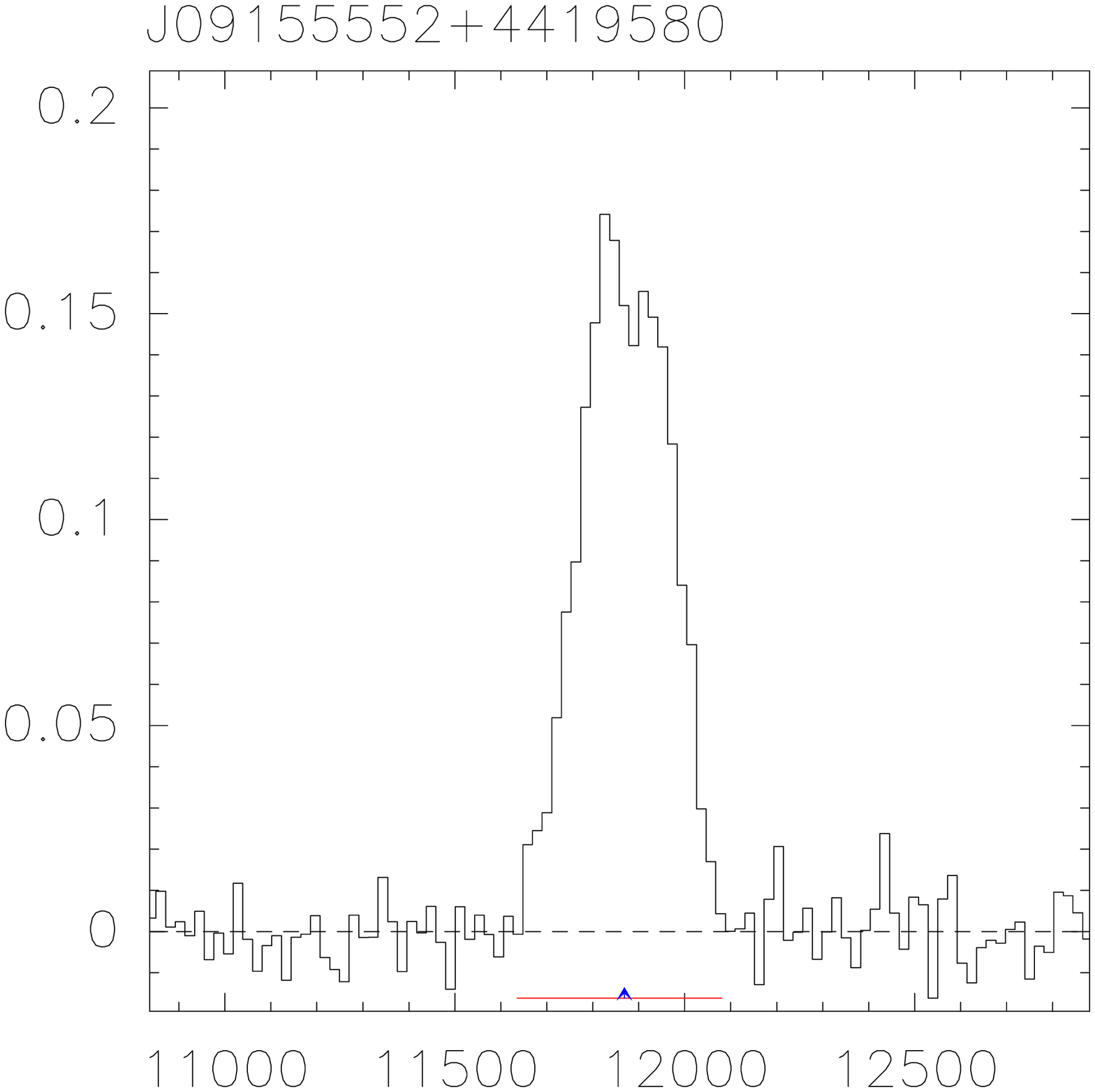}
\hspace{0.1cm}
\includegraphics[width=3.6cm,clip,trim = 0.cm 0.cm 0.cm 0.0cm, angle=-0]{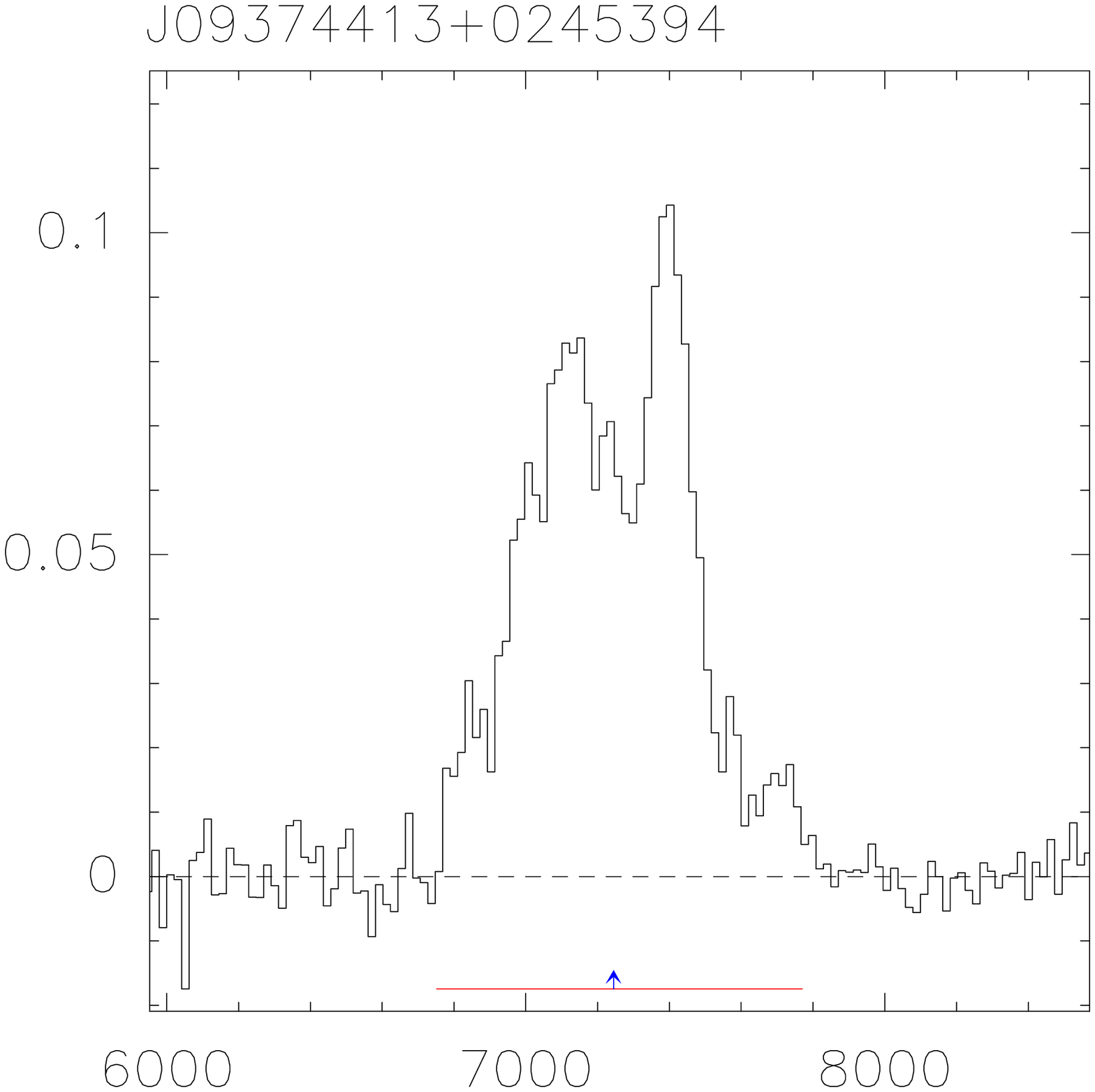}
}
\quad

\centerline{
\includegraphics[width=3.6cm,clip,trim = 0.cm 0.cm 0.cm 0.0cm, angle=-0]{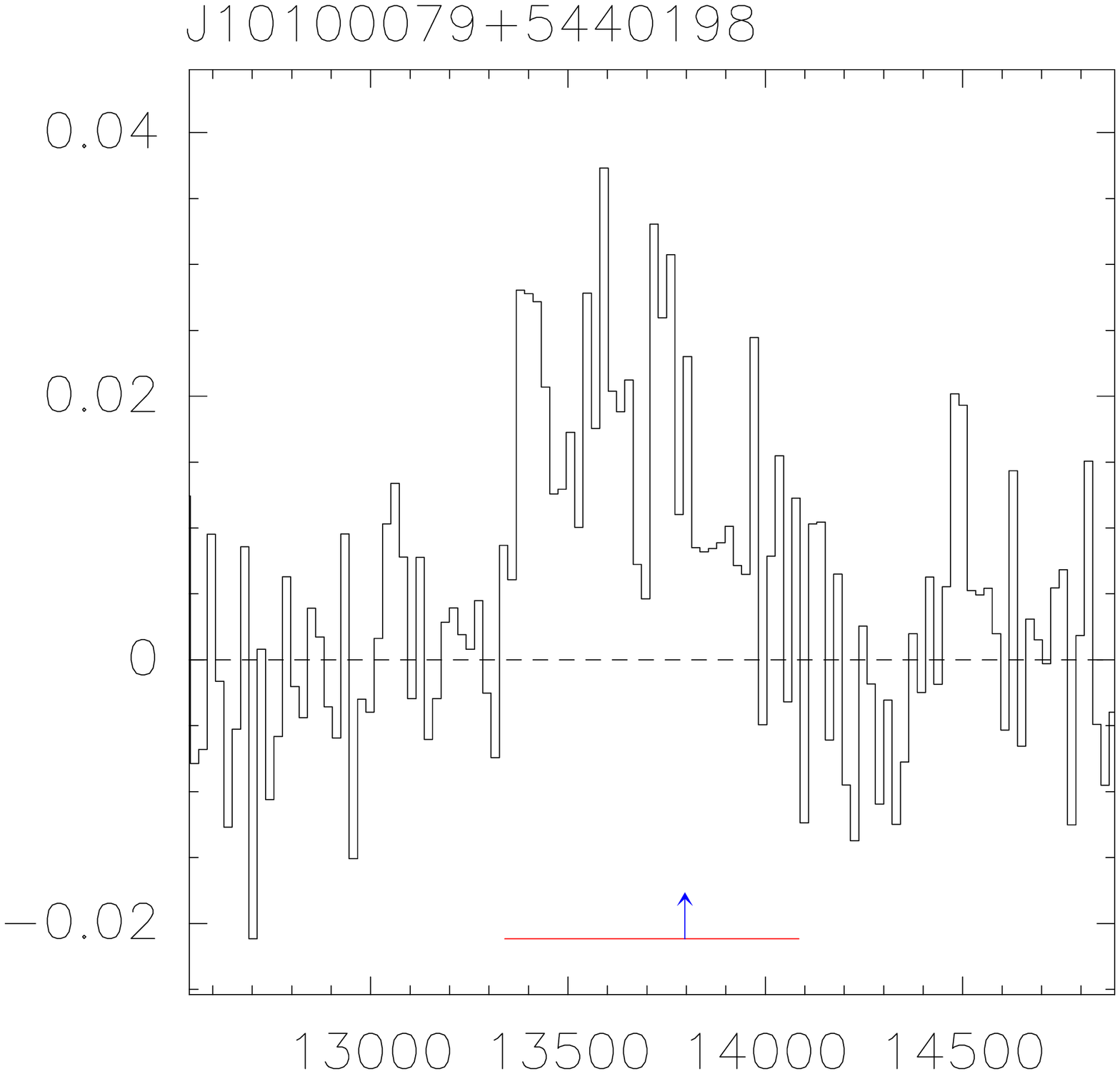}
\hspace{0.1cm}
\includegraphics[width=3.6cm,clip,trim = 0.cm 0.cm 0.cm 0.0cm, angle=-0]{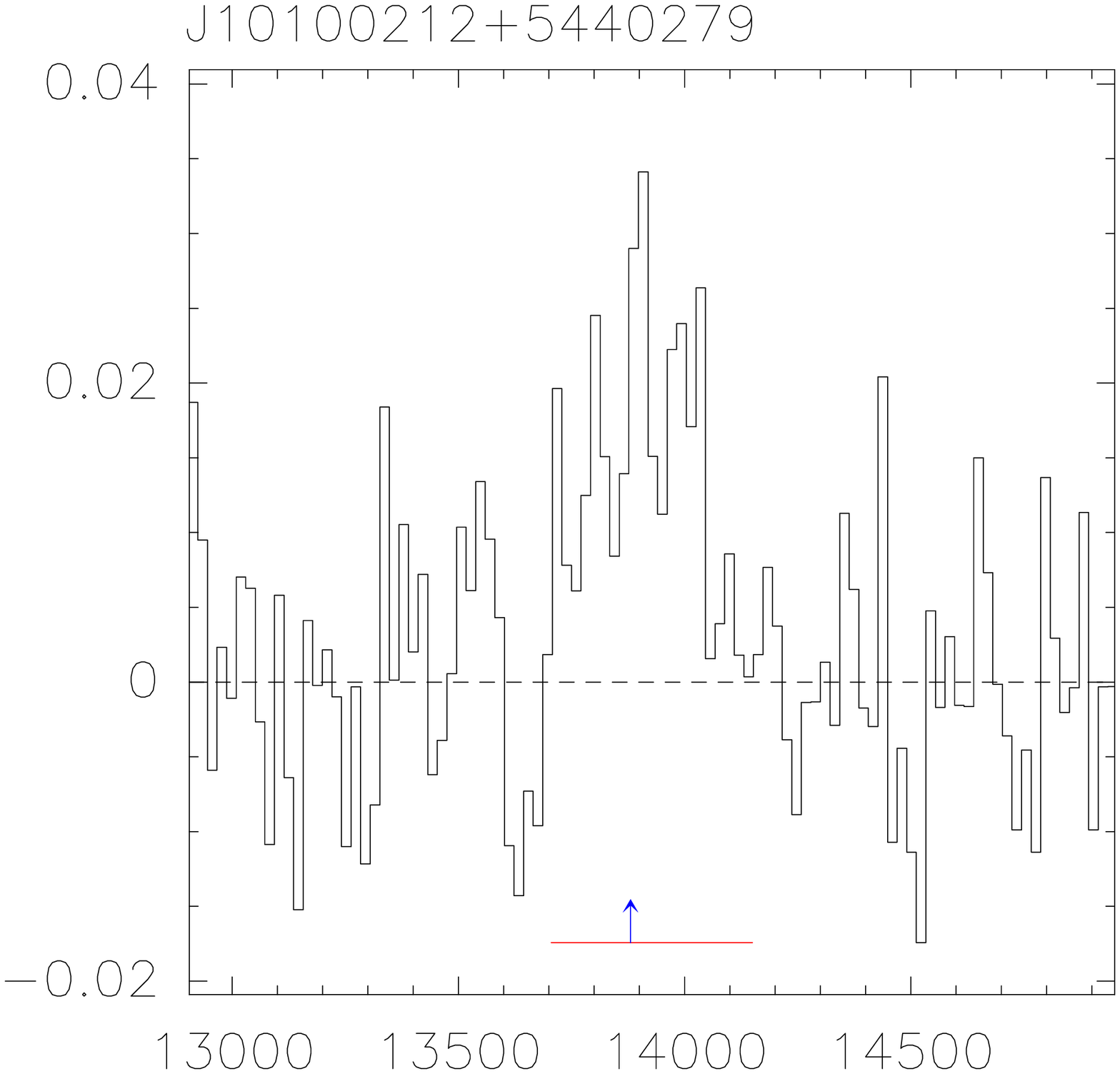}
\hspace{0.1cm}
\includegraphics[width=3.6cm,clip,trim = 0.cm 0.cm 0.cm 0.0cm,angle=-0]{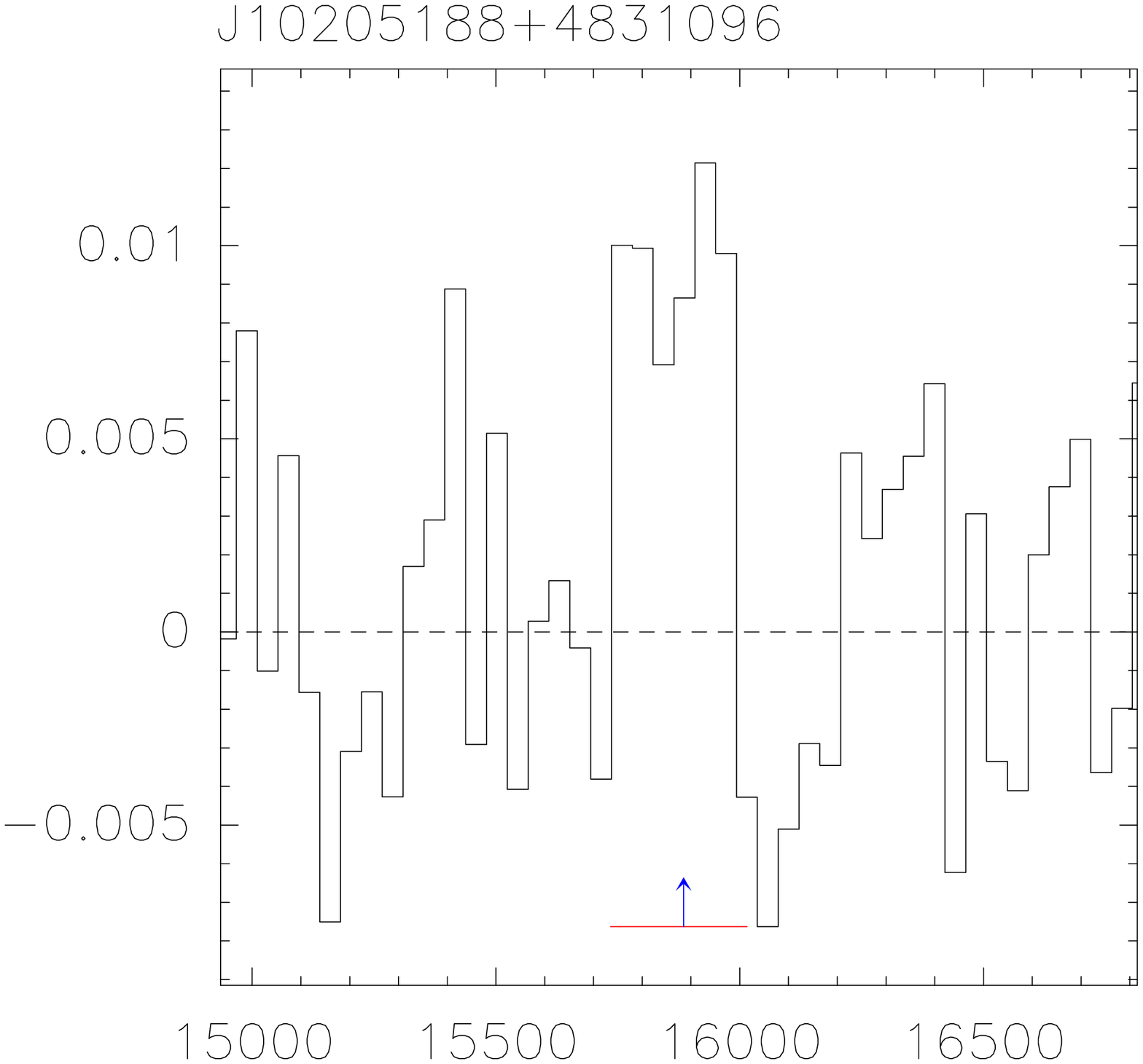}
\hspace{0.1cm}
\includegraphics[width=3.6cm,clip,trim = 0.cm 0.cm 0.cm 0.0cm, angle=-0]{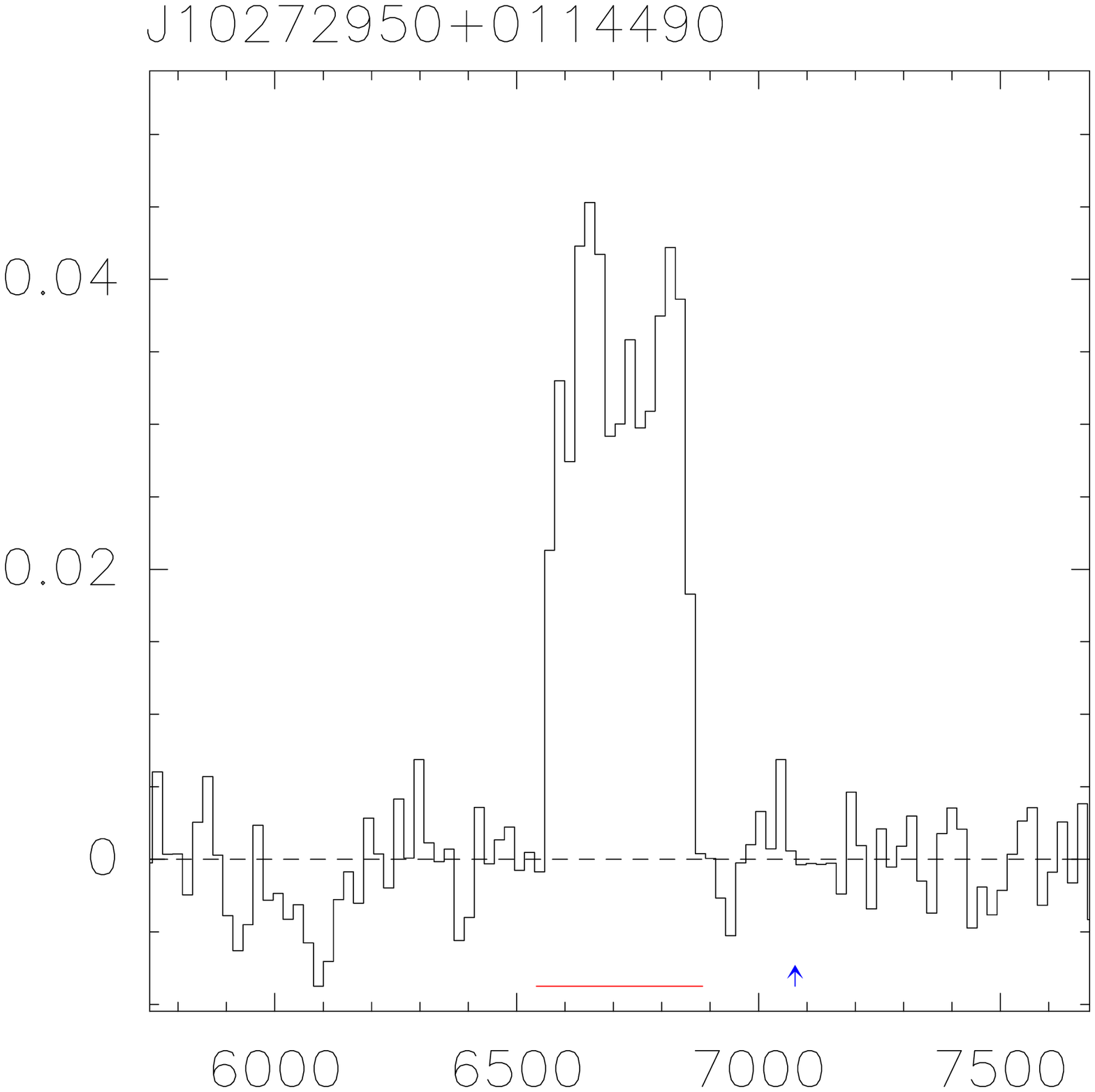}
}
\quad

\centerline{
\includegraphics[width=3.6cm,clip,trim = 0.cm 0.cm 0.cm 0.0cm, angle=-0]{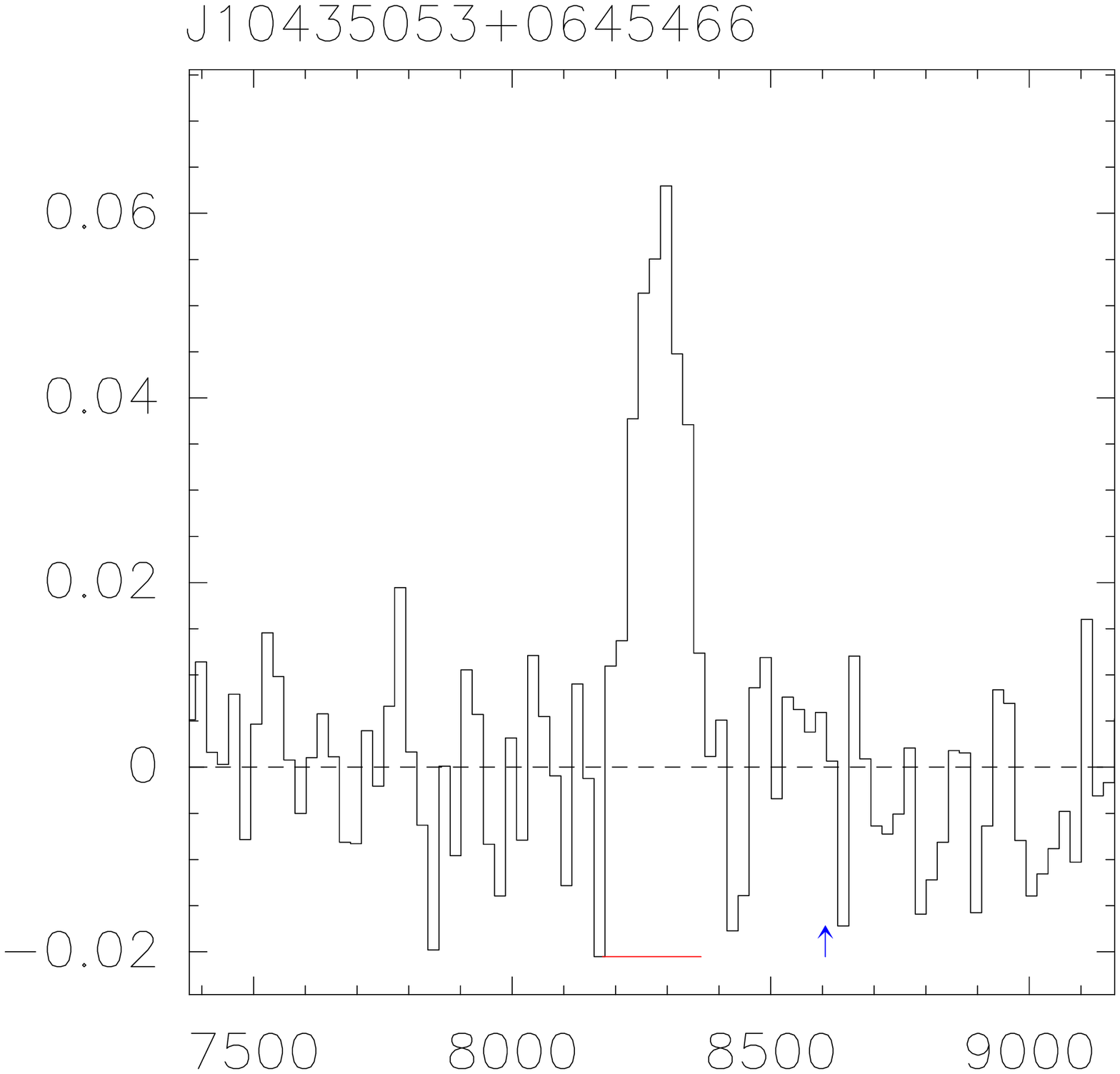}
\hspace{0.1cm}
\includegraphics[width=3.6cm,clip,trim = 0.cm 0.cm 0.cm 0.0cm, angle=-0]{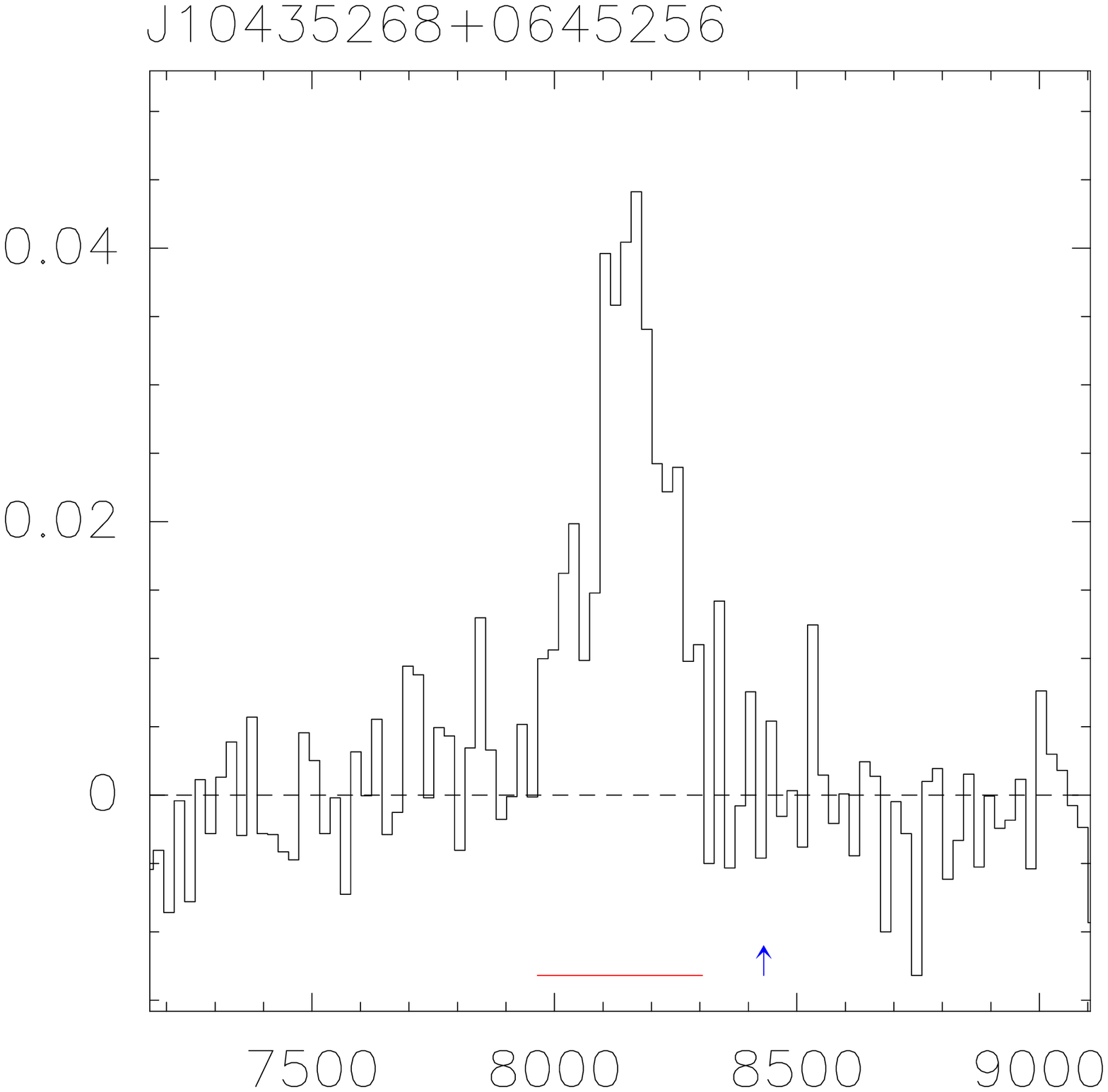}
\hspace{0.1cm}
\includegraphics[width=3.6cm,clip,trim = 0.cm 0.cm 0.cm 0.0cm,angle=-0]{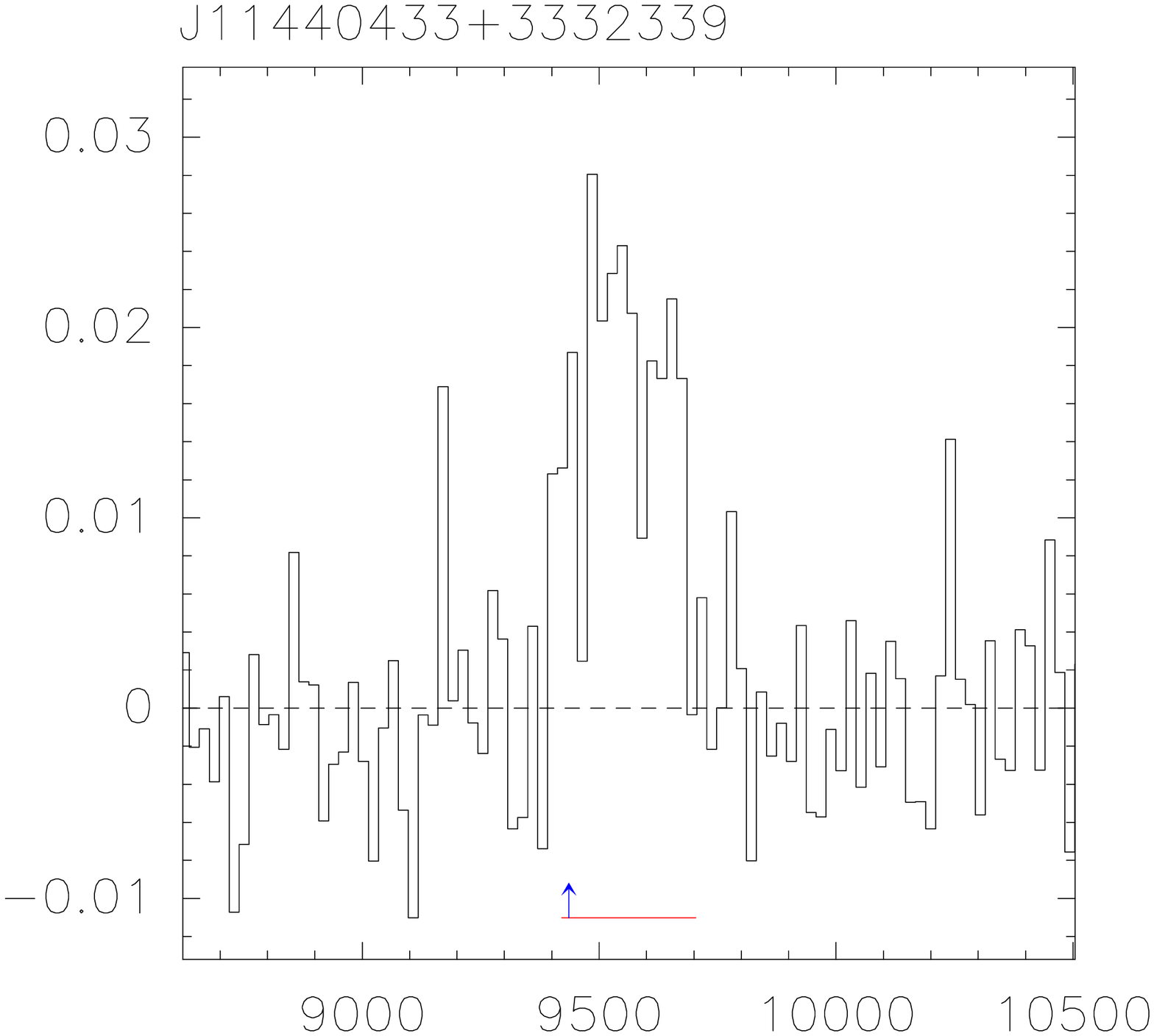}
\hspace{0.1cm}
\includegraphics[width=3.6cm,clip,trim = 0.cm 0.cm 0.cm 0.0cm, angle=-0]{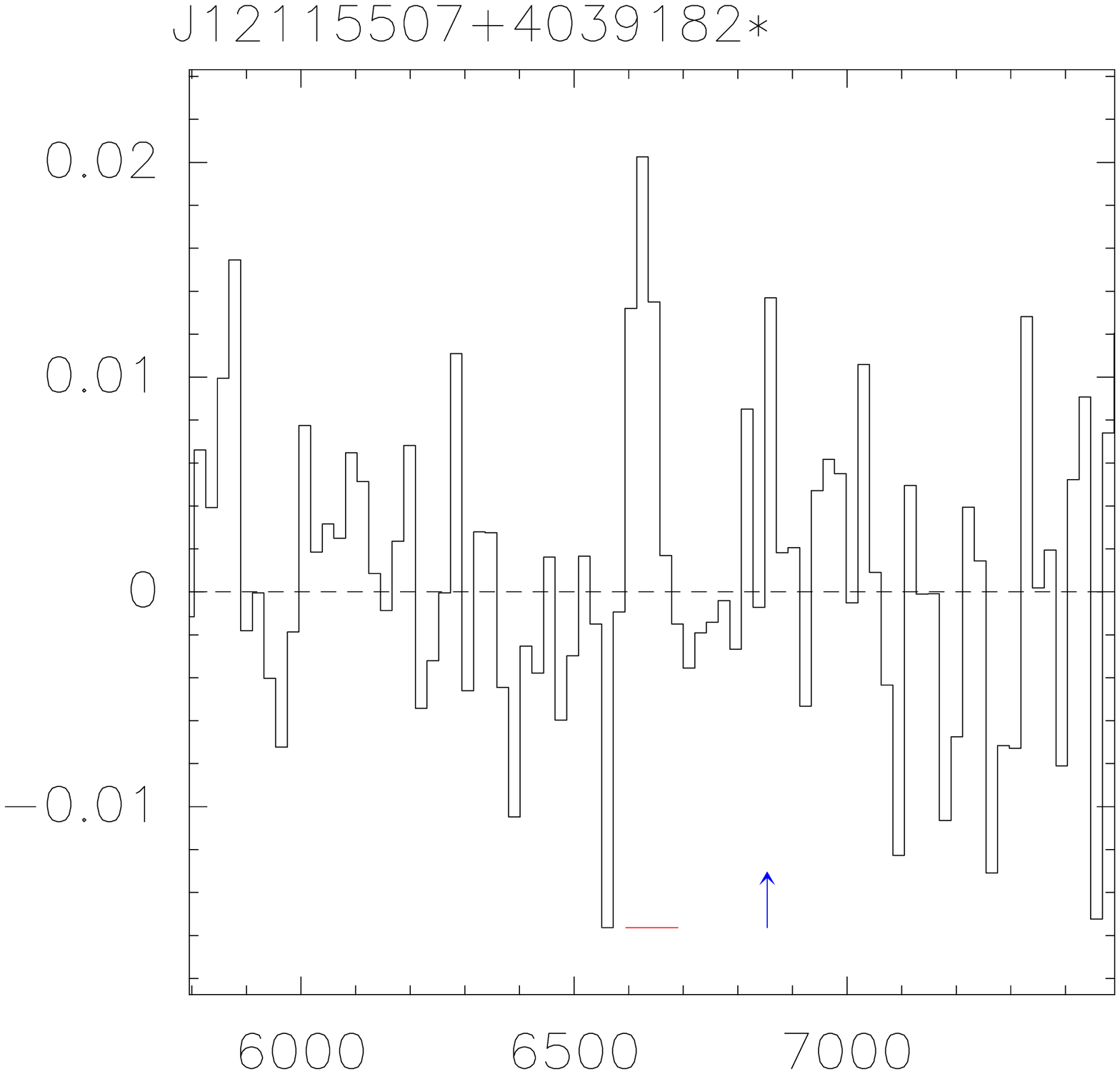}
}
\quad

\centerline{
\includegraphics[width=3.6cm,clip,trim = 0.cm 0.cm 0.cm 0.0cm, angle=-0]{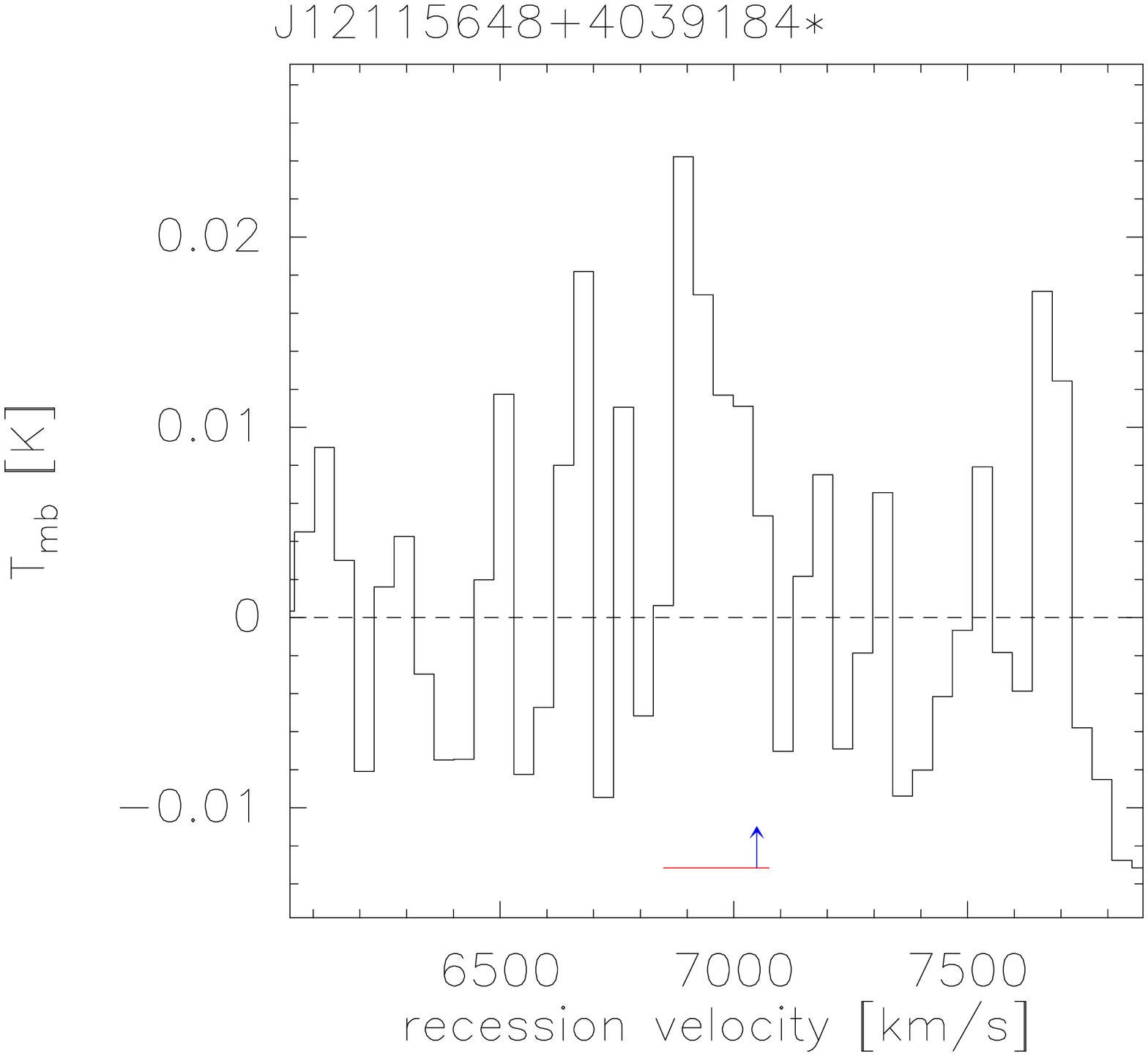}
\hspace{0.1cm}
\includegraphics[width=3.6cm,clip,trim = 0.cm 0.cm 0.cm 0.0cm, angle=-0]{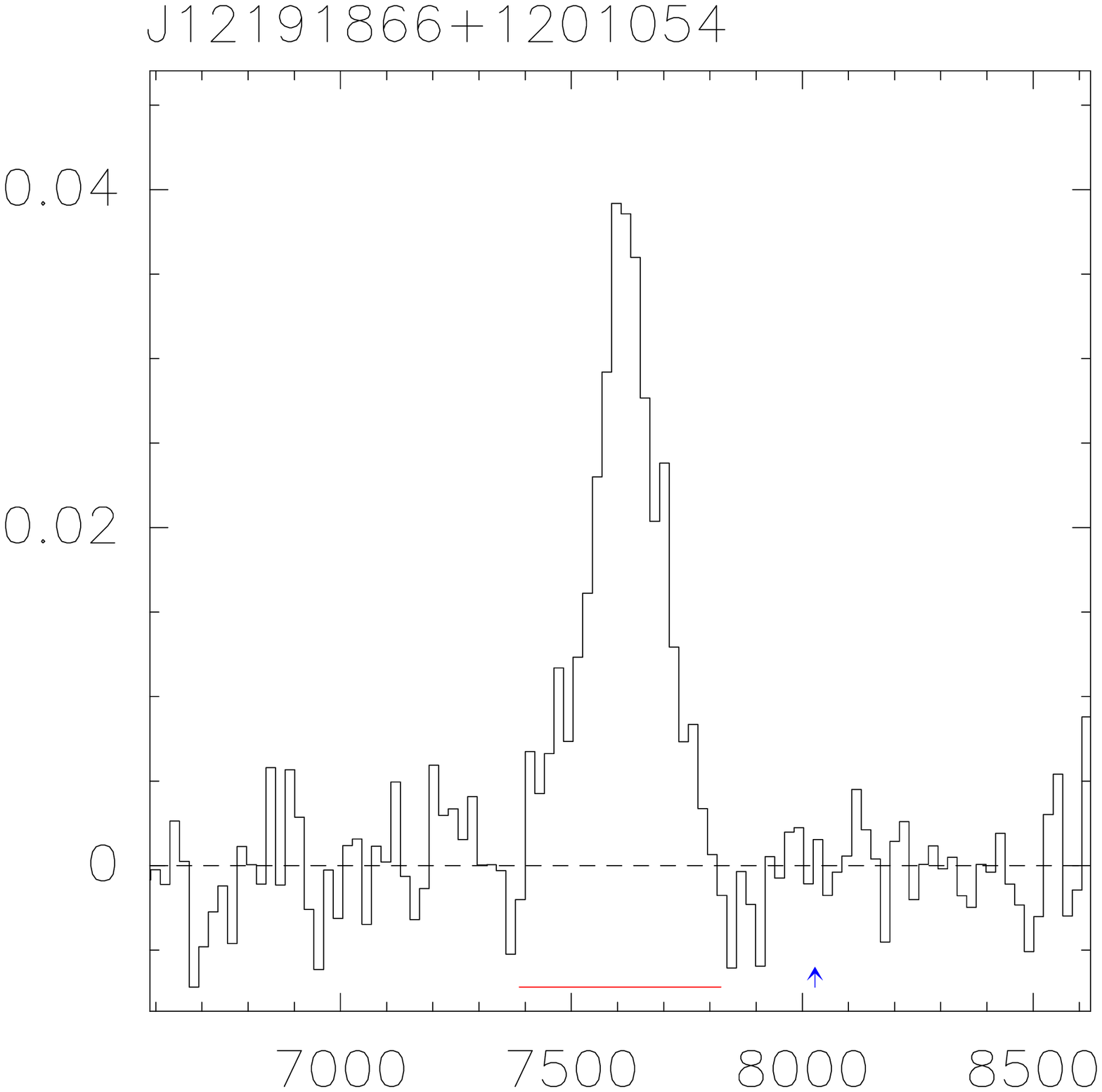}
\hspace{0.1cm}
\includegraphics[width=3.6cm,clip,trim = 0.cm 0.cm 0.cm 0.0cm,angle=-0]{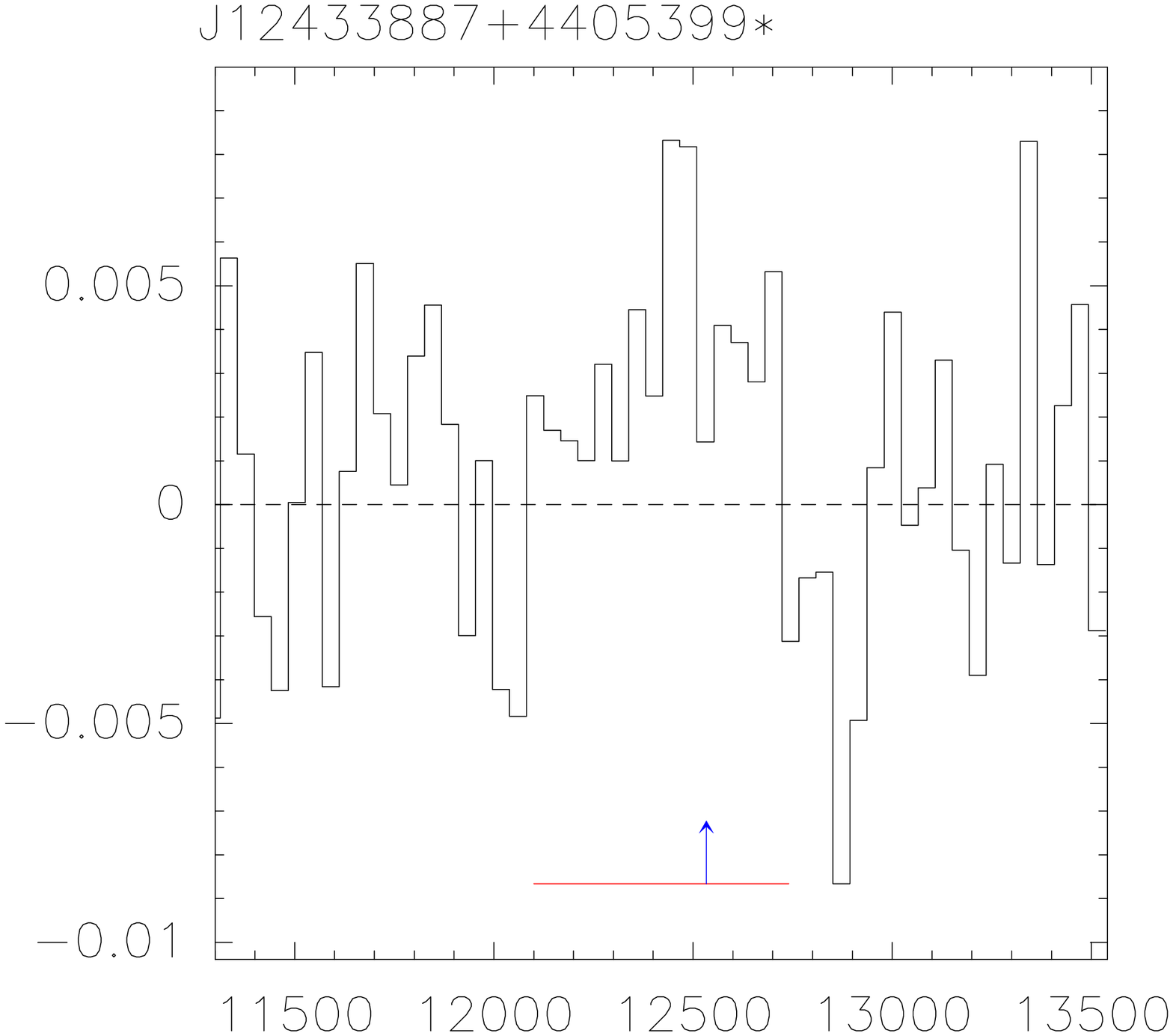}
\hspace{0.1cm}
\includegraphics[width=3.6cm,clip,trim = 0.cm 0.cm 0.cm 0.0cm, angle=-0]{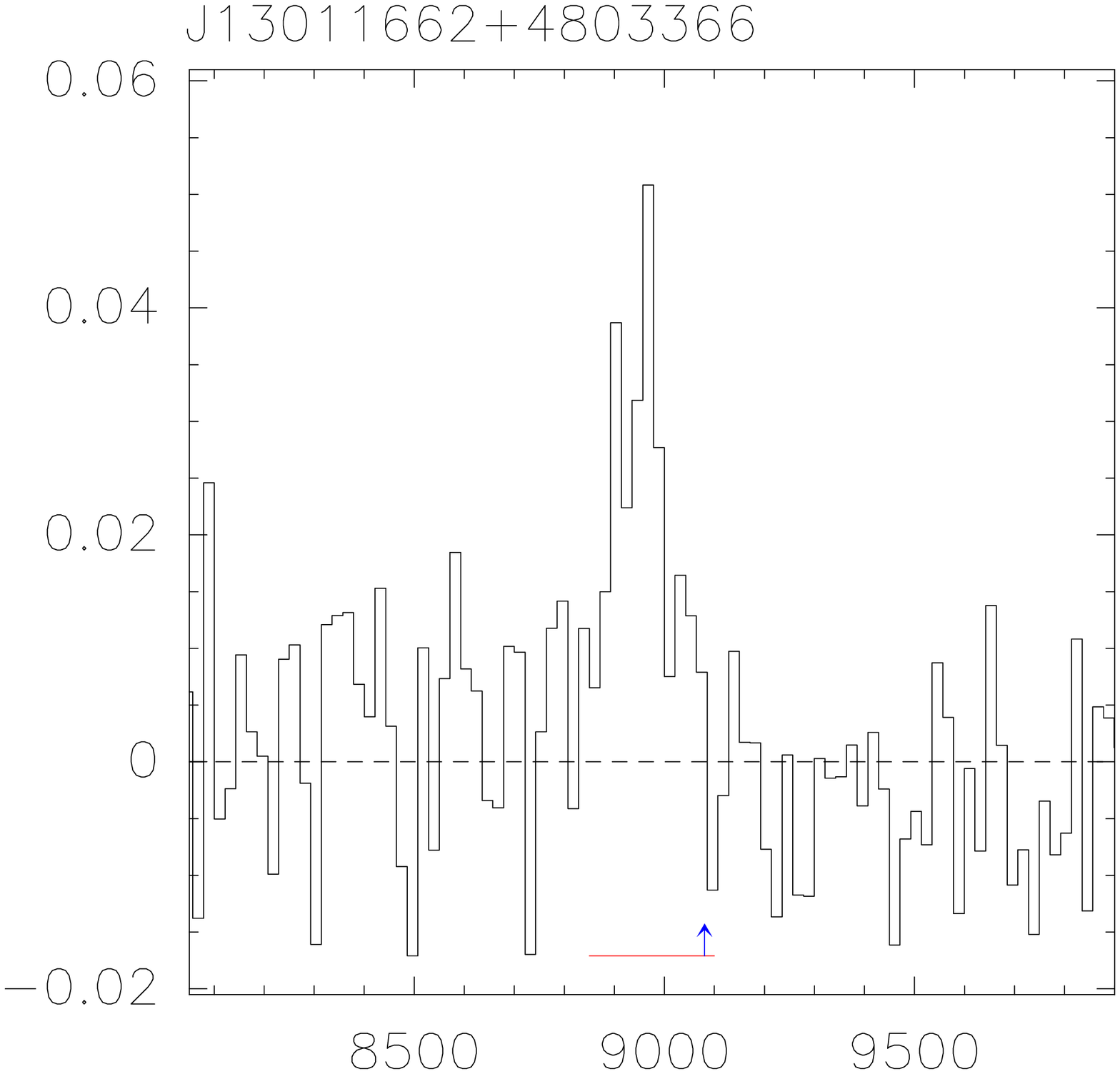}
}

\caption{CO(2-1) spectra of the detected  galaxies (including tentative detections). The velocity resolution
is $\sim$ 20 \kms\ for most spectra and $\sim$  40 \kms\ for some cases where a lower resolution was required
to clearly see the line. The red line segment shows the zero-level linewidth of the 
CO line adopted for the determination of the velocity integrated intensity. 
The blue upright arrow indicated the optical  heliocentric recession velocity. 
An asterisk next to the name indicates a tentative detection.
}
\label{fig:spectra_co10}
\end{figure*}

\setcounter{figure}{0} 

\begin{figure*}
\centerline{
\includegraphics[width=3.6cm,clip,trim = 0.cm 0.cm 0.cm 0.0cm, angle=-0]{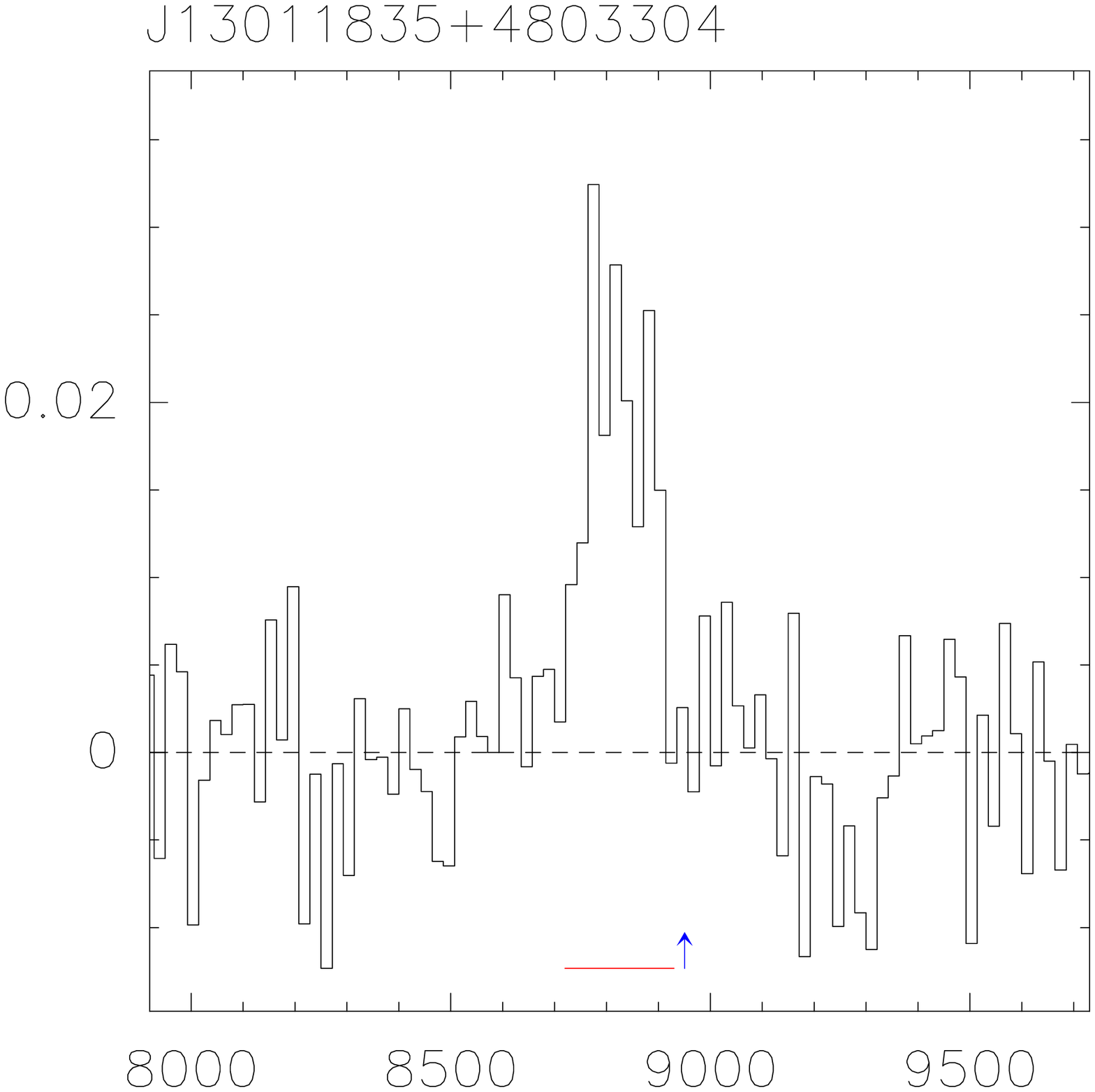}
\hspace{0.1cm}
\includegraphics[width=3.6cm,clip,trim = 0.cm 0.cm 0.cm 0.0cm, angle=-0]{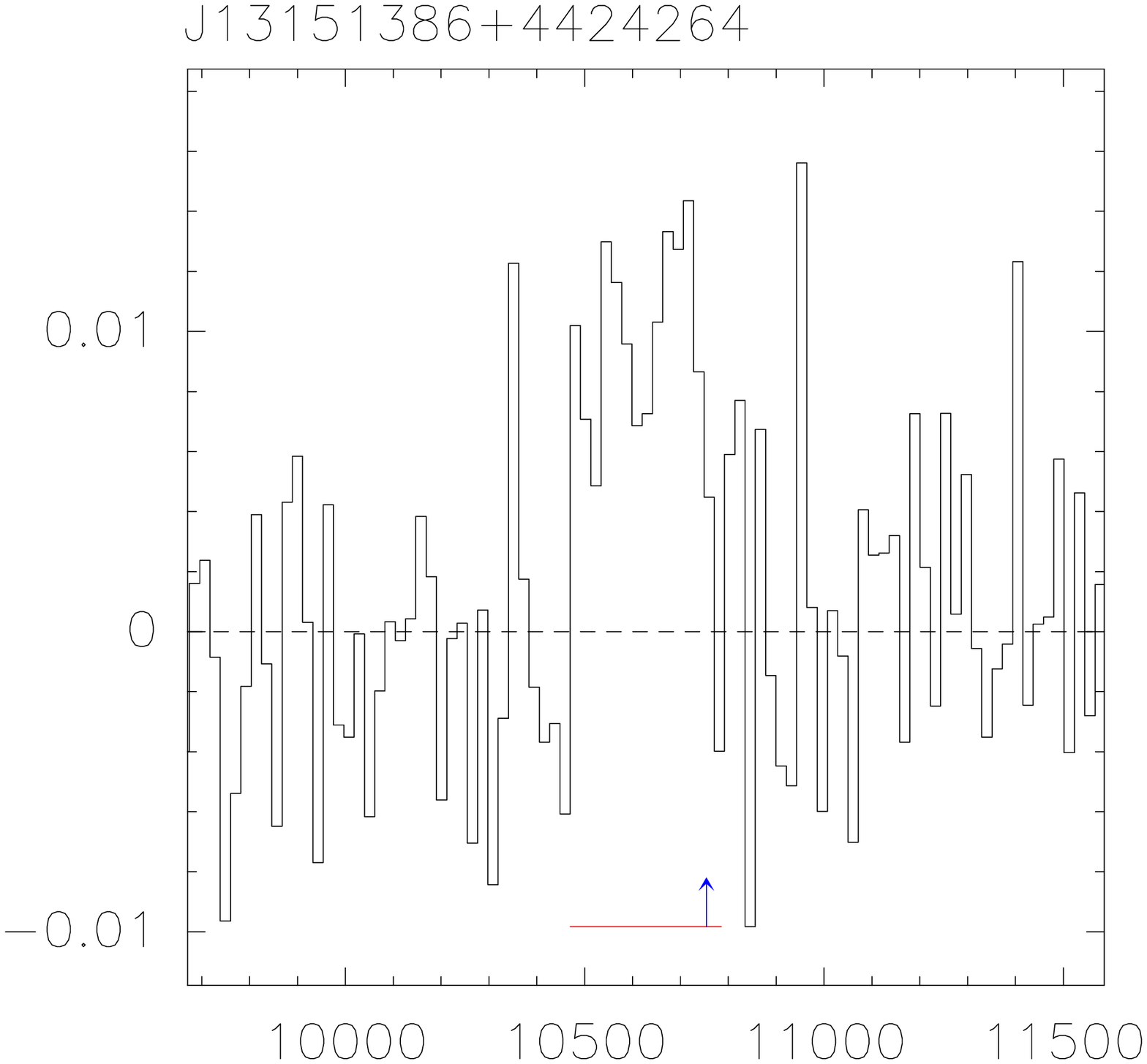}
\hspace{0.1cm}
\includegraphics[width=3.6cm,clip,trim = 0.cm 0.cm 0.cm 0.0cm,angle=-0]{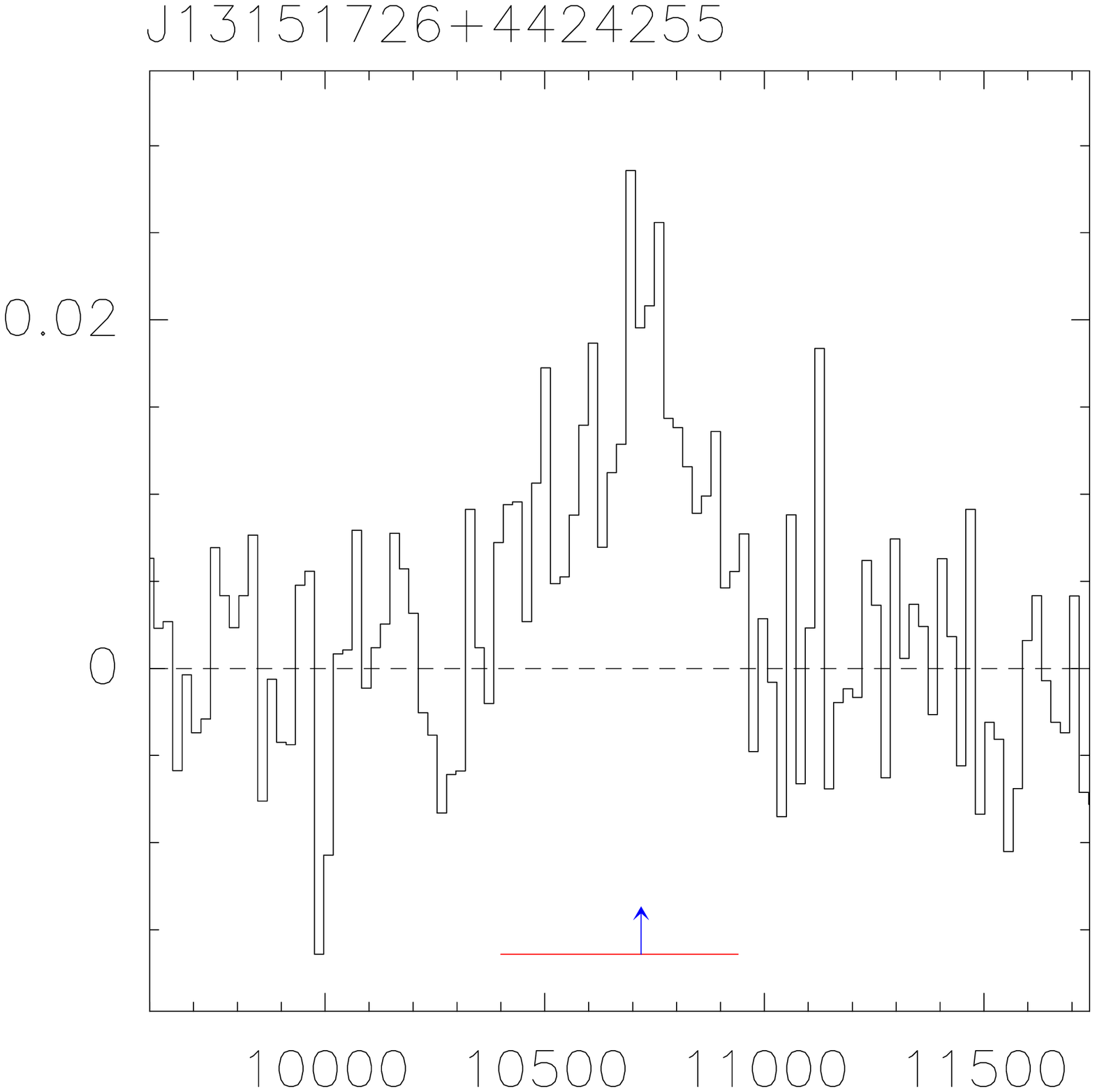}
\hspace{0.1cm}
\includegraphics[width=3.6cm,clip,trim = 0.cm 0.cm 0.cm 0.0cm, angle=-0]{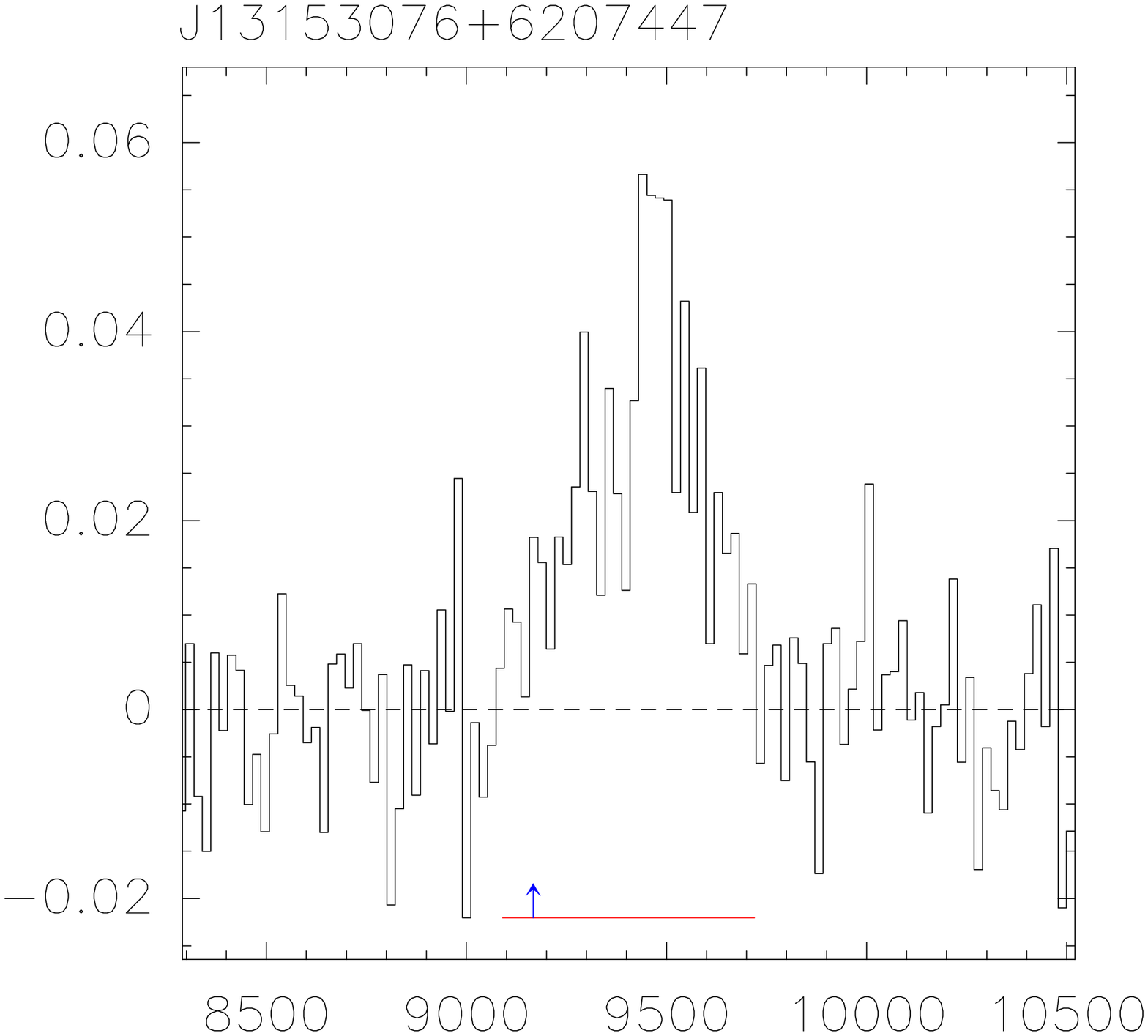}
}
\quad

\centerline{
\includegraphics[width=3.6cm,clip,trim = 0.cm 0.cm 0.cm 0.0cm, angle=-0]{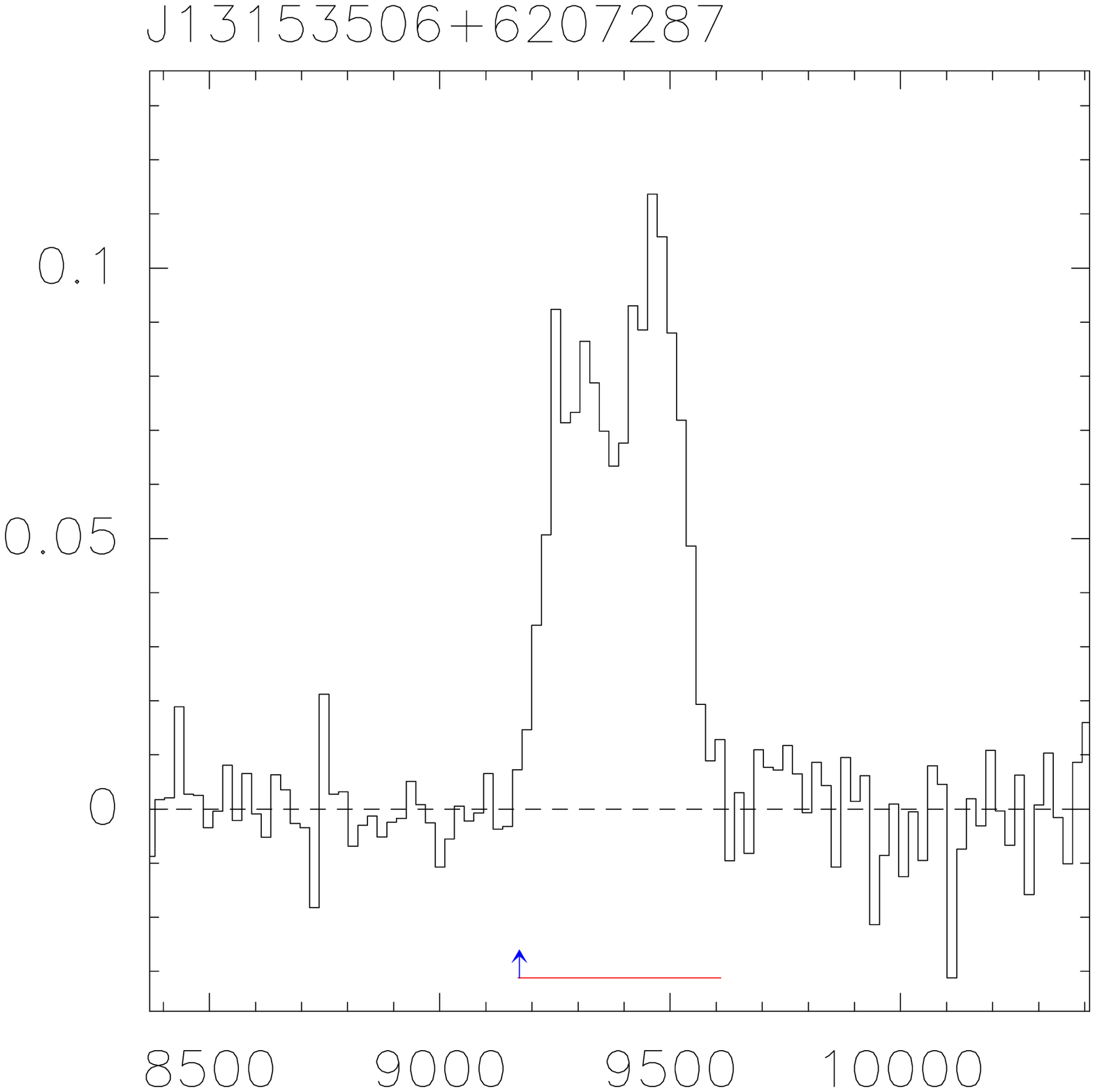}
\hspace{0.1cm}
\includegraphics[width=3.6cm,clip,trim = 0.cm 0.cm 0.cm 0.0cm, angle=-0]{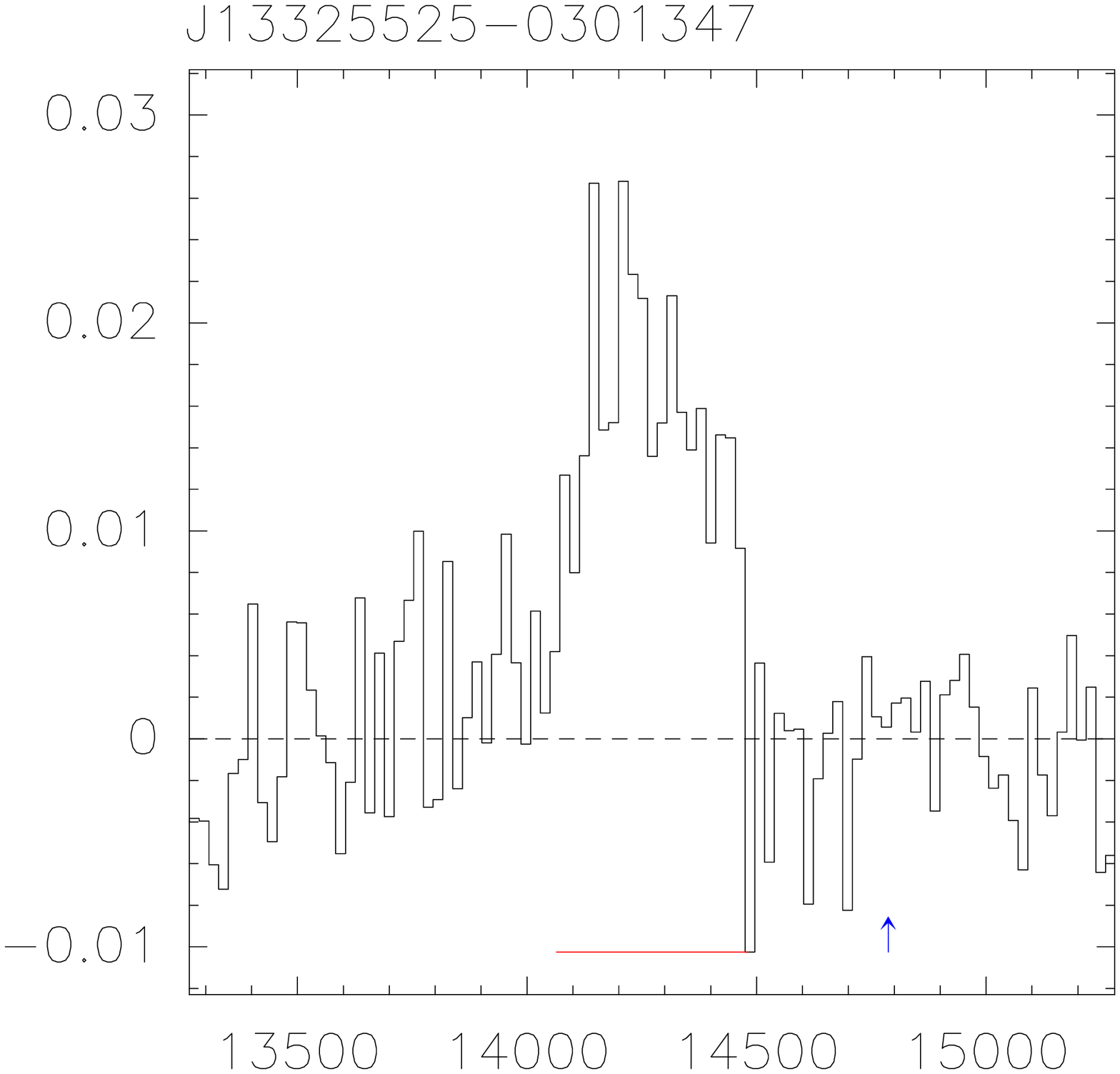}
\hspace{0.1cm}
\includegraphics[width=3.6cm,clip,trim = 0.cm 0.cm 0.cm 0.0cm, angle=-0]{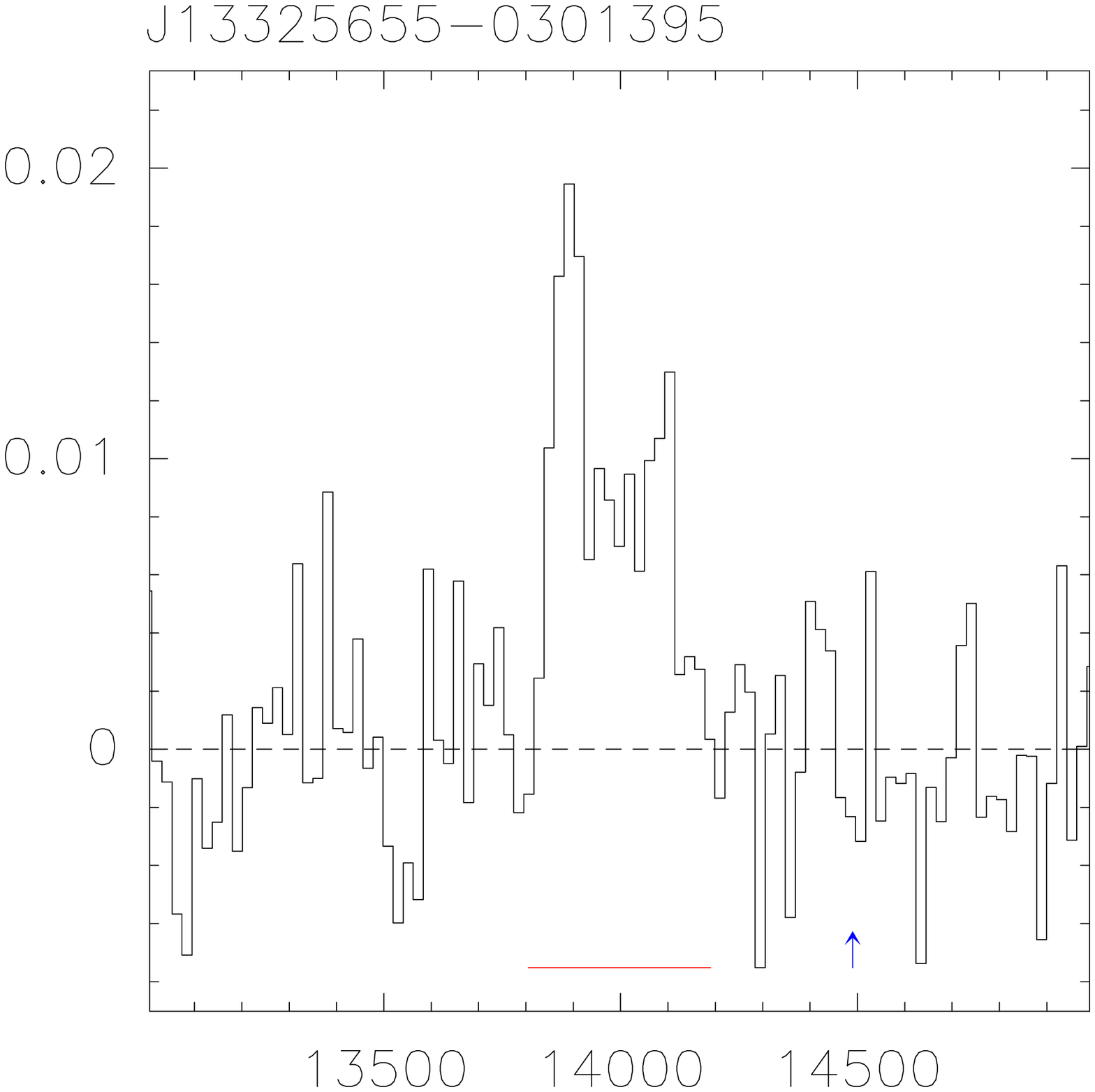}
\hspace{0.1cm}
\includegraphics[width=3.6cm,clip,trim = 0.cm 0.cm 0.cm 0.0cm, angle=-0]{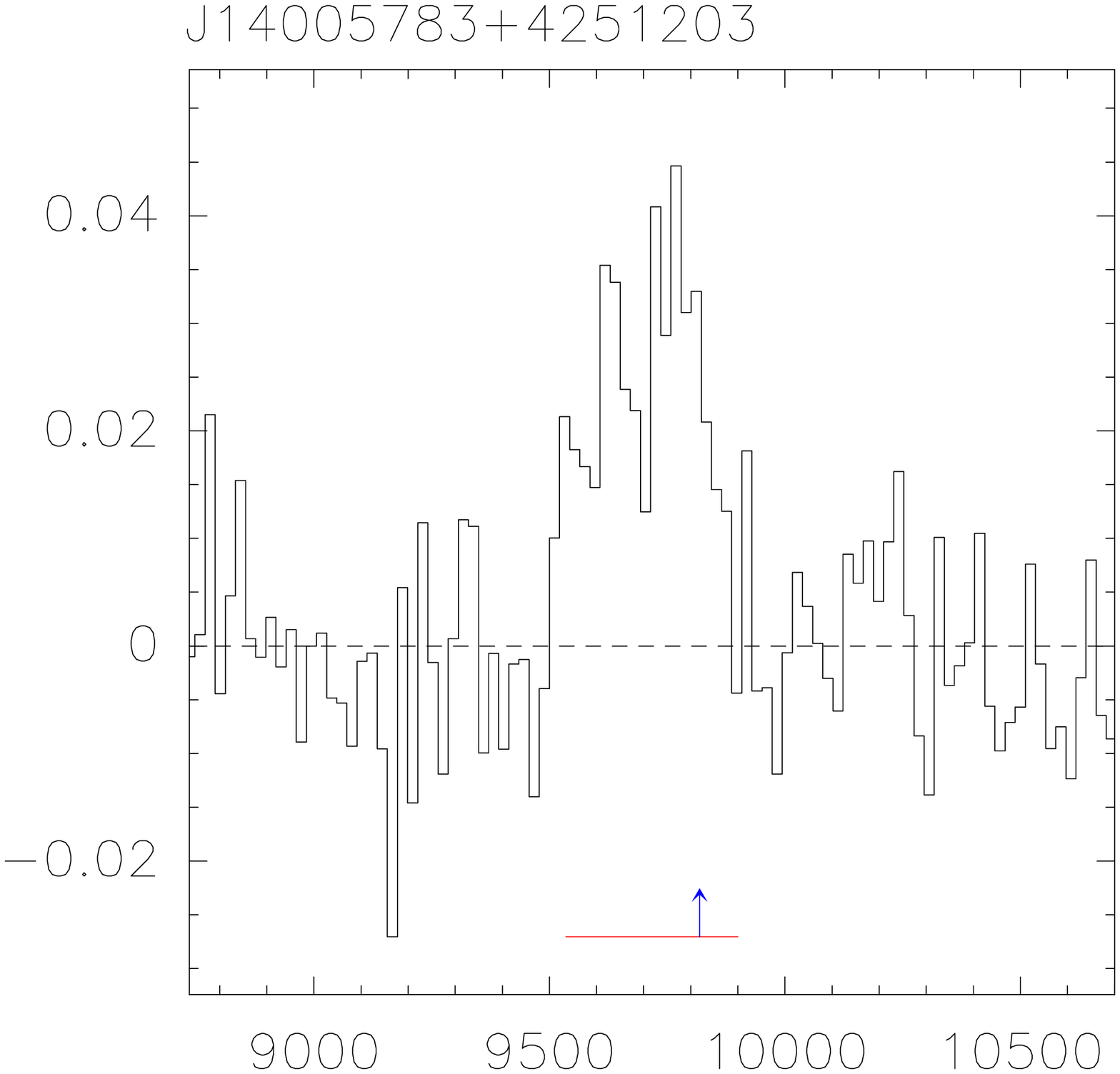}
}
\quad

\centerline{
\includegraphics[width=3.6cm,clip,trim = 0.cm 0.cm 0.cm 0.0cm, angle=-0]{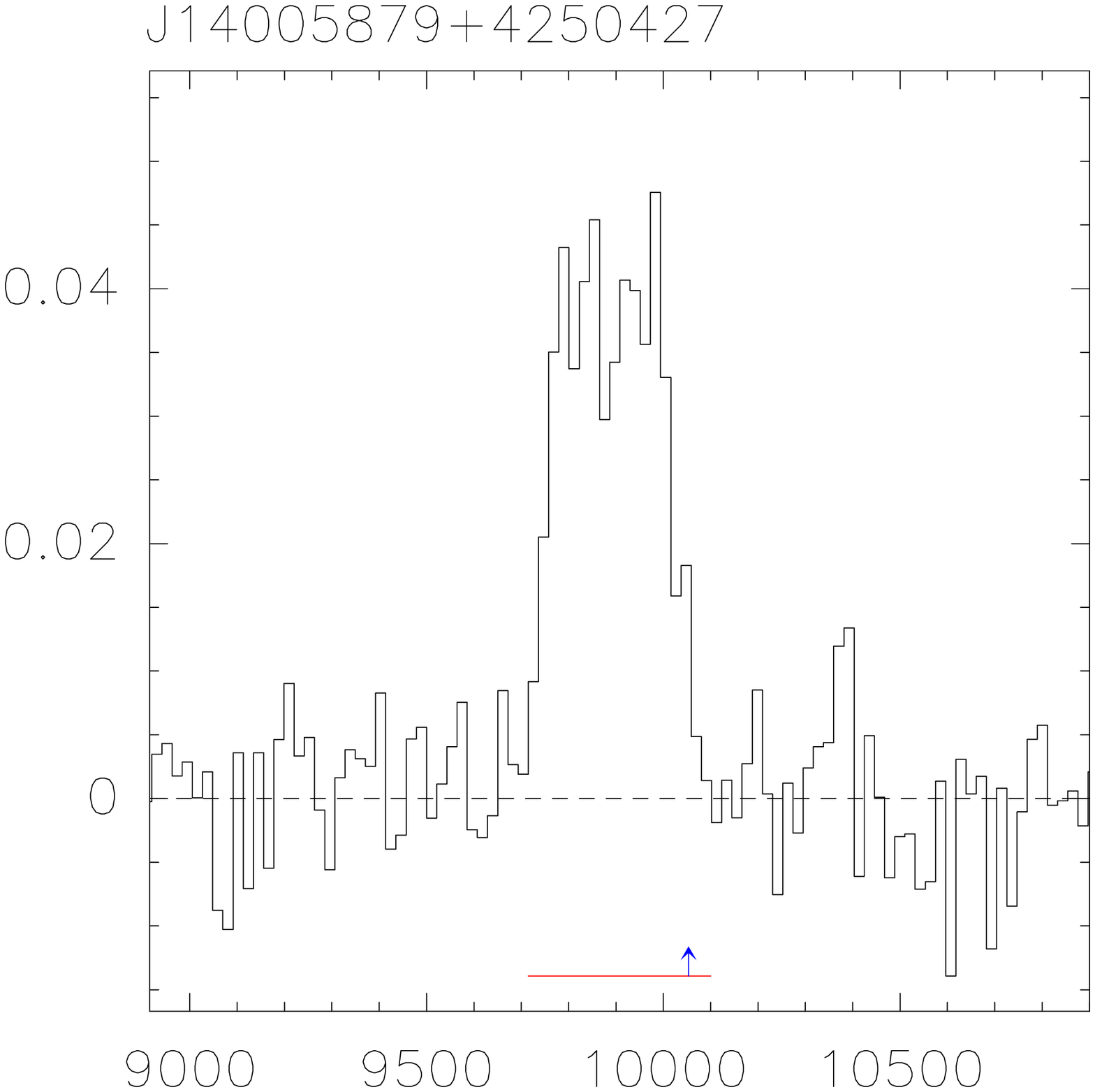}
\hspace{0.1cm}
\includegraphics[width=3.6cm,clip,trim = 0.cm 0.cm 0.cm 0.0cm, angle=-0]{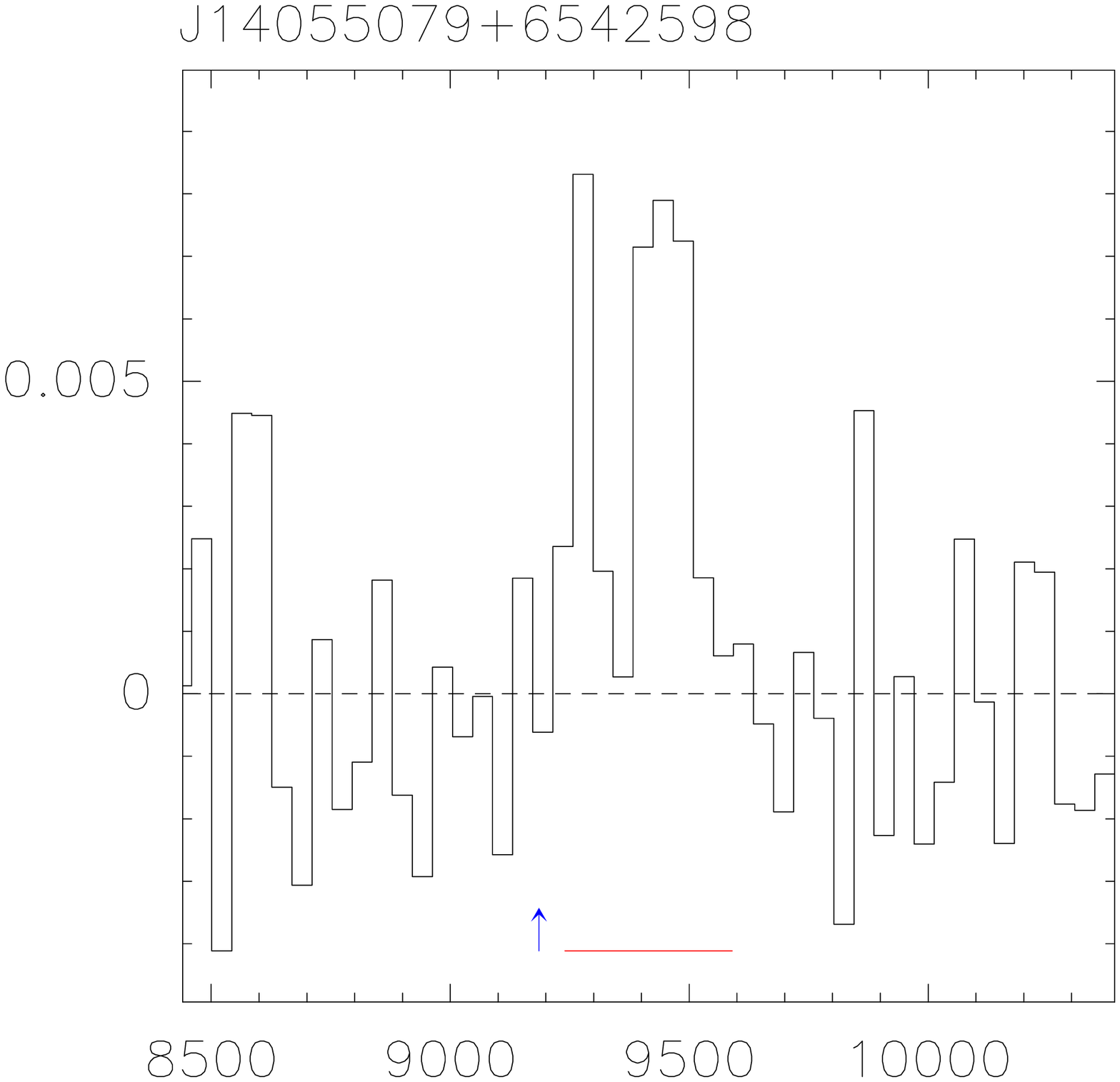}
\hspace{0.1cm}
\includegraphics[width=3.6cm,clip,trim = 0.cm 0.cm 0.cm 0.0cm,angle=-0]{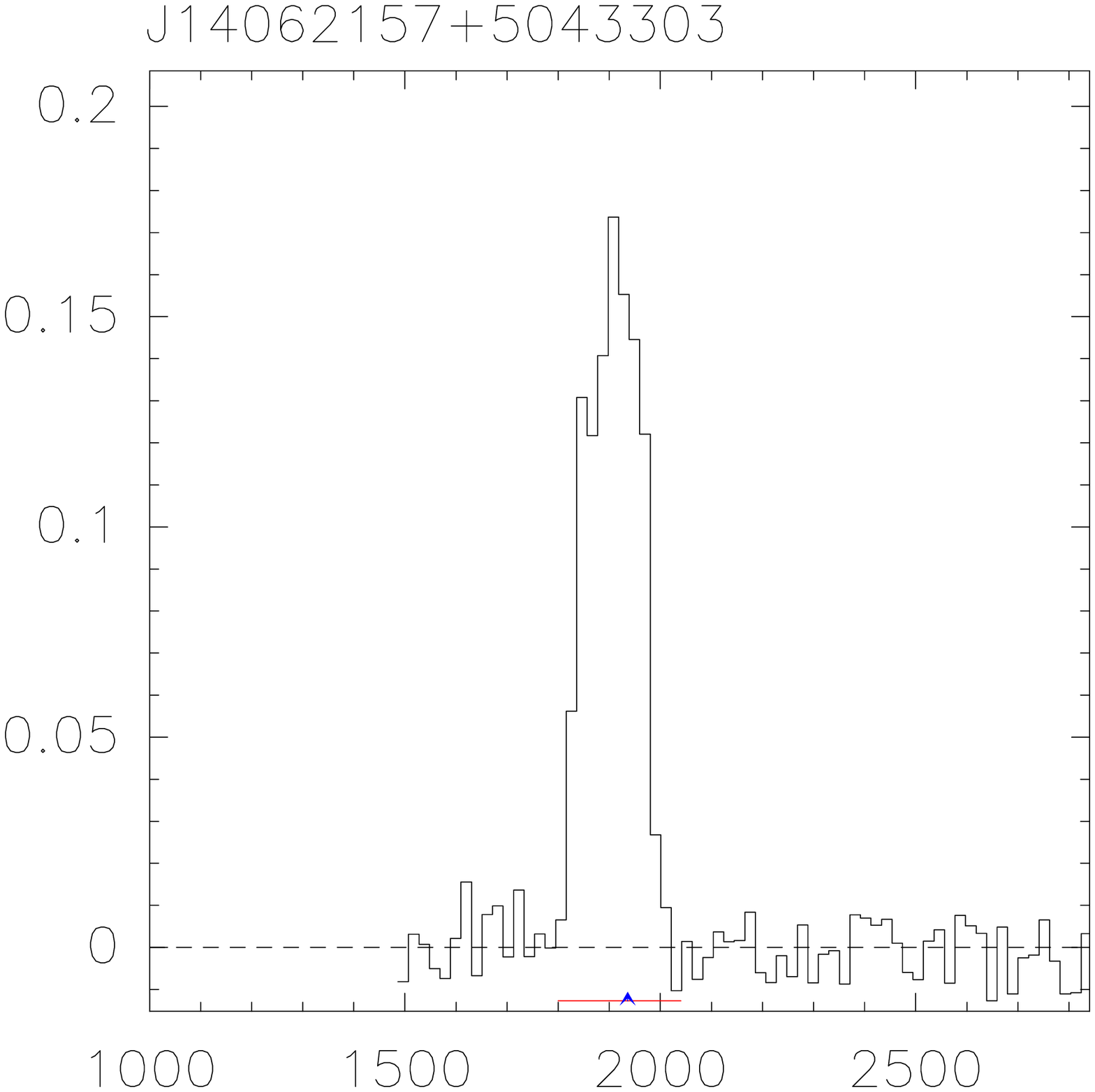}
\hspace{0.1cm}
\includegraphics[width=3.6cm,clip,trim = 0.cm 0.cm 0.cm 0.0cm, angle=-0]{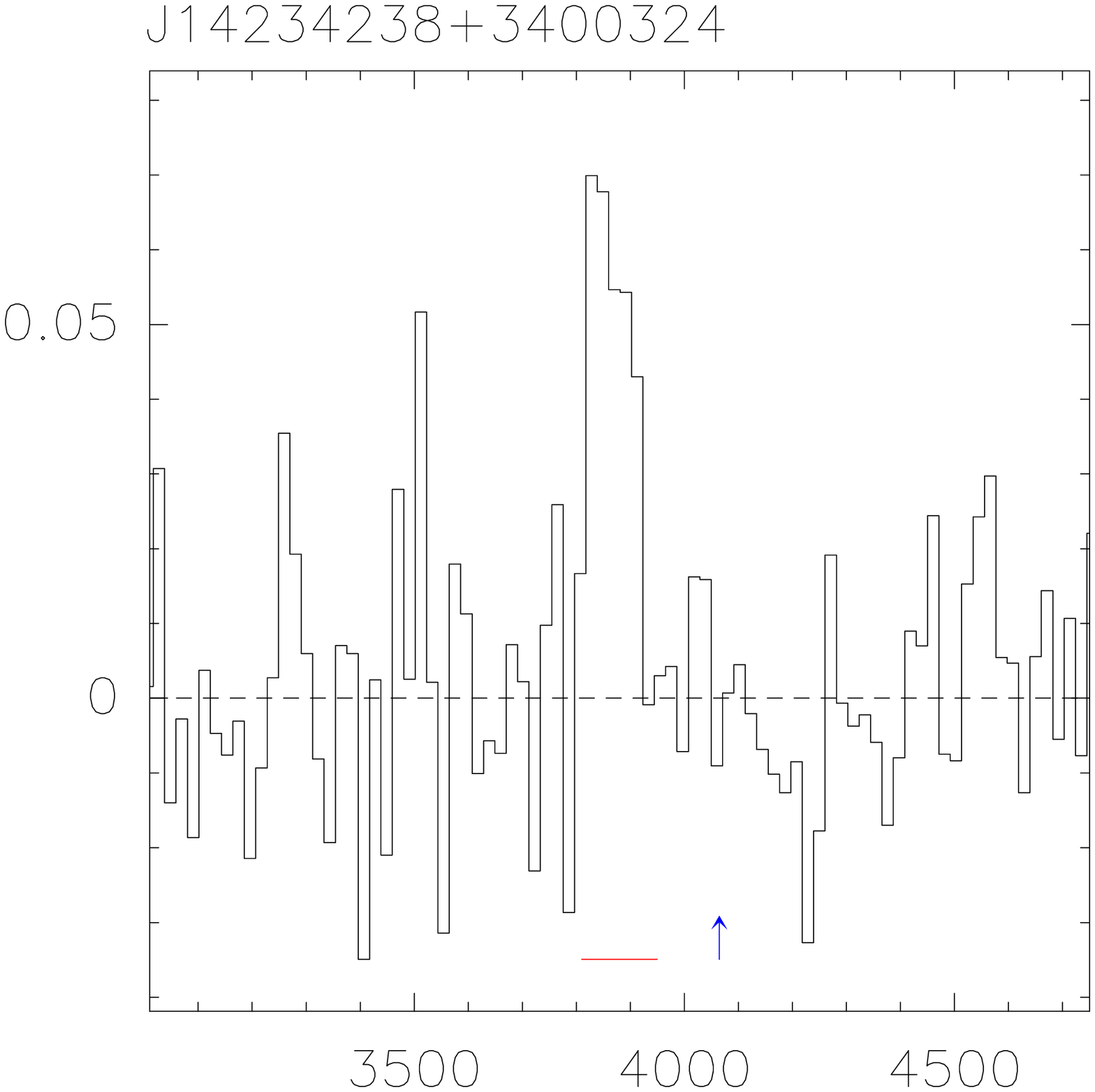}
}
\quad

\centerline{
\includegraphics[width=3.6cm,clip,trim = 0.cm 0.cm 0.cm 0.0cm, angle=-0]{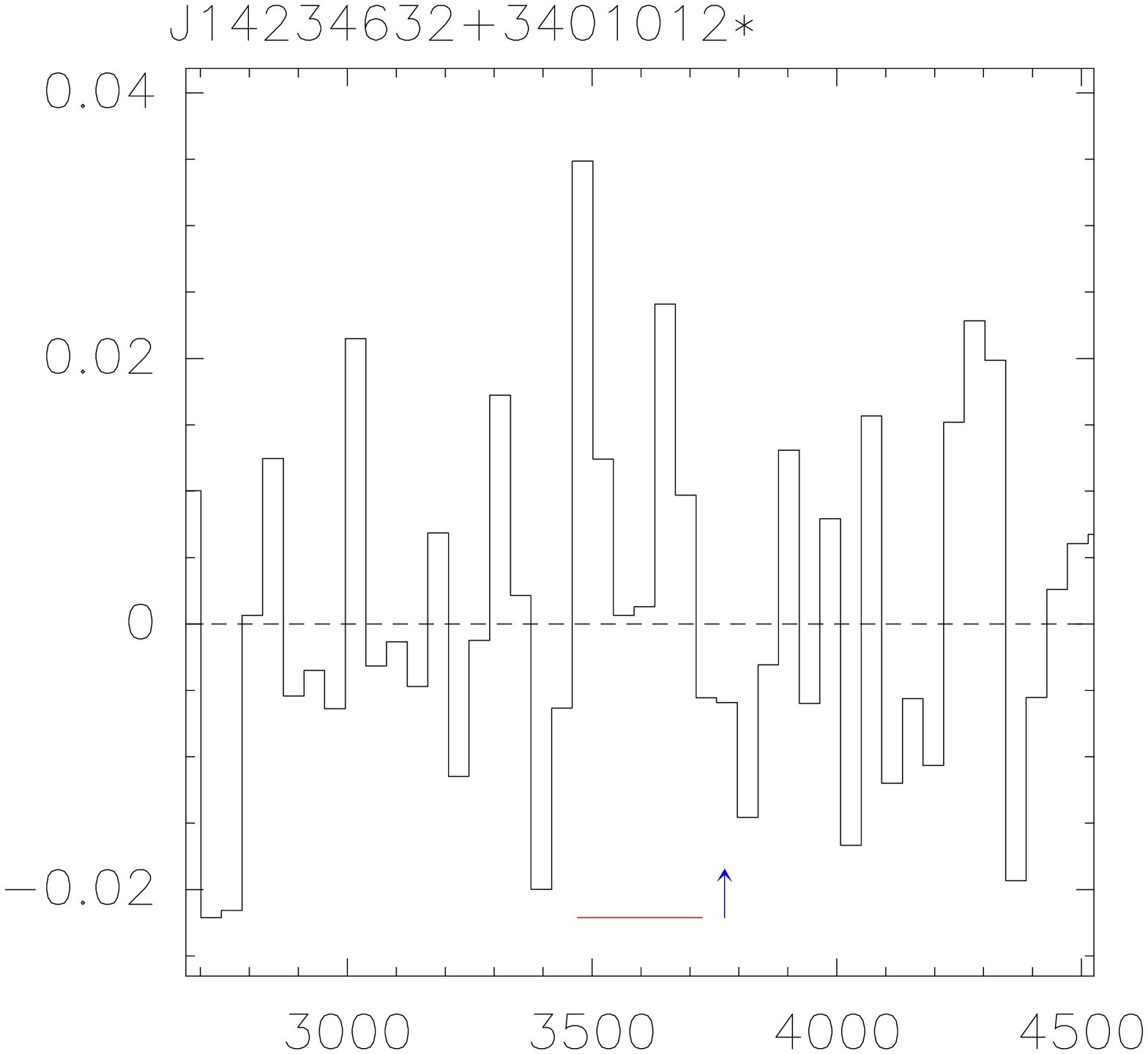}
\hspace{0.1cm}
\includegraphics[width=3.6cm,clip,trim = 0.cm 0.cm 0.cm 0.0cm, angle=-0]{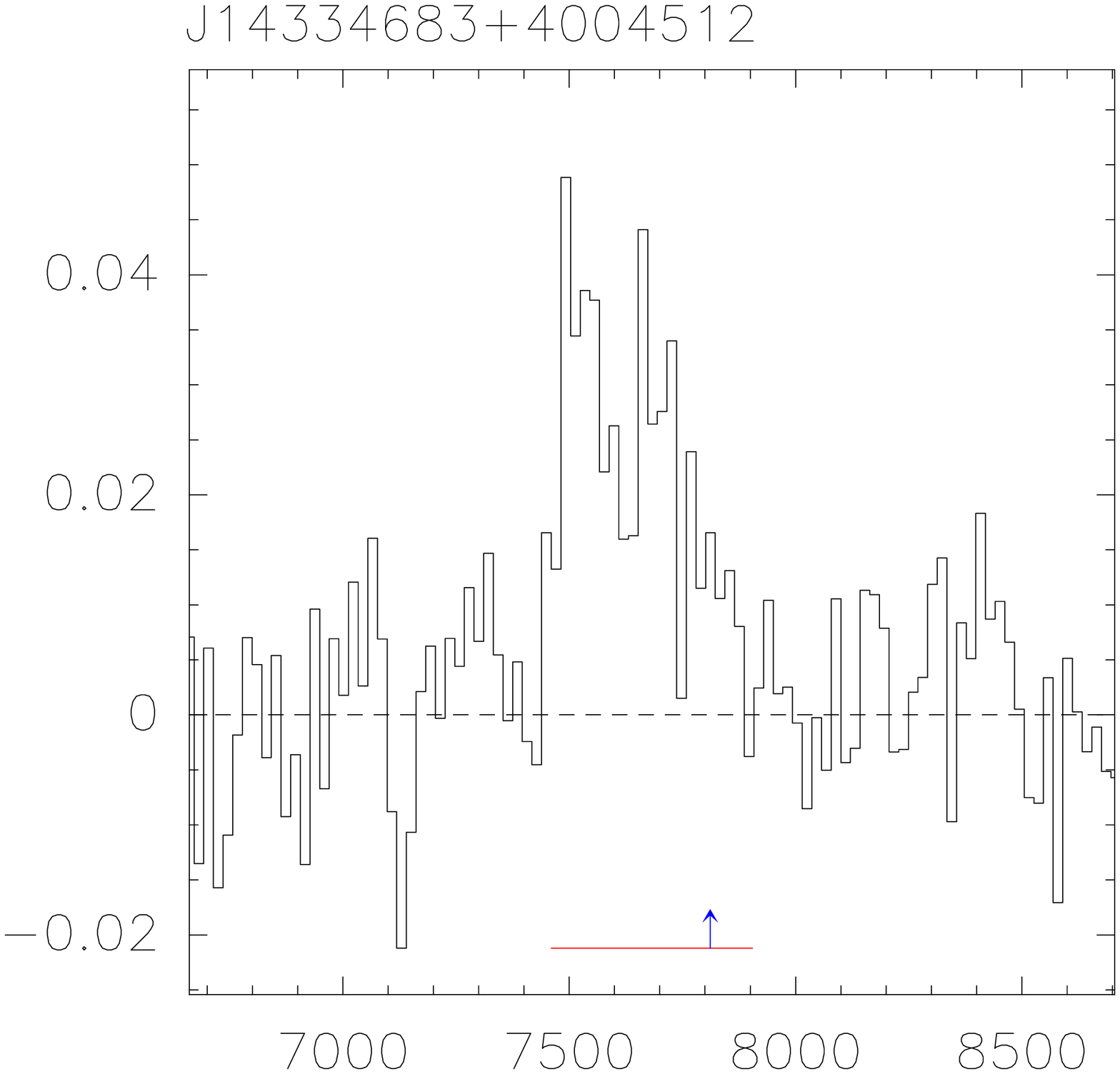}
\hspace{0.1cm}
\includegraphics[width=3.6cm,clip,trim = 0.cm 0.cm 0.cm 0.0cm,angle=-0]{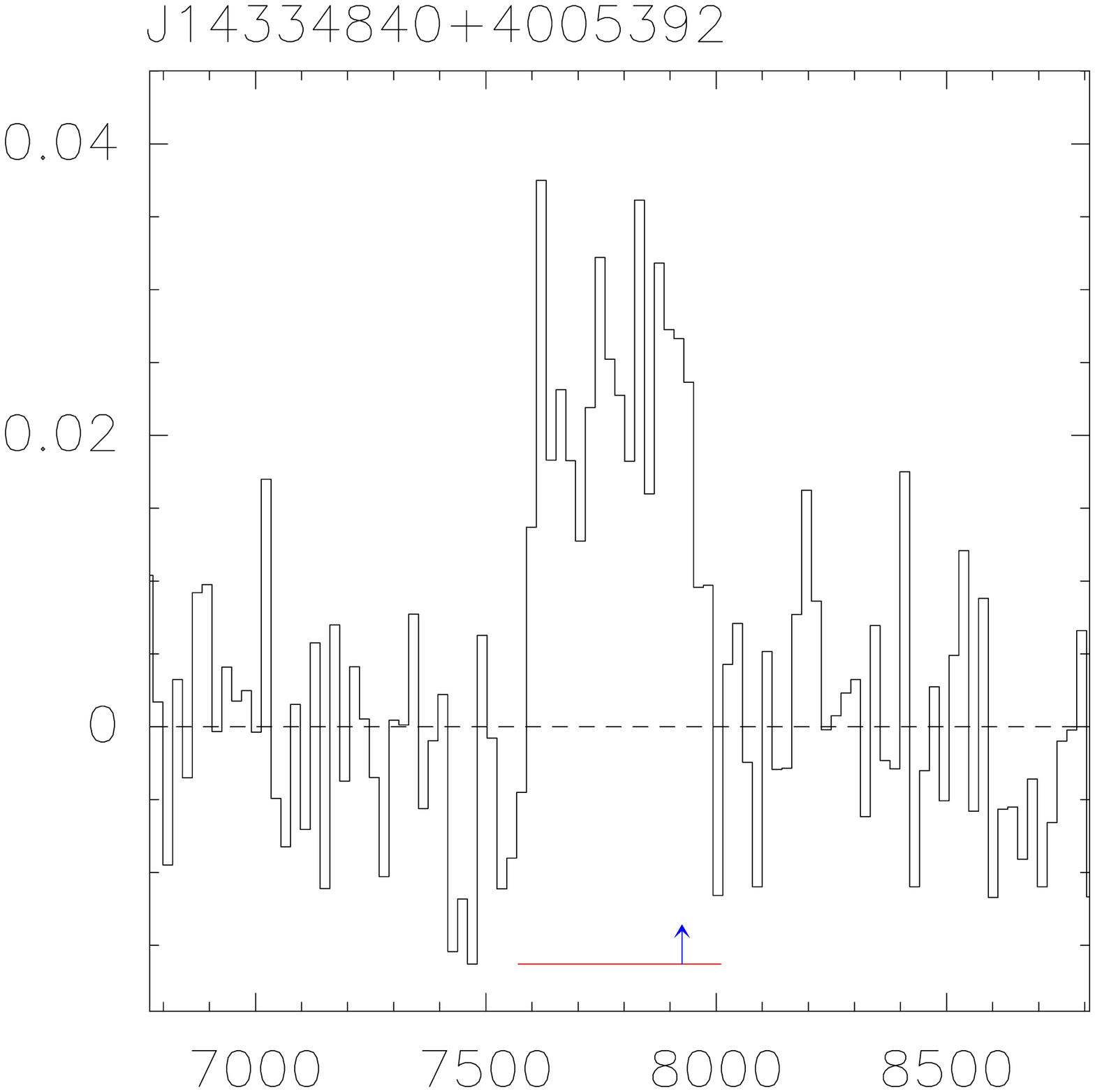}
\hspace{0.1cm}
\includegraphics[width=3.6cm,clip,trim = 0.cm 0.cm 0.cm 0.0cm, angle=-0]{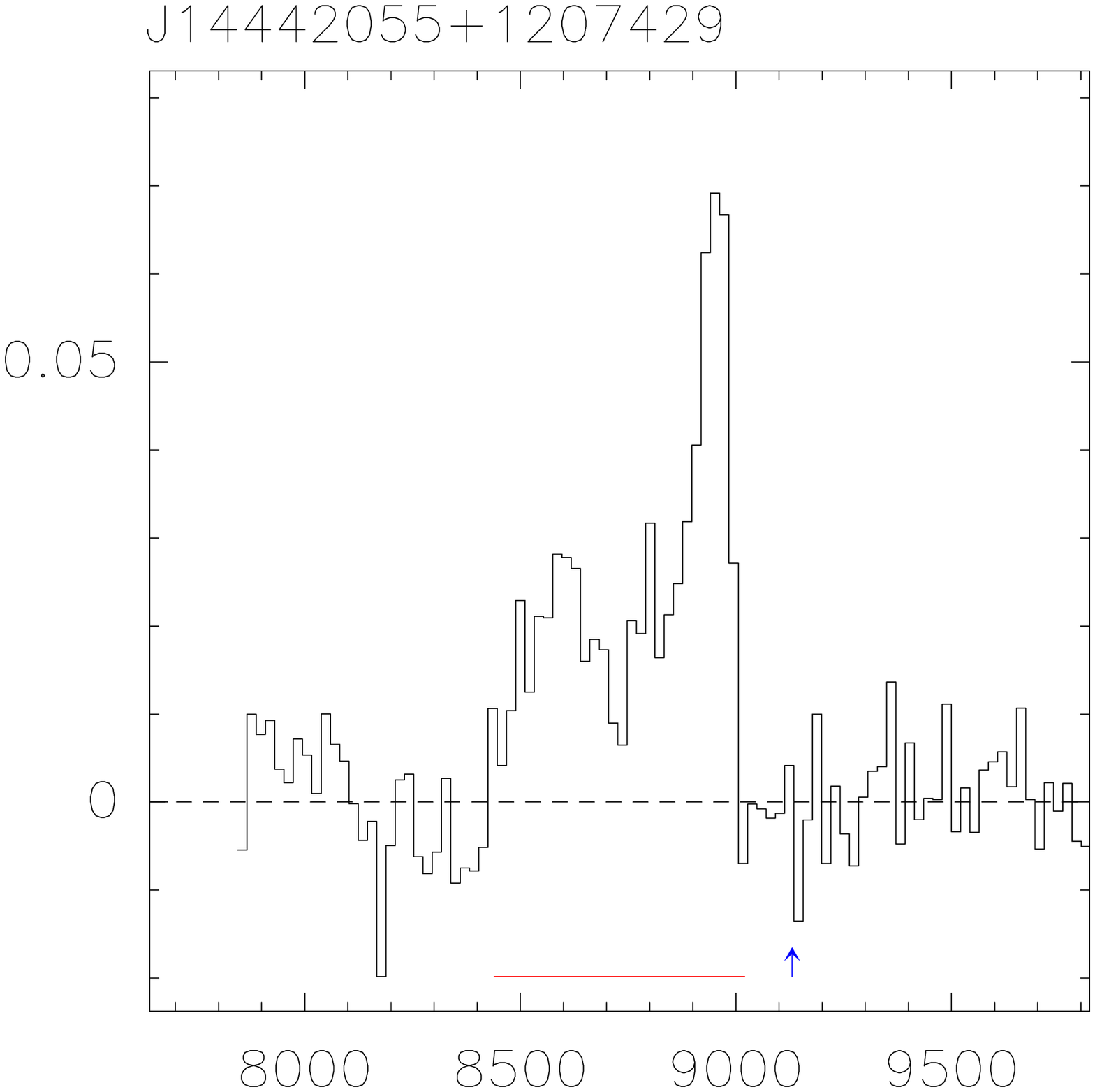}
}
\quad

\centerline{
\includegraphics[width=3.6cm,clip,trim = 0.cm 0.cm 0.cm 0.0cm, angle=-0]{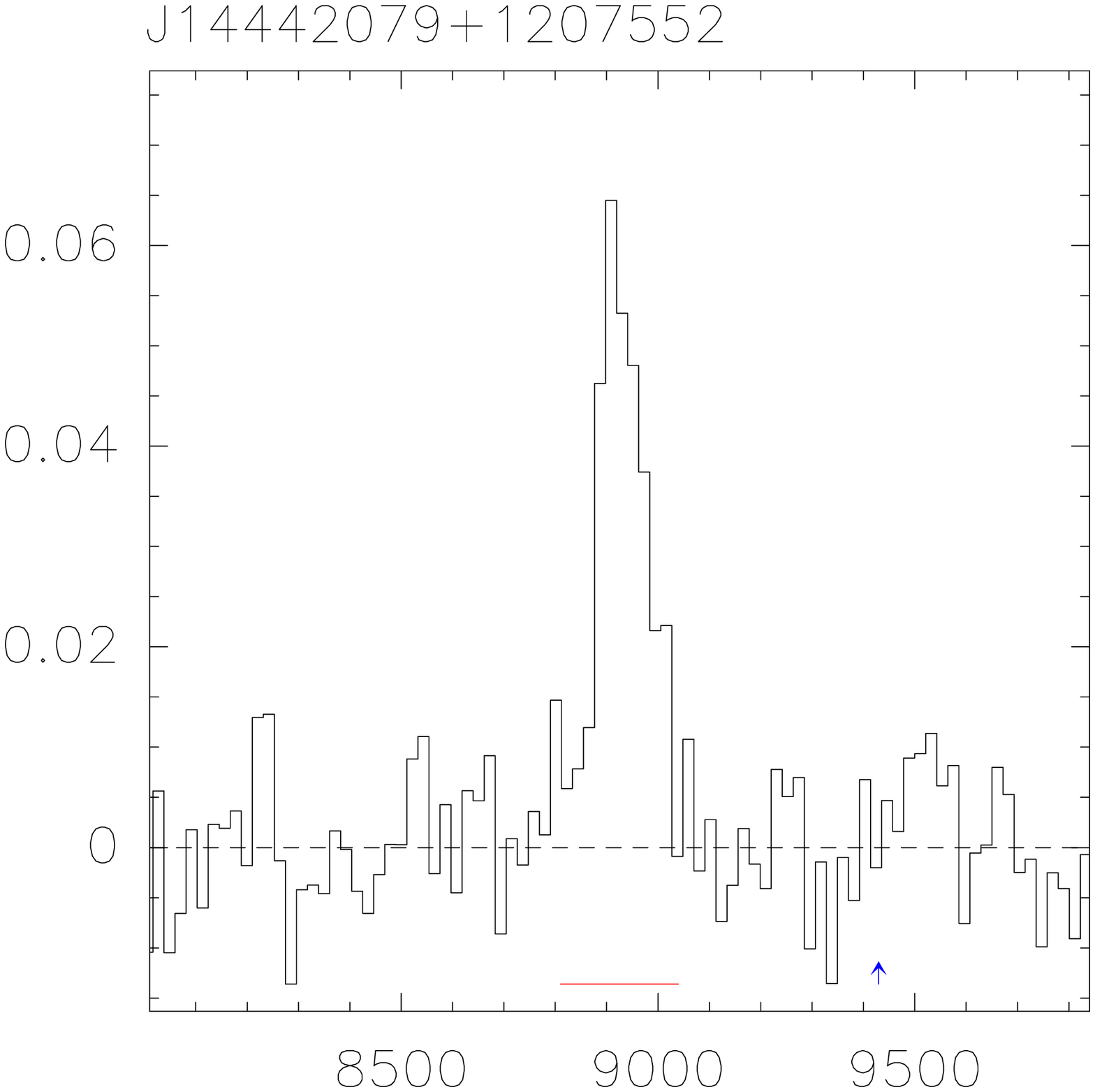}
\hspace{0.1cm}
\includegraphics[width=3.6cm,clip,trim = 0.cm 0.cm 0.cm 0.0cm, angle=-0]{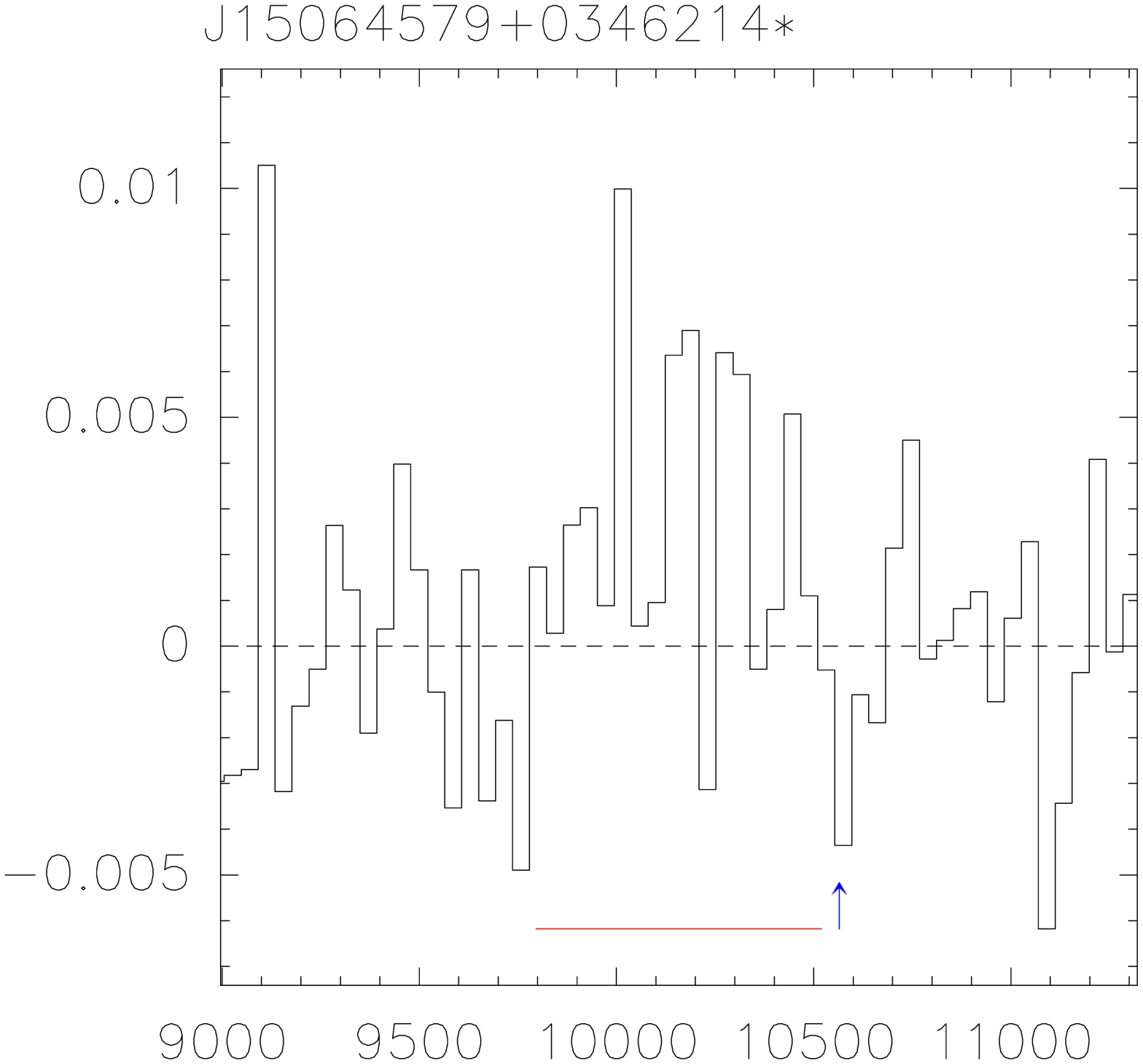}
\hspace{0.1cm}
\includegraphics[width=3.6cm,clip,trim = 0.cm 0.cm 0.cm 0.0cm,angle=-0]{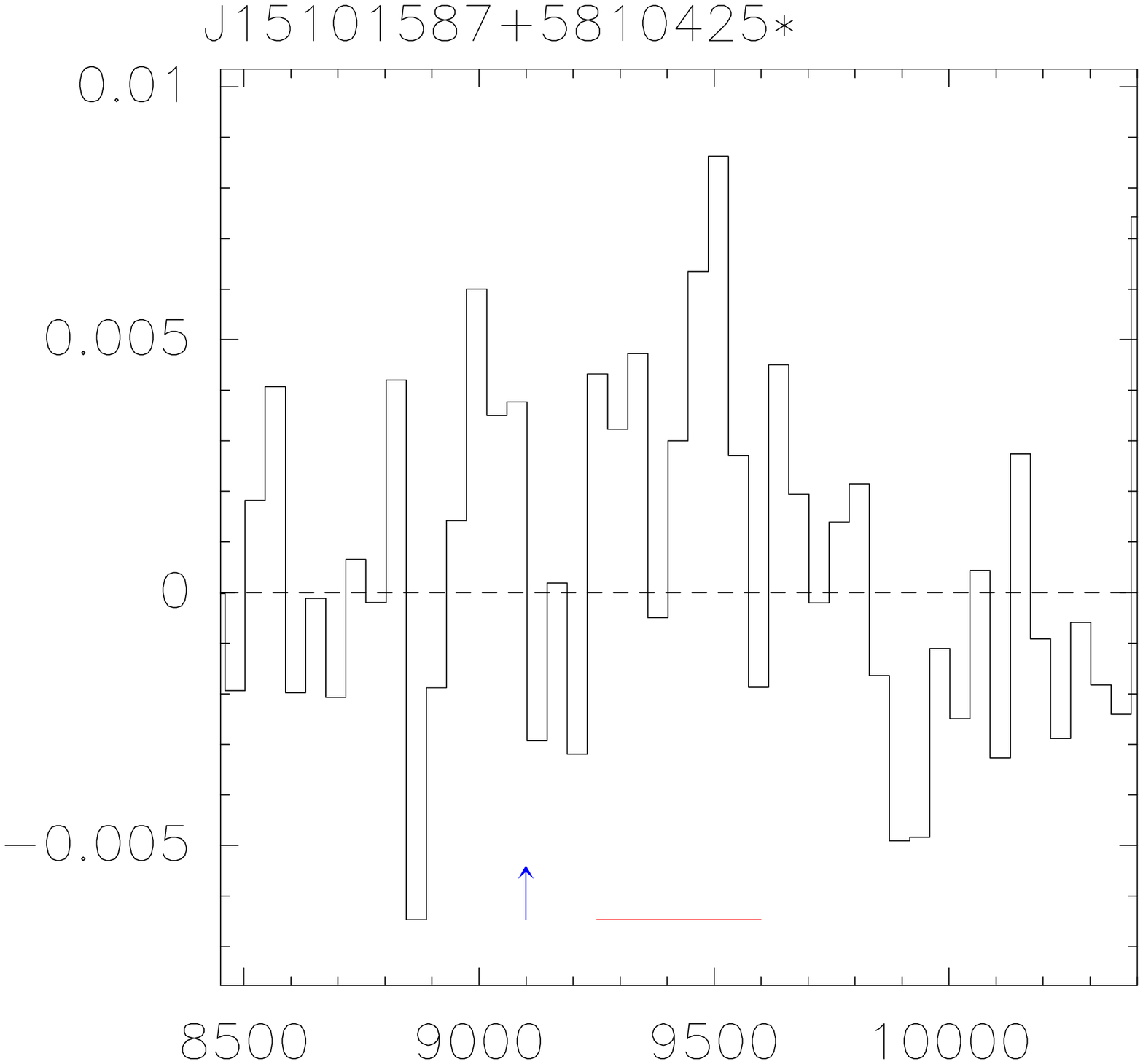}
\hspace{0.1cm}
\includegraphics[width=3.6cm,clip,trim = 0.cm 0.cm 0.cm 0.0cm,angle=-0]{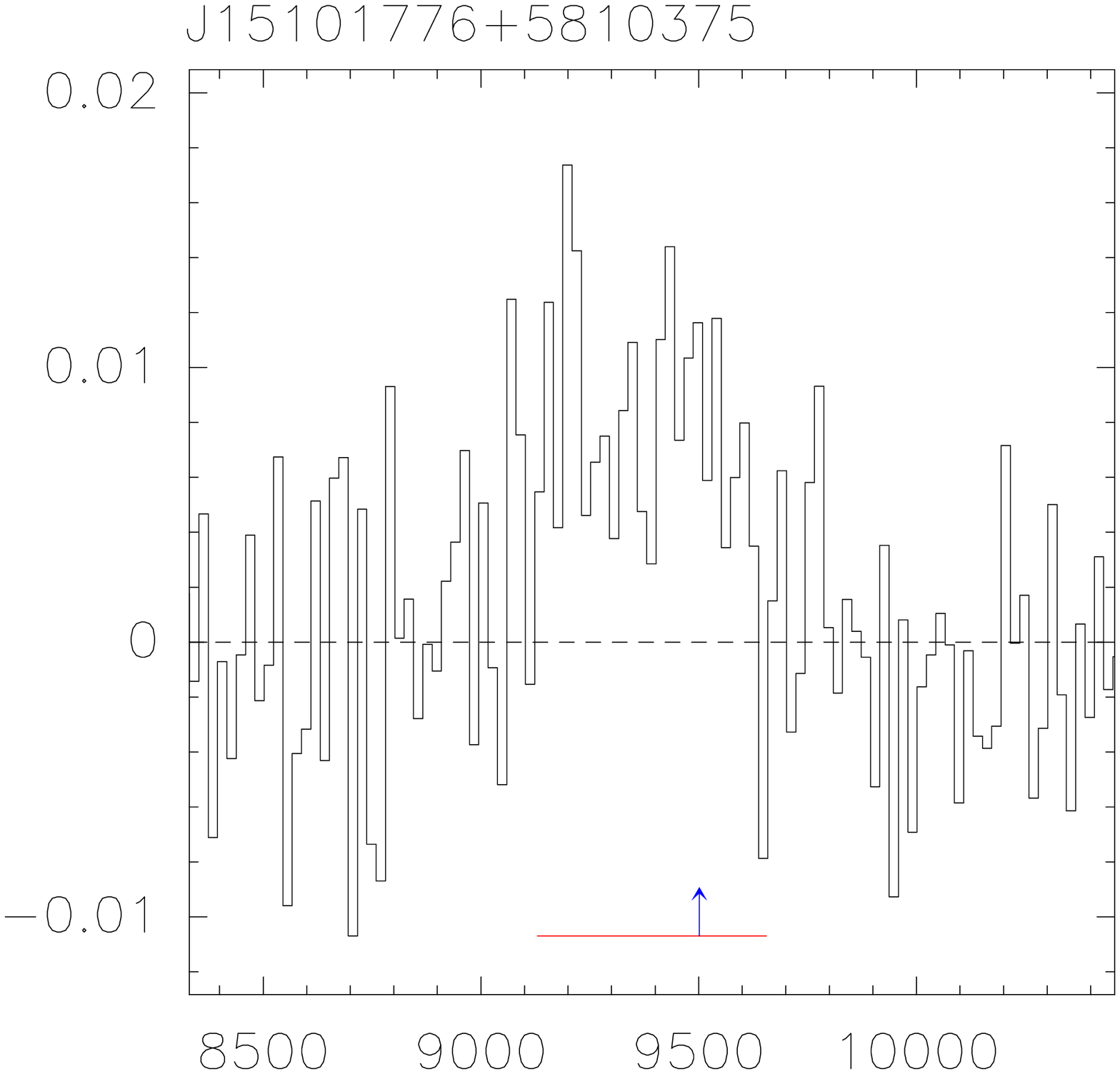}
}
\quad

\centerline{
\includegraphics[width=3.6cm,clip,trim = 0.cm 0.cm 0.cm 0.0cm, angle=-0]{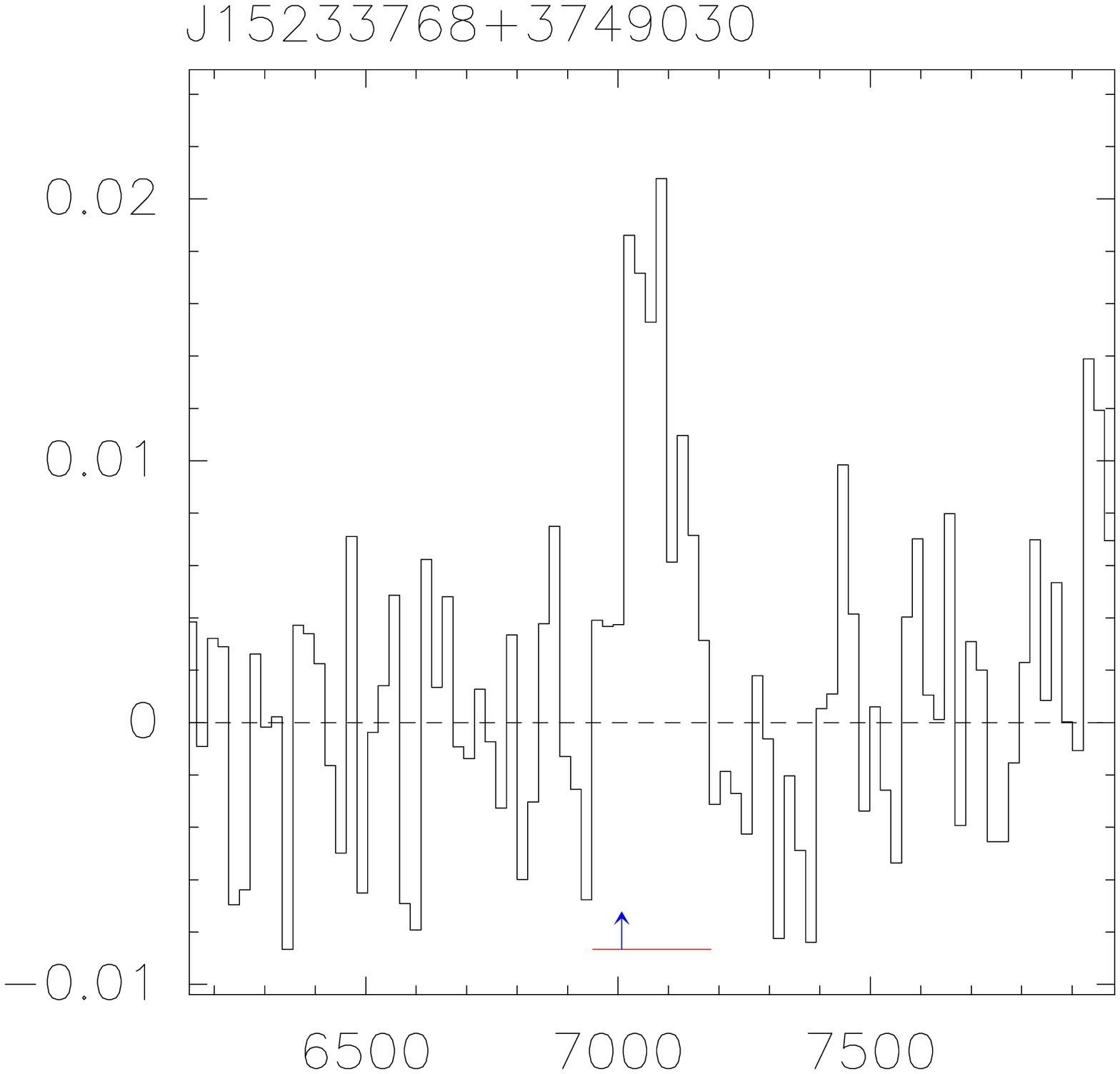}
\hspace{0.1cm}
\includegraphics[width=3.6cm,clip,trim = 0.cm 0.cm 0.cm 0.0cm, angle=-0]{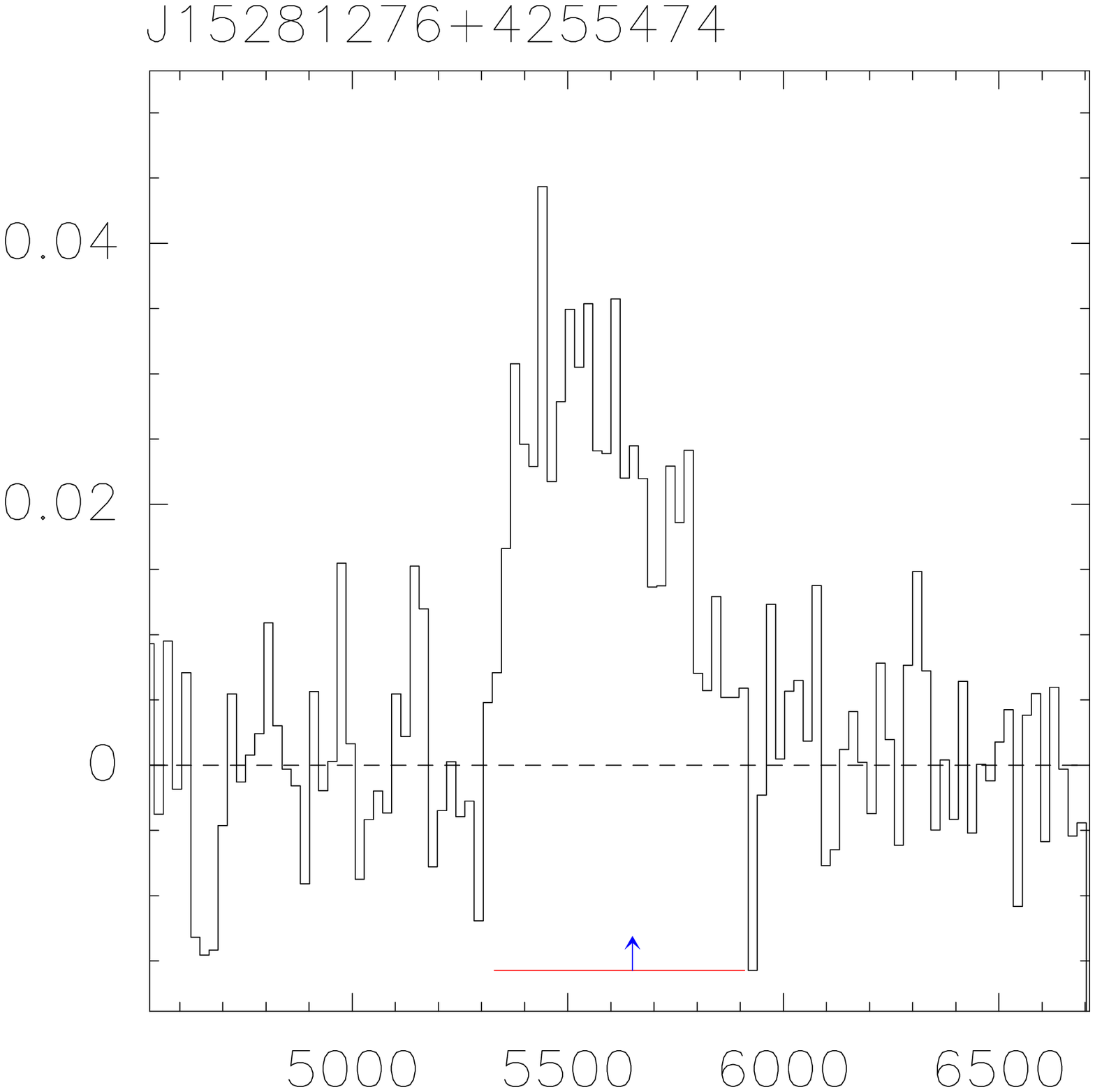}
\hspace{0.1cm}
\includegraphics[width=3.6cm,clip,trim = 0.cm 0.cm 0.cm 0.0cm, angle=-0]{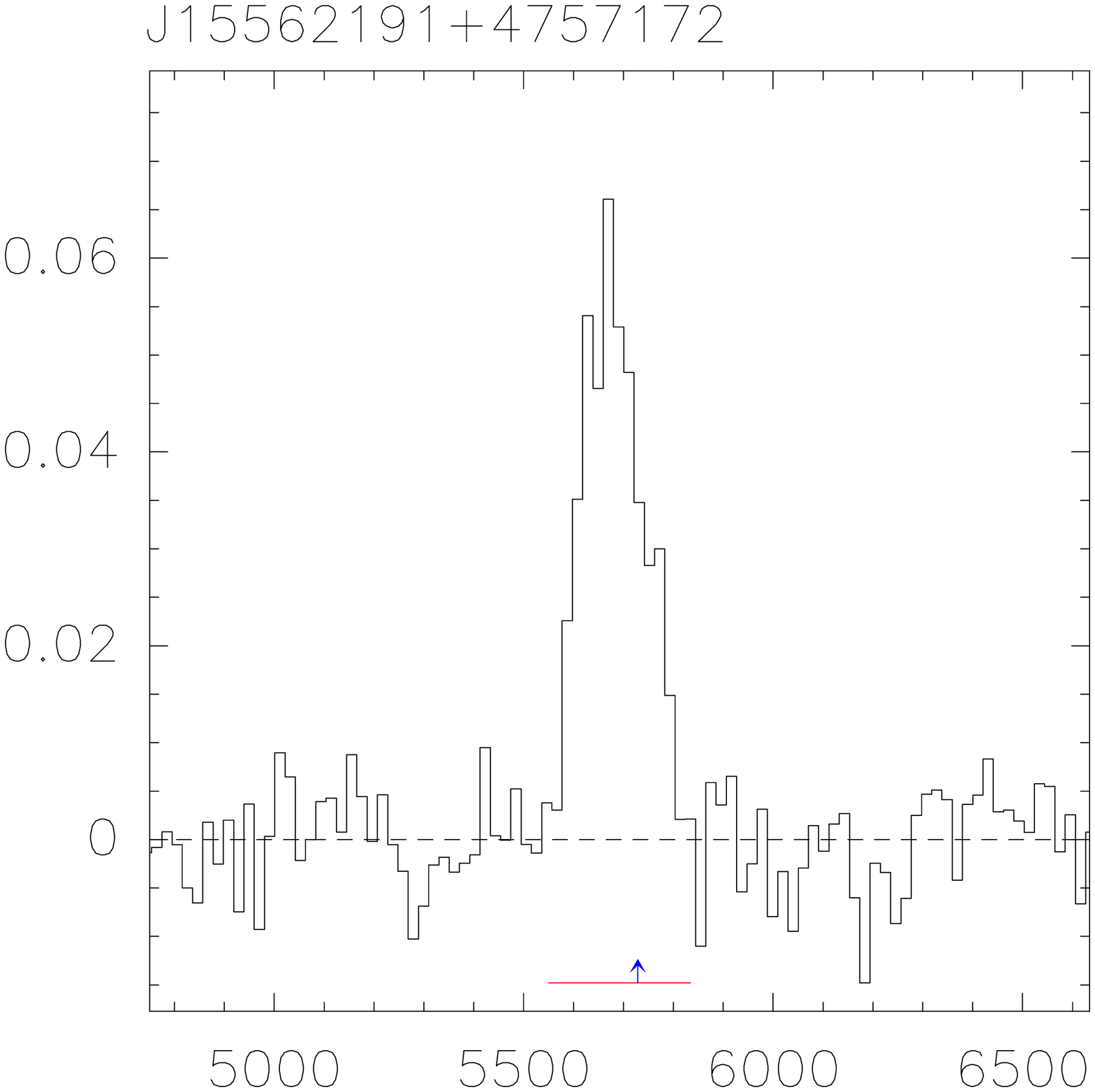}
\hspace{0.1cm}
\includegraphics[width=3.6cm,clip,trim = 0.cm 0.cm 0.cm 0.0cm,angle=-0]{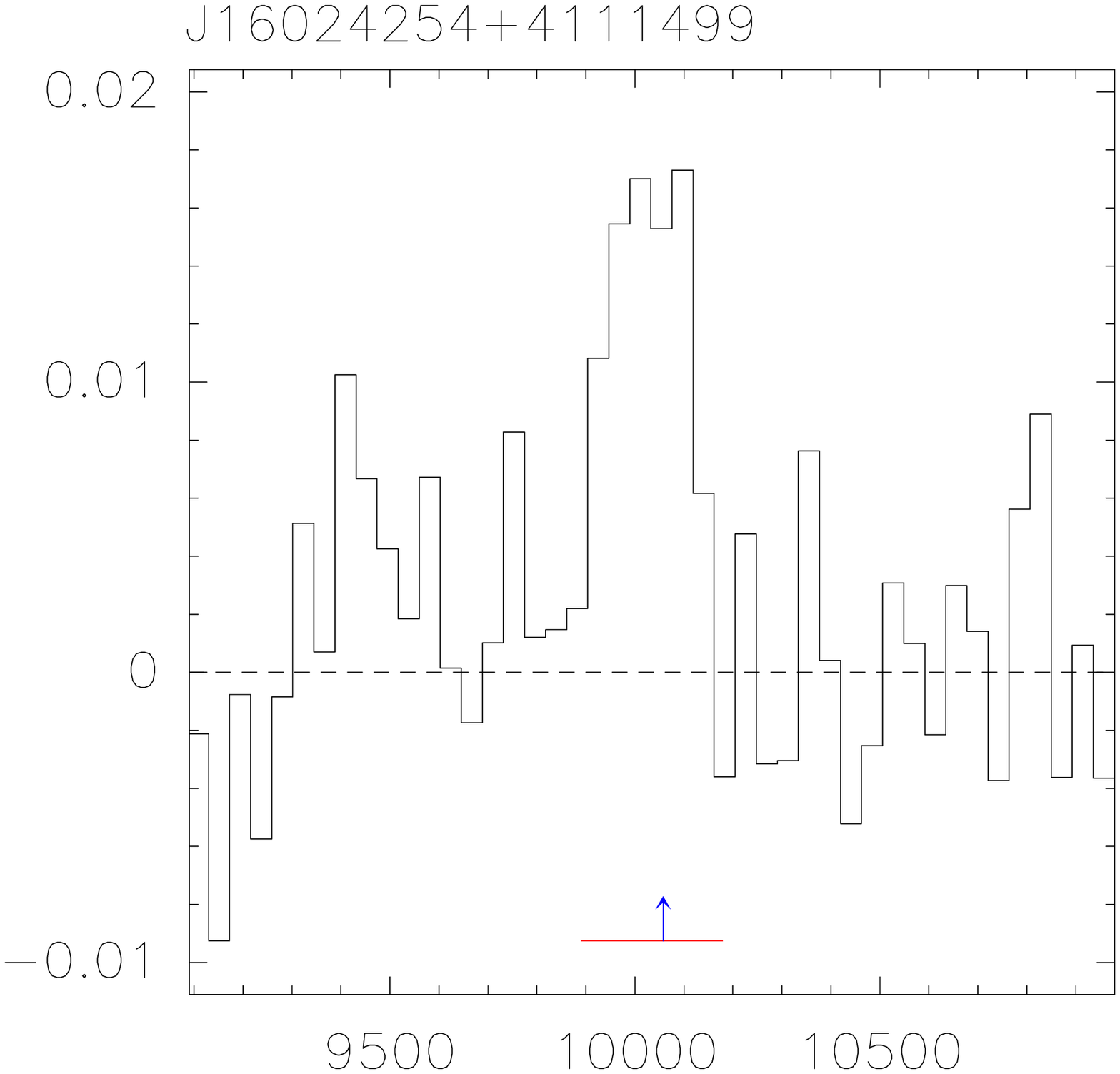}
}
\quad

\centerline{
\includegraphics[width=4cm,clip,trim = 0.cm 0.cm 0.cm 0.0cm, angle=-0]{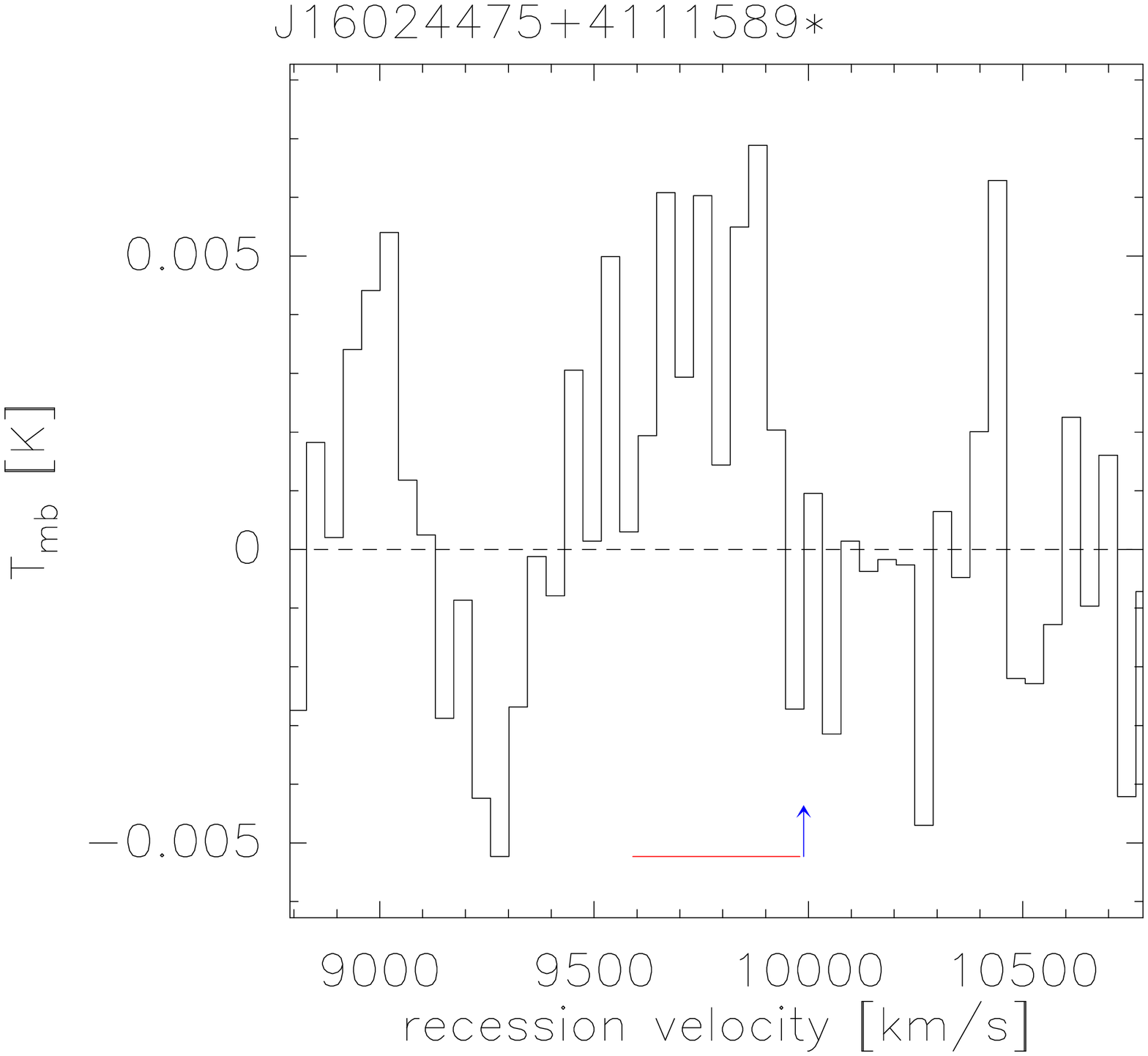}
\hspace{0.1cm}
\includegraphics[width=3.6cm,clip,trim = 0.cm 0.cm 0.cm 0.0cm, angle=-0]{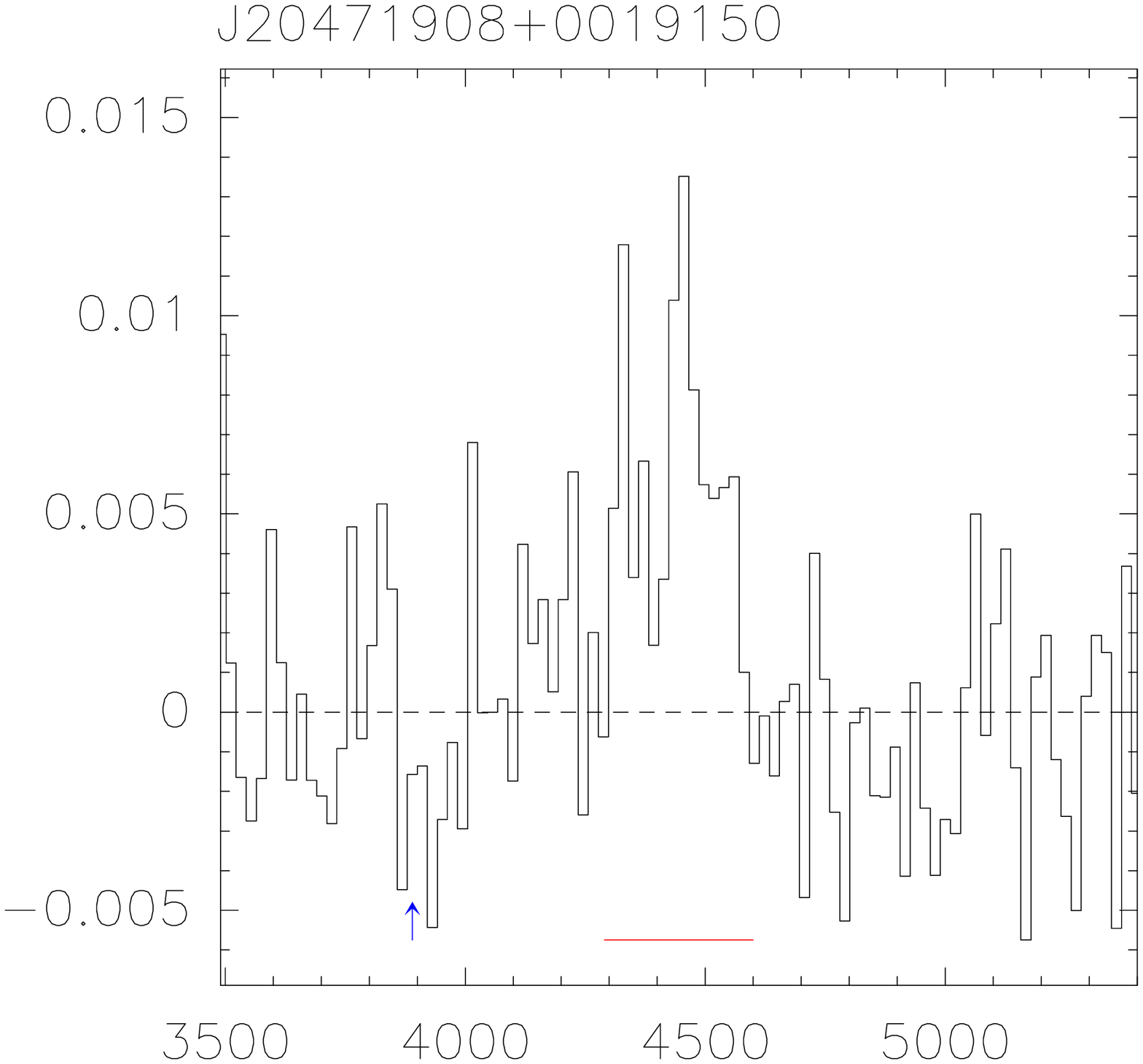}
}

\caption{Continued.
}
\end{figure*}

\end{document}